\documentclass[12pt,a4paper]{article}

\usepackage[margin=1in]{geometry}  
\usepackage{setspace}              
\usepackage{titlesec}              
\usepackage{titling}               
\usepackage{lipsum}                
\usepackage{tocloft}
\usepackage{amsmath,amssymb,amsfonts} 
\usepackage{graphicx}
\usepackage{bm}
\usepackage{xcolor}
\usepackage{color} 
\usepackage{dcolumn}
\usepackage{chngcntr}
\usepackage{hyperref}
\usepackage{mathtools}
\usepackage{amsmath}
\usepackage{amsfonts}
\usepackage{amsthm}
\usepackage{amssymb}
\usepackage{graphicx}
\usepackage{float}
\usepackage{multirow}
\usepackage{float}
\usepackage{placeins}
\usepackage{upgreek} 
\usepackage{mathrsfs}
\usepackage{cite}
\usepackage[font=footnotesize]{caption}
\usepackage{comment}
\usepackage{array}
\usepackage{multirow}

\renewcommand{\cftsecfont}{\normalfont}      
\renewcommand{\cftsecpagefont}{\normalfont}  
\renewcommand{\cftsubsecfont}{\normalfont}   
\renewcommand{\cftsubsecpagefont}{\normalfont} 
\renewcommand{\cftsubsubsecfont}{\normalfont}   
\renewcommand{\cftsubsubsecpagefont}{\normalfont}

\usepackage{longtable}

\titleformat{\section}
  {\normalfont\normalsize\bfseries}{\thesection.}{1em}{}           

\titleformat{\subsection}
  {\normalfont\normalsize\bfseries\itshape}{\thesubsection}{1em}{}  

\titleformat{\subsubsection}
  {\normalfont\normalsize\itshape}{\thesubsubsection}{1em}{}        

\begin{document}

\begin{center}
\textbf{\large{REVIEW ARTICLE}}
\end{center}

\begin{center}

\textbf{Thermal transport in crystals: from the quantum Dyson equation to mesoscopic phonon hydrodynamics}

\vspace{1em}
\normalsize
Enrico Di Lucente\textsuperscript{1,2}, 
Michele Simoncelli\textsuperscript{2}, 
Nicola Marzari\textsuperscript{1,3} \\
[0.5em]
\textsuperscript{1} Theory and Simulation of Materials (THEOS), École Polytechnique Fédérale de Lausanne, Lausanne 1015, Switzerland; 
\textsuperscript{2} Department of Applied Physics and Applied Mathematics, Columbia University, New York, NY 10027, USA; 
\textsuperscript{3} Theory of Condensed Matter, Cavendish Laboratory, University of Cambridge, Cambridge CB3 0US, United Kingdom
\end{center}

\begin{abstract}

Thermal transport in dielectric, non-magnetic crystals is mediated by quantized lattice vibrations, which drift and interact when driven out of equilibrium by a temperature gradient.
This phenomenon can be described at multiple theoretical levels, ranging from fully quantum descriptions to semiclassical and mesoscopic, continuum approaches. 
This review rigorously discusses the theoretical steps and approximations connecting these different levels of theory, bridging the gap between quantum phonon Dyson and Kadanoff-Baym equations and semiclassical Boltzmann transport formalism, and then discussing the coarse-graining procedures that yield the mesoscopic viscous heat equations for non-diffusive, hydrodynamic heat transport in devices.
We show how the established Guyer-Krumhansl and dual-phase-lag equations emerge as special linear-isotropic-band and inviscid limit of the viscous heat equations, respectively; most importantly, we demonstrate that the viscous heat equations predict not only the canonical Poiseuille flow and second sound, but also more exotic effects such as negative thermal resistance, steady-state thermal backflow and vortices.
We highlight how combining these frameworks with first-principles simulations enables us to transparently connect microscopic phonon physics to observable non-diffusive heat-transport phenomena, and to guide experimentalists in their detection, amplification, and control.
\noindent
We discuss how to recast the viscous heat equations in terms of Helmholtz and biharmonic equations that we solve analytically. We rely on this to discuss the similarities and differences between the macroscopic behavior of the “phonon fluid” and that of other hydrodynamic systems (such as classical fluids and electron fluids), discussing their hallmark on compressibility and vorticity, and their influence on  strength and nature of phonon hydrodynamics.
We conclude by providing a roadmap on how to generalize the tools used to describe phonon hydrodynamics to other quasiparticles, motivating future advances in the study of collective quantum transport phenomena in solids.
\end{abstract}

\noindent
\textbf{Keywords:} quantum heat conduction, phonon Boltzmann transport; viscous heat equations, thermal conductivity, thermal viscosity

\vspace{1em}

\begin{center}
\textbf{Table of Contents}
\end{center}
\vspace{-40pt}
\renewcommand{\contentsname}{}
\tableofcontents
\renewcommand{\contentsname}{Contents}

\newpage
\section{Introduction}

The phenomenon of heat flowing through a solid is of critical importance to many and diverse technological applications, including, e.g., miniaturization and efficiency of electronic devices, waste-heat harvesting, and even in therapy against cancer. 
All of these different applications aim to engineer heat flow via the chemical composition, the crystal structure, or the shape of the material through which heat flows. As a result, the variety of conditions and materials in which this phenomenon is exploited is very wide. 
For example, current research efforts in electronics aim to find materials with very high thermal conductivity \cite{qian_phonon-engineered_2021,kimExtremelyAnisotropicVan2021,liProbingLimitHeat2025}, 
since building electronic devices using materials with conductivity higher than silicon would facilitate the removal of heat generated by Joule effect, and consequently allow one to increase their efficiency and miniaturization \cite{1705144,balandin2011thermal}.
Conversely, in the field of thermoelectric energy harvesting, increasing the efficiency of the conversion of waste heat into electricity requires finding electrically conductive materials with a thermal conductivity as low as possible \cite{hanusThermalTransportDefective2021}. 
Finally, in medicine heat is employed in photothermal therapy \cite{liClinicalDevelopmentPotential2020,aramiRemotelyControlledNearinfraredtriggered2022} against cancer,  consisting in injecting nanoparticles into a tumor and then heating these with electromagnetic radiation up to a temperature that can destroy cancer cells.\\
Heat transport can display very different behavior depending on the material in which it takes place, and engineering such a phenomenon in everyday applications requires equations that allow us, e.g., to predict how the shape of a device affects its capability to exchange heat and thus its local temperature (which is a macroscopic, easily measurable quantity). 
This is particularly relevant in electronics, where integrated circuits have to be designed in ways that prevent the formation of hot spots, {i.e.}, regions where the temperature reaches values that cause damage or melting of the device.
On the other hand, innovating on current technologies that rely on heat transfer, or envisioning new applications,  
requires understanding heat transport from first principles --- \textit{i.e.}, in Aristoteles's words, understanding it in terms of \textit{“the first basis from which a thing is known”} \cite{Aristotele}. In practice, this corresponds to finding a  microscopic theoretical description in the context of quantum physics.

\paragraph{Fourier's macroscopic equation and Peierls Boltzmann's microscopic equation.} 

The first solution to the challenge of developing an equation capable of describing the macroscopic evolution of a temperature field $T(\bm{r},t)$ in space $(\bm{r})$ and time $(t)$ during heat transfer was proposed by Fourier in 1822 \cite{fourier1822theorie}, who introduced the celebrated “heat equation” 
\begin{equation}
C \frac{\partial T(\bm{r},t)}{\partial t} - \sum_{i,j=1}^3\kappa^{ij} \frac{\partial^2 T(\bm{r},t)}{\partial r^i\partial r^j} = 0 \;,
\end{equation}
where $C$ is the specific heat and $\kappa^{ij}$ the thermal conductivity tensor, both quantities that can be measured in experiments. Fourier's heat equation is  widely used nowadays in engineering applications involving macroscopic objects exchanging heat.\\
The first microscopic theory for thermal transport was developed by Peierls in 1929 \cite{peierls1929kinetischen}, who formulated a semiclassical theory and rationalized heat conduction in crystals in terms of atomic vibrational waves diffusing and colliding as if they were particles of a classical gas (\textit{i.e.} evolving according to an equation analogous to that developed by Boltzmann for classical gases). In insulating crystals, solids in which atoms are arranged in a periodic repeating, ordered structure called crystal lattices, heat is mainly carried by such quanta of lattice vibrations, called phonons. In fact, in insulators and semiconductors, electrons are not very free to move, and so, when subject to an external temperature gradient, phonons provide the main contribution to thermal conductivity. The quasiparticle (phonon) wave-packet distribution ${n}(\bm{r},\bm{q},t)_s$ is the central quantity of this semiclassical formulation. Here, $\bm{r}$ is the position of the wave packet in direct space, $\hbar\bm{q}$ is its quasimomentum (in a crystal, periodicity implies that the wavevector $\bm{q}$ can be restricted to the first Brillouin zone $\mathfrak{B}$), $t$ is time, and $s$ is a band index that specifies the energy of the wave packet and, equivalently, its polarization branch. The evolution of this semiclassical distribution function is ruled by the phonon Boltzmann transport equation (BTE) \cite{peierls1996quantum,ziman2001electrons}:
\begin{equation}  \label{PBoltzmann_BTE}
\frac{\partial n(\bm{r},\bm{q},t)_s}{\partial t} +\bm{v}(\bm{q})_s\cdot \nabla_{\bm{r}}n(\bm{r},\bm{q},t)_s =\frac{\partial n(\bm{r},\bm{q},t)_s}{\partial t}\bigg|_{\rm col} \!\!,
\end{equation}
where $\bm{v}(\bm{q})_s$ is the group velocity of the wave packet with wavevector $\bm{q}$ and in band $s$, and $\tfrac{\partial n(\bm{r},\bm{q},t)_s}{\partial t}\big|_{\rm col}$ a collision operator that is computed using the Fermi golden rule \cite{ziman2001electrons,peierls1996quantum}.
The relevance of the BTE \eqref{PBoltzmann_BTE} is its capability to predict the thermal conductivity of a crystal from its microscopic vibrational properties (described by the distribution ${n}(\bm{r},\bm{q},t)_s$).
In fact, Hardy \cite{hardy1963energy} showed that the heat flux $\bm{Q}$, which is by definition related to the conductivity tensor $\kappa^{ij}$ and temperature gradient $\nabla T$ via $Q^i=-\kappa^{ij}\nabla_j T$ (we employ Einstein's repeated index convention for the sum), is determined by the microscopic solution of Eq. \eqref{PBoltzmann_BTE} in the steady-state and homogeneous regime:
\begin{equation} \label{heat_flux_intro}
\boldsymbol{Q}=\frac{1}{(2\pi)^3}\int_{\mathfrak{B}}\sum_{s}\hbar \omega(\bm{q})_s \bm{v}(\bm{q})_s n(\bm{q})_s,
\end{equation}
where $\omega(\bm{q})_s$ is the frequency of the wave packet with wavevector $\bm{q}$ and in band $s$.
Therefore, solving the linearized BTE (LBTE), linear in the temperature gradient in Eq. \eqref{PBoltzmann_BTE}, in the steady-state and spatially homogeneous regime yields the heat flux \eqref{heat_flux_intro} and, consequently, determines the thermal conductivity tensor of the material through the relation $Q^i=-\kappa^{ij}\nabla_j T $.\\
Recent years have seen major theoretical \cite{lindsay2014phonon,cepellotti2015phonon,levitov2016electron,simoncelli2020generalization} and experimental \cite{bandurin2016negative,crossno2016observation,moll2016evidence,lee2015hydrodynamic,ding2022observation,huberman2019observation,goblot2024imaging} advances in electrical and thermal transport in fast conductors, especially with the emergence of non-diffusive and hydrodynamic regimes for electrons and phonons. In contrast to conventional heat conduction, dominated by momentum-relaxing interactions, phonon hydrodynamics involves fluid-like heat flow driven by momentum-conserving phonon-phonon scattering. First studied in the 1960s, it revealed Poiseuille-like flow \cite{mezhov1965measurement} and second sound \cite{ackerman1966second,guyer1966solution,gurzhi1968hydrodynamic,enz1968one,hardy1970phonon,gotze1967first} in solid helium \cite{ackerman1966second}, sodium fluoride \cite{jackson1970second,pohl1976observation}, bismuth \cite{narayanamurti1972observation}, sapphire \cite{danil1979observation}, and strontium titanate \cite{hehlen1995observation}, all at cryogenic temperatures. Theoretical efforts bridged microscopic and macroscopic heat transport: Sussmann and Thellung \cite{sussmann1963thermal} derived mesoscopic equations from the LBTE \cite{peierls1955quantum}, while Gurzhi \cite{gurzhi1964thermal,gurzhi1968hydrodynamic} and Guyer and Krumhansl \cite{guyer1966solution,guyer1966thermal} introduced weak momentum dissipation to model second sound and Poiseuille flow. Early models assumed specific phonon dispersions—linear-isotropic or power-law—but were later refined by Hardy's mesoscopic framework incorporating weak Umklapp (momentum-dissipating) scattering \cite{hardy1970phonon,hardy1974hydrodynamic}. Gurzhi’s work also paved the way for electron hydrodynamics, enabling recent observations of electron viscosity \cite{bandurin2016negative}. Subsequent theoretical developments, in particular the work of Gurevich \cite{gurevich1986transport} together with related contributions \cite{aronov1981boltzmann,gurevich1991intrinsic}, provided a detailed analysis of momentum-conserving scattering processes and their implications for collective phonon transport. These studies anticipated several key aspects of the modern hydrodynamic description, including the emergence of a phonon-viscosity concept. Related discussions can also be found in the monograph by Chen \cite{chen2005nanoscale}, which places these ideas in the broader context of nanoscale heat transport. First-principles simulations with the LBTE predicted hydrodynamic behavior in graphene and other 2D materials \cite{cepellotti2015phonon,lee2015hydrodynamic,cepellotti2017transport}, carbon nanotubes \cite{lee2017hydrodynamic}, and graphite \cite{ding2018phonon} at non-cryogenic temperatures. Experiments have since confirmed viscous thermal transport in such conductors \cite{schmidt2008pulse,balandin2011thermal,fugallo2014thermal,machida2020phonon}, including room-temperature second sound \cite{melis2021room,beardo2021observation,ding2022observation,huberman2019observation,xie2026room}, Poiseuille-like flow \cite{huang2023observation,li2022reexamination,cepellotti2017boltzmann,machida2018observation,sendra2022hydrodynamic}, and lattice cooling \cite{jeong2021transient}.\\
In the hydrodynamic regime, where phenomena such as Poiseuille flow or second sound occur, Fourier’s law breaks down \cite{huberman2019observation,machida2018observation,martelli2018thermal,khodusov2012second}, making this commonly used and convenient tool unreliable for predicting temperature profiles in a device. Despite the predictive power of the full LBTE, its mathematical and computational complexity makes it unsuitable for describing thermal transport in realistic devices with non-trivial geometries \cite{cepellotti2017boltzmann}. This has motivated the development of mesoscopic or coarse-grained models that retain the essential hydrodynamic and non-local effects captured by the LBTE, yet remain tractable at the continuum level. Several strategies have been explored. Some approaches simplify the LBTE by neglecting the repopulation of phonon modes due to scattering—the so-called single-mode relaxation-time approximation (RTA)—allowing for analytical \cite{hua2014analytical,vermeersch2015superdiffusive,maznev2011onset,yang2015heating} or asymptotic \cite{peraud2016extending} solutions. Based on the LBTE within the RTA, mesoscopic models have been developed that generalize Fourier’s law to account for ultrafast thermal processes or ballistic effects \cite{chen2001ballistic,anderson2006novel,ordonez2011constitutive,ramu2014enhanced,hua2019experimental}. Other works construct mesoscopic models independently of the LBTE \cite{cao2007equation}, or extend the Guyer-Krumhansl equation to include boundary effects on heat transport \cite{alvarez2009phonon,guo2015phonon,guo2018phonon,ziabari2018full,torres2018emergence}. Finally, a hydrodynamic transport model has been derived from the LBTE using the Callaway approximation, introducing a phonon viscosity that can be computed from atomistic simulations \cite{li2018role}. In passing, it is worth noting that analogous theoretical developments have been made in electron hydrodynamics, where the electronic BTE is coarse-grained into Navier–Stokes–like models for charge and momentum flow \cite{levitov2016electron,torre2015nonlocal,bandurin2016negative,moll2016evidence,scaffidi2017hydrodynamic}. However, while electronic hydrodynamics is often observed in semimetals and two-dimensional conductors, this review focuses on phonons. \\
Broadly speaking, current research in phonon hydrodynamics can be viewed as evolving along two complementary directions. The first seeks to identify and engineer materials that exhibit hydrodynamic behavior, involving both experimental efforts to synthesize new compounds and computational approaches to predict promising candidates, including those not yet realized in the laboratory. The second direction seeks to broaden the temperature window for observing hydrodynamic behavior, pushing it toward higher, ideally room-temperature, conditions to enhance its technological applicability. \\
Alongside experimental progress, theoretical and computational modeling of phonon hydrodynamics has undergone rapid and very recent growth. Within computational physics and materials science, such modeling has become particularly powerful when combined with first-principles (ab initio) simulations based on, i.e., density-functional theory (DFT). The latter is used to solve the underlying Schrödinger equation of quantum mechanics in crystalline solids using controlled approximations, offering both predictive accuracy and cost efficiency. By bridging fundamental physics with realistic modeling, such simulations not only complement experiments but also guide them by predicting phenomena that may not yet have been observed.\\
The purpose of this review is to present a unified and rigorous theoretical framework for thermal transport across all relevant length and time scales, from the microscopic quantum-mechanical description of phonons to the mesoscopic and macroscopic regimes characteristic of realistic devices. We aim to clarify how these regimes are connected, elucidate the transition from fully quantum and semiclassical descriptions to macroscopic hydrodynamic behavior, and provide both fundamental insight and practical guidance for future applications.

\paragraph{Organization of the review: beyond the Peierls-Boltzmann equation, and beyond Fourier's law.} 

\begin{figure}[h!]
\centering
\includegraphics[width=0.8\textwidth]{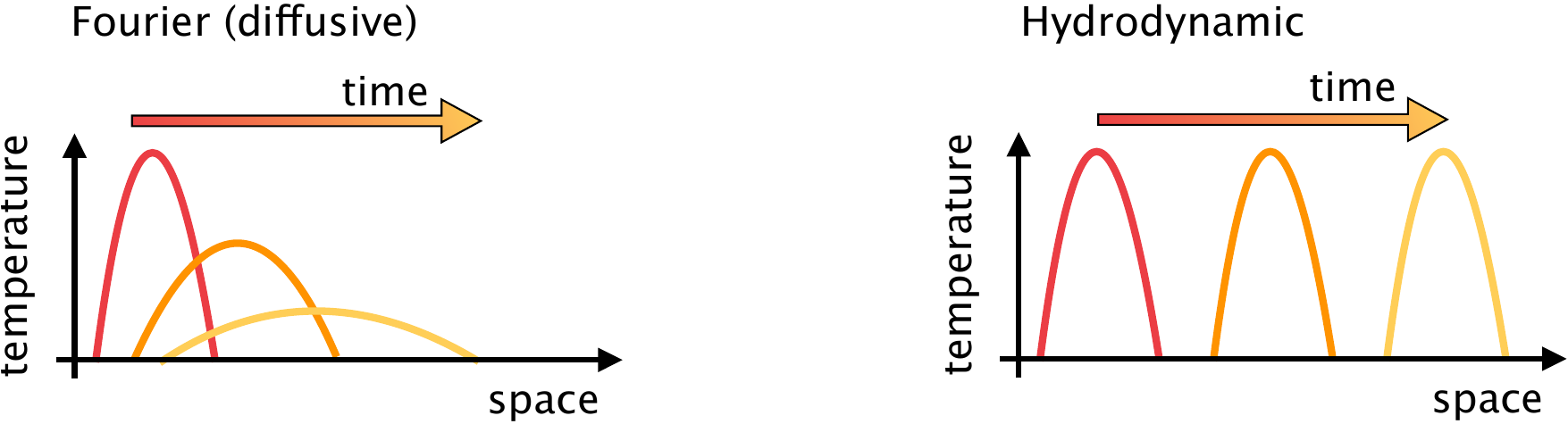}
\caption{\textbf{Qualitative difference between diffusive and hydrodynamic thermal transport.} (Left panel) Schematic representation of the temporal evolution of a temperature perturbation in the diffusive regime, where Fourier's equation is accurate. (Right panel) Schematic representation of heat propagation in the form of “second sound” in the hydrodynamic regime of thermal transport, where Fourier's equation fails. The color here is related to time, with red representing a time instant happened earlier than yellow (orange represents an intermediate time instant). }
\label{fig:Fourier_vs_second_sound}
\end{figure}

The purpose of this review is to summarize the theoretical framework to describe thermal transport beyond Fourier, and beyond Boltzmann. We provide a rigorous derivation that bridges the quantum-mechanical description of heat in terms of microscopic phonon excitations, and the mesoscopic and macroscopic regimes characteristic of realistic devices. We offer both fundamental insight and practical guidance for future applications. The review is organized as follows:\\
section 2 presents an overview of the nonequilibrium Green’s function (NEGF) formalism for phonons. This formalism was developed in the 1960s through a series of foundational works by Martin and Schwinger \cite{martin1959theory}, Kadanoff and Baym \cite{kadanoff2018quantum}, and Keldysh \cite{keldysh1964diagram}, which established a general framework for describing quantum systems driven out of equilibrium. 
We discuss the quantum Dyson equation \cite{dyson1949s}, which establishes a perturbative framework for relating Green’s functions by resumming infinite classes of Feynman diagrams. Then, we discuss the nonperturbative set of integral equations
formulated by Schwinger \cite{schwinger1951green} (now known as the Schwinger–Dyson equations) which allow us to study interacting quantum systems, and lead to the Kadanoff–Baym equation (KBE) describing the quantum kinetics of phonon quasiparticles \cite{kadanoff2018quantum}. \\
In section 3 we discuss how to simplify the KBE into the celebrated semiclassical Peierls-Boltzmann transport equation (the BTE) \cite{peierls1929kinetischen}, and how its linearized form (the LBTE) allows us to relate microscopic phonon properties to macroscopic, experimentally observable, thermal conductivity. \\
Section 4 discusses thermal transport at the mesoscopic level. It begins by recalling how the exact solution of the microscopic LBTE can be written as a linear combination of “relaxons” \cite{cepellotti2016thermal} (\textit{i.e.} collective phonon excitations that are eigenvectors of the LBTE's scattering matrix and have a well-defined mathematical parity), and such a solution determines a microscopic, closed-form expression for the thermal conductivity that receives contributions exclusively from odd relaxons. Then,  we show that the complementary set of even relaxons determines another quantity, thermal viscosity. We discuss how the thermal viscosity becomes especially relevant in the hydrodynamic regime of thermal transport \cite{simoncelli2020generalization}, where collisions between phonon wavepackets that conserve the crystal momentum (normal processes) are much more frequent than momentum-dissipating collisions (Umklapp processes) \cite{gurzhi1968hydrodynamic,ziman2001electrons}. Under these conditions, we show that the equilibrium distribution of phonon wave packets depends on two parameters, temperature $T$ (which emerges from the conservation of energy in collisions), and “drift-velocity” $\bm{u}$ (which emerges from the conservation of momentum in the most frequent normal collisions); $T$ and $\bm{u}$ play a role analogous to the pressure and velocity fields in fluids, and from this follows a prediction of fluid-like (hydrodynamic) behavior for heat \cite{gurzhi1968hydrodynamic}.\\
We discuss how microscopic conservation laws can be exploited to coarse-grain the microscopic  LBTE (which has a complex integro-differential form) into a set of mesoscopic “viscous heat equations” (VHE), partial differential equations that describe  both Fourier's heat diffusion and non-diffusive, hydrodynamic phenomena such as second sound at a much reduced complexity and more intuitively than the LBTE (see Fig. \ref{fig:Fourier_vs_second_sound}).
Importantly, the VHE constitute a general framework that naturally encompasses several well-known heat-transport models as special limits. In the inviscid limit, the VHE reduce to the dual-phase-lag equation (DPLE), which retains heat-flux inertia but eliminates viscous momentum transport and the associated spatial nonlocality. Enforcing a vanishing temperature phase lag further reduces the DPLE to Cattaneo’s equation, removing the distinction between drifting and diffusive heat-flux components. In contrast, the Guyer–Krumhansl equations are obtained from the VHE in the steady-state limit under the assumptions of linear and isotropic phonon dispersion and drift-dominated transport. While viscosity is retained in this case, the incompressibility constraint enforces a harmonic temperature field, thereby precluding the emergence of internal temperature extrema. We show that this novel mesoscopic formulation, combined with modern first-principles calculations for its parameters \cite{simoncelli2020generalization,dragavsevic2023viscous}, yields for graphite predictions of the temperatures and sizes at which hydrodynamic effects are largest in good agreement with very recent experimental findings \cite{huberman2019observation,ding2022observation}. 
\\
In section 5 we discuss how the steady-state VHE can be recast into two decoupled modified biharmonic equations—one for a drift-velocity potential and one for a stream function—enabling analytic solutions in reciprocal space. This procedure delivers a full closed-form expression for the temperature field and exposes its decomposition into compressible and vortical contributions. Compressibility, in particular, emerges as a defining property of viscous phonon flow, distinguishing it from most electronic hydrodynamic systems \cite{torre2015nonlocal,bandurin2016negative,levitov2016electron}, which are typically modeled as nearly incompressible \cite{levitov2016electron}. We identify the interplay of compressibility and vorticity as the microscopic source of thermal viscosity. Following the route of Ref. \cite{levitov2016electron}, we relate this mechanism to negative nonlocal thermal resistance in a two-dimensional strip—an unmistakable signature of hydrodynamic heat flow—arising from thermal vortices \cite{raya2022hydrodynamic,restuccia2023non,sykora2023multiscale,shang2020heat,zhang2021heat,tur2024microscopic,di2025vortices} or heat backflow \cite{dragavsevic2023viscous}.
\\
Finally, in section 6 we first discuss further latest developments in  theoretical and experimental signatures of phonon hydrodynamics across new materials and then provide an overview of possible future directions, and a roadmap on how to generalize and extend the theoretical tools presented to describe other quasiparticles, motivating future studies of non-diffusive quantum transport phenomena in solids.

\paragraph{Executive roadmap: thermal transport across scales}

Thermal transport in solids spans a wide hierarchy of length and time scales, ranging from fully quantum-mechanical dynamics at the atomic level to macroscopic heat diffusion in extended devices. No single theoretical description is able to capture all relevant regimes. Instead, different equations emerge as controlled limits of a more general framework, depending on which physical processes dominate and which degrees of freedom can be coarse-grained.\\
The purpose of this review is not to advocate a single “best” model, but to clarify how commonly used approaches---from Fourier’s law to hydrodynamic and microscopic transport equations---are connected, what physical assumptions underlie each of them, and under which conditions they succeed or fail. This Executive Roadmap provides a guide to the theoretical landscape explored in the remainder of the article, summarizing the regimes of validity of the different transport equations discussed and highlighting the physical effects that each description captures or neglects.
\\\\
Table \ref{tab:acronyms} summarizes the acronyms used throughout the text.
\begin{table}[h!]
\centering
\renewcommand{\arraystretch}{1.4}
\resizebox{0.85\textwidth}{!}{%
\begin{tabular}{>{\centering\arraybackslash}m{4.5cm}
                >{\centering\arraybackslash}m{10.5cm}}
\hline
\hline
\textbf{Acronym} & \textbf{Definition} \\
\hline
NEGF & Nonequilibrium Green's Function \\
KBE & Kadanoff--Baym Equation \\
BTE & Boltzmann Transport Equation \\
LBTE & Linearized Boltzmann Transport Equation \\
FGR & Fermi's Golden Rule \\
GKB (ansatz) & Generalized Kadanoff--Baym (ansatz) \\
SCPH & Self-consistent phonon \\
SCAILD & Self-consistent \textit{ab initio} lattice dynamics \\
SSCHA & Stochastic self-consistent harmonic approximation \\
AIMD & \textit{Ab initio} molecular dynamics \\
RTA & Relaxation time approximation \\
VHE & Viscous heat equations \\
SW (transformation) & Schrieffer--Wolff (transformation) \\
LLW & Lubrication layer width \\
FDN & Fourier deviation number \\
DPLE & Dual-phase-lag equation \\
SS & Second sound \\
GKE & Guyer--Krumhansl equations \\
LCS & Lattice cooling strength \\
\hline
\hline
\end{tabular}}
\caption{List of acronyms used throughout the manuscript.}
\label{tab:acronyms}
\end{table}
$\,$\\\\
Table \ref{tab:transport_regimes} summarizes the main transport regimes addressed in this review, the dominant physical processes at play, the corresponding governing equations, and their principal limitations. The regimes should be understood as overlapping rather than sharply separated, with smooth crossovers controlled by material properties, temperature, and device geometry.
\begin{table}[h!]
\centering
\renewcommand{\arraystretch}{1.4}
\resizebox{\textwidth}{!}{%
\begin{tabular}{>{\centering\arraybackslash}m{3.2cm}
                >{\centering\arraybackslash}m{4.0cm}
                >{\centering\arraybackslash}m{4.2cm}
                >{\centering\arraybackslash}m{4.4cm}
                >{\centering\arraybackslash}m{4.8cm}}
\hline
\hline
\textbf{Level/regime} &
\textbf{Transport behavior} &
\textbf{Temperature regime} &
\textbf{Governing equations} &
\textbf{Scope and limitations} \\
\hline
\textbf{Fully quantum} &
Quantum kinetic regime &
All $T$ and ultrafast (beyond quasiparticle) dynamics &
\textbf{Dyson equation} and \textbf{Kadanoff--Baym equation} for the phonon Green's functions &
Captures quantum coherence, correlations, and transient dynamics; lacks an intuitive continuum description and is extremely demanding to solve numerically \\
\hline
\textbf{Semiclassical microscopic} &
Mode-resolved phonon transport &
All $T$ (when phonon quasiparticles are well defined) &
\textbf{Boltzmann transport equation} (BTE, LBTE) for the phonon distribution function &
Predicts thermal conductivity from microscopic phonon properties; offers clear physical interpretation and straightforward implementation in numerical codes; computationally demanding for realistic geometries \\
\hline
\textbf{Unified mesoscopic framework} &
Diffusive + hydrodynamic (with ballistic corrections) &
Medium--low to high $T$ &
\textbf{Viscous heat equations} for temperature and phonon drift velocity (numerical); \textbf{modified biharmonic equations} for velocity potential and stream-function fields (analytical) &
Unified description retaining viscosity, compressibility, and spatial nonlocality. The dual-phase-lag equation (DPLE) emerges as a limiting case for vanishing viscosity, capturing finite-speed thermal waves but neglecting viscous momentum transport, vorticity, boundary effects, and nonlocal hydrodynamic heat flow \\
\hline
                    &
\textbf{Ballistic}: dominant extrinsic sources of scattering (finite domains, grain boundaries); plug-like heat-flux profiles &
Low $T$ &
\textbf{Phenomenological ballistic models}; generalized Fourier-type equations &
Boundary-dominated transport; captures finite-size effects but lacks microscopic mode resolution \\
\textbf{Mesoscopic}  & \textbf{Hydrodynamic}: dominant intrinsic momentum-conserving (normal) phonon--phonon scattering; Poiseuille (parabolic) heat-flux profile &
Intermediate $T$ &
\textbf{Guyer--Krumhansl equations} &
Assume linear and isotropic phonon dispersion, drift-dominated heat flux, and a harmonic steady-state temperature field \\
               &
\textbf{Kinetic/Diffusive}: dominant intrinsic momentum-nonconserving phonon--phonon scattering (Umklapp, impurities); flat heat-flux profile &
Medium--high $T$ &
\textbf{Fourier's law of heat conduction} &
Accurate for long length and time scales; fails for finite propagation speed, nonlocality, and hydrodynamic effects \\
\hline
\hline
\end{tabular}}
\caption{Hierarchical organization of thermal-transport theories across fully quantum, semiclassical microscopic, and mesoscopic descriptions. The mesoscopic level explicitly encompasses hydrodynamic and diffusive regimes within a unified continuum framework, and allow for ballistic corrections.}
\label{tab:transport_regimes}
\end{table}

\newpage
\section{Quantum regime: phonon Dyson and Kadanoff-Baym equations}

In this section, we present a concise overview of the nonequilibrium Green’s function (NEGF) formalism for phonons. This formalism was developed in the 1960s through a series of foundational works by Martin and Schwinger \cite{martin1959theory}, Kadanoff and Baym \cite{kadanoff2018quantum}, and Keldysh \cite{keldysh1964diagram}, which established a general framework for describing quantum systems driven out of equilibrium. These early developments introduced the contour-ordered Green’s functions and the associated diagrammatic techniques that underpin modern nonequilibrium many-body theory. Comprehensive treatments and extensions of this formalism have since been provided in influential reviews by Danielewicz \cite{danielewicz1984quantum1,danielewicz1984quantum2} and Mahan \cite{mahan1987quantum,mahan2000many}, which remain standard references for applications of NEGF to transport and dynamical phenomena in condensed matter systems.\\
In this review, we begin with the quantum Dyson equation \cite{dyson1949s}, which establishes a perturbative framework for relating Green’s functions by resumming infinite classes of Feynman diagrams within quantum electrodynamics. Building on this foundation, Schwinger formulated a nonperturbative set of integral equations—now known as the Schwinger–Dyson equations—which generalize Dyson’s relations and provide a powerful tool for studying interacting quantum systems \cite{schwinger1951green}. These equations are now widely used across quantum field theory, ranging from condensed-matter to high-energy physics, and they naturally lead to the Kadanoff–Baym equation (KBE) describing quasiparticle (phonons, in this case) quantum kinetics \cite{kadanoff2018quantum}. This provides a direct link to semiclassical transport, bridging to the Boltzmann transport equation (BTE) and enabling predictions of macroscopic thermal properties from first-principles simulations, typically based on density functional theory \cite{DFT_1,DFT_2,omini1995iterative,omini1997heat,broido_intrinsic_2007,lindsay_perspective_2019,mcgaughey2019phonon,lindsay2010flexural,fugallo2013ab,li2012thermal,li2015ultralow,fugallo2014thermal,lindsay2019perspective,lindsay2016first,togo2015distributions,togo2023first,chaput2013direct,cepellotti2016thermal,cepellotti2016relaxons,cepellotti2017boltzmann,cepellotti2017transport,bonini2012acoustic,di2023crossover,di2026spin,ponet2024energy}

\subsection{Nonequilibrium phonon Green's function theory}

The starting point is the full lattice Hamiltonian for a crystal with basis atoms:
\begin{equation} \label{start_hamiltonian}
H = \sum_{\bm{R}b} \frac{P_{\bm{R}b}^{2}}{2m_b} + V(\boldsymbol{r}_{\bm{R}_1 b_1}, \boldsymbol{r}_{\bm{R}_2 b_2}, \boldsymbol{r}_{\bm{R}_3 b_3}, \dots),
\end{equation}
where $\boldsymbol{r}_{\bm{R}b} = \boldsymbol{r}_{\bm{R}} + \boldsymbol{r}_b$ denotes the position of atom $b$ in unit cell $\bm{R}$, $m_b$ is its mass, and $V$ is the interatomic potential. Within the harmonic approximation, the potential energy is expanded to second order around equilibrium:
\begin{equation} 
V_{\text{harm}} = \frac{1}{2} \sum_{\substack{\bm{R}b\alpha \\ \bm{R}'b'\beta}} \upphi_{\bm{R}b\alpha, \bm{R}'b'\beta} \, u_{\bm{R}b\alpha} u_{\bm{R}'b'\beta}, 
\end{equation}  
with the real-space dynamical matrix defined as
\begin{equation} 
\upphi_{\bm{R}b\alpha, \bm{R}'b'\beta} = \left. \frac{\partial^2 V}{\partial u_{\bm{R}b\alpha} \, \partial u_{\bm{R}'b'\beta}} \right|_{\text{eq}}.
\end{equation}  
Its elements are the second-order interatomic force constants, and $\boldsymbol{u}_{\bm{R}b}$ denotes small displacements around equilibrium positions. By assuming plane-wave solutions, the Hamiltonian can be expressed in reciprocal space using momentum and position operators
\begin{equation}
P_{\bm{q}b} = \frac{1}{\sqrt{m_b}} \sum_{\bm{R}} P_{\bm{R}b} e^{-i \bm{q} \cdot (\bm{r}_{\bm{R}} + \bm{B}_b)}, 
\quad
X_{\bm{q}b} = \frac{1}{\sqrt{m_b}} \sum_{\bm{R}} \boldsymbol{u}_{\bm{R}b} e^{-i \bm{q} \cdot (\bm{r}_{\bm{R}} + \bm{B}_b)}.
\end{equation}
The harmonic Hamiltonian becomes
\begin{equation}
H_{\text{harm}} = \frac{1}{\mathcal{V}} \sum_{\bm{q}b} \frac{P_{\bm{q}b} P_{\bm{q}b}^{\dagger}}{2m_b} + \frac{1}{2\mathcal{V}} 
\sum_{\substack{\bm{q} \\ b\alpha \\ b'\beta}} 
\upphi_{b\alpha, b'\beta}(\bm{q}) \, 
X_{\bm{q} b\alpha}X_{\bm{q}b'\beta}^{\dagger},
\end{equation}
where $\mathcal{V}$ is the crystal volume defined as the unit cell volume times the number of unit cells that constitute the crystal $N_{c}$ ($N_{c}$ is also the number of wave vectors $\boldsymbol{q}$ used to sample the Brillouin zone. Here, $\upphi_{b\alpha, b'\beta; \bm{q}} = \sum_{\bm{R}'} \upphi_{\bm{0}b\alpha, \bm{R}'b'\beta} e^{-i \bm{q} \cdot r_{\bm{R}'}}$ is the Fourier-transformed dynamical matrix. Diagonalizing $\upphi(\bm{q})$ at each $\bm{q}$-point yields phonon eigenvectors $\upepsilon_{b\alpha; \bm{q}s}$ and frequencies $\omega_{\bm{q}s}$:
\begin{equation}
\sum_{b'\beta}\mathcal{D}_{b\alpha, b'\beta; \bm{q}} \, \upepsilon_{b'\beta; \bm{q}s} = \omega_{\bm{q}s}^2 \, \upepsilon_{b\alpha; \bm{q}s},
\end{equation}
where, choosing one lattice vector as the origin,
\begin{equation}
\mathcal{D}_{b\alpha,\, b'\beta}(\bm{q}) =
\frac{1}{\sqrt{m_b m_{b'}}}
\sum_{\bm{R}}
\upphi_{0b\alpha,\, \bm{R}b'\beta}
\, e^{i \bm{q}\cdot \bm{R}}.
\end{equation}
The harmonic Hamiltonian in terms of phonon normal modes reads
\begin{equation} \label{harmonic_hamiltonian} 
H_{\text{harm}} = \sum_{\nu} \hbar \omega_{\nu} \left( \frac{1}{2} + a_{\nu}^{\dagger} a_{\nu} \right),
\end{equation}
where $a_\nu^\dagger$ and $a_\nu$ are phonon creation and annihilation operators, and $\pm\nu = (\pm \bm{q}, s)$ combines wavevector and branch index. Anharmonic interactions are included through the leading third-order perturbation term:
\begin{equation} 
H = H_{\text{harm}} + \Delta H_{\text{3rd}} = H_{\text{harm}} + \frac{1}{3!} \sum_{\substack{\bm{R}\bm{R}'\bm{R}'' \\ bb'b'' \\ \alpha\beta\eta}} \uppsi_{\bm{R}b\alpha, \bm{R}'b'\beta, \bm{R}''b''\eta} \, u_{\bm{R}b\alpha} u_{\bm{R}'b'\beta} u_{\bm{R}''b''\eta},
\end{equation}
where the third-order interatomic force constants are 
\begin{equation} 
\uppsi_{\bm{R}b\alpha,\bm{R}'b'\beta,\bm{R}''b''\eta}=\left.\frac{\partial^{3}V}{\partial u_{\bm{R}b\alpha}\partial u_{\bm{R}'b'\beta}\partial u_{\bm{R}''b''\eta}}\right|_{\text{eq}}. 
\end{equation}  
In reciprocal space, the anharmonic contribution can be written as
\begin{equation} \label{phonon_hamiltonina_with_anharmonic_term} 
\begin{split} 
\Delta H_{\text{3rd}} &= \frac{1}{3!}\sum_{\nu\nu'\nu''} \uppsi_{\nu\nu'\nu''} 
\delta_{\boldsymbol{q}+\boldsymbol{q}'+\boldsymbol{q}'',\boldsymbol{K}}
(a_{\nu}+a^{\dagger}_{-\nu})(a_{\nu'}+a^{\dagger}_{-\nu'})(a_{\nu''}+a^{\dagger}_{-\nu''})\\ 
&= \frac{1}{3!}\sum_{\nu\nu'\nu''} \uppsi_{\nu\nu'\nu''} 
\delta_{\boldsymbol{q}+\boldsymbol{q}'+\boldsymbol{q}'',\boldsymbol{K}} \, A_{\nu} A_{\nu'} A_{\nu''},
\end{split} 
\end{equation}  
where
\begin{equation} 
\uppsi_{\nu\nu'\nu''} = \sqrt{\frac{\hbar^{3}}{\mathcal{V}}} 
\sum_{\substack{bb'b''\\ \alpha\beta\eta}} 
\frac{\upepsilon_{b\alpha;\nu} \, \upepsilon_{b'\beta;\nu'} \, \upepsilon_{b''\eta;\nu''} \, 
\uppsi_{b\alpha;\boldsymbol{q},b'\beta;\boldsymbol{q}',b''\eta;\boldsymbol{q}''}} 
{\sqrt{8 m_{b} m_{b'} m_{b''} \, \omega_{\nu} \omega_{\nu'} \omega_{\nu''}}},
\end{equation}
and $\boldsymbol{K}$ is a reciprocal lattice vector. In Eq. \eqref{phonon_hamiltonina_with_anharmonic_term} we have already introduced the vibronic phonon operators:
\begin{equation}
\begin{split}
A_{\nu} &= a_{\nu} + a^{\dagger}_{-\nu},\\
A^{\dagger}_{\nu} &= A_{-\nu} = a^{\dagger}_{\nu} + a_{-\nu},
\end{split}
\end{equation}
which allow the anharmonic interactions to be written compactly in terms of collective phonon excitations. They naturally capture the mixed-mode interactions that arise in anharmonic systems and, as evident from Eq. \eqref{phonon_hamiltonina_with_anharmonic_term}, enter directly into the anharmonic part of the phonon Hamiltonian. This representation also provides a clear connection to physical observables, making it particularly convenient when constructing the phonon Green's function.\\
In the vibronic representation, the single-phonon Green's function is defined using the operators $A$ and $A^{\dagger}$ as
\begin{equation} \label{green's_function}
G_{\nu_{1}\nu_{2}}(t_{1},t_{2}) = -i\,{\rm Tr}\left(\rho \, \mathcal{T}\Big(A_{\nu_{1}}(t_{1}) A^{\dagger}_{\nu_{2}}(t_{2})\Big)\right) 
= -i \langle \mathcal{T} A_{\nu_{1}}(t_{1}) A^{\dagger}_{\nu_{2}}(t_{2}) \rangle,
\end{equation}
where $\mathcal{T}$ is the time-ordering operator, which arranges operators such that the one with the earlier time argument appears to the right. Following the standard conventions in many-body theory, we define the “lesser” and “greater” Green's functions as
\begin{equation} \label{G<_corr_function}
G^{<}_{\nu_{1}\nu_{2}}(t_{1},t_{2}) = -i \, {\rm Tr}\Big(\rho \, A^{\dagger}_{\nu_{2}}(t_{2}) A_{\nu_{1}}(t_{1})\Big) = -i \langle A^{\dagger}_{\nu_{2}}(t_{2}) A_{\nu_{1}}(t_{1}) \rangle,
\end{equation}
\begin{equation} \label{G>_corr_function}
G^{>}_{\nu_{1}\nu_{2}}(t_{1},t_{2}) = -i \, {\rm Tr}\Big(\rho \, A_{\nu_{1}}(t_{1}) A^{\dagger}_{\nu_{2}}(t_{2})\Big) = -i \langle A_{\nu_{1}}(t_{1}) A^{\dagger}_{\nu_{2}}(t_{2}) \rangle.
\end{equation}
Using these definitions, the full time-ordered Green's function can be expressed compactly as
\begin{equation} \label{green's_function_bis}
G_{\nu_{1}\nu_{2}}(t_{1},t_{2}) = \theta(t_{2}-t_{1}) G^{<}_{\nu_{1}\nu_{2}}(t_{1},t_{2}) + \theta(t_{1}-t_{2}) G^{>}_{\nu_{1}\nu_{2}}(t_{1},t_{2}).
\end{equation}
For practical calculations, it is useful to introduce the retarded and advanced Green's functions:
\begin{equation} \label{GR_commutation}
G^{R}_{\nu_{1}\nu_{2}}(t_{1},t_{2}) = -i \theta(t_{1}-t_{2}) \langle [A_{\nu_{1}}(t_{1}), A^{\dagger}_{\nu_{2}}(t_{2})] \rangle 
= \theta(t_{1}-t_{2}) \left[ G^{>}_{\nu_{1}\nu_{2}}(t_{1},t_{2}) - G^{<}_{\nu_{1}\nu_{2}}(t_{1},t_{2}) \right],
\end{equation}
\begin{equation} \label{GA_commutation}
G^{A}_{\nu_{1}\nu_{2}}(t_{1},t_{2}) = i \theta(t_{2}-t_{1}) \langle [A_{\nu_{1}}(t_{1}), A^{\dagger}_{\nu_{2}}(t_{2})] \rangle 
= -\theta(t_{2}-t_{1}) \left[ G^{>}_{\nu_{1}\nu_{2}}(t_{1},t_{2}) - G^{<}_{\nu_{1}\nu_{2}}(t_{1},t_{2}) \right],
\end{equation}
where $[A,B] = AB - BA$ follows from the bosonic commutation relations \cite{mahan2000many}. The retarded Green's function $G^{R}$ is nonzero only for $t_{1} \ge t_{2}$, describing the system's response at time $t_{1}$ to a perturbation applied at an earlier time $t_{2}$. Conversely, the advanced Green's function $G^{A}$ is nonzero only for $t_{1} \le t_{2}$. It is important to distinguish between the non-interacting Green's function $G^0$ and the full interacting Green's function $G$. The free Green's function $G^0$, determined solely by the harmonic Hamiltonian $H_{\rm harm}$, describes the amplitude for a phonon created in mode $\nu_2$ at time $t_2$ to be found in mode $\nu_1$ at time $t_1$ in the absence of interactions. In contrast, the full Green's function $G$ accounts for the actual propagation including all anharmonic interactions described by $V$, capturing phonon scattering, decay, and renormalization effects. Physically, $G$ represents the complete propagation amplitude of a phonon, including all possible interactions with other phonons.\\
Compared to approaches based on the one-body density matrix, whose time evolution of the density matrix is governed by the von Neumann equation \cite{neumann1955mathematical,neumann1927wahrscheinlichkeitstheoretischer,dirac1981principles} (often referred to as the Liouville–von Neumann equation by analogy with classical Liouville dynamics), this formalism offers a transparent framework for understanding and controlling the approximations involved in modeling quantum phonon interactions (see also Refs. \cite{bonitz2016quantum,stefanucci2013nonequilibrium,stefanucci2024semiconductor}).

\subsection{From the Dyson equation to the Kadanoff-Baym equation}

In the quantum description of lattice vibrations, phonon Green’s functions play a central role, capturing the dynamics and interactions of quantized vibrational modes in solids. They encode comprehensive information about phonon propagation, scattering, and correlations, forming the foundation of modern theoretical approaches to quantum thermal transport. Within this framework, the Dyson equation provides a self-consistent relation that links the interacting phonon Green’s function to its non-interacting counterpart through a self-energy term. This term can account for all many-body interaction effects, including phonon–phonon scattering arising from anharmonicity, which constitutes the primary focus of the present review.\\
Historically, the Dyson equation emerged in the mid-20th century as part of the development of quantum field theory, introduced by Freeman Dyson in 1949 to provide a systematic perturbative framework for quantum electrodynamics. Dyson’s formulation revolutionized the understanding of interacting quantum systems by introducing a diagrammatic representation, now known as Feynman diagrams, which allows the infinite series of perturbative corrections to be organized in a compact and intuitive manner. \\
For phonons, the Dyson equation arises naturally from the perturbative expansion of the phonon Green's function in terms of the anharmonic potential $V$ in Eq. \eqref{start_hamiltonian}:
\begin{equation}
G = G^0 + G^0 V G^0 + G^0 V G^0 V G^0 + \dots
\end{equation}
Rather than summing all terms explicitly, we introduce the phonon self-energy $\Sigma_{\nu_1 \nu_2}(t_1, t_2)$, which encompasses all one-phonon-irreducible (1PI) diagrams. These are diagrams connecting $t_1$ and $t_2$ that cannot be separated by cutting a single phonon line:
\begin{equation}
\Sigma_{\nu_1 \nu_2}(t_1, t_2) = \sum_{n=1}^{\infty} i^n 
\int dt_3 \dots dt_{n+2} \, 
\langle \mathcal{T} \, V(t_1) \dots V(t_n) \, X_{\nu_1}(t_1) X_{\nu_2}(t_2) \rangle_{\text{1PI, connected}},
\end{equation}
where “1PI” denotes one-phonon-irreducible and “connected” indicates that disconnected diagrams are excluded. Intuitively, $V$ represents the bare interaction from the Hamiltonian, while $\Sigma$ is the effective interaction experienced by a phonon due to all interaction processes. In this way, the full phonon Green's function satisfies the Dyson equation (where we explicitly show the integration over intermediate times and phonon modes):
\begin{equation} \label{dyson_eq}
G_{\nu_1 \nu_2}(t_1, t_2) = G^0_{\nu_1 \nu_2}(t_1, t_2) + 
\int d\boldsymbol{q}' \int d\boldsymbol{q}'' \int dt' \int dt'' 
\, G^0_{\nu_1 \nu'}(t_1, t') \, \Sigma_{\nu' \nu''}(t', t'') \, G_{\nu'' \nu_2}(t'', t_2).
\end{equation}
In what follows, we restrict our attention to intraband phonon dynamics ($s_{1}=s_{2}$), thereby neglecting interband tunneling-like processes \cite{simoncelli2019unified,simoncelli2022wigner,di2023crossover,shin2024thermodynamics}. Extending the formalism to include such interband effects is, however, straightforward.  
When dealing with the Dyson equation, one encounters products involving multiple Green’s functions and self-energies. In this context, Langreth’s analytic continuation rules \cite{langreth1976linear,haug2008quantum} (see also page 71 of Ref. \cite{haug2008quantum}) are particularly useful. For a product of two quantities $C=AB$, these rules read:
\begin{equation} \label{Langreth_theorem_1}
C^{\lessgtr}_{\nu_{1}\nu_{2}}(t_{1},t_{2})=\int dt'\Big[A^{R}_{\nu_{1}\nu'}(t_{1},t')B^{\lessgtr}_{\nu'\nu_{2}}(t',t_{2})+A^{\lessgtr}_{\nu_{1}\nu'}(t_{1},t')B^{A}_{\nu'\nu_{2}}(t',t_{2})\Big].
\end{equation}
To simplify notation, it is convenient to treat products as operator multiplications over internal indices and time variables. Within this compact formalism, Eq. \eqref{Langreth_theorem_1} can be generalized to the case of three factors, $D = ABC$, yielding \cite{haug2008quantum}
\begin{equation} \label{Langreth_theorem_2}
D^{\lessgtr}=A^{R}B^{R}C^{\lessgtr}+A^{R}B^{\lessgtr}C^{A}+A^{\lessgtr}B^{A}C^{A}.
\end{equation}
Using this compact notation, the Dyson equation in Eq. \eqref{dyson_eq} takes the concise form
\begin{equation} \label{dyson_eq_compact}
G = G^{0} + G^{0}\Sigma G,
\end{equation}
which can equivalently be rewritten as
\begin{equation} \label{double_dyson_eq_for_KB}
\begin{split}
G^{0^{-1}}G = 1 + \Sigma G,\\
GG^{0^{-1}} = 1 + G\Sigma.
\end{split}
\end{equation}
Applying Langreth’s theorem \eqref{Langreth_theorem_1} to these equations gives
\begin{equation}
\begin{split}
G^{0^{-1}}G^{<} = \Sigma^{R}G^{<} + \Sigma^{<}G^{A},\\
G^{<}G^{0^{-1}} = G^{R}\Sigma^{<} + G^{<}\Sigma^{A}.
\end{split}
\end{equation}
Subtracting the two relations above, we obtain
\begin{equation} \label{differential_KB_equation}
\Big[G^{0^{-1}}\,,\,G^{<}\Big]=\Sigma^{R}G^{<}+\Sigma^{<}G^{A}-G^{R}\Sigma^{<}-G^{<}\Sigma^{A},
\end{equation}
where $[A,B]=AB-BA$ denotes a commutator. This expression corresponds to the differential form of the KBE \cite{kadanoff2018quantum}. The overall structure of Eq. \eqref{differential_KB_equation} closely parallels that of the nonequilibrium BTE, once the lesser Green’s function $G^{<}$ is mapped into a generalized quasiparticle distribution function. The commutator on the left-hand side represents a driving term, while the right-hand side yields a quantum collision term, with positive (scattering-in) and negative (scattering-out) contributions corresponding to phonon repopulation and depopulation processes, respectively. The following relations from the NEGF formalism are also useful \cite{kadanoff2018quantum,ryndyk2016theory}:
\\
\begingroup
\setlength{\abovedisplayskip}{1.75pt}
\setlength{\belowdisplayskip}{1.75pt}
\begin{equation} \label{spectral_function_def}
\mathrm{A}=-2\,\text{Im}\{G^{R}\}=i\,(G^{R}-G^{A})=i\,(G^{>}-G^{<}),
\end{equation}
\begin{equation} \label{gamma_def_NEGF}
\Gamma=-2\,\text{Im}\{\Sigma^{R}\}=i\,(\Sigma^{R}-\Sigma^{A})=i\,(\Sigma^{>}-\Sigma^{<}),
\end{equation}
\begin{equation} \label{G_from_GR_and_GA}
G=\tfrac{1}{2}(G^{R}+G^{A}),
\end{equation}
\begin{equation} \label{Resigma_as_sum_sigma_R_A}
\text{Re}\{\Sigma_{\nu}(\omega)\}=\tfrac{1}{2}\big(\Sigma_{\nu}^{R}(\omega)+\Sigma_{\nu}^{A}(\omega)\big),
\end{equation}
\endgroup
\\
where $\mathrm{A}$ is the spectral function and $\Gamma$ represents the phonon linewidth, i.e., the inverse of twice the phonon lifetime.  
To symmetrize the retarded and advanced components, we make use of the following decompositions:
\begin{equation}
\begin{split}
&f^{R}=\frac{1}{2}\left(f^{R}+f^{A}\right)+\frac{1}{2}\left(f^{R}-f^{A}\right),\\
&f^{A}=\frac{1}{2}\left(f^{A}+f^{R}\right)+\frac{1}{2}\left(f^{A}-f^{R}\right).
\end{split}
\end{equation}
Substituting these relations into Eq. \eqref{differential_KB_equation}, we obtain
\begin{equation}
\begin{split}
\Big[G^{0^{-1}}\,,\,G^{<}\Big]=&\,\frac{1}{2}\left(\Sigma^{R}+\Sigma^{A}\right)G^{<}+\frac{1}{2}\left(\Sigma^{R}-\Sigma^{A}\right)G^{<}+\frac{1}{2}\Sigma^{<}\left(G^{A}+G^{R}\right)+\\
&+\frac{1}{2}\Sigma^{<}\left(G^{A}-G^{R}\right)-\frac{1}{2}\left(G^{R}+G^{A}\right)\Sigma^{<}-\frac{1}{2}\left(G^{R}-G^{A}\right)\Sigma^{<}-\\
&-\frac{1}{2}G^{<}\left(\Sigma^{A}+\Sigma^{R}\right)-\frac{1}{2}G^{<}\left(\Sigma^{A}-\Sigma^{R}\right).
\end{split}
\end{equation}
By factorizing terms and employing Eqs. \eqref{G_from_GR_and_GA} and \eqref{Resigma_as_sum_sigma_R_A}, one obtains
\begin{equation}
\begin{split}
\Big[G^{0^{-1}}\,,\,G^{<}\Big]=\Big[\text{Re}\{\Sigma\},G^{<}\Big]+\Big[\Sigma^{<},G\Big]+\frac{1}{2}\Big\{\left(\Sigma^{R}-\Sigma^{A}\right),G^{<}\Big\}-\frac{1}{2}\Big\{\left(G^{R}-G^{A}\right),\Sigma^{<}\Big\},
\end{split}
\end{equation}
where $\lbrace A,B\rbrace = AB + BA$ denotes the anticommutator. Making use of Eqs. \eqref{spectral_function_def} and \eqref{gamma_def_NEGF}, this leads to
\begin{equation} \label{KB_equation}
\big[G^{0^{-1}}-\text{Re}\{\Sigma\},G^{<}\big]-\big[\Sigma^{<},G\big]=\frac{1}{2i}\big\lbrace\Gamma,G^{<}\big\rbrace-\frac{1}{2i}\big\lbrace \mathrm{A},\Sigma^{<}\big\rbrace=\frac{1}{2}\big\lbrace\Sigma^{>},G^{<}\big\rbrace-\frac{1}{2}\big\lbrace G^{>},\Sigma^{<}\big\rbrace.
\end{equation}
In this form of the KBE, the linewidths $\Gamma$ represent scattering-out processes, the term $\lbrace \mathrm{A},\Sigma^{<}\rbrace$ accounts for scattering-in contributions, while $G^{0^{-1}}$ and $-\text{Re}\{\Sigma\}$ describe renormalization of the quasiparticle energy levels, associated with phonon frequency shifts \cite{maradudin1962scattering}. This formulation establishes an explicit connection with the structure of the BTE, while extending it by incorporating full spectral resolution through the spectral function $\mathrm{A}$. This makes it possible to go beyond the conventional Fermi’s golden rule (FGR) picture of phonon scattering, which is also considered here, where each scattering event is strictly energy-conserving (see also later in the text). In Ref. \cite{di2025broadening}, this condition was relaxed through a physically derived, anharmonic, self-consistent, and phonon-resolved treatment of collisional broadening. Furthermore, Eq. \eqref{KB_equation} contains an additional term, $[\Sigma^{<},G]$, which lacks a direct semiclassical analogue. This contribution arises from out-of-pole propagation \cite{vspivcka1994quasiparticle,lipavsky2001kinetic,di2025theory}—transport processes mediated by satellite features of the spectral function \cite{lipavsky2001kinetic}. In the weak-interaction regime (small scattering rates, as assumed in Landau’s quasiparticle picture \cite{landau1987statistical_part2}), this term does not influence the driving term of the kinetic equation \cite{vspivcka1995quasiparticle}. Nevertheless, it lies outside the scope of the conventional BTE, which only describes transport around the phonon frequency pole. A detailed and rigorous discussion of this out-of-pole term and its impact on phonon transport is provided in Ref. \cite{di2025theory}. \\
Neglecting the out-of-pole term for simplicity, the KBE can be rewritten as \cite{reggiani1987quantum}:
\begin{equation} \label{differential_KB_equation_final}
2\left[G^{0^{-1}},G^{<}\right]=\Sigma^{>}G^{<}+G^{<}\Sigma^{>}-\Sigma^{<}G^{>}-G^{>}\Sigma^{<},
\end{equation}
which, in compact notation, becomes
\begin{equation} \label{KBE_skeleton}
{\rm DT}=2\left[G^{0^{-1}},G^{<}\right]=\left\lbrace\Sigma^{>},G^{<}\right\rbrace-\left\lbrace\Sigma^{<},G^{>}\right\rbrace={\rm CT},
\end{equation}
where “DT” and “CT” denote, respectively, the driving and collision terms. As evident from Eqs. \eqref{differential_KB_equation_final} and \eqref{KBE_skeleton}, the present review neglects frequency-shift effects and focuses exclusively on phonon linewidths and collisional broadening.

\subsection*{Takeaways}

In this section we have introduced the fully quantum description of phonon transport based on the nonequilibrium Green’s function formalism. Starting from the lattice Hamiltonian with anharmonic interactions, we defined phonon Green’s functions and derived the Dyson equation, which encodes all interaction effects through the phonon self-energy. By applying Langreth’s rules, we obtained the Kadanoff–Baym equation, which governs the quantum kinetics of phonons and provides a formally exact description of driving, scattering, and renormalization processes. We showed that the structure of the Kadanoff–Baym equation closely parallels that of the Boltzmann transport equation, while extending it by retaining full spectral information and allowing for quantum corrections to the semiclassical regime, such as collisional broadening and out-of-pole propagation effects. This establishes the quantum-mechanical foundation upon which semiclassical transport theories are built, clarifying both the regime of validity and the limitations of the Boltzmann picture discussed in the following sections.

\section{Semiclassical regime: phonon Boltzmann transport equation}

The first quantitative theory describing thermal conduction in insulating crystals in terms of phonon dynamics was introduced by Peierls in 1929 \cite{peierls1955quantum,peierls1996quantum,ziman2001electrons}, who formulated a microscopic model based on the BTE. His construction of the BTE, however, lacked a rigorous derivation from quantum thermal transport; instead, it combined quantum lattice dynamics with classical kinetic theory, treating phonons as gas-like quasiparticles whose scattering processes were introduced phenomenologically rather than obtained from a microscopic phonon self-energy.\\
In this section, we derive the BTE starting from the KBE that can be solved numerically. The procedure begins with the Dyson equation for the Green’s function, from which the quantum kinetic equation, namely the KBE, is derived. As will become clear in the following discussion, the KBE is written in a fully notation-independent form, meaning that it remains valid for any type of Green’s function and for any choice of phase-space variables, such as $(\boldsymbol{q},t)$, $(\boldsymbol{R},t)$, or even $(\boldsymbol{q},\omega,\boldsymbol{R},t)$ in the framework of Wigner’s mixed representation \cite{moyal1949quantum,hillery1984distribution,imre1967wigner}.\\
The next step consists in rewriting the KBE within the Wigner representation, beginning with the analysis of the driving term. In this representation, the driving term takes a particularly compact form, since it naturally appears as the commutator between the Green’s function and its noninteracting counterpart. The collision term, by contrast, involves commutators between products of the Green’s function and the phonon self-energy $\Sigma$. Hence, our objective is to express $\Sigma$ as a functional of the Green’s function, retaining the lowest-order diagrams in perturbation theory, specifically within the Wigner framework.\\
Subsequently, we perform a formal mapping of the KBE onto the semiclassical transport equation presented earlier. This correspondence is achieved through the generalized Kadanoff–Baym (GKB) ansatz \cite{haug2008quantum,reggiani1987quantum,vspivcka1995quasiparticle,lipavsky1986generalized,stefanucci2023and,stefanucci2024semiconductor}, followed by the application of the gradient expansion, analogous to the generalized Moyal product formalism \cite{moyal1949quantum,simoncelli2021thermal}, to both the driving and collision terms. This approximation simplifies the commutators in the driving term, introducing gradient corrections that give rise to the time and spatial derivative terms of the BTE. Similarly, it simplifies the anticommutators in the collision term, reducing them to direct products between $\Sigma$ and $G$. Importantly, the collision integral appearing in the generalized BTE obtained from the KBE is intrinsically nonlinear, providing a natural foundation for extending the formalism beyond the linear-response regime \cite{yamada1968nonlinear}.\\
By means of the GKB ansatz, we establish a coherent framework that connects the KBE with the semiclassical BTE. This formulation explicitly exposes the approximations involved, elucidates the treatment of energy conservation—here maintained within the FGR picture, where each phonon scattering event conserves energy exactly through the spectral function and the phonon scattering matrix—and highlights possible extensions that can be developed within this unified theoretical scheme.

\subsection{Gradient approximation} \label{gradient_approximation_main}

A rigorous approach to derive the BTE from the KBE is to formulate the theory within the Wigner mixed representation \cite{reggiani1987quantum,vspivcka1995quasiparticle}. This formalism conveniently allows the phonon self-energy functional to be expressed while consistently retaining both spatial and temporal dependencies. \\
In this framework, we outline how the BTE can be obtained from the KBE by means of a gradient approximation, which is valid under the assumption that variations in the center-of-mass coordinates occur much more slowly than those in the relative coordinates. Since the BTE inherently describes transport regimes characterized by smooth spatial and temporal variations, it becomes crucial to distinguish the “fast” microscopic quantum fluctuations from the “slow” macroscopic changes. This separation is achieved by introducing the Wigner coordinates and systematically expanding the theory in gradients, keeping only the leading-order contributions.\\
The Wigner mixed representation of the phonon Green’s function $G_{\nu_{1}\nu_{2}}(t_{1},t_{2})$ is defined as \cite{haug2008quantum}:
\begin{equation} \label{Wigner's_mixed_representation}
\tilde{G}_{\nu}^{\lessgtr}(\omega;\boldsymbol{R},t)=\iint d\tau \, d\boldsymbol{q}' \, e^{i\omega\tau+i\boldsymbol{q}'\cdot\boldsymbol{R}}G_{s}^{\lessgtr}\!\left(\boldsymbol{q}+\frac{\boldsymbol{q}'}{2},t+\frac{\tau}{2};\boldsymbol{q}-\frac{\boldsymbol{q}'}{2},t-\frac{\tau}{2}\right),
\end{equation}
where the integration over reciprocal space covers the entire Brillouin zone (BZ). In the near-equilibrium regime, boundary effects at the edges of the BZ can be neglected \cite{simoncelli2022wigner}. Equation \eqref{Wigner's_mixed_representation} is expressed in terms of the center-of-mass and relative variables introduced above:
\begin{equation} \label{center-of-mass_difference_variables}
\begin{split}
&\boldsymbol{q}'=\boldsymbol{q}_1-\boldsymbol{q}_2,\hspace{1cm}\tau=t_{1}-t_{2},\\
&\boldsymbol{q}=\frac{1}{2}(\boldsymbol{q}_1+\boldsymbol{q}_2),\hspace{1cm}t=\frac{1}{2}(t_{1}+t_{2}).
\end{split}
\end{equation}
For clarity of notation, the tilde symbol in the Wigner representation will be omitted from this point onward, as the dependence on the center-of-mass variables makes the representation self-evident. Importantly, $\boldsymbol{q}'$ and $\tau$ describe fast, microscopic variations and must be treated exactly (after Fourier transforms, $\boldsymbol{q}'\leftrightarrow\boldsymbol{R}$, $\tau\leftrightarrow\omega$), whereas $\boldsymbol{q}$ and $t$ represent slow, macroscopic quantities with weak gradients, and are thus treated approximately. By inverting the relations between the center-of-mass and relative coordinates, one obtains:
\begin{equation} \label{times_expressions}
\begin{split}
&\boldsymbol{q}_1=\boldsymbol{q}+\frac{\boldsymbol{q}'}{2},\hspace{1cm}t_{1}=t+\frac{\tau}{2},\\
&\boldsymbol{q}_2=\boldsymbol{q}-\frac{\boldsymbol{q}'}{2},\hspace{1cm}t_{2}=t-\frac{\tau}{2},
\end{split}
\end{equation}
and, from definition \eqref{green's_function}, we find:
\begin{equation} \label{Wigner's_mixed_representation_bis}
G^{<}_{\nu_{1}\nu_{2}}(t_{1},t_{2})=G_{s}^{<}\!\left(\boldsymbol{q}+\frac{\boldsymbol{q}'}{2},t+\frac{\tau}{2};\boldsymbol{q}-\frac{\boldsymbol{q}'}{2},t-\frac{\tau}{2}\right).
\end{equation}
Using this formalism (see Ref. \cite{di2025broadening} for a detailed derivation), one can define the Wigner distribution function $\tilde{n}_{s}(\boldsymbol{q},\boldsymbol{R},t)$ as:
\begin{equation} \label{wigner_distr_function}
\begin{split}
\tilde{n}_{s}(\boldsymbol{q},\boldsymbol{R},t)=\frac{1}{2\pi}\int d\boldsymbol{q}'\,e^{i\boldsymbol{q}'\cdot\boldsymbol{R}}\left\langle a_{s}^{\dagger}\!\left(\boldsymbol{q}-\frac{\boldsymbol{q}'}{2},t\right) a_{s}\!\left(\boldsymbol{q}+\frac{\boldsymbol{q}'}{2},t\right)\right\rangle,
\end{split}
\end{equation}
which plays the role of the quasiparticle distribution function in the extended transport equation.  
In this representation, matrix products take a nontrivial form \cite{haug2008quantum,lipavsky2001kinetic,vspivcka1995quasiparticle}, for instance:
\begin{equation} 
C_{\nu_{1}\nu_{2}}(t_{1},t_{2})=\int d\boldsymbol{q}'dt'\,A_{\nu_{1}\nu'}(t_{1},t')B_{\nu'\nu_{2}}(t',t_{2}).
\end{equation}
Expressed in terms of the variables $(\boldsymbol{q},\omega,\boldsymbol{R},t)$, this becomes \cite{haug2008quantum}:
\begin{equation} 
C_{s}(\boldsymbol{q},\omega,\boldsymbol{R},t)=A_{s}(\boldsymbol{q},\omega,\boldsymbol{R},t)\mathsf{G}(\boldsymbol{q},\omega,\boldsymbol{R},t)B_{s}(\boldsymbol{q},\omega,\boldsymbol{R},t),
\end{equation}
where the gradient operator $\mathsf{G}$ is defined as:
\begin{equation} \label{gradient_operator}
\mathsf{G}=\exp\!\left[\frac{1}{2i}\!\left(\frac{\partial^{A}}{\partial t}\frac{\partial^{B}}{\partial\omega}-\frac{\partial^{A}}{\partial\omega}\frac{\partial^{B}}{\partial t}-\frac{\partial^{A}}{\partial\boldsymbol{R}}\!\cdot\!\frac{\partial^{B}}{\partial\boldsymbol{q}}+\frac{\partial^{A}}{\partial\boldsymbol{q}}\!\cdot\!\frac{\partial^{B}}{\partial\boldsymbol{R}}\right)\!\right],
\end{equation}
where the superscripts $(A,B)$ indicate which function in the product is differentiated. Retaining the lowest nonvanishing order yields the gradient approximation, in which the commutators and anticommutators appearing in the KBE \eqref{KB_equation} reduce to \cite{haug2008quantum,vspivcka1995quasiparticle}: \\
\begingroup
\setlength{\abovedisplayskip}{1.75pt}
\setlength{\belowdisplayskip}{1.75pt}
\begin{equation} \label{gradient_expansion_1} 
\left\lbrace A_{s},B_{s}\right\rbrace_{(\boldsymbol{q},\omega,\boldsymbol{R},t)}=2A_{s}(\boldsymbol{q},\omega,\boldsymbol{R},t)B_{s}(\boldsymbol{q},\omega,\boldsymbol{R},t),
\end{equation}
\begin{equation} \label{gradient_expansion_2}
\left[A_{s},B_{s}\right]_{(\boldsymbol{q},\omega,\boldsymbol{R},t)}=-i\!\left(\frac{\partial A}{\partial t}\frac{\partial B}{\partial\omega}-\frac{\partial A}{\partial\omega}\frac{\partial B}{\partial t}-\frac{\partial A}{\partial\boldsymbol{R}}\!\cdot\!\frac{\partial B}{\partial\boldsymbol{q}}+\frac{\partial A}{\partial\boldsymbol{q}}\!\cdot\!\frac{\partial B}{\partial\boldsymbol{R}}\right).
\end{equation}
\endgroup
\\
For compactness, dependencies are left implicit in the above commutators. A straightforward generalization to higher-order derivatives in Eq. \eqref{gradient_operator} highlights the flexibility of the KBE framework in extending transport theory beyond standard linear-response formulations.  
It is important to emphasize once more that the KBE is fundamentally notation-independent, meaning it is universally valid for any kind of Green’s function $G$ and for any phase-space variable set. This includes $(\boldsymbol{r},t)$, $(\boldsymbol{q},t)$, or the mixed representation $(\boldsymbol{q},\omega,\boldsymbol{R},t)$ as introduced in Wigner’s formalism. Within this framework, we can rewrite the KBE \eqref{KBE_skeleton} as:
\begin{equation} \label{KBE_to_use_for_derivation}
\begin{split}
&\text{DT}_{\nu}(\omega,\boldsymbol{R},t)=2\left[G^{0^{-1}}_{\nu}(\omega)\,,\,G^{<}_{\nu}(\omega,\boldsymbol{R},t)\right]=\\
=&\,4\Big[\Sigma_{\nu}^{>}(\omega,\boldsymbol{R},t)G_{\nu}^{<}(\omega,\boldsymbol{R},t)-\Sigma_{\nu}^{<}(\omega,\boldsymbol{R},t)G_{\nu}^{>}(\omega,\boldsymbol{R},t)\Big]=\text{CT}_{\nu}(\omega,\boldsymbol{R},t),
\end{split}
\end{equation}
where the gradient approximation \eqref{gradient_expansion_1} has already been applied to the collision term.

\subsection{Generalized Kadanoff-Baym ansatz} \label{GKB_ansatz_section}

The final step in connecting the KBE, expressed in terms of the Green’s function $G$, to the BTE, formulated in terms of the (phonon) Wigner distribution function $n$, requires defining an explicit correspondence between these two quantities. This relationship is established through what is known as the generalized Kadanoff–Baym (GKB) ansatz. Originally developed in the study of electronic transport \cite{lipavsky1986generalized}, the ansatz takes the form $g^{<}=iAf$ and $g^{>}=-iA[1-f]$, where $g$ and $f$ denote the nonequilibrium electronic Green’s and distribution functions, respectively. This formulation emerged in the early development of nonequilibrium Green’s function (NEGF) theory, particularly in the context of describing intracollisional field effects under strong electric fields \cite{haug2008quantum,reggiani1987quantum,jauho1982rigorous,ciancio2004gauge}.\\
The phonon analogue of this ansatz has been analyzed in several works \cite{niklasson1968theory,kwok1966unified,meier1969green,horie1964boltzmann,stefanucci2023and}. However, it is known that the BTE obtained through this route leads to an incomplete expression for the collision integral (see, for example, Eqs. 3.3$'$, 3.12, and 3.13 of Ref. \cite{kwok1966unified}), where only half of the scattering contributions appearing in the conventional BTE are recovered. \\
A general, non-Lorentzian spectral function of a phonon system can be written as:
\begin{equation} \label{spectral_function_d_bosonic_main}
\begin{split}
\mathrm{d}_{\nu}(\omega)=\frac{1}{\pi}\frac{\gamma_{\nu}(\omega)}{[\omega-\omega_{\nu}-\text{\scriptsize{$\Delta$}}_{\nu}(\omega)]^{2}+\gamma_{\nu}^{2}(\omega)}.
\end{split}
\end{equation}
Within the vibronic representation, the GKB ansatz reads:
\begin{equation} \label{GKB_ansatz_equation}
\begin{split}
G^{<}_{\nu}(\omega,\boldsymbol{R},t)&=-2\pi i\Big[n_{\nu}(\omega,\boldsymbol{R},t)\mathrm{d}_{\nu}(\omega)+\big(n_{\nu}(-\omega,\boldsymbol{R},-t)+1\big)\mathrm{d}_{\nu}(-\omega)\Big],\\
G^{>}_{\nu}(\omega,\boldsymbol{R},t)&=-2\pi i\Big[n_{\nu}(-\omega,\boldsymbol{R},-t)\mathrm{d}_{\nu}(-\omega)+\big(n_{\nu}(\omega,\boldsymbol{R},t)+1\big)\mathrm{d}_{\nu}(\omega)\Big].
\end{split}
\end{equation}
In the above expression, the frequency shift $\text{\scriptsize{$\Delta$}}_{\nu}(\omega) = \text{Re}\,\sigma_{\nu}(\omega)$ is not considered further in this work and will be neglected for simplicity. The ansatz itself is completely general: it replaces the unknown Green’s functions $g^{\lessgtr}$ with the function $n$, which serves as the phonon distribution, while still fulfilling the exact NEGF relation $\mathrm{a}=i(g^{>}-g^{<})$. In Ref. \cite{di2025broadening}, the validity of the GKB ansatz is demonstrated both at thermal equilibrium—where it can be physically interpreted through the fluctuation-dissipation theorem—and out of equilibrium, provided that the phonon distribution entering the ansatz is the Wigner distribution function introduced in Eq. \eqref{wigner_distr_function}. \\
In the so-called quasiparticle approximation, the spectral function is replaced by a Dirac delta distribution,
\begin{equation} \label{quasiparticle_approximation_main}
\mathrm{d}_{\nu}(\pm\omega)\rightarrow\delta(\omega\mp\omega_{\nu}),
\end{equation}
leading to the standard quasiparticle form of the GKB ansatz \cite{di2025broadening,di2025phonon,lipavsky1986generalized}:
\begin{equation} \label{quasiparticle_ansatz}
\begin{split}
G^{<}_{\nu}(\omega,\boldsymbol{R},t)&=-2\pi i\Big[n_{\nu}(\omega,\boldsymbol{R},t)\delta(\omega-\omega_{\nu})+\big(n_{\nu}(-\omega,\boldsymbol{R},-t)+1\big)\delta(\omega+\omega_{\nu})\Big],\\
G^{>}_{\nu}(\omega,\boldsymbol{R},t)&=-2\pi i\Big[n_{\nu}(-\omega,\boldsymbol{R},-t)\delta(\omega+\omega_{\nu})+\big(n_{\nu}(\omega,\boldsymbol{R},t)+1\big)\delta(\omega-\omega_{\nu})\Big].
\end{split}
\end{equation}
This approximation reflects the physical picture in which the system’s excitations behave as well-defined quasiparticles with negligible broadening and no significant frequency shifts. As a result, the spectral weight is entirely localized at the quasiparticle energy, yielding a delta-like form consistent with Fermi’s Golden Rule (FGR). The approximation holds when quasiparticle lifetimes are long enough that individual phonon scattering events conserve energy exactly. Under these conditions, frequency shifts can be safely ignored at the level of the conventional BTE, since the leading-order diagrams in the perturbative expansion of the self-energy—namely, the “loop” and “bubble” diagrams \cite{maradudin1962scattering,cowley1963lattice} (see the diagrams in the first row of Fig. \ref{fig:self_energy_diagrams})—contribute at the same order. The loop diagram affects only the real part of the self-energy (producing frequency shifts), while the bubble diagram contributes to both the real and imaginary parts (corresponding to frequency shifts and linewidths) \cite{maradudin1962scattering}. If the real part of the self-energy is neglected, the loop contribution can therefore be omitted entirely, leaving only the imaginary part responsible for the linewidths (and thus the phonon lifetimes) through the bubble diagram.\\
This reasoning underlies the standard BTE formulation: since the spectral function is approximated by a Dirac delta (excluding broadening), it is consistent to neglect line shifts within the same level of approximation. Even at the stage of the quasiparticle approximation \eqref{quasiparticle_approximation_main}, one can observe that higher-order renormalization effects are also neglected. These effects are characterized by the pole (or wavefunction) renormalization factor:
\begin{equation} \label{z_pole_renormalization}
z_{\nu}(\omega)=\frac{1}{1-\frac{\partial\Delta_{\nu}(\omega)}{\partial\omega}},
\end{equation}
which is directly linked to the real part of the self-energy. Hence, when the real part of the self-energy is excluded—meaning the energy renormalization effects are ignored—the use of either the quasiparticle form \eqref{quasiparticle_ansatz} or the general GKB ansatz \eqref{GKB_ansatz_equation} (possibly with a broadening description \cite{di2025broadening}) remains fully justified. \\
It is also important to note that an additional leading-order “tad-pole” diagram \cite{SSCHA_0,calandra2007anharmonic,maradudin1962scattering,paulatto2015first} should, in principle, be included to account for the modification of atomic equilibrium positions due to anharmonicity; at zero temperature, this is attributed to zero-point motion. Furthermore, this diagram is not relevant for a high-symmetry crystal because it only provides corrections at the Gamma point in the Brillouin zone \cite{lazzeri2003anharmonic}. The loop diagram depends on $\uppsi_{4}$ once, while the bubble depends on $\uppsi_{3}$ and $\Gamma_{3}$. If the vertex $\Gamma_{3}$ is not dressed, then the bubble depends on $\uppsi_{3}$ twice. Although the tad-pole also depends on $\uppsi_{3}$ twice, it is diagrammatically distinct from the bubble at zero-order in the vertex corrections. In the perturbative treatment above, we have assumed that the atoms oscillate around their harmonic positions, and under this assumption, the tad-pole is absent, as is the case in high-symmetry systems where the atomic positions are fixed by the symmetry group \cite{lazzeri2003anharmonic}.

\subsection{Driving term of the Boltzmann transport equation}

At this stage we are fully equipped to derive the extended phonon transport equation from Eq. \eqref{KBE_to_use_for_derivation}. We begin by applying the gradient approximation \eqref{gradient_expansion_2} to the driving term:
\begin{equation}
\begin{split}
\text{DT}_{\nu}(\omega,\boldsymbol{R},t)=&2\left[G^{0^{-1}}_{\nu}(\omega)\,,\,G^{<}_{\nu}(\omega,\boldsymbol{R},t)\right]=\\
=&-2i\frac{\partial G^{0^{-1}}_{\nu}(\omega)}{\partial t}\frac{\partial G^{<}_{\nu}(\omega,\boldsymbol{R},t)}{\partial\omega}+2i\frac{\partial G^{0^{-1}}_{\nu}(\omega)}{\partial\omega}\frac{\partial G^{<}_{\nu}(\omega,\boldsymbol{R},t)}{\partial t}+\\
&+2i\frac{\partial G^{0^{-1}}_{\nu}(\omega)}{\partial\boldsymbol{R}}\cdot\frac{\partial G^{<}_{\nu}(\omega,\boldsymbol{R},t)}{\partial\boldsymbol{q}}-2i\frac{\partial G^{0^{-1}}_{\nu}(\omega)}{\partial\boldsymbol{q}}\cdot\frac{\partial G^{<}_{\nu}(\omega,\boldsymbol{R},t)}{\partial\boldsymbol{R}}.
\end{split}
\end{equation}
Substituting the expression for $G_{\nu}^{0}(\omega)=\frac{2\omega_{\nu}}{\omega^{2}-\omega_{\nu}^{2}}$ (see i.e. Refs. \cite{mahan2000many,ryndyk2016theory}) and noting that $G^{0^{-1}}_{\nu}(\omega)$ is time- and position-independent (only $-\frac{\partial A}{\partial\omega}\frac{\partial B}{\partial t}$ and $\frac{\partial A}{\partial\boldsymbol{q}}\cdot\frac{\partial B}{\partial\boldsymbol{R}}$ derivatives contribute) we get
\begin{equation}
\text{DT}_{\nu}(\omega,\boldsymbol{R},t)=2i\frac{\omega}{\omega_{\nu}}\frac{\partial G^{<}_{\nu}(\omega,\boldsymbol{R},t)}{\partial t}+2i\frac{\boldsymbol{v}_{\nu}}{2}\left(\frac{\omega^{2}}{\omega_{\nu}^{2}}+1\right)\cdot\frac{\partial G^{<}_{\nu}(\omega,\boldsymbol{R},t)}{\partial\boldsymbol{R}},
\end{equation}
where $\boldsymbol{v}_{\nu}=\frac{\partial\omega_{\nu}}{\partial\boldsymbol{q}}$ is the phonon group velocity. Substituting the GKB ansatz \eqref{GKB_ansatz_equation} for the phonon Green's function we then get
\begin{equation} \label{DT_on_going}
\begin{split}
\text{DT}_{\nu}(\omega,\boldsymbol{R},t)=&\,\,4\pi\frac{\omega}{\omega_{\nu}}\mathrm{d}_{\nu}(\omega)\frac{\partial n_{\nu}(\omega,\boldsymbol{R},t)}{\partial t}+4\pi\frac{\omega}{\omega_{\nu}}\mathrm{d}_{\nu}(-\omega)\frac{\partial\big[n_{\nu}(-\omega,\boldsymbol{R},-t)+1\big]}{\partial t}+\\
&+4\pi\left(\frac{\omega^{2}}{\omega_{\nu}^{2}}+1\right)\mathrm{d}_{\nu}(\omega)\frac{\boldsymbol{v}_{\nu}}{2}\cdot\frac{\partial n_{\nu}(\omega,\boldsymbol{R},t)}{\partial\boldsymbol{R}}+\\
&+4\pi\left(\frac{\omega^{2}}{\omega_{\nu}^{2}}+1\right)\mathrm{d}_{\nu}(-\omega)\frac{\boldsymbol{v}_{\nu}}{2}\cdot\frac{\partial\big[n_{\nu}(-\omega,\boldsymbol{R},-t)+1\big]}{\partial\boldsymbol{R}}.
\end{split}
\end{equation}
The second time derivative in Eq. \eqref{DT_on_going} can be written as
\begin{equation}
\begin{split}
\frac{\partial\big[n_{\nu}(-\omega,\boldsymbol{R},-t)+1\big]}{\partial t}=-\frac{\partial n_{\nu}(-\omega,\boldsymbol{R},t)}{\partial t},
\end{split}
\end{equation}
while the second space derivative reads
\begin{equation} \label{time_reversal_invariance}
\frac{\partial\big[n_{\nu}(-\omega,\boldsymbol{R},-t)+1\big]}{\partial\boldsymbol{R}}=\frac{\partial n_{\nu}(-\omega,\boldsymbol{R},-t)}{\partial\boldsymbol{R}}=\frac{\partial n_{\nu}(-\omega,\boldsymbol{R},t)}{\partial\boldsymbol{R}};
\end{equation}
the latter equality stems from the time-reversal invariance of quantum kinetic equations in the NEGF formalism \cite{bonitz2016quantum}. Indeed, as the KBE can be directly derived from the equations of motion of the field operators in second quantization, which are time-reversal invariant, it can be shown that the KBE shares the same symmetry properties \cite{bonitz2016quantum,scharnke2017time}. So in the end the driving term reads
\begin{equation} \label{DT}
\begin{split}
&\text{DT}_{\nu}(\omega,\boldsymbol{R},t)=\\
=&\,\,4\pi\Bigg[\frac{\omega}{\omega_{\nu}}\mathrm{d}_{\nu}(\omega)\frac{\partial n_{\nu}(\omega,\boldsymbol{R},t)}{\partial t}-\frac{\omega}{\omega_{\nu}}\mathrm{d}_{\nu}(-\omega)\frac{\partial n_{\nu}(-\omega,\boldsymbol{R},t)}{\partial t}+\\
&\hspace{0.75cm}+\left(\frac{\omega^{2}}{\omega_{\nu}^{2}}+1\right)\mathrm{d}_{\nu}(\omega)\frac{\boldsymbol{v}_{\nu}}{2}\cdot\frac{\partial n_{\nu}(\omega,\boldsymbol{R},t)}{\partial\boldsymbol{R}}+\left(\frac{\omega^{2}}{\omega_{\nu}^{2}}+1\right)\mathrm{d}_{\nu}(-\omega)\frac{\boldsymbol{v}_{\nu}}{2}\cdot\frac{\partial n_{\nu}(-\omega,\boldsymbol{R},t)}{\partial\boldsymbol{R}}\Bigg].
\end{split}
\end{equation}
Finally, we can obtain the quasiparticle approximation \eqref{quasiparticle_approximation_main} of the driving term \eqref{DT} by simply integrating over all frequencies, term by term (so we bear in mind that the same operation is to be performed on the collision term,):
\begin{equation} \label{DT_qp}
\begin{split}
\text{DT}_{\nu}&=8\pi\Bigg[\frac{\partial n_{\nu}(\boldsymbol{R},t)}{\partial t}
+\boldsymbol{v}_{\nu}\frac{\partial n_{\nu}(\boldsymbol{R},t)}{\partial\boldsymbol{R}}\Bigg].
\end{split}
\end{equation}
By having the Dirac delta functions act on the driving term, the general dependence on $\omega$ is reduced to a dependence on the wave vector $\boldsymbol{q}$ and band index $s$ through phonon frequencies $\omega_{\nu}$. Additionally, the phonon distribution $n$ in the driving term corresponds directly to the Wigner distribution function $\tilde{n}$. This demonstrates clearly that, when applying the quasiparticle ansatz \eqref{quasiparticle_ansatz} within the gradient approximation scheme, the driving term of the KBE reduces to that of the BTE.

\subsection{Collision term of the Boltzmann transport equation}

The collision operator involves anticommutators containing products of the Green’s function $G$ and the self-energy $\Sigma$ (see the right-hand side of Eq. \eqref{KBE_skeleton}). Evaluating it requires an additional step, namely expressing $\Sigma$ as a functional of $G$. Since we employ Wigner’s mixed representation to establish a consistent framework that simultaneously accounts for both the driving and scattering dynamics, we need to recast the self-energy functional in this representation \cite{di2025broadening}.\\
In the present review, the self-energy is considered up to third order in the perturbative Feynman expansion.
\begin{figure}[!htb]
\centering
\includegraphics[width=0.75\textwidth]{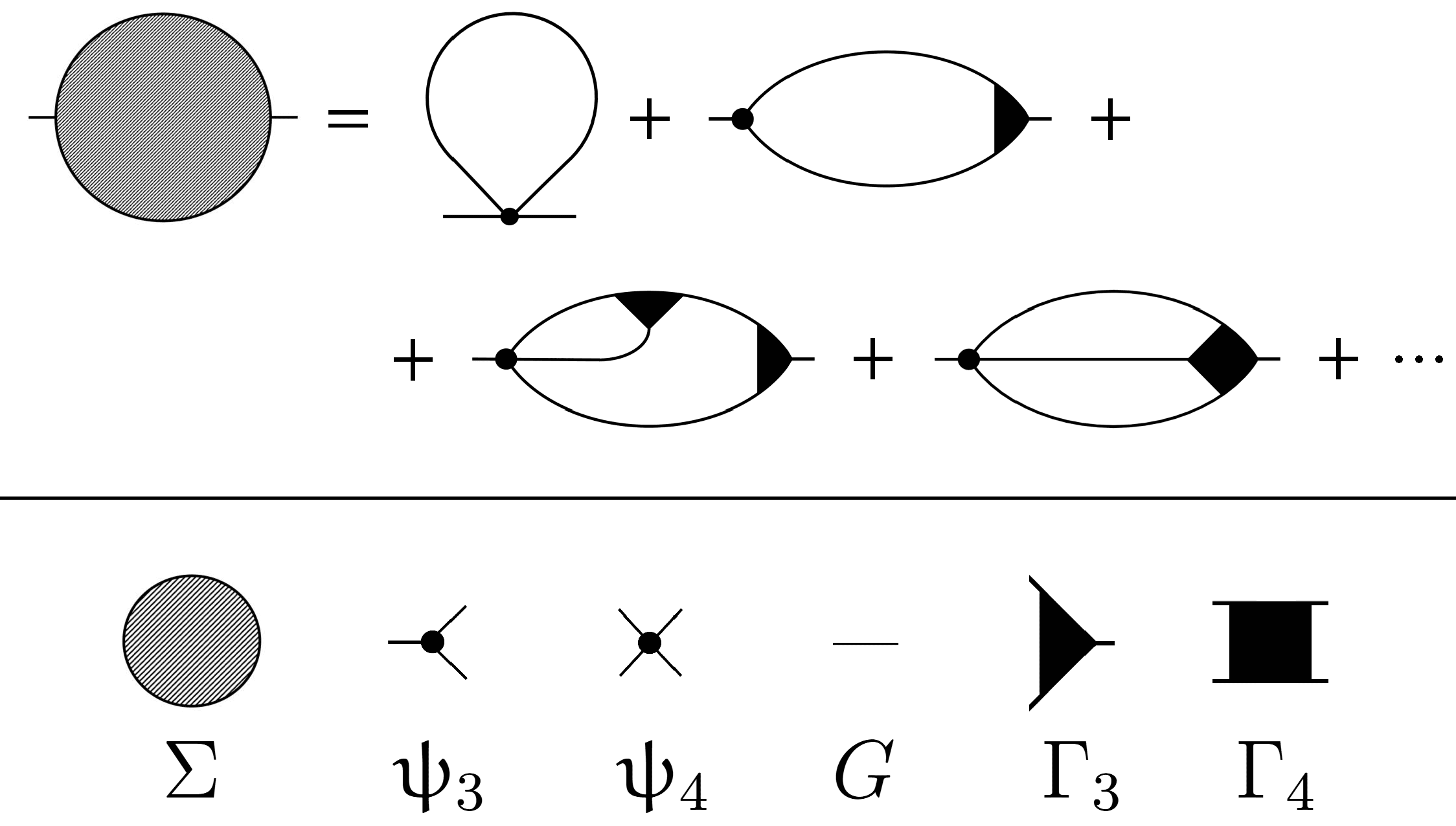}
\caption{\textbf{Diagrammatic representation of the leading contributions to the phonon self-energy expressed through Feynman diagrams.} The two lowest-order terms proportional to the two-point correlation function $G$ correspond to the standard “loop” ($\uppsi_{4}G$) and “bubble” ($\uppsi_{3}G\Gamma_{3}G$) diagrams. The meaning of the symbols and line styles are also clarified: the symbol $\Gamma$ denotes interaction vertices that should, in principle, be renormalized and dressed by higher-order processes (see Fig. \ref{fig:vertex_diagrams} and Eq. \eqref{gamma_3_integral_eq}), whereas the symbol $\psi$ represents the bare interaction. See also Fig. \ref{fig:self_energy_diagrams_appendix2} in appendix \ref{self-energy_section} and Eqs. \eqref{G_is_f_2_eq1} and \eqref{G_is_f_2_eq2}.}
\label{fig:self_energy_diagrams}
\end{figure}
This can be rigorously derived using the functional method of Martin and Schwinger \cite{wehner1967phonon,wehner1966infra,martin1959theory}, which was also employed by Kadanoff \cite{kadanoff2018quantum}. A full derivation of the phonon self-energy within this framework is provided in appendix \ref{self-energy_section}. \\
Moreover, as discussed previously, we disregard the frequency renormalization stemming from the real part of $\Sigma$, thereby reducing the self-energy to its bubble-diagram form (i.e., retaining only the second diagram in the first row of Fig. \ref{fig:self_energy_diagrams}):
\begin{equation} \label{sigma_eq}
\begin{split}
\Sigma_{\nu}(\omega)=\frac{2\pi}{\hbar^{2}\mathcal{V}}\sum_{\nu'\nu''}\delta_{\boldsymbol{q}-\boldsymbol{q}'-\boldsymbol{q}'',\boldsymbol{K}}\uppsi_{\nu-\nu'-\nu''}\int&\frac{d\omega_{1}}{2\pi}\int\frac{d\omega_{2}}{2\pi}(2i\hbar)2\pi\delta(\omega-\omega_{1}-\omega_{2})\\
&G_{\nu'}(\omega_{1})G_{\nu''}(\omega_{2})\Gamma_{3,\nu\nu'\nu''}(\omega,\omega_{1},\omega_{2}),
\end{split}
\end{equation}
where the third-order vertex function in the bubble diagram, $\Gamma_{3}$, is diagrammatically defined to be beyond the simple bare three phonon interaction $\Gamma_{3}^{0}\propto\uppsi_{3}$ \cite{klein1969derivation,wehner1967phonon} (see appendix \ref{vertex_corrections_section}). When dealing with vertex corrections, we are mainly interested in the contributions to the self-energy which correspond to so called ladder diagrams \cite{sham1967equilibrium,sham1967temperature,kadanoff2018quantum,danielewicz1984quantum1,danielewicz1984quantum2,botermans1990quantum}. The self-consistent equation for the vertex function $\Gamma_{3}$ is given by \cite{klein1969derivation} (see also Eqs. \eqref{final_Gamma_3} and \eqref{gamma_3_integral_eq_appendix} in appendix \ref{vertex_corrections_section})
\begin{equation} \label{gamma_3_integral_eq}
\begin{split}
\Gamma_{3}(1,2,3)=&\,\,\Gamma_{3}^{0}(1,2,3)-i\,12\,\uppsi_{4}(1,2,4,5)G(4,4')G(5,5')\Gamma_{3}(4',5',4)+\\
&+\Gamma_{3}^{0}(1,4,5)G(4,4')G(5,5')G(6,6')\Gamma_{3}(4',6,2)\Gamma_{3}(5',6',3)+\cdots,
\end{split}
\end{equation}
which is an integral equation requiring integration over all internal variables, frequencies, and wave vectors. The equation is self-consistent because the fully dressed vertex $\Gamma_{3}$ appears on both sides. An analogous Bethe--Salpeter–type equation can be derived for the fourth-order vertex $\Gamma_{4}$ \cite{wehner1967phonon,klein1969derivation,klein1968linear}. In the compact notation used here, these integrations are encoded by numerical labels for brevity. The diagram form of this equation is shown in Fig. \ref{fig:vertex_diagrams}. Note that the relationship between the equations governing the vertex parts or ladder diagrams and transport equations has been extensively explained by Abrikosov, Gorkov, and Dzyaloshinskii in their famous book \cite{abrikosov1965quamtum}.
\begin{figure}[!htb]
\centering
\includegraphics[width=0.75\textwidth]{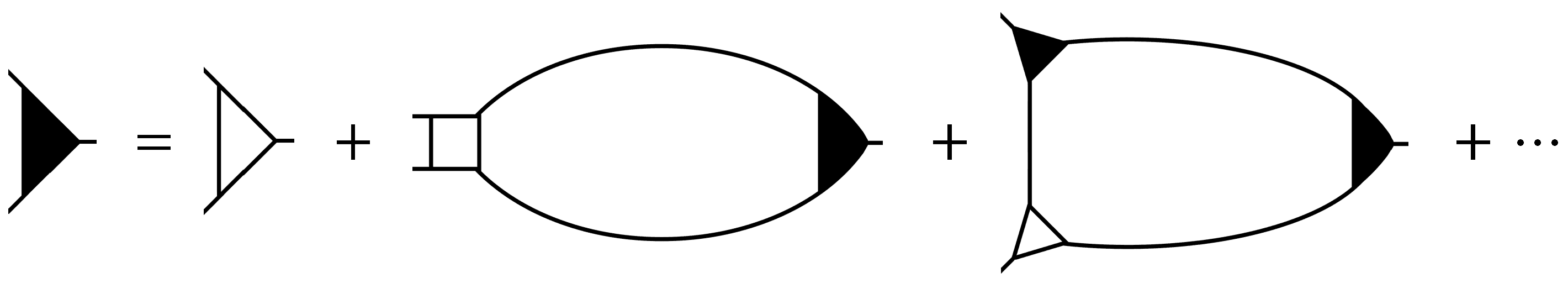}
\caption{\textbf{Diagrammatic illustration of the nonlinear integral (Bethe-Salpeter–type) equation \eqref{gamma_3_integral_eq}) for the vertex function.} The diagram emphasizes the ladder corrections that renormalize, or “dress” the vertex $\Gamma_{3}$ appearing in the bubble contribution of Fig. \ref{fig:self_energy_diagrams}. The meaning of all symbols and lines is identical to that used in Fig. \ref{fig:self_energy_diagrams}.}
\label{fig:vertex_diagrams}
\end{figure}
It can be seen that the iteration of the third term on the RHS of Eq. \eqref{gamma_3_integral_eq} (or Fig. \ref{fig:vertex_diagrams}) leads to the ladder diagrams where the vertex corrections providing the second and third term on the RHS contain four- and three-phonon processes respectively. The ladder diagrams or vertex corrections can be interesting for their impact on the self-energy, affecting various crystal properties governed by the one-phonon Green function, such as sound absorption \cite{klein1967ultrasonic}, infrared radiation absorption \cite{wehner1966infra}, and also scattering properties \cite{cochran1967phonons}. Secondly, they can play a crucial role in deriving a BTE-like equation (in the sense of a partial differential equation that mathematically describes the time evolution of the distribution function) for the vertex function $\Gamma_{3}$ \cite{sham1967equilibrium}. In fact, as discussed by Sham \cite{sham1967equilibrium,sham1967temperature}, it can be shown that starting from the non-linear integral equation \eqref{gamma_3_integral_eq} it is possible to obtain a Bethe-Salpeter equation for the vertex function $\Gamma_{3}$. Taking the latter as the starting point, one can derive an equation with the same structure of the BTE, where the vertex function takes the place of the phonon distribution and the term on the RHS has the structure of the Peierls' collision operator \cite{klein1969derivation}. Finally, Eq. \eqref{gamma_3_integral_eq} can be further simplified for some specific cases. Specifically, vertex corrections (i.e. ladder diagrams) become important whenever the external phonon frequency is on the same scale as the linewidth of the thermal phonons it couples to, as first discussed by Klein and Sham \cite{klein1969derivation,sham1967equilibrium}. This situation arises in the hydrodynamic regime, namely for long-wavelength acoustic phonons entering and leaving the self-energy diagram, where the associated products of propagators acquire almost singular “pinching” configurations. By contrast, the phonons inside the bubble diagram remain thermal excitations with characteristic energies on the order of $k_{\mathrm{B}}T$. Therefore, the relevant kinematics correspond to the case in which one of the three phonons in $\Gamma_{3}(1,2,3)$ has a frequency much smaller than $k_{\mathrm{B}}T$, while the other two are thermal phonons whose wave vectors have nearly the same magnitude and opposite directions. Taking into account this separation of scales, together with the distinct momentum dependence of the different $\Gamma_{3}$ contributions in Eq. \eqref{gamma_3_integral_eq}, it is sufficient to treat self-consistently only the vertex associated with the external soft phonon. In contrast, the second $\Gamma_{3}$ appearing on the RHS of Eq. \eqref{gamma_3_integral_eq}, which involves only thermal phonons, may be approximated by its lowest-order expression, $\Gamma_{3}\equiv\Gamma_{3}^{0}\propto\uppsi_{3}$. As a result, the equation for the vertex $\Gamma_3$ reduces to an effectively linearized form \cite{klein1968linear}:
\begin{equation} \label{gamma_3_integral_eq_bis}
\begin{split}
\Gamma_{3}(1,2,3)=\Gamma_{3}^{0}(1,2,3)-\Big[&i12\uppsi_{4}(1,2,5,6)\\
&-\Gamma_{3}^{0}(1,4,5)\Gamma_{3}^{0}(2,4',6)G(4,4')\Big]G(5,5')G(6,6')\Gamma_{3}(5',6',3).
\end{split}
\end{equation}
This allows to take into account vertex corrections explicitly since, when iterated, it simply leads to the sum of a geometric series \cite{di2025vertex}:
\begin{equation} 
\Gamma_{3}^{{\rm{N}}}\simeq\Gamma_{3}^{0}+\frac{\Gamma_{3}^{0}}{1+\big(i12\uppsi_{4}-\Gamma_{3}^{0}\Gamma_{3}^{0}G\big)GG}.
\end{equation}
However, the treatment of vertex corrections on top of the bare three-phonon interaction is beyond the scope of this review.\\
In summary, here we use the bubble approximation with the 0th-order approximation of the vertex function $\Gamma_{3}$, which translates into considering Eq. \eqref{sigma_eq} (bubble diagram in Fig. \ref{fig:self_energy_diagrams}) and only the first term in Eq. \eqref{gamma_3_integral_eq} (empty triangular vertex in Fig. \ref{fig:vertex_diagrams}). In this way we are consistent in neglecting the second term (proportional to $\uppsi_{4}$, loop diagram), because we are neglecting $\uppsi_{4}$ terms also in the vertex function $\Gamma_{3}$. \\
The phonon self-energy could be better approximated by following two different strategies. The first is to retain higher orders in the expansion given Fig. \ref{fig:self_energy_diagrams}; this would mean obtaining vertices of 4th order and higher. The second instead refers to a better approximation of the vertex part itself, as we already mentioned, i.e., not limiting the derivation to $\Gamma_{3}\equiv\Gamma_{3}^{0}$, but including the additional terms in its definition of the functional derivative. \\
By setting $\Gamma_{3}\equiv\Gamma_{3}^{0}$ we get the bubble self-energy in the form usually considered for the calculation of various anharmonic effects \cite{xu2008nonequilibrium,volz2020quantum,maradudin1962scattering,cowley1966anharmonic,semwal1972thermal}:
\begin{equation} \label{bubble_self_energy_final_main}
\Sigma^{\lessgtr}_{\nu}(\omega)=\sum_{\nu'\nu''}\int\frac{d\omega_{1}}{2\pi}\int\frac{d\omega_{2}}{2\pi}(2i\hbar)2\pi\delta(\omega-\omega_{1}-\omega_{2})|\mathcal{F}_{\nu-\nu'-\nu''}|^{2}G^{\lessgtr}_{\nu'}(\omega_{1})G^{\lessgtr}_{\nu''}(\omega_{2}),
\end{equation}
where
\begin{equation}
\mathcal{F}_{\nu-\nu'-\nu''}=\delta_{\boldsymbol{q}-\boldsymbol{q}'-\boldsymbol{q}'',\boldsymbol{K}}|\uppsi_{\nu-\nu'-\nu''}|,
\end{equation}
with $\boldsymbol{K}$ a reciprocal lattice vector. The Dirac delta $\delta(\omega-\omega_{1}-\omega_{2})$ arises because the time dependence only enters through the time difference $t_1-t_2$.\\
Within Wigner’s mixed representation, the bubble self-energy functional of the Green’s function reads \cite{di2025broadening} (see appendix \ref{self_energy_wigner} for a rigorous derivation):
\begin{equation} \label{final_selfenergy_wigner_main} 
\tilde{\Sigma}_{\nu}^{\lessgtr}(\omega;\boldsymbol{R},t)=4i\pi\hbar\sum_{\nu'\nu''}|\mathcal{F}_{\nu-\nu'-\nu''}|^{2}\int\frac{d\omega_{1}}{2\pi}\int\frac{d\omega_{2}}{2\pi}\delta(\omega-\omega_{1}-\omega_{2})\tilde{G}_{\nu'}^{\lessgtr}(\omega_{1};\boldsymbol{R},t)\tilde{G}_{\nu''}^{\lessgtr}(\omega_{2};\boldsymbol{R},t).
\end{equation}
Applying the quasiparticle ansatz \eqref{quasiparticle_ansatz} to the Green’s functions in both Eq. \eqref{KBE_to_use_for_derivation} and the self-energy \eqref{final_selfenergy_wigner_main} yields the standard BTE collision term \cite{di2025broadening}:
\begin{equation} \label{final_BTE_scattering_qp_non_homo_main}
\resizebox{\textwidth}{!}{$
\begin{split}
{\rm{CT}}_{\nu}=8\pi\hbar\sum_{\nu'\nu''}\Bigg\lbrace&\frac{1}{2}|\mathcal{F}_{\nu-\nu'-\nu''}|^{2}\Bigg[\big(n_{\nu}+1\big)n_{\nu'}n_{\nu''}-n_{\nu}\big(n_{\nu'}+1\big)\big(n_{\nu''}+1\big)\Bigg]\delta(\omega_{\nu}-\omega_{\nu'}-\omega_{\nu''})+\\
&+|\mathcal{F}_{\nu\nu'-\nu''}|^{2}\Bigg[\big(n_{\nu}+1\big)\big(n_{\nu'}+1\big)n_{\nu''}-n_{\nu}n_{\nu'}\big(n_{\nu''}+1\big)\Bigg]\delta(\omega_{\nu}+\omega_{\nu'}-\omega_{\nu''})\Bigg\rbrace,
\end{split}$}
\end{equation}
where the three-phonon processes, including both decays and coalescences (with the $1/2$ factor avoiding double counting), are distinguished and multiplied by Dirac delta functions, ensuring exact energy conservation according to FGR. This expression is obtained after integrating over all frequencies $\omega$, consistent with the treatment of the driving term in Eq. \eqref{DT_qp}, which selects $\omega=\omega_\nu$ within the collision operator. Eq. \eqref{final_BTE_scattering_qp_non_homo_main} represents a key result of this work. While the electron case has been extensively treated in the literature \cite{lipavsky1986generalized,ponce2020first}, previous phonon derivations were partial due to ambiguities in the choice of the ansatz in a vibronic formulation \cite{kwok1966unified,niklasson1968theory}. Importantly, here ${\rm{CT}}_{\nu}\equiv{\rm{CT}}_{\nu}(\boldsymbol{R},t)$ and $n_{\nu}\equiv n_{\nu}(\boldsymbol{R},t)$, as follows directly from the protocol we adopted: both the phonon self-energy in Eq. \eqref{final_selfenergy_wigner_main} and the GKB ansatz in Eq. \eqref{quasiparticle_ansatz} explicitly retain space and time dependence. This is a consequence of employing Wigner's mixed representation, which provides a rigorous mathematical framework for incorporating these dependencies. In this way, this derivation resolves the long-standing issue \cite{spohn2006phonon} of rigorously obtaining the BTE collision term with explicit time- and space-dependence, typically introduced via empirical approximations \cite{vasko2006quantum}. It also demonstrates that the space-time structure of the BTE naturally emerges from Wigner’s mixed representation applied to the KBE, clarifying the correct approach for solving time-dependent and non-homogeneous BTEs, often tackled via Monte Carlo simulations \cite{carrete2017almabte,raya2022bte}.\\
Even with Eq. \eqref{final_BTE_scattering_qp_non_homo_main}, the BTE remains a complicated integro-differential equation. In most practical scenarios, phonons are close to equilibrium \cite{ziman2001electrons,fugallo2013ab}. To simplify, we assume small deviations from a local temperature $T(\boldsymbol{R},t)$, such that the phonon distribution $n_\nu(\boldsymbol{R},t)$ can be expressed as a small perturbation around the local Bose-Einstein equilibrium $\bar{n}_\nu(\boldsymbol{R},t)$:
\begin{equation} \label{expansion_n_bar_n_delta_n_main}
n_{\nu}=\bar{n}_{\nu}+\Delta n_{\nu}=\bar{n}_{\nu}+\bar{n}_{\nu}(\bar{n}_{\nu}+1)h_{\nu},
\end{equation}
where $h_{\nu}=\dfrac{\hbar\omega_{\nu}}{k_{{\rm B}}T^{2}}\,\nabla T\cdot\boldsymbol{{\rm f}}_{\nu}$, 
and $\boldsymbol{{\rm f}}_{\nu}$ is the solution of the LBTE, recast in matrix form as a linear algebra problem. Inserting Eq. \eqref{expansion_n_bar_n_delta_n_main} into Eq. \eqref{final_BTE_scattering_qp_non_homo_main} results in the linearized Boltzmann transport equation (LBTE), in which the collision operator can be written in the following matrix form:
\begin{equation} \label{final_scattering_linerized_qp_matrix_main}
{\rm{CT}}_{\nu}=\sum_{\nu'}A_{\nu\nu'}h_{\nu'}=\underbrace{-A^{\text{out}}_{\nu\nu}h_{\nu}}_{\text{diagonal}}+\underbrace{\sum_{\nu'\ne\nu}A^{\text{in}}_{\nu\nu'}h_{\nu'}}_{\text{non-diagonal}},
\end{equation}
with
\begin{equation} \label{final_A_in_A_out_qp_main}
\begin{split}
A^{\text{out}}_{\nu\nu}&=8\pi\hbar\sum_{\nu'\nu''}\Big(\mathcal{L}_{\nu\nu'}^{\nu''}{\rm{R}}_{\nu\nu'}^{\nu''}+\frac{1}{2}\mathcal{L}_{\nu}^{\nu'\nu''}{\rm{R}}_{\nu}^{\nu'\nu''}\Big),\\
A^{\text{in}}_{\nu\nu'}&=8\pi\hbar\sum_{\nu''}\Big(\mathcal{L}_{\nu\nu''}^{\nu'}{\rm{R}}_{\nu\nu''\nu'}-\mathcal{L}_{\nu\nu'}^{\nu''}{\rm{R}}_{\nu\nu'}^{\nu''}+\mathcal{L}_{\nu}^{\nu'\nu''}{\rm{R}}_{\nu}^{\nu'\nu''}\Big),
\end{split}
\end{equation}
where
\begin{equation} \label{pieces_in_collision_term_qp}
\begin{split}
&\mathcal{L}_{\nu}^{\nu'\nu''}=\delta(\omega_{\nu}-\omega_{\nu'}-\omega_{\nu''})|\mathcal{F}_{\nu-\nu'-\nu''}|^{2},\\
&\mathcal{L}_{\nu\nu'}^{\nu''}=\delta(\omega_{\nu}+\omega_{\nu'}-\omega_{\nu''})|\mathcal{F}_{\nu\nu'-\nu''}|^{2},\\
&{\rm{R}}_{\nu\nu'}^{\nu''}=(\bar{n}_{\nu}+1)(\bar{n}_{\nu'}+1)\bar{n}_{\nu''},\\
&{\rm{R}}_{\nu}^{\nu'\nu''}=(\bar{n}_{\nu}+1)\bar{n}_{\nu'}\bar{n}_{\nu''}.
\end{split}
\end{equation}
This decomposition makes explicit that the scattering matrix comprises two contributions: $A^{\text{out}}_{\nu\nu}$, describing the depopulation of phonon $\nu$ due to scattering, and $A^{\text{in}}_{\nu\nu'}$, describing the repopulation of $\nu$ from incoming phonons. Thus, the BTE is naturally understood as a specific instance of the KBE, whose collision term can generally be expressed as ${\rm CT}_{\rm in}-{\rm CT}_{\rm out}$. Furthermore, it can be shown \cite{fugallo2013ab} that the diagonal elements of the matrix $A^{\text{out}}_{\nu\nu}$ are directly linked to the phonon linewidths $\Gamma_{\nu}$—the latter being inversely proportional to the phonon lifetimes $\tau_{\nu}=\frac{1}{2\Gamma_{\nu}}$—through
\begin{equation} \label{A_out_as_Gamma}
A^{\text{out}}_{\nu\nu}=\bar{n}_{\nu}(\bar{n}_{\nu}+1)\Gamma_{\nu}.
\end{equation}
This identity naturally follows from the derivation of the scattering matrix $A^{\text{out}}_{\nu\nu}$ and is independently confirmed within the NEGF framework when Eq. \eqref{gamma_def_NEGF} is employed \cite{di2025broadening}. For comparison with experimental analyses, the spectral function of the system is then computed starting from the phonon self-energy, written as
\begin{equation} \label{self_energy_general_def}
\Sigma_{\nu}(\omega)=\Delta_{\nu}(\omega)+i\Gamma_{\nu}(\omega).
\end{equation}
It should be noted that in frameworks based on self-consistent phonon (SCPH) theories \cite{werthamer1970self}, this expression may be modified to incorporate additional renormalization effects. SCPH-based methods, including self-consistent \textit{ab initio} lattice dynamics (SCAILD) \cite{souvatzis2008entropy} and the stochastic self-consistent harmonic approximation (SSCHA) \cite{errea2014anharmonic}, explicitly account for anharmonic contributions to phonon frequencies. These approaches originate from Dyson’s equation and establish a self-consistent procedure to obtain renormalized frequencies. The renormalization is typically driven by the fourth-order interaction term (loop diagram), which has no external legs and can therefore be solved self-consistently to capture anharmonic corrections \cite{tadano2015self}. Furthermore, these methods routinely incorporate temperature-induced modifications of phonon bands, often through \textit{ab initio} molecular dynamics (AIMD) or within the SSCHA formalism itself \cite{SSCHA_0,SSCHA_1,SSCHA_2,SSCHA_3}. More recently, Tadano and co-workers have extended this approach by including lineshift effects originating from the bubble diagram in perturbation theory \cite{tadano2022first}. This review does not explicitly examine the renormalization of phonon frequencies induced by anharmonicity or temperature, as our primary aim is to establish a connection with mesoscopic models of thermal transport. However, we emphasize that much remains open for further discussion and extension at the level of quantum microscopic transport theories \cite{di2025broadening,di2025theory,di2026ephvirtual,di2025vertex,di2025theoretical}.\\
The separation into diagonal and off-diagonal parts is particularly important in the hydrodynamic regime. Properly capturing phonon hydrodynamics requires the full scattering matrix, since, as will be seen later in the text, conservation of crystal momentum is encoded in special eigenvectors of the matrix associated with normal scattering events, which appear in both diagonal and off-diagonal components. Consequently, as we will see, simplified approaches such as the single-mode relaxation time approximation (RTA), which reduce the scattering matrix to its diagonal entries to extract a single-mode relaxation time, fail to describe hydrodynamic phonon transport (see also Ref. \cite{chiloyan2021green}).

\subsection{Linearized driving term and thermal conductivity}

In many cases, we are primarily interested in solving the steady state problem, where a constant temperature gradient is applied to the crystal. Under these conditions, the time derivative in the BTE can be set to zero, allowing the driving term to be further simplified:
\begin{equation}
\boldsymbol{v}_{\nu}\frac{\partial n_{\nu}(\boldsymbol{R},t)}{\partial\boldsymbol{R}}=\boldsymbol{v}_{\nu}\cdot\nabla T\frac{\partial n_{\nu}(\boldsymbol{R},t)}{\partial T}\approx\boldsymbol{v}_{\nu}\cdot\nabla T\frac{\partial\bar{n}_{\nu}(\boldsymbol{R},t)}{\partial T}
\end{equation}
The first equality relies on the assumption that the system is spatially homogeneous, and that any variation in the distribution arises solely from the presence of a temperature gradient. The subsequent approximation is valid in the regime of small deviations from thermal equilibrium—precisely where the linearization of the scattering operator also applies. In this limit, the change in the out-of-equilibrium distribution due to temperature is approximately the same as that of the Bose-Einstein distribution. As a result, the steady-state, homogeneous BTE simplifies considerably and takes the form:
\begin{equation} \label{homo_BTE_solving_preliminar}
\boldsymbol{v}_{\nu}\cdot\nabla T\frac{\partial\bar{n}_{\nu}(\boldsymbol{R},t)}{\partial T}=\sum_{\nu'}A_{\nu\nu'}h_{\nu'},
\end{equation}
where the collision operator is linearized as discussed in Eq. \eqref{expansion_n_bar_n_delta_n_main}. When using Eq. \eqref{expansion_n_bar_n_delta_n_main} to look for solution that are linear in the temperature gradient, we get $h_{\nu}\approx\nabla Tf_{\nu}$, so Eq. \eqref{homo_BTE_solving_preliminar} becomes
\begin{equation} \label{homo_BTE_solving}
v_{\nu}\frac{\partial\bar{n}_{\nu}(\boldsymbol{R},t)}{\partial T}=\sum_{\nu'}A_{\nu\nu'}f_{\nu'},
\end{equation}
where we selected the component of the phonon group velocity parallel to $\nabla T$. The advantage of using matrix notation is that it reveals the underlying mathematical structure of the equation, reducing it to a straightforward linear algebra problem
\begin{equation} \label{linear_algebra_problem}
Af=b    
\end{equation}
with $b=v_{\nu}\frac{\partial\bar{n}_{\nu}(\boldsymbol{R},t)}{\partial T}$. The solution to this equation can, in principle, be obtained by directly inverting the scattering matrix: $f=A^{-1}b$. Assuming this equation is solvable, determining the deviation function $f_{\nu}$ effectively resolves the most challenging part of the transport problem. Once $f_{\nu}$ is known, the lattice heat flux can be evaluated, to leading (harmonic) order, as \cite{simoncelli2022wigner,hardy1963energy}
\begin{equation}
Q=\frac{1}{\mathcal{V}}\sum_{\nu}\hbar\omega_{\nu}v_{\nu}n_{\nu}=\frac{1}{\mathcal{V}k_{{\rm{B}}}T^{2}}\sum_{\nu}\bar{n}_{\nu}(\bar{n}_{\nu}+1)\hbar\omega_{\nu}v_{\nu}f_{\nu}\nabla T.
\end{equation}
This equation can be directly compared with the definition of thermal conductivity $Q=-\kappa\nabla T$, obtaining
\begin{equation} \label{thermal_conductivity_definition}
\kappa=-\frac{1}{\mathcal{V}k_{{\rm{B}}}T^{2}}\sum_{\nu}\bar{n}_{\nu}(\bar{n}_{\nu}+1)\hbar\omega_{\nu}v_{\nu}f_{\nu}.
\end{equation}
Together with the thermal viscosity (see section \ref{thermal_viscosity}), this quantity is one of the central transport coefficients governing the mesoscopic behavior of a phonon system.

\subsection{Normal and Umklapp scattering}

We now examine the type of microscopic phonon scattering required for hydrodynamic behavior to emerge. Consider the scattering of a phonon in the state $n_{\nu}$. Both decay and coalescence processes, such as those described in Eq. \eqref{final_BTE_scattering_qp_non_homo_main}, involve two additional phonons in the states $\nu'$ and $\nu''$. As discussed in the previous section, within the lowest-order quasiparticle approximation \eqref{quasiparticle_ansatz}, phonon–phonon interactions are constrained by energy conservation \cite{di2025broadening}, which restricts the frequencies of the three phonons involved. In addition, there is also a constraint on their momenta. The anharmonic term of the phonon Hamiltonian (see Eq. \eqref{phonon_hamiltonina_with_anharmonic_term}) contains a factor $\delta_{\boldsymbol{q}+\boldsymbol{q}'+\boldsymbol{q}'',\boldsymbol{K}}$, which naturally emerges from the Fourier transform when moving from real to reciprocal space \cite{cepellotti2016thermal}. This factor encodes the periodicity of the Bravais lattice and expresses crystal momentum conservation. Accordingly, the momenta of the three interacting phonons must satisfy
\begin{equation}
\boldsymbol{q} + \boldsymbol{q}' + \boldsymbol{q}'' = \boldsymbol{K}.
\end{equation}
Thus, in a three-phonon process, crystal momentum is conserved only modulo a reciprocal lattice vector $\boldsymbol{K}$; the mismatch between initial and final momentum may correspond to any such vector. Taken together, energy and crystal-momentum conservation impose strong restrictions on the allowed scattering channels for a phonon $(\boldsymbol{q}, s)$ \cite{di2025theory,feng2017four}.
\begin{figure}[h!]
\centering
\includegraphics[width=0.75\textwidth]{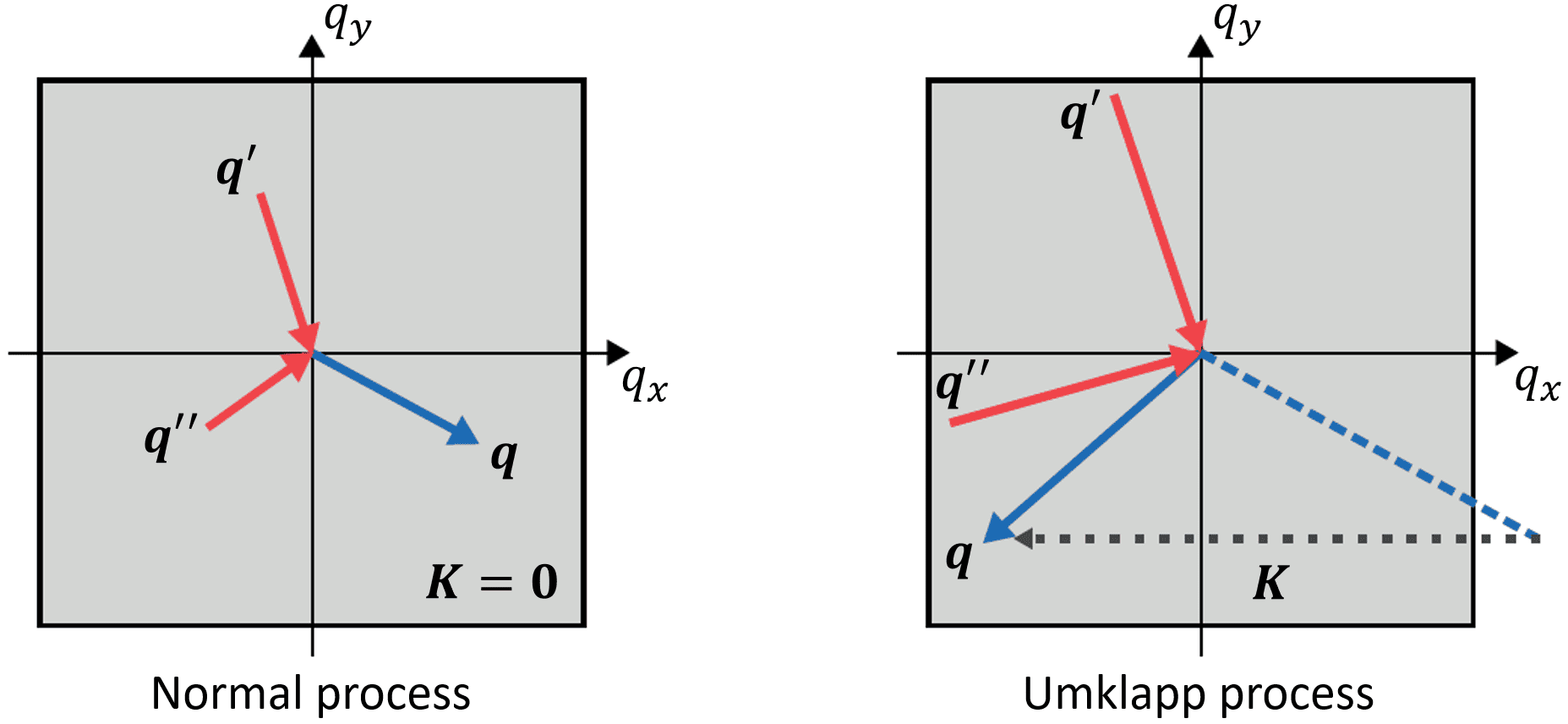}
\caption{\textbf{Normal and Umklapp phonon scattering processes.} Normal and Umklapp processes during a coalescence event involving two phonons with wave vectors $\boldsymbol{q}'$ and $\boldsymbol{q}''$ merging into a phonon with wave vector $\boldsymbol{q}$. In a normal process (left), the sum of the initial momenta lies inside the first Brillouin zone (shaded square). In an Umklapp process (right), the sum lies outside the first Brillouin zone but can be folded back inside by subtracting a reciprocal lattice vector.}
\label{normal_Umklapp}
\end{figure}
Peierls introduced a classification of three-phonon scattering events \cite{peierls2014bird}: processes that conserve crystal momentum exactly ($\boldsymbol{K}=0$) are called \emph{normal} (left panel of Fig. \ref{normal_Umklapp}), whereas processes for which momentum conservation requires a nonzero reciprocal lattice vector ($\boldsymbol{K}\neq 0$) are known as \emph{Umklapp}. The term traces back to Lenz, who used \emph{Umklappen}---German for “turn over”---to describe the flipping of a magnetic moment in early formulations of the Ising model \cite{lenz1920beitrag}. In phonon physics, Umklapp events can effectively reverse the direction of momentum (right panel of Fig. \ref{normal_Umklapp}). In practice, one distinguishes normal from Umklapp processes by computing the third wave vector as $\boldsymbol{q}'' = \boldsymbol{q} \pm \boldsymbol{q}'$. If $\boldsymbol{q}''$ lies inside the first Brillouin zone, the process is normal; if instead it falls outside and must be brought back by adding a reciprocal lattice vector $\boldsymbol{K}$, the process is classified as Umklapp.

\subsection{Solution of the Boltzmann transport equation}

With the advent of modern computational techniques, the LBTE can now be solved exactly using iterative \cite{omini1995iterative,broido_intrinsic_2007,carrete2017almabte}, variational \cite{fugallo2013ab}, or exact diagonalization schemes \cite{chaput2013direct,cepellotti2017transport}. These advances have enabled parameter-free, first-principles predictions of phonon transport properties in bulk crystals with genuine predictive power.\\
In the previous section we showed that the phonon BTE can be cast as a linear algebra problem whose formal solution is obtained by inverting the scattering matrix (see Eq. \eqref{linear_algebra_problem}). Although this provides a compact mathematical expression, it is of limited practical value: directly inverting the full collision operator is computationally prohibitive, and even when such an inversion is achieved, the resulting thermal conductivity offers little intuitive connection to the underlying microscopic physics. Here we summarize a set of strategies for solving the LBTE to obtain both the nonequilibrium phonon distribution and the resulting thermal conductivity, with a final emphasis on the relaxon formalism \cite{cepellotti2016relaxons}, which underpins the macroscopic description of heat transport that we aim to develop.\\
We begin with two classical approximations—the single-mode relaxation time approximation (RTA) and Callaway’s model—which yield closed-form solutions and an appealingly simple physical interpretation. These approaches, however, rely on assumptions about the structure of the scattering operator that are not universally valid, particularly in low-dimensional systems such as 2D crystals \cite{fugallo2014thermal}. We then introduce three approaches capable of solving the LBTE exactly, without any assumptions on the form of the collision operator. The iterative method \cite{omini1995iterative} was the first practical algorithm to treat the full phonon–phonon scattering problem and remains widely used today. It has been applied successfully with first-principles scattering rates—for instance, the first fully \textit{ab initio} solution was reported for silicon \cite{broido_intrinsic_2007}, and later extended to alloys \cite{garg2011role}. However, the iterative scheme may exhibit convergence issues in 2D materials, including graphene \cite{cepellotti2016thermal}, where strong mode-coupling prevents the geometric series from converging. A more recent approach takes advantage of the variational formulation of thermal transport to efficiently obtain the exact solution of the Boltzmann equation \cite{fugallo2013ab}. This method is particularly robust in systems where the iterative method fails to converge. Finally, we discuss a direct and exact diagonalisation of the scattering operator. Although this is by far the most computationally demanding approach and thus rarely attempted, it yields a closed-form solution to the transport problem and provides a rigorous physical interpretation of thermal transport in terms of collective excitations—here named relaxons \cite{cepellotti2016relaxons}.

\subsubsection{Single-mode relaxation-time approximation}

We begin with the RTA, which represents the most elementary attempt at approximating the solution of the LBTE. In this approach, the full collision operator is replaced by its diagonal part,
\begin{equation}
A_{\nu\nu'} \approx A^{\mathrm{out}}_{\nu\nu'}=\frac{\bar{n}_\nu(\bar{n}_\nu+1)}{\tau_\nu}\,\delta_{\nu\nu'},
\end{equation}
where $\tau_\nu$ is interpreted as the relaxation time (or lifetime) of phonon mode $\nu$. Historically, the RTA derives from Boltzmann’s early work on classical gases and predates modern phonon transport theory. The LBTE in the RTA can then be written as (spatial and temporal dependencies are omitted for brevity)
\begin{equation}
\boldsymbol{v}_\nu \cdot \nabla T
\frac{\partial \bar{n}_\nu}{\partial T}
=
-\bar{n}_\nu(\bar{n}_\nu+1)
\frac{1}{\tau_\nu}
\frac{\hbar\omega_\nu}{k_{\mathrm B}T^2}
\nabla T\cdot\boldsymbol f_\nu .
\end{equation}
Using
\begin{equation}
\frac{\partial \bar{n}_\nu}{\partial T}
=
\bar{n}_\nu(\bar{n}_\nu+1)
\frac{\hbar\omega_\nu}{k_{\mathrm B}T^2},
\end{equation}
one obtains
\begin{equation}
\nabla T\cdot\boldsymbol f_\nu
=
-\tau_\nu\,\nabla T\cdot\boldsymbol{v}_\nu,
\end{equation}
and therefore the solution is
\begin{equation}
\boldsymbol{f}_\nu = -\tau_\nu \boldsymbol{v}_\nu.
\end{equation}
Inserting this expression into the thermal conductivity formula yields
\begin{equation}
\kappa_{\mathrm{RTA}} = \frac{1}{\mathcal{V}}
\sum_\nu C_\nu v_\nu^2\tau_\nu
= \frac{1}{\mathcal{V}}\sum_\nu C_\nu v_\nu \Lambda_\nu,
\end{equation}
where $\Lambda_\nu = v_\nu\tau_\nu$ is the mean free path. This expression mirrors that of kinetic gas theory, although its physical interpretation in the phonon context is more subtle. The limitations of the RTA follow from the fact that it neglects all off-diagonal elements of the scattering matrix. These terms describe mode-coupling processes that redistribute population between phonons and are essential for correctly capturing normal processes (which conserve momentum). As emphasized by Ziman \cite{ziman2001electrons}, the RTA relaxation time characterizes the decay of individual modes, not the decay of the heat flux itself.

\subsubsection{Callaway’s approximation} \label{callaway_section}

At sufficiently low temperatures, Umklapp processes are strongly suppressed \cite{ashcroft2022solid} and normal processes dominate phonon scattering. In this regime, the RTA often breaks down, as observed in, e.g., thermal conductors at low temperature \cite{cepellotti2015phonon,fugallo2014thermal}. This inadequacy motivates Callaway’s refinement \cite{callaway1959model} in which the full scattering operator is explicitly separated into normal (N) and resistive (R) components (like Umklapp and isotopic scattering, for example), and assumes that resistive processes relax the system toward equilibrium, while normal processes relax it toward the drifting distribution $\bar{n}^{\mathrm{D}}_{\nu}$ (see Eq. \eqref{drifting_distribution}). Within this approximation, the LBTE becomes
\begin{equation}
\boldsymbol{v}_{\nu}\cdot\nabla T\frac{\partial\bar{n}_{\nu}}{\partial T}=-\frac{n_\nu-\bar{n}_\nu}{\tau_\nu^{\mathrm{R}}}-\frac{n_\nu-\bar{n}^{\mathrm{D}}_{\nu}}{\tau_\nu^{\mathrm{N}}}.
\end{equation}
After a Taylor expansion of $\bar{n}^{\mathrm{D}}_{\nu}$ and algebraic manipulation, one finds
\begin{equation}
f_\nu =
-\frac{\hbar\omega_\nu v_\nu\tau_\nu}{k_B T^2}
+(\boldsymbol{q}\cdot\boldsymbol{u})
\frac{\tau_\nu}{\tau_\nu^{\mathrm{N}}}\frac{1}{k_B T},
\end{equation}
where $\tau_\nu^{-1} = (\tau_\nu^{\mathrm{N}})^{-1}+(\tau_\nu^{\mathrm{R}})^{-1}$ and $\boldsymbol{u}$ is the phonon drift velocity (see section \ref{hydro_equilibrium}). To determine the phonon drift velocity, Callaway imposes momentum conservation under normal processes:
\begin{equation}
\frac{d\boldsymbol{P}}{dt}\bigg|_{\mathrm{N}} = 0.
\end{equation}
This condition determines $\boldsymbol{u}$ and leads to Callaway’s final expression for the
thermal conductivity,
\begin{equation}
\kappa = \kappa_{\mathrm{RTA}}
+ \kappa_{\mathrm{Callaway}}^{\mathrm{corr}},
\end{equation}
where the second term that captures the contribution of normal processes is
\begin{equation}
\kappa_{\mathrm{Callaway}}^{\mathrm{corr}} =\frac{1}{k_B T^2 \mathcal{V}}\frac{\left( \sum_{\nu} \bar{n}_\nu (\bar{n}_\nu + 1)\hbar \omega_\nu v_\nu q_\parallel \frac{\tau_\nu}{\tau^\mathrm{N}_\nu} \right)^2}{\sum_{\nu} \bar{n}_\nu (\bar{n}_\nu + 1) (q_\parallel)^2 \frac{\tau_\nu}{\tau^\mathrm{N}_\nu \tau^\mathrm{R}_\nu}},
\end{equation}
where $q_\parallel$ is the component of the phonon wave vector parallel to the temperature gradient. Although approximate, Callaway’s model introduces the crucial insight that a single relaxation time per phonon mode is insufficient and has inspired later developments concerning hydrodynamic transport and second sound; nevertheless, it remains quantitatively limited because it neglects mode-to-mode coupling and the full structure of the scattering matrix, which are essential for accurately describing collective phonon dynamics beyond the weakly hydrodynamic regime. Moreover, it is also shown to underestimate the suppression of normal processes in relaxing thermal current \cite{allen2013improved}.

\subsubsection{Iterative method}

The first exact solution of the phonon LBTE was obtained using the iterative scheme introduced in Ref. \cite{omini1995iterative}. In matrix form, the method begins
with the RTA solution $f^{(0)} = A_{\mathrm{out}}^{-1}b$ and iteratively generates
\begin{equation}
f^{(i)} = \sum_{j=0}^i \left[-A_{\mathrm{out}}^{-1}A_{\mathrm{in}}\right]^j A_{\mathrm{out}}^{-1}b.
\end{equation}
This expansion is a geometric series in the operator $-A_{\mathrm{out}}^{-1}A_{\mathrm{in}}$. Convergence is guaranteed only if all eigenvalues of this operator have magnitude $<1$. When this condition is met, the method yields the exact solution.\\
While extremely successful in many bulk materials, the iterative approach often fails in two-dimensional systems (like e.g. graphene at room temperature \cite{cepellotti2016thermal}), where off-diagonal scattering elements can be large. In such cases, the iterative operator may have eigenvalues with magnitude greater than unity, causing the geometric series, and therefore the iterative solution for the distribution function, to diverge.

\subsubsection{Variational method}

The thermal conductivity can be written as the maximizer of the
functional \cite{cepellotti2016thermal,fugallo2013ab}
\begin{equation}
\kappa[f] = \frac{1}{\mathcal{V}}\left(2\langle b | f \rangle - \langle f | A | f \rangle \right),
\end{equation}
where $b$ and $f$ verify Eq. \eqref{linear_algebra_problem}. Maximizing this functional is equivalent to solving the LBTE exactly. Traditional approaches \cite{hamilton1969variational} used trial functions with adjustable parameters. While such methods improve over RTA, they provide no guarantee of reaching the exact solution and depend sensitively on the choice of ansatz. The modern formulation introduced in Ref. \cite{fugallo2013ab} applies a conjugate-gradient algorithm to maximize $\kappa[f]$. Since $A$ is symmetric and semidefinite, the functional has a unique maximum; the algorithm therefore converges to the exact solution without requiring trial functions. To improve convergence, it is advantageous to precondition the functional using the diagonal part of the scattering matrix:
\begin{align}
\tilde{f} &= A_{\mathrm{out}} f,\\
\tilde{A} &= A_{\mathrm{out}}^{-1} A A_{\mathrm{out}}^{-1}.
\end{align}
Despite their efficiency, both the iterative and variational methods share a conceptual limitation: the phonon distribution $f_\nu$ they produce does not yield a clean definition of relaxation times or mean free paths. Although one may formally define $\tau_\nu^{(\mathrm{eff})}=\frac{f_\nu}{v_\nu \hbar\omega_\nu}$ and $\Lambda_\nu^{(\mathrm{eff})}=\frac{f_\nu}{\hbar\omega_\nu}$, these quantities lack a rigorous physical interpretation. In particular, such mean free paths can be negative or diverge when $v_\nu = 0$. In the next section we will show how diagonalising the scattering operator resolves these issues and leads to well-defined microscopic relaxation times.

\subsubsection{Exact diagonalisation in terms of relaxons} \label{relaxons_section}

The final method we consider in this review is the explicit diagonalization of the scattering operator. While computationally demanding, this approach leads to a powerful and physically transparent interpretation of thermal transport \cite{cepellotti2016relaxons}. Starting from the time-dependent BTE
\begin{equation}
\frac{\partial n_\nu}{\partial t}
+ \boldsymbol{v}_\nu\cdot\nabla n_\nu
= -\sum_{\nu'} \Omega_{\nu\nu'} n_{\nu'},
\end{equation}
where 
\begin{equation}
{\rm{\Omega}}_{\nu\nu'}=\frac{A_{\nu\nu'}}{\bar{n}_{\nu'}(\bar{n}_{\nu'}+1)},
\end{equation}
we first symmetrize the scattering matrix via the transformation \cite{hardy1965lowest,hardy1970phonon,chaput2013direct,krumhansl1965thermal} 
\begin{equation} \label{symmetric_transformation}
\tilde{\Omega}_{\nu\nu'} = 
\Omega_{\nu\nu'}\sqrt{\frac{\bar{n}_{\nu'}( \bar{n}_{\nu'} +1)}
{\bar{n}_\nu( \bar{n}_\nu +1)}},
\qquad
\tilde{n}_\nu = [\bar{n}_\nu(\bar{n}_\nu+1)]^{-1/2} n_\nu,
\end{equation}
so that $\tilde{\Omega}$ is real, symmetric and positive semidefinite. 
\begin{figure}[h!]
\centering
\includegraphics[width=\textwidth]{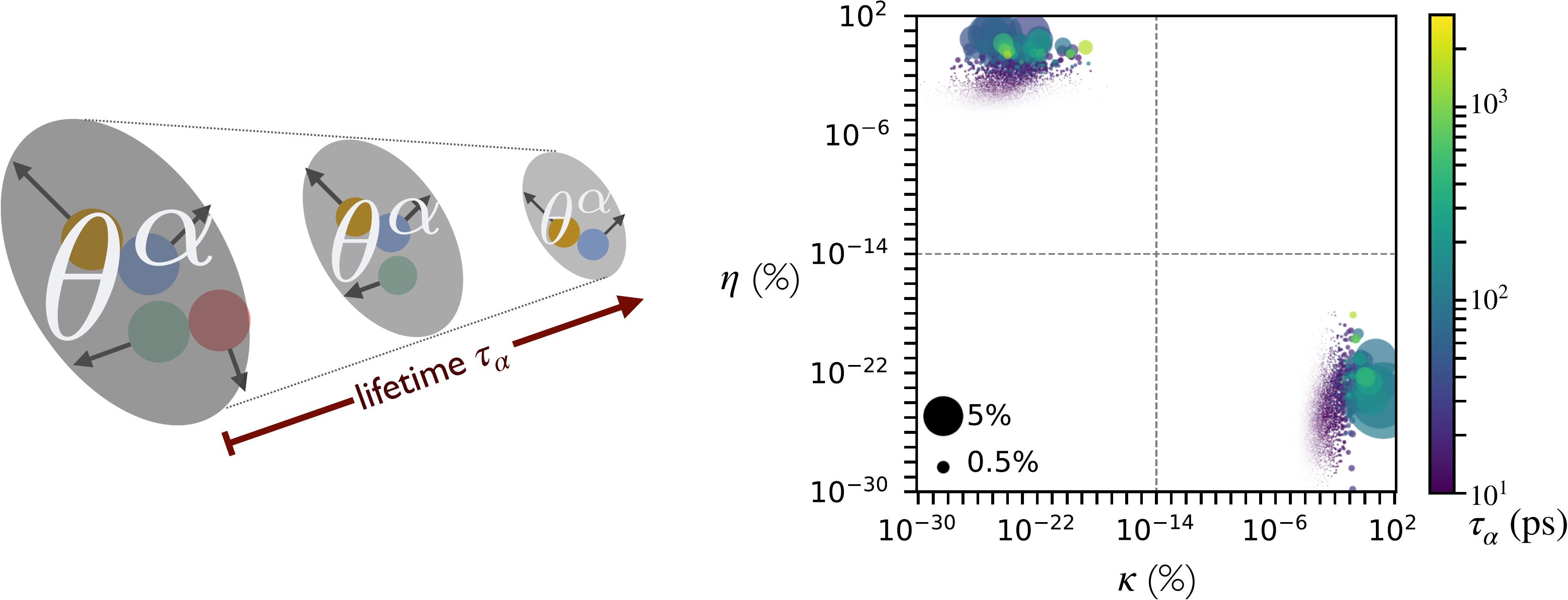}
\caption{\textbf{Relaxon picture of lattice thermal transport.} (Left panel) Schematic of lattice-vibration equilibration following a thermal perturbation. Each relaxon is a collective phonon excitation that evolves independently and relaxes exponentially toward equilibrium with a characteristic lifetime. (Right panel) Contributions of individual relaxons to the bulk thermal conductivity and thermal viscosity of graphite (in-plane components) at 300 K. Each point represents a relaxon; its color denotes the relaxation time, and its area is proportional to its combined contribution to conductivity and viscosity. Odd relaxons govern thermal conductivity and contribute negligibly to viscosity, while even relaxons dominate the thermal viscosity and contribute negligibly to conductivity (see also section \ref{thermal_viscosity}). Dashed lines are guides to the eye highlighting the decoupling between the two sets of eigenvectors. \textit{Figure reproduced with permission from Ref. \cite{simoncelli2020generalization}.}}
\label{fig:relaxon_figure}
\end{figure}
Diagonalising $\tilde{\Omega}$,
\begin{equation}
\sum_{\nu'} \tilde{\Omega}_{\nu\nu'} \theta^{i}_{\nu'}
= \frac{1}{\tau_i}\,\theta^{i}_\nu,
\end{equation}
yields a complete orthonormal basis $\{\theta^{(i)}\}$ of eigenvectors. These eigenvectors as collective excitations—relaxons—whose occupation numbers are defined by \cite{cepellotti2016relaxons}
\begin{equation}
\tilde{n}_\nu = \sum_i f_i\,\theta^{i}_\nu.
\end{equation}
In this representation, the BTE becomes
\begin{equation}
\frac{\partial f_i}{\partial t}
+ \sum_\alpha \boldsymbol{w}_{i\alpha}\cdot\nabla f_\alpha
= -\frac{f_i}{\tau_i},
\end{equation}
where
\begin{equation} \label{relaxon_velocity}
\boldsymbol{w}_{i\alpha} = \frac{1}{\mathcal{V}}\sum_{\nu}\theta^{i}_{\nu}\boldsymbol{v}_{\nu}\theta^{\alpha}_{\nu} = \langle i|\boldsymbol{v}|\alpha\rangle
\end{equation}
derives from the action of the scattering operator on the deviation from equilibrium. The relaxation of such collective excitations is purely exponential:
\begin{equation}
f_i(t) = f_i(0)e^{-t/\tau_i}.
\end{equation}
Thus, relaxons possess true relaxation times, unlike phonons, whose relaxation is a superposition of many exponentials. In steady state, the thermal conductivity becomes
\begin{equation}
\kappa_{ij} = C\sum_\alpha w_{\alpha}^{i}w_{\alpha}^{j}\tau_\alpha= C \sum_\alpha w_{\alpha}^{i}\Lambda_{\alpha}^{j},
\end{equation}
where $\Lambda_{\alpha}^{j}=w_{\alpha}^{j}\tau_\alpha$ is the relaxon mean free path, $C$ denotes the volumetric specific heat and 
\begin{equation}
\boldsymbol{w}_{\alpha}\equiv\boldsymbol{w}_{0\alpha} = \frac{1}{\mathcal{V}}\sum_{\nu}\theta^{0}_{\nu}\boldsymbol{v}_{\nu}\theta^{\alpha}_{\nu} = \langle 0|\boldsymbol{v}|\alpha\rangle
\end{equation}
derives from the action of the scattering operator on the equilibrium distribution (the meaning of $\theta^{0}_{\nu}$ will be clear later, see Eq. \eqref{Bose_Einstein_eigenvector}). The specific heat is defined as the derivative of the total energy with respect to temperature, evaluated at equilibrium, and it can equivalently be expressed as a sum over the modal specific heats $C_\nu$:
\begin{equation} \label{specific_heat}
C = \left. \frac{\partial E}{\partial T} \right|_{\text{eq}}
= \frac{1}{\mathcal{V}} \sum_\nu C_\nu
= \frac{1}{\mathcal{V}} \sum_\nu \frac{(\hbar \omega_\nu)^2}{k_{\mathrm{B}} \bar{T}^{2}} \,\bar{n}_\nu(\bar{n}_\nu + 1).
\end{equation}
This constitutes a kinetic theory for the exact solution of the BTE: heat is carried not by individual phonons but by relaxons, each with a well-defined lifetime, velocity and mean free path.\\
Finally, we note that the idea of diagonalizing a scattering operator to simplify transport equations has also been explored in other fields. For example, in radiative transfer theory, the discrete–ordinate formulation of the transport equation leads to a system of coupled equations that can be solved by diagonalizing the scattering matrix, thereby decoupling the angular modes of the radiation field \cite{stamnes1988numerically}.

\subsection*{Takeaways}

In this section we derived the phonon Boltzmann transport equation starting from the fully quantum Kadanoff--Baym equation, thereby establishing the semiclassical regime as a controlled approximation to quantum phonon kinetics. We showed how the gradient expansion and the generalized Kadanoff--Baym ansatz map quantum Green’s functions onto a phase-space phonon distribution, clarifying the assumptions underlying the quasiparticle picture and Fermi’s golden rule. This derivation makes explicit the origin and structure of the Boltzmann transport equation driving and collision terms, including their space--time dependence and possible extensions to the nonlinear regime. We discussed how different intrinsic scattering mechanisms—normal and Umklapp processes—govern the relaxation of phonon populations and thermal currents, and how their interplay determines the emergence of hydrodynamic behavior. Finally, by examining exact solution strategies for the linearized Boltzmann transport equation, we showed that diagonalizing the scattering operator leads naturally to the relaxon picture, in which heat is carried by collective excitations with well-defined lifetimes and mean free paths. This formulation provides the conceptual bridge between microscopic phonon kinetics and the mesoscopic hydrodynamic theories developed in the following sections.

\section{Mesoscopic treatment: viscous heat equations}

In this section, we focus on the viscous heat equations (VHE) \cite{simoncelli2020generalization}, which, as anticipated in the introduction, provide a general and unified approach for extending Fourier’s law into the hydrodynamic regime. Starting from the LBTE, the VHE derives two novel coupled mesoscopic heat transport equations that consistently capture Fourier diffusion, hydrodynamic propagation, and all intermediate transport regimes on an equal footing. To introduce the VHE, we begin with the microscopic description of thermal transport provided by the LBTE, which governs the evolution of nonequilibrium phonon populations. In its time-dependent form, the LBTE reads:
\begin{equation}
\frac{\partial\Delta n_{\nu}(\boldsymbol{R},t)}{\partial t} 
+ \boldsymbol{v}_{\nu} \cdot  \frac{\partial\Delta n_{\nu}(\boldsymbol{R},t)}{\partial \boldsymbol{R}}= -\frac{1}{\mathcal{V}} \sum_{\nu'} \Omega_{\nu\nu'} \Delta n_{\nu'}(\boldsymbol{R},t),
\label{LBTE}
\end{equation}
where $\Delta_{\nu}$ is defined in Eq. \eqref{expansion_n_bar_n_delta_n_main} and, as usual, $\nu$ labels a phonon state, i.e., the combined index $\nu \equiv (\boldsymbol{q}, s)$ running over phonon wave vectors $\boldsymbol{q}$ and branches $s$, and $\Omega_{\nu\nu'}$ is the phonon scattering matrix \cite{cepellotti2016relaxons}. Eq. \eqref{LBTE} describes the evolution of the deviation, $\Delta n_{\nu}(\boldsymbol{R},t)$, of the phonon occupation from its thermal equilibrium value, given by the Bose-Einstein distribution at the equilibrium temperature $\bar{T}$:
\begin{equation}
\bar{n}_{\nu} = \left(e^{\beta\hbar\omega_{\nu}} - 1\right)^{-1},
\end{equation}
where $\beta=(k_{\text{B}}\bar{T})^{-1}$. From the solution of Eq. \eqref{LBTE} one can compute the local lattice energy, $E$, and the total crystal momentum, $\boldsymbol{P}$ \cite{hardy1970phonon}:
\begin{equation}
\begin{split}
E(\boldsymbol{R},t)&= \frac{1}{\mathcal{V}} \sum_{\nu} \hbar \omega_{\nu} \bar{n}_{\nu}(\boldsymbol{R},t),\\
\boldsymbol{P}(\boldsymbol{R},t)&= \frac{1}{\mathcal{V}} \sum_{\nu} \hbar \boldsymbol{q} \bar{n}_{\nu}(\boldsymbol{R},t),
\end{split}
\end{equation}
respectively relevant for thermal conductivity \cite{cepellotti2016relaxons} and for the hydrodynamic regime of heat transport \cite{sussmann1963thermal,gurzhi1968hydrodynamic,klemens1951thermal}. The hydrodynamic behavior typically emerges only in simple crystals—those where phonon branch separations are much larger than their linewidths \cite{simoncelli2019unified,simoncelli2022wigner,di2023crossover}.

\subsection{Hydrodynamic thermal equilibrium} \label{hydro_equilibrium}

In a phonon system where only normal scattering processes occur, the total crystal momentum is conserved. Because of this conservation law, a system that initially carries momentum cannot relax to the standard Bose-Einstein equilibrium distribution. Instead, the phonon population evolves toward a different stationary state known as the drifting distribution:
\begin{equation} \label{drifting_distribution}
\bar{n}^{\mathrm{D}}_{\nu}
= \left(e^{\beta(\hbar\omega_\nu - \boldsymbol{u}\cdot\boldsymbol{q})}-1\right)^{-1},
\end{equation}
where $\boldsymbol{u}$ is the phonon drift velocity. This velocity emerges from maximizing the entropy of the system while enforcing not only energy conservation but also crystal momentum conservation; the corresponding Lagrange multiplier, when multiplied by $k_{\mathrm B}\bar{T}$, naturally acquires the dimension of a velocity. Formally, $T$ and $\boldsymbol{u}$ correspond to the Lagrange multipliers enforcing the conservation of local energy and momentum, respectively. The drifting distribution is therefore similar to the Bose-Einstein distribution but “shifted’’ by the drift velocity. It represents the equilibrium state of the hydrodynamic regime of phonon transport, since---as will become clear later---the phonon drift velocity behaves analogously to the velocity field of a classical fluid and can be interpreted as the velocity of a phonon fluid \cite{simoncelli2020generalization}. It is important to note that the drifting distribution \eqref{drifting_distribution} differs from the one used in electron hydrodynamics, not only because of the different statistics involved (Fermi–Dirac–like rather than Bose–Einstein–like), but more fundamentally because phonons lack a chemical potential. In fact, the total phonon number is not a constant of motion (for example, phonon-coalescence events reduce the number of phonons in the system), and therefore no continuity equation for a phonon number density can be defined—its very definition lacks a rigorous physical meaning, in contrast to electronic systems. This feature, combined with the fact that the phonon BTE is typically treated in linear response—where nonlinear terms that would prevent the derivation of closed-form hydrodynamic equations are neglected—restricts current models of phonon hydrodynamics to the laminar regime only (as discussed later in the text). As a consequence, mesoscopic phonon turbulence in solids has not yet been modeled within a first-principles hydrodynamic framework.\\
Differing from the Bose-Einstein distribution, the drifting distribution explicitly depends on the phonon wave vector $\boldsymbol{q}$, which breaks the symmetry between modes with $\boldsymbol{q}$ and $-\boldsymbol{q}$. As a consequence, the drifting distribution carries a finite total momentum and a non-zero heat flux. Indeed, for the Bose-Einstein distribution one has $\bar{n}_{\boldsymbol{q}s} = \bar{n}_{-\boldsymbol{q}s}$ because it depends only on the frequency, and $\omega_{\boldsymbol{q}s} = \omega_{-\boldsymbol{q}s}$. At the same time, the phonon group velocity is an odd function of the wave vector, $\boldsymbol{v}_{-\boldsymbol{q}s} = -\boldsymbol{v}_{\boldsymbol{q}s}$. Thus, the heat flux (evaluated here to the lowest harmonic order \cite{hardy1963energy}),
\begin{equation}
\boldsymbol{Q}
= \frac{1}{\mathcal{V}} \sum_{\nu} \hbar \omega_{\nu} \boldsymbol{v}_{\nu} n_{\nu},
\end{equation}
cancels pairwise between opposite wave vectors for the Bose-Einstein distribution, yielding zero net momentum and zero heat flux. In contrast, for the drifting distribution $\bar{n}^{\mathrm{D}}_{\boldsymbol{q}s}\ne \bar{n}^{\mathrm{D}}_{-\boldsymbol{q}}$, so the cancellation does not occur and a finite heat flux remains. Because the drifting distribution carry finite heat flux, a material in which only normal scattering events take place would exhibit infinite thermal conductivity as there would not be any mechanism to relax momentum and dissipate the heat flux. This situation is analogous to a perfectly harmonic crystal, where the absence of scattering prevents momentum dissipation, again leading to divergent thermal conductivity. Therefore, anharmonicity---and in particular Umklapp processes---is essential for producing a finite thermal conductivity in ordinary crystals, as these processes constitute the only mechanism capable of relaxing the total phonon momentum.\\
Depending on which type of scattering dominates, different transport regimes emerge \cite{cepellotti2015phonon}.
\begin{figure}[h!]
\centering
\includegraphics[width=0.6\textwidth]{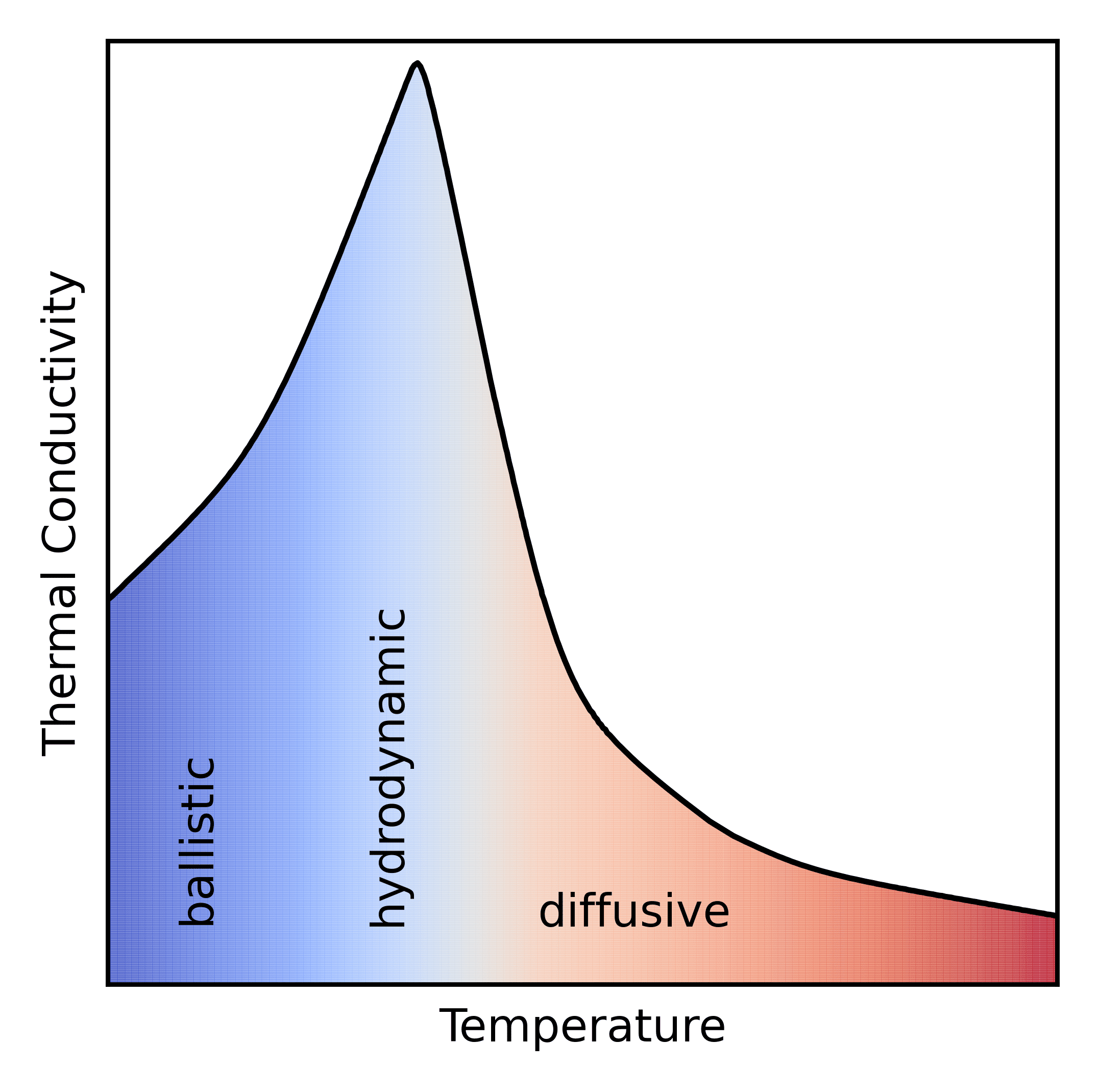}
\caption{\textbf{Thermal transport regimes in a generic three-dimensional crystal as a function of temperature.} At low temperatures, transport is ballistic, with extrinsic scattering much weaker than both normal and resistive processes. At intermediate temperatures, the hydrodynamic regime emerges, which can be further divided into the Poiseuille regime—where normal processes dominate and extrinsic scattering is stronger than resistive scattering—and the Ziman regime—where normal processes remain dominant but resistive mechanisms exceed extrinsic ones. At high temperatures, resistive scattering prevails over both normal and extrinsic processes, leading to the kinetic (diffusive) regime.}
\label{fig:regimes}
\end{figure}
These regimes can be understood, for example, by examining the temperature dependence of the thermal conductivity of a generic three-dimensional crystal (see Fig. \ref{fig:regimes}). At low temperatures, atomic vibrations are weak because ions carry little energy. Phonons are therefore more likely to undergo boundary and defect scattering, such as grain-boundary scattering in polycrystalline samples \cite{klemens1994phonon,fugallo2013ab}, point-defect scattering \cite{walker1963phonon}, or scattering from the finite system size. In this regime, extrinsic scattering dominates, and the system exhibits ballistic transport. As temperature increases, intrinsic phonon-phonon interactions become more frequent. The first to dominate are normal processes, giving rise to the hydrodynamic regime. The temperature window where this regime appears is typically narrow, but significant effort in thermal-transport physics, quantum transport, and materials science is devoted to identifying materials where hydrodynamic behavior is present, and especially to pushing this behavior toward room temperature for technological applications. Within the hydrodynamic regime, crystal momentum is conserved, heat flux is not dissipated, and a peak in the thermal conductivity is typically observed. At even higher temperatures, the system enters the diffusive or kinetic regime, where intrinsic dissipative processes---such as Umklapp scattering and isotope scattering due to mass fluctuations in the lattice \cite{omini1997heat,fugallo2013ab}---dominate. These processes relax momentum, dissipate heat flux, and lead to a decrease in thermal conductivity with increasing temperature. In passing, we note that in some conductors, electron-phonon coupling can also contribute non-negligibly to thermal conductivity \cite{liao2015significant}.\\
The energy flux generated in response to a temperature gradient determines the thermal conductivity, while the crystal-momentum flux generated by a drift-velocity gradient defines the thermal viscosity (for the analogous quantity in electronic hydrodynamics, see Ref. \cite{rice1967theory}). To formalize this connection, we consider a crystal in the hydrodynamic regime, carrying a finite amount of crystal momentum. The corresponding local-equilibrium phonon distribution is the drifting distribution \eqref{drifting_distribution}. Expanding the deviation from equilibrium \eqref{expansion_n_bar_n_delta_n_main} in proximity of the local equilibrium \eqref{drifting_distribution} yields \cite{simoncelli2020generalization,hardy1970phonon}:
\begin{equation} \label{expansion_of_deviations_from_eq}
\begin{split}
\Delta n_{\nu}(\boldsymbol{R},t)
&= 
\left(\frac{\partial \bar{n}^{D}_{\nu}}{\partial T}\right)_{\mathrm{eq}} [T(\boldsymbol{R},t) - \bar{T}]
+ 
\left(\frac{\partial \bar{n}^{D}_{\nu}}{\partial \boldsymbol{u}}\right)_{\mathrm{eq}} \cdot \boldsymbol{u}(\boldsymbol{R},t)
+\Delta n^{\delta}_{\nu}(\boldsymbol{R},t) \\
&=\Delta{n}^{T}_{\nu}(\boldsymbol{R},t) +\Delta n^{D}_{\nu}(\boldsymbol{R},t) +\Delta n^{\delta}_{\nu}(\boldsymbol{R},t),
\end{split}
\end{equation}
where $\Delta{n}^{T}_{\nu}$ arises from temperature variations, $\Delta{n}^{D}_{\nu}$ from the local drift velocity, and $\Delta n^{\delta}_{\nu}$ represents the residual non-equilibrium contribution that cannot be mapped onto a local-equilibrium state. For simplicity, the subscript “eq” indicates evaluation at equilibrium ($T=\bar{T}$, $\boldsymbol{u}=0$), and is omitted henceforth.\\
Considering the steady-state regime and linearizing the LBTE around small, constant gradients of $T$ and $\boldsymbol{u}$, we substitute Eq. \eqref{expansion_of_deviations_from_eq} into Eq. \eqref{LBTE}, keeping only linear terms in the temperature and drift-velocity gradients:
\begin{equation}
\boldsymbol{v}_{\nu}\cdot
\left[
\frac{\partial \bar{n}_{\nu}}{\partial T} \nabla T
+
\frac{\partial \bar{n}^{D}_{\nu}}{\partial \boldsymbol{u}}\cdot\nabla\boldsymbol{u}
\right]
=
-\frac{1}{\mathcal{V}}\sum_{\nu'}\Omega_{\nu\nu'}
\left[\Delta n^{T}_{\nu'} + \Delta n^{D}_{\nu'} + \Delta n^{\delta}_{\nu'}\right].
\label{LBTE_linearized}
\end{equation}
Following Refs. \cite{hardy1965lowest,hardy1970phonon,chaput2013direct,krumhansl1965thermal,cepellotti2016relaxons} we use the symmetrized scattering matrix introduced in Eq. \eqref{symmetric_transformation},
\begin{equation}
\tilde{\Omega}_{\nu\nu'} =
\Omega_{\nu\nu'}
\sqrt{\frac{\bar{n}_{\nu'}(\bar{n}_{\nu'} + 1)}
{\bar{n}_{\nu}(\bar{n}_{\nu} + 1)}},
\end{equation}
and define the symmetrized distribution $\Delta\tilde{n}_{\nu} = \Delta n_{\nu}/\sqrt{\bar{n}_{\nu}(\bar{n}_{\nu}+1)}$, which makes $\tilde{\Omega}$ Hermitian and, hence, diagonalizable in the relaxon basis \cite{cepellotti2016relaxons} (see also section \ref{relaxons_section}). Using parity symmetry, even functions satisfy $f_{\nu}=f_{-\nu}$ (e.g., for $\hbar\omega_{\nu}$), while odd functions satisfy $f_{\nu}=-f_{-\nu}$ (e.g., for $\boldsymbol{v}_{\nu}$). It follows that $\Delta\tilde{n}^{T}_{\nu}$ is even, $\Delta\tilde{n}^{D}_{\nu}$ is odd. Moreover, since the eigenvectors of the scattering matrix have a well-defined parity \cite{hardy1970phonon} (note that $\Omega_{\nu\nu'}=\Omega_{-\nu-\nu'}$), we can also decompose $\Delta\tilde{n}^{\delta}_{\nu}$ into even and odd parts: $\Delta\tilde{n}^{\delta}_{\nu} = \Delta\tilde{n}^{\delta,E}_{\nu} + \Delta\tilde{n}^{\delta,O}_{\nu}$. At steady state, Eq. \eqref{LBTE_linearized} separates into two independent equations for the two parities. The odd part describes the response to a temperature gradient:
\begin{equation} \label{odd_part}
\boldsymbol{v}_{\nu}\sqrt{\bar{n}_{\nu}(\bar{n}_{\nu}+1)}
\cdot\frac{\partial \bar{n}_{\nu}}{\partial T}\nabla T
= -\frac{1}{\mathcal{V}}\sum_{\nu'}\tilde{\Omega}_{\nu\nu'}\Delta\tilde{n}^{\delta,O}_{\nu'},
\end{equation}
where $\Delta\tilde{n}^{\delta,O}_{\nu}$ represents the odd non-equilibrium phonon population generated in response to a temperature gradient. In writing Eq. \eqref{odd_part} we used the fact that $\Delta \tilde{n}^{D}_{\nu}$ is an eigenvector with zero eigenvalue of the scattering matrix including only normal scattering processes, $\tilde{\Omega}^{N}$ and that $\frac{1}{\mathcal{V}}\sum_{\nu'}\tilde{\Omega}_{\nu\nu'}^{U}\Delta \tilde{n}_{\nu}^{D}\simeq0$ at both high and low temperatures ($\tilde{\Omega}^{U}$ being the scattering matrix including only Umklapp processes), because at high temperatures the strong crystal-momentum dissipation ensures $\Delta \tilde{n}_{\nu}^{D}\propto\boldsymbol{u}\approx0$ and at low temperatures Umklapp processes are much less frequent than normal processes. The even part describes the response to a drift-velocity gradient:
\begin{equation}
\boldsymbol{v}_{\nu}\sqrt{\bar{n}_{\nu}(\bar{n}_{\nu}+1)}
\cdot\frac{\partial \bar{n}^{D}_{\nu}}{\partial \boldsymbol{u}}\cdot\nabla\boldsymbol{u}
= -\frac{1}{\mathcal{V}}\sum_{\nu'}\tilde{\Omega}_{\nu\nu'}\Delta\tilde{n}^{\delta,E}_{\nu'},
\end{equation}
where $\Delta\tilde{n}^{\delta,E}_{\nu}$ is the even component generated by the drift-velocity gradient. Similarly, in this case we used the fact that $\Delta \tilde{n}^{T}_{\nu}$ is related to a special eigenvector with zero eigenvalue of the total scattering matrix (see also Eq. \eqref{Bose_Einstein_eigenvector}), also referred to as the Bose-Einstein eigenvector \cite{simoncelli2020generalization}.

\subsection{Thermal viscosity} \label{thermal_viscosity}

As anticipated above, hydrodynamic thermal transport arises when most collisions between phonon wave packets conserve crystal momentum. This occurs, for example, when the mean free path for normal collisions $\Lambda_N$ is much smaller than both the boundary scattering length $L_S$ (the sample size for a single crystal or the grain size for a polycrystal) and the mean free path for Umklapp collisions $\Lambda_U$: $\Lambda_N \ll L_S, \Lambda_U$ \cite{gurzhi1968hydrodynamic,cepellotti2015phonon}. Under these conditions, the local equilibrium can be described in terms of the local fields $T(\boldsymbol{R},t)$ and $\boldsymbol{u}(\boldsymbol{R},t)$, along with four special eigenvectors \cite{simoncelli2020generalization}: $\phi^0_\nu$, the eigenvector of the full LBTE scattering matrix with zero eigenvalue, and $\phi^i_\nu$ ($i=1,2,3$), the eigenvectors with zero eigenvalue of the LBTE scattering matrix including only normal scattering processes. For the discussion that follows, it is useful to recall the notation: we denote by $\theta_{\nu}$ the eigenvectors of the full scattering matrix $\tilde{\Omega}_{\nu\nu'} = \tilde{\Omega}^{U}_{\nu\nu'} + \tilde{\Omega}^{N}_{\nu\nu'}$, and by $\phi_{\nu}$ the eigenvectors of the scattering matrix including only normal processes, $\tilde{\Omega}^{N}_{\nu\nu'}$. Because energy is conserved by both normal and Umklapp processes, the energy eigenvector $\phi^{0}_{\nu}$ is also an eigenvector of the full scattering matrix, so that $\phi^{0}_{\nu} \equiv \theta^{0}_{\nu}$. These four special eigenvectors are part of the eigenvectors of the LBTE scattering matrix: four special relaxons \cite{cepellotti2016relaxons} (see section \ref{relaxons_section}). They define collective phonon excitations with well-defined relaxation times. Recently, thermal conductivity was reformulated as a sum over relaxons, showing that only odd relaxons contribute to conductivity \cite{cepellotti2016relaxons,simoncelli2020generalization}, where odd refers to the parity of the linearized scattering matrix $\tilde{\Omega}_{\nu\nu'}$. In contrast, as will become clearer later in the text, even relaxons define thermal viscosity in the hydrodynamic regime \cite{simoncelli2020generalization} (see Fig. \ref{fig:relaxon_figure}). These four special relaxons are distinguished by their zero eigenvalue, representing conserved quantities in the system: $\phi^0_\nu$ corresponds to energy conservation (satisfying $\tilde{\Omega}\phi^{0}_{\nu}=0$), while $\phi^i_\nu$ ($i=1,2,3$) corresponds to crystal momentum conservation in the three Cartesian directions (satisfying $\tilde{\Omega}^{N}\phi^{i}_{\nu}=0$). In particular, the explicit expression of the Bose-Einstein eigenvector is \cite{simoncelli2020generalization}
\begin{equation} \label{Bose_Einstein_eigenvector}
\phi^0_\nu =  
\sqrt{\frac{\bar{n}_\nu (\bar{n}_\nu + 1)}{k_{\text{B}} \bar{T}^2 C}} \, \hbar \omega_\nu
= \sqrt{\frac{k_{\text{B}} \bar{T}^2}{C \bar{n}_\nu (\bar{n}_\nu + 1)}} \, \frac{\partial \bar{n}_\nu}{\partial T},
\end{equation}
where the specific heat $C$ is defined in Eq. \eqref{specific_heat}, while the three eigenvectors of the normal scattering matrix $\tilde{\Omega}^N_{\nu\nu'}$ with zero eigenvalues \cite{simoncelli2020generalization} are
\begin{equation}
\phi^i_\nu = 
\sqrt{\frac{k_{\text{B}} \bar{T} }{A_i\bar{n}_\nu (\bar{n}_\nu + 1)}} \frac{\partial \bar{n}^{D}_\nu}{\partial u_i} 
= \sqrt{\frac{\bar{n}_\nu (\bar{n}_\nu + 1)}{k_{\text{B}} \bar{T} A_i}} \, \hbar q_i,
\end{equation}
where $i=1,2,3$, and 
\begin{equation}
A_{i} = \left(\frac{\partial P_{i}}{\partial u_{i}}\right)_{\mathrm{eq}}
= \frac{1}{k_{B}\bar{T}\mathcal{V}}\sum_{\nu} \bar{n}_{\nu}(\bar{n}_{\nu}+1)(\hbar q_{i})^{2}
\end{equation}
is the specific momentum \cite{simoncelli2020generalization}.\\
The deviation $\Delta\tilde{n}^{\delta,E}_{\nu}$ gives rise to a crystal-momentum flux \cite{gurzhi1968hydrodynamic,hardy1970phonon}:
\begin{equation}
\Pi^{\delta E}_{ij} = \frac{1}{\mathcal{V}}\sum_{\nu} \hbar q_{i} v^{j}_{\nu}
\sqrt{\bar{n}_{\nu}(\bar{n}_{\nu}+1)}\Delta\tilde{n}^{\delta,E}_{\nu}.
\end{equation}
Analogously to viscous electronic flow \cite{rice1967theory}, the local relationship between the crystal-momentum flux and the gradient of the drift velocity defines the thermal viscosity tensor:
\begin{equation} \label{muijkl}
\eta_{ijkl}=\sqrt{{A_{i}}{A_{k}}}
\sum_{\alpha>0}
{w^{j}_{i\alpha} w^{l}_{k\alpha} }\, \tau_{\alpha},
\end{equation}
where $\tau_{\alpha}$ is the relaxation time associated with relaxon $\alpha$, and $w^{j}_{i\alpha} = (1/\mathcal{V})\sum_{\nu}\phi^{i}_{\nu}v^{j}_{\nu}\phi^{\alpha}_{\nu}$ is the velocity tensor for relaxon $\alpha$ with eigenvectors $\phi^{i}_{\nu}$ and $\phi^{\alpha}_{\nu}$ (see Eq. \eqref{relaxon_velocity}). The viscosity tensor defined in Eq. \eqref{muijkl} has the dimensions of a dynamic viscosity (Pa·s), and obeys the symmetries discussed in Ref.~\cite{dragavsevic2023viscous}. It quantifies the dissipative transfer of crystal momentum induced by gradients of the phonon drift velocity, and plays the same role for phonon hydrodynamics as the shear viscosity in real fluids. \\
Having established a microscopic definition of the thermal viscosity and its connection to the relaxon spectrum of the scattering matrix, we can now proceed to formulate the corresponding mesoscopic VHE. These equations couple the evolution of the local temperature and phonon drift velocity fields, providing a continuum-level model that captures both diffusive and hydrodynamic regimes within a unified theoretical framework.

\subsection{LBTE projection onto the special relaxons subspace}

Ref. \cite{simoncelli2020generalization} shows how the LBTE dynamics can be projected onto the basis of eigenvectors (relaxons) of the normal scattering matrix $\tilde{\Omega}^N_{\nu\nu'}$, which can be diagonalized as
\begin{equation} \label{normal_diag}
\frac{1}{\mathcal{V}} \sum_{\nu'} \tilde{\Omega}^N_{\nu\nu'} \phi^\beta_{\nu'} = \frac{1}{\tau^N_\beta} \phi^\beta_\nu, \quad \beta \ge 0,
\end{equation}
where the first four eigenvectors ($\beta=0,1,2,3$) correspond to the special eigenvectors introduced above. The deviation from equilibrium can then be expressed as a linear combination of these eigenvectors:
\begin{equation}
\Delta\tilde{n}_\nu(\boldsymbol{R},t) = \sum_\beta z_\beta(\boldsymbol{R},t) \phi^\beta_\nu.
\label{LBTE_expansion}
\end{equation}
This representation allows one to exploit the analytical knowledge of the first four eigenvectors to derive mesoscopic transport equations. In this basis, the LBTE reads
\begin{equation}
\sum_\beta \frac{\partial z_\beta(\boldsymbol{R},t)}{\partial t} \phi^\beta_\nu + \boldsymbol{v}_\nu \cdot \sum_\beta \nabla z_\beta(\boldsymbol{R},t) \phi^\beta_\nu
= -\sum_{\beta>3} \frac{z_\beta(\boldsymbol{R},t)}{\tau^N_\beta} \phi^\beta_\nu - \frac{1}{\mathcal{V}} \sum_{\nu',\beta>0} z_\beta(\boldsymbol{R},t) \tilde{\Omega}^U_{\nu\nu'} \phi^\beta_{\nu'}.
\label{LBTE_normal_basis}
\end{equation}
The first step to move from the microscopic to the mesoscopic domain is to project Eq. \eqref{LBTE_normal_basis} onto the special Bose-Einstein eigenvector $\phi^0_\nu$, which is equivalent to calculating the energy moment of the LBTE. Conceptually, this involves (i) expressing the deviation from equilibrium in terms of temperature and drift-velocity contributions associated with the first four eigenvectors, plus out-of-equilibrium terms for subsequent eigenvectors, (ii) taking the scalar product with $\phi^0_\nu$ to obtain the energy balance equation, and (iii) expressing the heat flux as contributions from the drift velocity and from deviations from local equilibrium, relating the latter to the thermal conductivity. Following this procedure, one obtains the first mesoscopic viscous heat equation for the temperature field:
\begin{equation} \label{VHE1}
C \frac{\partial T(\boldsymbol{R},t)}{\partial t} 
+ \sum_{i,j=1}^{3} W^i_{0j} \sqrt{T \bar{A}_jC} \frac{\partial u_j(\boldsymbol{R},t)}{\partial R_{i}} 
- \sum_{i,j=1}^{3} \kappa^{D}_{ij} \frac{\partial^2 T(\boldsymbol{R},t)}{\partial R_{i} \partial R_{j}} = 0,
\end{equation}
where $W^0_{ij}$ is the velocity tensor and $\kappa^{D}_{ij}$ is the thermal conductivity contribution from diffusion-damped relaxons \cite{dragavsevic2023viscous}, whose derivation will be presented in detail in section \ref{diffusion_damped_relaxons}. We emphasize that the right-hand side of Eq. \eqref{VHE1} is zero because the system considered here does not exchange energy with any localized external heat source \cite{allen2018temperature}. This equation can be seen as a natural extension of Fourier’s law of heat conduction, 
\begin{equation} \label{FourieR_{l}aw}
Q_{i} = -\sum_j \kappa_{ij} \nabla_j T
\end{equation}
with $\boldsymbol{Q}$ being the heat flux. For instance, in the steady-state regime, it reduces to the Laplace equation for the temperature — i.e., the Laplacian of $T$ equals zero — which is exactly what one obtains if the term associated with the phonon drift velocity $\boldsymbol{u}$ in Eq. \eqref{VHE1} is neglected.\\
Similarly, projecting the LBTE onto the three momentum eigenvectors $\phi^i_\nu$ ($i=1,2,3$) is equivalent to calculating the momentum moment of the LBTE. This requires (i) expressing the momentum flux in terms of contributions from the local equilibrium temperature and the out-of-equilibrium drift, (ii) taking the scalar product with each $\phi^i_\nu$ to derive a set of three coupled equations for the components of $\boldsymbol{u}(\boldsymbol{R},t)$, (iii) relating the out-of-equilibrium contributions to the viscosity tensor and the spatial derivatives of the drift velocity, and (iv) including weak Umklapp dissipation which vanishes in the ideal hydrodynamic limit. This procedure yields the vectorial viscous heat equation for the drift velocity components:
\begin{equation} \label{VHE2}
A_i \frac{\partial u_i(\boldsymbol{R},t)}{\partial t} 
+ \sqrt{\frac{C A_i}{\bar{T}}} \sum_{j=1}^{3}  W^j_{i0} \frac{\partial T(\boldsymbol{R},t)}{\partial R_{j}} 
- \sum_{j,k,l=1}^{3} \eta_{ijkl} \frac{\partial^2 u_k(\boldsymbol{R},t)}{\partial R_{j} \partial R_{l}} 
= - \sum_{j=1}^{3} \sqrt{A_iA_j} D^U_{ij} u_j(\boldsymbol{R},t),
\end{equation}
where $D^U_{ij}$ is the momentum dissipation rate caused by Umklapp processes, which can also include contributions from boundary scattering. For example, if boundary scattering is included at a first approximation as in Ref. \cite{fugallo2014thermal} over a characteristic length $L_S$, then $D^U_{ij} = D^U_{ij;\mathrm{bulk}} + D^U_{ij;\mathrm{boundary}}(L_S)$; a more detailed discussion of finite-size effects is provided in section \ref{finite_size_effects}. Note that the velocity tensor $W^j_{\alpha\beta}$ arising from the nondiagonal form of the diffusion operator in the basis of the eigenvectors of the normal part of the scattering matrix is indeed different from the general velocity tensor $w^j_{\alpha\beta}$ which is obtained considering the eigenvectors of the full scattering matrix, including both Umklapp and normal scattering processes. \\
Eqs. \eqref{VHE1} and \eqref{VHE2} together are known as the viscous heat equations (VHE) \cite{simoncelli2020generalization}. In particular, Eq. \eqref{VHE2} closely resembles the Navier-Stokes momentum equation (in its convective form) in the linear, laminar regime—namely, the regime in which the nonlinear inertial term $(\boldsymbol{u}\!\cdot\!\nabla)\boldsymbol{u}$ is absent. As a consequence, the VHE describe a phonon fluid with zero Reynolds number, i.e. an ideal, perfectly laminar flow. The VHE obtained by coarse-graining the LBTE are mesoscopic coupled partial differential equations that, besides temperature, include a phonon fluid drift velocity, enabling a direct analogy with the pressure and velocity of classical fluids. The VHE describe hydrodynamic, diffusive, and intermediate regimes, greatly reduce computational cost versus full LBTE, and provide transparent physical interpretation. LBTE's limitations in handling complex geometries hinder practical predictions, whereas the VHE accommodate shape and boundary effects efficiently and have been also benchmarked against spatially resolved LBTE results in micrometer-scale devices \cite{dragavsevic2023viscous}. Consequently, the problem is recast as a set of mesoscopic partial differential equations that can be solved to simulate devices with complex geometries, provided that physically meaningful boundary conditions are specified for the unknown temperature and drift-velocity fields. Appropriate boundary conditions are required: for temperature, standard Dirichlet or Neumann conditions \cite{logan2014applied} can be employed; for the drift velocity, one may impose slip conditions ($u_{\parallel}\neq 0$, $u_{\perp}=0$), which allow only longitudinal drifting flux at the boundaries, or no-slip conditions ($u_{\parallel}=u_{\perp}=0$), which enforce zero drifting heat flux. Most importantly, the VHE retain the predictive and quantum-mechanical character of the BTE because they are a coarse-grained form of it, and all transport coefficients appearing in Eqs. \eqref{VHE1} and \eqref{VHE2} can be computed from first-principles, without relying on experimental inputs or fitting procedures. In particular, the thermal conductivity, the thermal viscosity, and the remaining coefficients are obtained from an ab initio solution of the LBTE in the relaxon formalism (see e.g. Refs. \cite{simoncelli2020generalization,dragavsevic2023viscous}), computed once for the bulk material. These coefficients are then used to parameterize the VHE, which can subsequently describe different equilibrium temperatures and device geometries at the mesoscopic scale.

\subsubsection{Momentum and diffusion-damped conductivity contributions} \label{diffusion_damped_relaxons}

It has been shown in Refs. \cite{cepellotti2016relaxons,simoncelli2020generalization} that only odd relaxons contribute to thermal conductivity. Going more into details, the thermal conductivity that parametrizes the macroscopic treatment, the first VHE (see Eq. \eqref{VHE1}), is actually coming only from a subset of diffusion-damped relaxons \cite{dragavsevic2023viscous}. To do this we rewrite the expression for the thermal conductivity written in the relaxon basis, originally introduced in Ref. \cite{cepellotti2016relaxons}. Relaxons $\{\theta_{l\nu}\}$ are the eigenvectors of the full LBTE scattering operator, $\tilde{\Omega}_{\nu\nu'} = \tilde{\Omega}^{U}_{\nu\nu'} + \tilde{\Omega}^{N}_{\nu\nu'}$. In this representation, the conductivity tensor reads
\begin{equation} \label{kappa_relaxon}
\kappa_{ij}
= C \sum_{l=1}^{\mathcal{N}}
\bigl\langle \theta_{\nu}^{0} \big| v_{\nu}^{i} \big| \theta_{\nu}^{l} \bigr\rangle
\tau^{l}
\bigl\langle \theta_{\nu'}^{l} \big| v_{\nu'}^{j} \big| \theta_{\nu'}^{0} \bigr\rangle
= C \bigl\langle \theta_{\nu}^{0} \big| v_{\nu}^{i}\, \breve{\Omega}_{\nu\nu'}^{-1} v_{\nu'}^{j} \big| \theta_{\nu'}^{0} \bigr\rangle ,
\end{equation}
where $\mathcal{N}$ is the total number of relaxons, $\tau^{l}$ their relaxation times, and $\breve{\Omega}_{\nu\nu'}^{-1}=\sum_{l=1}^{\mathcal{N}}\big| \theta_{\nu}^{l} \bigr\rangle\tau^{l}\bigl\langle \theta_{\nu'}^{l} \big|$ is the inverse of the scattering matrix restricted to the non-null space, i.e., restricted to the subspace orthogonal to the Bose-Einstein eigenvector \eqref{Bose_Einstein_eigenvector} ($l$ starts from 1 in the summation). The exclusion of this energy eigenvector from the conductivity expression is well established \cite{simoncelli2020generalization,spohn2006phonon}, as it represents a locally equilibrated distribution that cannot carry heat. In fact, it can be directly verified that 
\begin{equation}
\langle \theta^{0}_{\nu} | v^{i}_{\nu} | \theta^{0}_{\nu} \rangle \propto 
\sum_{s} \int_{\mathrm{BZ}} \omega^{2}_{\boldsymbol{q}s}\, v^{i}_{\boldsymbol{q}s}\, d^{3}q = 0,
\end{equation}
since the integrand is an odd function of $\boldsymbol{q}$ and the integral is taken over the symmetric Brillouin zone BZ. Eq. \eqref{kappa_relaxon} has been shown to reproduce the thermal conductivity obtained from variational and other exact LBTE solvers \cite{fugallo2013ab,cepellotti2016relaxons,dragavsevic2023viscous}, validating the relaxon formalism. For clarity, in what follows we write explicitly the sums over relaxon indices. Both the eigenvectors of the full collision matrix $\{\theta^{l}_{\nu}\}$ and those of the normal-only operator $\{\phi^{m}_{\nu}\}$ form complete bases for the space of phonon distributions. Thus, the identity operator admits the two equivalent decompositions
\begin{equation} 
\hat{1}_{\nu\nu'} =
|\theta^{0}_{\nu}\rangle\langle\theta^{0}_{\nu'}| + \sum_{l=1}^{\mathcal{N}}|\theta^{l}_{\nu}\rangle\langle\theta^{l}_{\nu'}|
= 
|\phi^{0}_{\nu}\rangle\langle\phi^{0}_{\nu'}| + \sum_{\ell=1}^{3}|\phi^{\ell}_{\nu}\rangle\langle\phi^{\ell}_{\nu'}|
+ \sum_{\iota=4}^{\mathcal{N}}|\phi^{\iota}_{\nu}\rangle\langle\phi^{\iota}_{\nu'}|.
\end{equation}
Here $\mathcal{N}=3N_{\rm at}N_{q}$, with $N_{q}$ the number of sampled wavevectors and $N_{\rm at}$ the number of atoms in the unit cell. The Bose-Einstein eigenvector responsible for energy conservation represents local equilibrium \cite{simoncelli2020generalization,allen2018temperature,di2025broadening} and so it does not contribute to heat conduction or viscosity. To obtain the linear-response solution, the LBTE is projected onto the subspace orthogonal to the energy mode. We therefore introduce the projector onto this subspace,
\begin{equation} \label{projector}
\hat{P}_{\nu\nu'}^{\perp 0} = 
\sum_{l=1}^{\mathcal{N}}|\theta^{l}_{\nu}\rangle\langle\theta^{l}_{\nu'}|
= 
\underbrace{\sum_{\ell=1}^{3}|\phi^{\ell}_{\nu}\rangle\langle\phi^{\ell}_{\nu'}|}_{\hat{P}_{\nu\nu'}^{M}}
+
\underbrace{\sum_{\iota=4}^{\mathcal{N}}|\phi^{\iota}_{\nu}\rangle\langle\phi^{\iota}_{\nu'}|}_{\hat{P}_{\nu\nu'}^{D}},
\end{equation}
which separates the three “momentum” eigenvectors of normal processes ($\hat{P}_{\nu\nu'}^{M}$) from the remaining $(\mathcal{N}-4)$ eigenvectors ($\hat{P}_{\nu\nu'}^{D}$), hereafter referred to as diffusion-damped modes because, as it will become clear later, they generate the Laplacian (diffusive) contributions in the VHE. Inserting the decomposition \eqref{projector} into the conductivity \eqref{kappa_relaxon}, we obtain
\begin{equation} \label{kappa_projected}
\begin{split}
\kappa_{ij} &=C \langle \theta_{\nu}^{0} | v_{\nu}^{i} \hat{P}_{\nu\nu'}^{\perp0}\breve{\Omega}^{-1}\hat{P}_{\nu''\nu'''}^{\perp0} v_{\nu}^{j} | \theta_{\nu'''}^{0} \rangle \\
&\approx
C \langle \theta_{\nu}^{0} | v_{\nu}^{i} \hat{P}_{\nu\nu'}^{M}\breve{\Omega}^{-1}\hat{P}_{\nu''\nu'''}^{M} v_{\nu}^{j} | \theta_{\nu'''}^{0} \rangle +C \langle \theta_{\nu}^{0} | v_{\nu}^{i} \hat{P}_{\nu\nu'}^{D}\breve{\Omega}^{-1}\hat{P}_{\nu''\nu'''}^{D} v_{\nu}^{j} | \theta_{\nu'''}^{0} \rangle 
\end{split}
\end{equation}
where mixed blocks ($MD$ and $DM$) can be neglected. This follows from the fact that (i) in the LBTE framework, zero eigenvalues correspond to conserved quantities; (ii) because the normal-scattering operator conserves only energy and momentum, the projected matrix  $\breve{\Omega}^{N}_{\perp} = \hat{P}^{\perp 0}\tilde{\Omega}^{N}\hat{P}^{\perp 0}$ possesses a finite spectral gap $g$ separating the three momentum modes (subspace $M$) from all remaining modes (subspace $D$); otherwise, additional conserved quantities would exist; (iii) Umklapp scattering acts as a small perturbation relative to this gap, i.e.\ $\lambda/g \ll 1$, where $\lambda = \|\breve{\Omega}^{U}\|$. In this regime, a unitary Schrieffer-Wolff (SW) transformation \cite{schrieffer1966relation,bravyi2011schrieffer} can be applied to systematically block-diagonalize the full collision operator \cite{dragavsevic2023viscous}. After the SW transformation of the collision matrix, and noting that it is symmetric, positive definite, and defined in partitioned form, its inverse can be determined using the expression for the inverse of a partitioned matrix \cite{horn1994topics}, obtaining:
\begin{equation}
\breve{\Omega}'^{-1}=
\begin{pmatrix}
\left[\breve{\Omega}'_{MM}\right]^{-1} + \mathcal{O}\left(\frac{\lambda^{2}}{g^{3}}\right) & \mathcal{O}\left(\frac{\lambda}{g^{2}}\right) \\
\mathcal{O}\left(\frac{\lambda}{g^{2}}\right) & \left[\breve{\Omega}'_{DD}\right]^{-1} +\mathcal{O}\left(\frac{\lambda^{2}}{g^{4}}\right)
\end{pmatrix}
\end{equation}
where $\breve{\Omega}'_{MM}=[D^{U}]^{-1}+\mathcal{O}(g^{-1})$ and $\breve{\Omega}'_{DD}=\mathcal{O}(g^{-1})$. In summary, the SW procedure effectively replaces the inverse collision matrix $\Omega^{-1}$ with a block-diagonal form containing only its momentum ($MM$) and diffusion-damped ($DD$) blocks, namely $[\Omega']^{-1}$, with corrections confined to order $\mathcal{O}(\lambda/g^{2})$. This yields a clean decomposition of the conductivity into momentum and diffusion-damped contributions. The momentum contribution follows from Eq. \eqref{kappa_projected},
\begin{equation}
\begin{split}
\kappa^{MM}_{ij}&=C \langle \theta_{\nu}^{0} | v_{\nu}^{i} \hat{P}_{\nu\nu'}^{M}\breve{\Omega}^{-1}\hat{P}_{\nu''\nu'''}^{M} v_{\nu}^{j} | \theta_{\nu'''}^{0} \rangle = \sum_{k,l=1}^{3} C\, W^{i}_{0k} W^{j}_{l0}\Big([D^{U}_{kl}]^{-1}+\mathcal{O}(g^{-1})_{kl}\Big)=\\
&=\kappa_{ij}^{M}+\Delta\kappa_{ij}^{M}.
\end{split}
\end{equation} 
The leading term scales as $\lambda^{-1}$, whereas subleading corrections scale as $g^{-1}$; since $\lambda \ll g$, the former dominates and we get \cite{dragavsevic2023viscous}
\begin{equation}
\kappa^{MM}_{ij} = \kappa^{M}_{ij}+\mathcal{O}\left(\frac{\lambda}{g}\right)
\end{equation}
It follows that when the strength of Umklapp process $\lambda$ is weak compared to the spectral gap $g$, one can approximate
\begin{equation} \label{kappa_momentum}
\kappa^{MM}_{ij}\approx\kappa^{M}_{ij}=C\sum_{k,l=1}^{3}W_{0k}^{i}W_{l0}^{j}\left[D^{U}_{kl}\right]^{-1}
\end{equation}
The remaining part of the conductivity comes from the diffusion-damped modes,
\begin{equation} \label{diffusion_damped_kappa}
\kappa^{D}_{ij} = \kappa_{ij} - \kappa^{M}_{ij},
\end{equation}
and accounts for the contribution of all relaxons outside the momentum subspace. This decomposition is essential for correctly identifying the coefficients that appear in the VHE. Although the structure of the derivation is unchanged whether one works with the total conductivity $\kappa_{ij}$ or with its diffusion-damped component $\kappa^{D}_{ij}$, it is important to emphasize that only in the limit of strong Umklapp scattering—where $D^{U}$ dominates and viscous contributions become negligible—do the VHE reduce exactly to Fourier’s heat equation, with the total conductivity given by $\kappa = \kappa^{M} + \kappa^{D}$.\\
Crucially, this refinement resolves the discrepancy noted in Ref. \cite{simoncelli2020generalization}, where the effective conductivity appearing in the VHE differed numerically from the Fourier conductivity even in the fully diffusive limit. To illustrate this, let us examine the second VHE \eqref{VHE2} in the high-temperature regime, where Umklapp scattering dominates and the term proportional to $D^{U}_{ij}$ overwhelms viscous contributions. In this limit one finds the relation
$\frac{\sqrt{C A_{i}}}{T}\, W^{\,j}_{i0}\frac{\partial}{\partial\boldsymbol{R}_{j}}T(\boldsymbol{R},t)= -\,\sqrt{A_{i} A_{j} D^{U}_{ij}}\,u_{j}(\boldsymbol{R},t)$. Substituting this expression for $u_{j}$ back into the first VHE \eqref{VHE1} yields
\begin{equation}
C\,\frac{\partial T(\boldsymbol{R},t)}{\partial t}-\sum_{i,j=1}^{3}\left[C\sum_{k,l=1}^{3}W^{i}_{0k} W^{j}_{l0}\,[D^{U}_{kl}]^{-1}\right]\frac{\partial^{2}T(\boldsymbol{R},t)}{\partial R_{i}\partial R_{j}}-\sum_{i,j=1}^{3}\kappa^{D}_{ij}\frac{\partial^{2}T(\boldsymbol{R},t)}{\partial R_{i}\partial R_{j}}=0.
\end{equation}
In fact, the term in square brackets is precisely the momentum contribution $\kappa^{M}_{ij}$ defined in Eq. \eqref{kappa_momentum}. Adding the diffusion-damped part $\kappa^{D}_{ij}$ then reconstructs the full thermal conductivity $\kappa_{ij}$ of Fourier’s law, as in Eq. \eqref{diffusion_damped_kappa}.

\subsection{Poiseuille heat flow}

\begin{figure}[!htb]
\centering
\includegraphics[width=\textwidth]{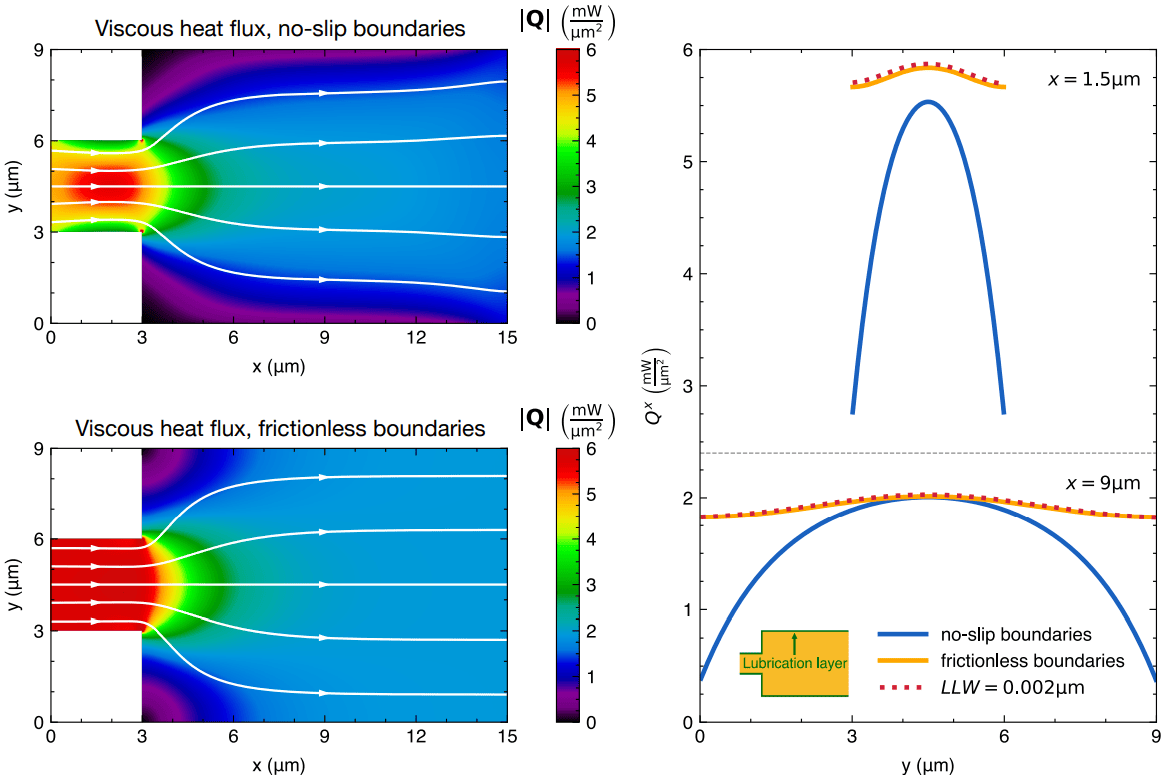}
\caption{\textbf{Influence of boundary conditions and Poiseuille heat profile of the in-plane heat-flow obtained by solving the VHE.} The panels show the $x$-$y$ heat flux in graphite, computed by solving the VHE with fixed temperatures of 80 K at $x=0 \mu$m and 60 K at $x=15 \mu$m, while all remaining edges are treated as adiabatic. The upper-left panel uses no-slip boundary conditions for the drift velocity ($\boldsymbol{u}=0$ on the boundary), whereas the lower-left panel illustrates the case of frictionless boundaries (infinite slip length $b$ in Eq. \eqref{slip_lenght}). The right panel compares the heat-flux profiles extracted along the vertical cuts at $x=1.5 \mu$m and $x=9 \mu$m for the different boundary conditions. The results demonstrate that the lubrication-layer method reproduces, to numerical precision (that is for sufficiently small lubrication layer width, $LLW$), the profiles obtained using \texttt{Mathematica}’s built-in perfectly slipping conditions in simple geometries. The insets schematically depict the placement of the lubrication layer near the boundary. \textit{Figure reproduced with permission from Ref. \cite{dragavsevic2023viscous}.}}
\label{fig:boundary_conditions_effects}
\end{figure}

Building on the analogy with classical fluid dynamics, the local energy $E(\boldsymbol{R},t)$ and crystal momentum $P_i(\boldsymbol{R},t)$ can be expressed in terms of the temperature and drift velocity via $E(\boldsymbol{R},t) = CT(\boldsymbol{R},t)$ and $P_i(\boldsymbol{R},t) = A_i u_i(\boldsymbol{R},t)$. With these identifications, the VHE can be recast in the form of balance equations:
\begin{equation}
\frac{\partial E(\boldsymbol{R},t)}{\partial t} + \nabla \cdot [\boldsymbol{Q}^\delta(\boldsymbol{R},t) + \boldsymbol{Q}^D(\boldsymbol{R},t)] = 0,
\end{equation}
\begin{equation} \label{momentum_balance}
\frac{\partial P_i(\boldsymbol{R},t)}{\partial t} + \sum_j \frac{\partial \Pi^{T}_{ij}(\boldsymbol{R},t)}{\partial R_{j}} + \sum_j \frac{\partial \Pi^{\delta E}_{ij}(\boldsymbol{R},t)}{\partial R_{j}} = - \left( \frac{\partial P_i}{\partial t} \right)_{\mathrm{Umklapp}},
\end{equation}
where, following the phonon expansion in Eq. \eqref{expansion_of_deviations_from_eq} the heat flux splits into contributions. The first is related to the temperature gradient,
\begin{equation} \label{delta_flux}
Q^{\delta}_{i}(\boldsymbol{R},t) = -\sum_j \kappa^{D}_{ij} \nabla_j T(\boldsymbol{R},t),
\end{equation}
and has the same mathematical form as Fourier's law but a distinct physical meaning. This is because since the temperature entering Eq. \eqref{delta_flux} is the viscous temperature obtained from the VHE and so coupled to a drift velocity rather than a diffusive temperature. The second contribution is a drifting flux associated with the drift velocity,
\begin{equation} \label{drifting_flux}
Q^{D}_{i}(\boldsymbol{R},t) = \sum_j W^i_{0j} \sqrt{\bar{T} A_j C} \, u_j(\boldsymbol{R},t).
\end{equation}
The total viscous heat flux is then given by $\boldsymbol{Q}(\boldsymbol{R},t)=\boldsymbol{Q}^{D}(\boldsymbol{R},t)+\boldsymbol{Q}^{\delta}(\boldsymbol{R},t)$. A simple validation of the VHE’s ability to capture the hydrodynamic regime—and to highlight the differences between the two heat-flux components—is provided in Ref. \cite{simoncelli2020generalization}. The simulation consists of solving the VHE for graphite at an equilibrium temperature of $\bar{T} = 70\,\mathrm{K}$, using the geometry shown in the left panels of Fig. \ref{fig:boundary_conditions_effects}, a planar configuration commonly employed as an illustrative example in fluid-dynamics textbooks. The VHE are solved numerically with a finite-element method implemented in \texttt{Mathematica} \cite{mathematica}, imposing boundary thermal baths of $80\,\mathrm{K}$ at the left edge ($x = 0\,\mu\mathrm{m}$) and $60\,\mathrm{K}$ at the right edge ($x = 15\,\mu\mathrm{m}$); all remaining boundaries are treated as adiabatic, and a no-slip condition is applied to the drift velocity $\boldsymbol{u}$ on every boundary.\\
Fig. \ref{fig:boundary_conditions_effects} shows the total viscous heat flux $\boldsymbol{Q}^\delta_\mathrm{viscous} + \boldsymbol{Q}^D_\mathrm{viscous}$ (upper-left panel of Fig. \ref{fig:boundary_conditions_effects}), as well its $x$-component along two transversal sections (right panel of Fig. \ref{fig:boundary_conditions_effects}). We emphasize here that Fourier’s law neglects the contribution to the heat flux arising from the local drift velocity.
Consequently, Fourier’s law fails to capture both qualitative and quantitative features of the heat flux profile, especially near spatial inhomogeneities such as boundaries or corners. In particular, $\boldsymbol{Q}^D$ rapidly increases (decreases) near the thermal reservoir on the hot (cold) side of the sample, while $\boldsymbol{Q}^\delta$ behaves in the opposite way. Microscopically, these variations arise from the transition of the phonon distribution from the Bose-Einstein equilibrium imposed at the boundaries to an out-of-equilibrium distribution carrying nonzero crystal momentum (i.e., a finite drift velocity) inside the sample.\\ 
The right panel of Fig. \ref{fig:boundary_conditions_effects} highlights the total heat flux along two transversal sections of the sample. Fourier’s law predicts a flat profile (not shown), whereas the VHE yield a Poiseuille-like shape, a characteristic signature of hydrodynamic transport. These results closely mimic the spatially-resolved solution of the LBTE, either in the frequency-dependent SMA approximation \cite{peraud2011efficient}, or considering the full scattering operator \cite{lindsay2016first}, which exhibits a minimum flux at the surfaces and maximum at the center. Notably, in contrast to classical fluid dynamics, the total heat flux does not vanish at the boundaries: the no-slip condition sets $\boldsymbol{Q}^D=0$, but the temperature-driven component $\boldsymbol{Q}^\delta$ can still be nonzero.\\
Ref. \cite{simoncelli2020generalization} shows that deviations from Fourier’s diffusive description are found to be negligible in silicon over the entire temperature range considered, whereas they become significant in high–thermal-conductivity materials such as graphite and diamond. In these systems, non-diffusive effects are maximized at intermediate temperatures—around $70\,\mathrm{K}$ for graphite and $225\,\mathrm{K}$ for diamond—for characteristic sample dimensions on the order of $10\,\mu\mathrm{m}$. In graphite, these predictions are consistent with experimental observations of second sound \cite{huberman2019observation} (see also section \ref{second_sound_section_after_DPLE}), indicating that the strongest deviations from Fourier behavior occur in the same temperature window where temperature waves dynamics is observed. \\
Overall, this section demonstrates that numerical solutions of the viscous heat
equations provide an efficient and predictive framework for identifying regimes where Fourier’s law breaks down and hydrodynamic transport emerges. In the following sections, this picture is further refined by constructing an analytical solution of the VHE, which offers deeper physical insight and enables a transparent interpretation of counterintuitive hydrodynamic phenomena, such as thermal backflow.

\subsubsection{Lubrication layer and frictionless boundaries} \label{lubrication}

\begin{figure}[!htb]
\centering
\includegraphics[width=\textwidth]{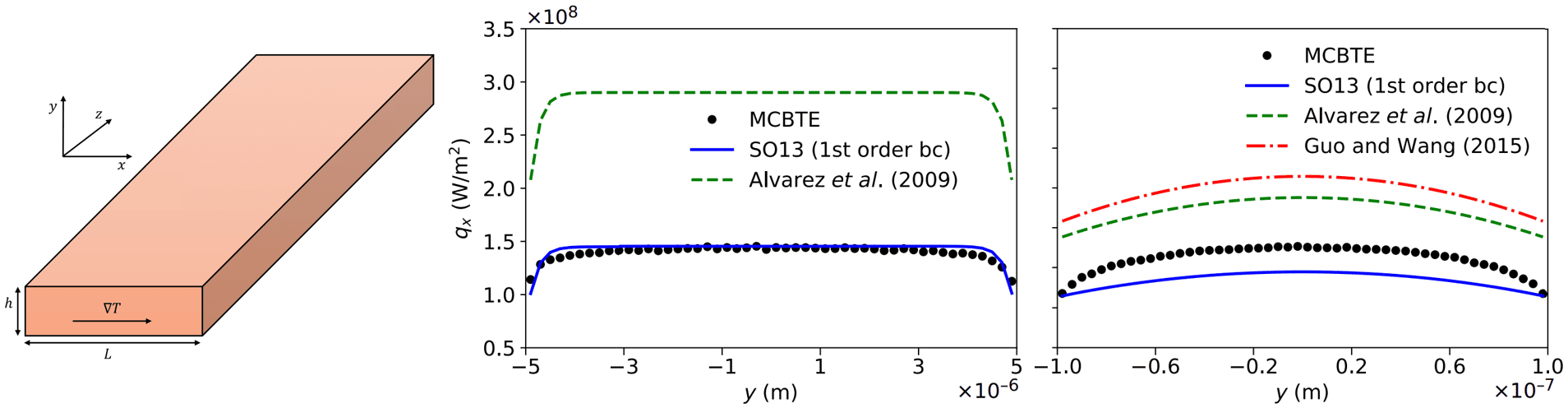}
\caption{\textbf{Comparison of the heat-flux predictions obtained from different theoretical models as discussed in Ref. \cite{yadav2025derivation}.} (Left panel) Schematic of the silicon thin-film geometry considered in Ref. \cite{yadav2025derivation}. The top and bottom surfaces (the $x$–$z$ planes at $y = \pm h/2$) are modeled as diffusively scattering boundaries ($C_1 = 1$) to account for boundary effects. A uniform temperature gradient is imposed along the $x$ direction, driving heat transport parallel to the film. The coordinate origin is located at the center of the film thickness, which extends a distance $L$ along the $x$ direction and is assumed infinite along $z$. (Central panel) Comparison of the longitudinal heat-flux profile $q_x$ obtained in Ref. \cite{yadav2025derivation} with the analytical models of Ref. \cite{alvarez2009phonon} (\textit{Alvarez et al.}) and Monte Carlo BTE (MCBTE) with $h = 10\,000\,\mathrm{nm}$ (and Knudsen number $\mathrm{Kn} = 0.02$). (Right panel) Analogous comparison to that of (central panel) including also Ref. \cite{guo2015phonon} (\textit{Guo and Wang}) with $h = 200\,\mathrm{nm}$ ($\mathrm{Kn} = 0.9$). \textit{Figure reproduced with permission from Ref. \cite{yadav2025derivation}.}}
\label{fig:yadav_figure}
\end{figure}

As anticipated above, the results discussed in the top-left panel of Fig. \ref{fig:boundary_conditions_effects} were obtained by imposing adiabatic boundary conditions for the temperature (except at the two thermal baths) together with no-slip boundary conditions for the phonon drift velocity. In this section, we extend the analysis of the same device by considering boundary configurations in which the no-slip assumption is relaxed, thereby exploring how partially or fully slipping boundaries influence the hydrodynamic features of thermal transport \cite{dragavsevic2023viscous}.\\
We begin by outlining the numerical strategy used to simulate different boundary conditions for the drift velocity. To model boundary behaviors other than the standard no-slip condition ($\boldsymbol{u}=0$ at the boundary), which corresponds to complete dissipation of crystal momentum, we introduce a lubrication layer \cite{bocquet2007flow} (see also the inset in the right panel of Fig. \ref{fig:boundary_conditions_effects}). As a general rule, the thickness of this layer, $w_{\mathrm{ll}}$, is chosen to be much smaller than the smallest characteristic length of the device \cite{dragavsevic2023viscous}. To quantify slipping effects, a finite slip length $b$ is introduced between the physical boundary of the device and its interior. The resulting auxiliary region is assigned reduced shear and rotational viscosity components (with $\eta_{\mathrm{shr}} = (\eta_{ijij} + \eta_{ijji})/2$ and $\eta_{\mathrm{rot}} = (\eta_{ijij} - \eta_{ijji})/2$). Partial slip is then modeled by  (i) reducing the shear and rotational viscosities inside the lubrication layer according to the $b$-dependent factor \cite{bocquet2007flow},
\begin{equation} \label{slip_lenght}
\frac{\eta_{\mathrm{shr}}}{\eta_{\mathrm{shr,ll}}}=\frac{\eta_{\mathrm{rot}}}{\eta_{\mathrm{rot,ll}}}=\frac{b}{w_{\mathrm{ll}}} + 1,
\end{equation}
and (ii) applying standard no-slip boundary conditions on the outer interface of the lubrication layer, i.e., the side facing the exterior of the device. By tuning the viscosity reduction and the thickness of this layer, one can continuously interpolate between the no-slip and frictionless (infinite slip-length) boundary regimes.\\
Fig. \ref{fig:boundary_conditions_effects} illustrates the effect of this procedure. As in conventional fluid dynamics, enforcing no-slip boundary conditions (upper-left panel of Fig. \ref{fig:boundary_conditions_effects}) produces a Poiseuille-like heat-flow profile as discussed above. In contrast, applying frictionless boundaries—corresponding to an infinite slip length $b$—yields a much flatter heat-flux distribution (lower-left panel of Fig. \ref{fig:boundary_conditions_effects}). The right panel of Fig. \ref{fig:boundary_conditions_effects} compares these results with \texttt{Mathematica}'s built-in perfectly slipping boundary conditions, demonstrating excellent agreement when a sufficiently small lubrication layer width (LLW $=0.002\,\mu\mathrm{m}$) is used. This validates the lubrication-layer approach as an effective tool for imposing frictionless boundary conditions in complex geometries, where a direct implementation can be significantly more challenging.\\
A comparison of the heat-flux predictions obtained from different theoretical models is discussed in Ref. \cite{yadav2025derivation} and presented in Fig. \ref{fig:yadav_figure} for film thicknesses
$h = 10\,000\,\mathrm{nm}$ and $h = 200\,\mathrm{nm}$.
The results obtained using a generalized heat-transport framework that could include linear, nonlinear, nonlocal, and relaxation terms, developed in Ref. \cite{yadav2025derivation}, are compared with the analytical solutions derived in Refs. \cite{alvarez2009phonon} and \cite{guo2015phonon} (referred to as \textit{Alvarez et al.} and \textit{Guo and Wang} in Fig. \ref{fig:yadav_figure}, respectively). In the latter two works, the heat flux in the thin film is expressed as the sum of two contributions, $q = q_b + q_w$, where the bulk component $q_b$ is obtained analytically and the wall-induced component $q_w$ is determined using first-order boundary conditions. Since both works rely on first-order boundary conditions, the heat-transport equation of Ref. \cite{yadav2025derivation} is also evaluated under the same assumptions in order to enable a consistent comparison.\\
As shown in the central panel of Fig. \ref{fig:yadav_figure}, the solution of Ref. \cite{alvarez2009phonon} captures the qualitative features of the $q_x$ profile, but exhibits quantitative deviations even in the regime $h \gg \lambda$, where the linear approach is expected to provide a reasonable description. For the thinner film, $h = 200\,\mathrm{nm}$ (right panel of Fig. \ref{fig:yadav_figure}), the predictions of both Refs. \cite{alvarez2009phonon} and \cite{guo2015phonon} yield larger heat-flux values compared to the solution of Ref. \cite{yadav2025derivation}. \\
Moreover, the equation developed in Ref. \cite{yadav2025derivation}, when supplemented with second-order boundary conditions, reproduces the heat-flux profiles obtained from Monte Carlo solutions of the BTE across both weakly and strongly nonlocal regimes. In the thick-film limit, all approaches recover the bulk Fourier heat flux in the interior, while differences emerge near the boundaries: the formulation of Ref. \cite{yadav2025derivation} accurately captures boundary-induced resistive effects. In the thin-film regime, it remains in close agreement with Monte Carlo results, yielding heat-flux profiles that are neither purely diffusive nor purely hydrodynamic.

\subsection{Deviations from diffusive behavior}

To quantify the conditions under which hydrodynamic heat conduction emerges, Ref. \cite{simoncelli2020generalization} introduces a descriptor known as the Fourier deviation number (FDN). This quantity is obtained by recasting the VHE into a dimensionless form via the Buckingham--Pi theorem \cite{buckingham1914physically}. All tensors entering the VHE are rescaled by extracting their largest components, yielding adimensional transport coefficients, while position, drift velocity, and temperature are nondimensionalized using a characteristic sample length $L$, a representative drift velocity $u_{0}$, and a temperature perturbation $\Delta T$. Inserting these into the steady-state VHE produces nondimensional equations from which three key parameters naturally arise: $\pi_{1}$, which compares drift-driven to diffusive (Fourier) heat flux; $\pi_{2}$, which measures the relative strength of temperature--drift coupling versus Umklapp momentum relaxation; and $\pi_{3}$, which compares viscous momentum transport to Umklapp dissipation and incorporates the effect of system size. These parameters determine the regimes in which the VHE reduce to Fourier’s law. The two most relevant are \cite{simoncelli2020generalization}:
\begin{equation}
\pi_{1}=\frac{\sqrt{\bar{T}A_{\mathrm{max}}C}W_{\mathrm{max}}u_{0}L}{\kappa^{D}_{\mathrm{max}}\Delta T}
\end{equation}
and
\begin{equation}
\pi_{3}=\frac{\eta_{\mathrm{max}}}{D^{U}_{\mathrm{max}}L^{2}A_{\mathrm{max}}},
\end{equation}
where each transport coefficient is represented by its maximum tensor component. When $\pi_{1}\ll1$, the coupling between drift velocity and temperature is weak, leading to Fourier-like behavior even when viscosity dominates. When $\pi_{3}\ll1$, strong Umklapp scattering or large system size suppresses viscous effects, again favoring Fourier-like transport. On the contrary, when both $\pi_{1}$ and $\pi_{3}$ are large, deviations from Fourier’s law become substantial and hydrodynamic behavior is enhanced. The FDN is defined from these two parameters as 
\begin{equation} \label{FDN}
\mathrm{FDN}=\left(\frac{1}{\pi_{1}}+\frac{1}{\pi_{3}}\right)^{-1},   
\end{equation}
and increases monotonically with the degree of departure from diffusive (Fourier) transport, providing a direct measure of hydrodynamic strength. Later on in the text, we will discuss the definition of the FDN within the analytical modeling of phonon hydrodynamics.

\subsection{Dual-phase-lag equation}

In this section we provide an analytical demonstration that the dual-phase-lag equation (DPLE) \cite{joseph1989heat,tzou1995unified}—a model frequently employed to describe thermal waves \cite{xu2002thermal,ordonez2010exact,kang2017method,gandolfi2019accessing,xu2021thermal,mazza2021thermal}—arises as a limiting case of the VHE when viscosity is neglected. To streamline the notation used in the VHE, we introduce the following compact definitions for the transport parameters, which will also be useful in the sections that follow:
\begin{equation}
\alpha_{ij} = W^{i}_{0j}\sqrt{T A_{j} C}, \qquad
\beta_{ij} = \sqrt{\frac{A_{i}C}{T}}\,W^{j}_{i0}, \qquad
\gamma_{ij} = \sqrt{A_{i} A_{j}} D^{U}_{ij},
\end{equation}
To connect the DPLE with the VHE, we start from the viscosity-free, source-free form of the VHE:
\begin{equation} \label{VHE_T_no_visc}
C\,\frac{\partial T}{\partial t}
+ \alpha_{ij}\,\frac{\partial u_{j}}{\partial x_{i}}
- \kappa^{D}_{ij}\,\frac{\partial^{2}T}{\partial x_{i}\partial x_{j}} = 0,
\end{equation}
\begin{equation} \label{VHE_u_no_visc}
A_{k}\,\frac{\partial u_{k}}{\partial t}
+ \gamma_{ik} u_{i}
+ \beta_{ik}\,\frac{\partial T}{\partial x_{i}} = 0,
\end{equation}
where \(i,k\) run over Cartesian coordinates. For simplicity, we consider a configuration where all tensors are effectively isotropic and diagonal—as occurs, for example, for in-plane transport in layered materials such as graphite or hBN \cite{simoncelli2020generalization,dragavsevic2023viscous}. In two dimensions, Eqs. \eqref{VHE_T_no_visc} and \eqref{VHE_u_no_visc} reduce to:
\begin{equation} \label{T_iso}
C\,\frac{\partial T}{\partial t} 
+ \alpha\left(\frac{\partial u_{x}}{\partial x} + \frac{\partial u_{y}}{\partial y}\right)
- \kappa^{D}\left(\frac{\partial^{2}T}{\partial x^{2}} + \frac{\partial^{2}T}{\partial y^{2}}\right)=0,
\end{equation}
\begin{equation} \label{ux_iso}
A\,\frac{\partial u_{x}}{\partial t} + \gamma\,u_{x} + \beta\,\frac{\partial T}{\partial x}=0,
\end{equation}
\begin{equation} \label{uy_iso}
A\,\frac{\partial u_{y}}{\partial t} + \gamma\,u_{y} + \beta\,\frac{\partial T}{\partial y}=0.
\end{equation}
Taking $x$- and $y$-derivatives Eqs. \eqref{ux_iso} and \eqref{uy_iso} respectively, gives:
\begin{equation} \label{duxdx}
\left(A\frac{\partial}{\partial t} + \gamma\right)\frac{\partial u_{x}}{\partial x} = -\beta\,\frac{\partial^{2}T}{\partial x^{2}}, \end{equation}
\begin{equation} \label{duydy}
\left(A\frac{\partial}{\partial t} + \gamma\right)\frac{\partial u_{y}}{\partial y} = -\beta\,\frac{\partial^{2}T}{\partial y^{2}}.
\end{equation}
Adding Eqs. \eqref{duxdx} and \eqref{duydy}, we obtain:
\begin{equation} \label{divu}
\left(A\frac{\partial}{\partial t} + \gamma\right)
\left(\frac{\partial u_{x}}{\partial x} + \frac{\partial u_{y}}{\partial y}\right)=-\beta\left(\frac{\partial^{2}T}{\partial x^{2}} + \frac{\partial^{2}T}{\partial y^{2}}\right).
\end{equation}
We now act on Eq. \eqref{T_iso} with the operator $\hat{O} = A\frac{\partial}{\partial t} + \gamma$, and use Eq. \eqref{divu} together with the identity $\kappa = \kappa^{D} + \frac{\alpha\beta}{\gamma}$ to eliminate the drift velocity, producing a closed equation for $T$:
\begin{equation} \label{DPLE_from_VHE}
\frac{C A}{\gamma}\,\frac{\partial^{2}T}{\partial t^{2}}+C\,\frac{\partial T}{\partial t}-\kappa\left(\frac{\partial^{2}T}{\partial x^{2}} + \frac{\partial^{2}T}{\partial y^{2}}\right)-\frac{\alpha\beta A}{\gamma^{2}}\frac{\partial}{\partial t}\left(\frac{\partial^{2}T}{\partial x^{2}} + \frac{\partial^{2}T}{\partial y^{2}}\right)=0.
\end{equation}
Eq. \eqref{DPLE_from_VHE} is identical to the DPLE introduced in Refs. \cite{joseph1989heat,tzou1995unified}. By comparison with the standard form of the DPLE, we identify the characteristic delay times: $\tau_{T} = \frac{A \kappa^{D}}{\kappa\gamma}$ and $\tau_{Q} = \frac{A}{\gamma}$, and the thermal diffusivity $\alpha_{E} = \frac{\kappa}{C}$. Here, $\tau_{Q}$ can be understood as the lag between the establishment of a temperature gradient and the resulting onset of heat flux, while $\tau_{T}$ quantifies the delay required for a temperature gradient to build up in response to an imposed heat flux \cite{dragavsevic2023viscous} (see also section \ref{lattice_cooling_section}). In the limits $\tau_{T}=\tau_{Q}$ or under steady-state conditions, Eq. \eqref{DPLE_from_VHE} reduces to Fourier's law of heat conduction. When $\tau_{T}=0$, it recovers Cattaneo’s hyperbolic heat equation \cite{tzou1995unified}.\\
Overall, this derivation shows that the DPLE is not an independent hydrodynamic theory, but rather represents the inviscid limit of the VHE. The temperature waves solutions predicted by the DPLE thus differ fundamentally from the viscous temperature waves characteristic of the full VHE.

\subsection{Second sound} \label{second_sound_section_after_DPLE}

In this section we show that the VHE are capable of capturing transient thermal behavior in which temperature propagates in a manner analogous to an acoustic wave. This phenomenon, known as second sound, corresponds to the coherent transport of temperature perturbations. From a macroscopic perspective, second sound arises when the temperature field satisfies a damped wave equation rather than a purely diffusive one. As anticipated above, when $\tau_{T}=0$, Eq. \eqref{DPLE_from_VHE} reduces to Cattaneo’s second sound equation \cite{tzou1995unified,cattaneo1958form}: 
\begin{equation} 
\frac{\partial^{2}T(x,y,t)}{\partial t^{2}}+\frac{\gamma}{A}\,\frac{\partial T(x,y,t)}{\partial t}-\frac{\gamma\kappa}{CA}\left(\frac{\partial^{2}T(x,y,t)}{\partial x^{2}} + \frac{\partial^{2}T(x,y,t)}{\partial y^{2}}\right)=0.
\end{equation}
As discussed in Ref. \cite{simoncelli2020generalization}, choosing $x$ as the propagation direction, the temperature obeys 
\begin{equation} \label{temperature_damped_wave}
\frac{\partial^{2}T(x,t)}{\partial t^{2}}
+ \frac{1}{\tau_{\mathrm{ss}}}\frac{\partial T(x,t)}{\partial t}
- v_{\mathrm{ss}}^{2}\frac{\partial^{2}T(x,t)}{\partial x^{2}}
= 0,
\end{equation}
where $\tau_{\mathrm{ss}}$ and $v_{\mathrm{ss}}$ are the relaxation time and the propagation velocity of the thermal wave. In contrast to Fourier’s law, which predicts an instantaneous adjustment of the heat flux to a temperature gradient, this equation ensures that a localized thermal disturbance spreads at a finite speed. Second sound can be obtained, e.g., through an approach inspired by Ref. \cite{cepellotti2017transport}. The idea is to look directly for conditions under which the microscopic fields entering the VHE behave as damped waves. To this end, one assumes for both temperature and drift velocity the ansatz of a solution of a damped-wave equation:
\begin{equation} \label{ansatz_ss_T_and_U}
\begin{split}
T(x,t) &= \bar{T} + \Delta T\, e^{i[kx - \hat{\omega}(k)t]} e^{-t/(2\hat{\tau}_{\mathrm{ss}})}, \\
u(x,t) &= u_{0}\, e^{i[kx - \hat{\omega}(k)t]} e^{-t/(2\hat{\tau}_{\mathrm{ss}})}, 
\end{split}
\end{equation}
where the complex amplitudes $\Delta T$ and $u_{0}$ allow for a phase shift between the two waves. This ansatz imposes that both fields oscillate with the same frequency and decay rate. Introducing the complex frequency  
\begin{equation} \label{tilde_omega}
\tilde{\omega}(k) = \hat{\omega}(k) - \frac{i}{2\hat{\tau}_{\mathrm{ss}}}
\end{equation}
and substituting Eqs. \eqref{ansatz_ss_T_and_U} into the VHE \eqref{VHE1} and \eqref{VHE2}, one can obtain the pair of coupled equations  
\begin{equation} \label{coupled_eq_ss}
\begin{split}
-iC\,\tilde{\omega}(k)\,\Delta T + W\sqrt{\bar{T}AC}\, ik\, u_{0} + \kappa^{D} k^{2}\Delta T &= 0, \\
-iA\,\tilde{\omega}(k)\,u_{0} + \sqrt{\frac{CA}{\bar{T}}}\, Wik\, \Delta T + \eta k^{2}u_{0} &= -A D^{U} u_{0}.
\end{split}
\end{equation}
The second of Eqs. \eqref{coupled_eq_ss} can be rearranged to express $u_{0}$ in terms of $\Delta T$:  
\begin{equation}
u_{0} = -\Delta T \frac{ik \sqrt{\frac{CA}{\bar{T}}}\, W}{\eta k^{2} + A D^{U} - iA \tilde{\omega}(k)}.
\end{equation}
Substituting this result into the first of Eqs. \eqref{coupled_eq_ss} yields a single equation for $\tilde{\omega}(k)$:  
\begin{equation} \label{quadratic}
[-iC\tilde{\omega}(k) + \kappa^{D} k^{2}]\,[A D^{U} + \eta k^{2} - iA\tilde{\omega}(k)]
+ CAW^{2}k^{2} = 0.
\end{equation}
Eq. \eqref{quadratic} is quadratic in $\tilde{\omega}(k)$ and its solution reads  
\begin{equation} \label{omega_full}
\begin{split}
\tilde{\omega}(k)=&-\frac{i}{2}\left(\frac{\eta}{A}k^{2} + D^{U} + \frac{\kappa^{D}}{C}k^{2}\right) \nonumber \\
&\mp \frac{1}{2}\sqrt{
\left(\frac{\eta}{A}k^{2} + D^{U} + \frac{\kappa^{D}}{C}k^{2}\right)^{2}
- 4\left(W^{2}k^{2} + \frac{D^{U}\kappa^{D}}{C}k^{2} + \frac{\kappa^{D} \eta}{AC}k^{4}\right)}. 
\end{split}
\end{equation} 
Since the semiclassical description applies in the long-wavelength limit, one can expand Eq. \eqref{omega_full} to lowest order in $k$, obtaining  
\begin{equation} \label{omega_simplified}
\tilde{\omega}(k) \approx -\frac{i D^{U}}{2}
\pm \sqrt{k^{2}\left(W^{2} + \frac{D^{U}\kappa^{D}}{2C} - \frac{\eta D^{U}}{2A}\right)- \left(\frac{D^{U}}{2}\right)^{2}}. 
\end{equation}
Comparing this expression with Eq. \eqref{tilde_omega} one gets  
\begin{equation}
\hat{\omega}(k) = 
\sqrt{
k^{2}\left(W^{2} + \frac{D^{U}\kappa^{D}}{2C} - \frac{\eta D^{U}}{2A}\right)
- \left(\frac{D^{U}}{2}\right)^{2}
},
\qquad
\hat{\tau}_{\mathrm{ss}} = \frac{1}{D^{U}},
\end{equation}
meaning that the decay time of second sound is controlled by the crystal-momentum dissipation rate. In the hydrodynamic limit $D^{U} \to 0$, second sounds disperses linearly with $k$, $\hat{\omega}(k) \approx W k$ and has a velocity $\hat{v}_{g}(k) = \frac{\partial \hat{\omega}}{\partial k} \approx W$ which confirms that the second-sound velocity approaches the drift-temperature coupling coefficient $W$ when momentum-relaxing processes vanish. \\
We will see that the approach developed in this section is fully consistent with the procedure that identifies the conditions under which the damped-wave equation \eqref{temperature_damped_wave} emerges directly from the VHE \eqref{VHE1} and \eqref{VHE2} (this will be discussed in detail in section \ref{second_sound_analytical}). When these conditions are met, the temperature field satisfies Eq. \eqref{temperature_damped_wave}, whose solution is given by the first of Eqs. \eqref{ansatz_ss_T_and_U}. Substituting this solution into the damped-wave equation yields the dispersion relation $\tilde{\omega}(k) = \sqrt{v_{\mathrm{ss}}^{2}k^{2} - (2\tau_{\mathrm{ss}})^{-2}}$, which allows one to express the relaxation time and propagation velocity of second sound in terms of the transport coefficients entering the VHE. In particular, \begin{equation}
\tau_{\mathrm{ss}} = \frac{CW^{2}}{\kappa^{D} D^{U^{2}} + D^{U}CW^{2}}, \qquad v_{\mathrm{ss}} = \frac{\kappa^{D} D^{U} + CW^{2}}{C W}.
\end{equation}
Because second sound is a damped phenomenon, its effective propagation speed depends on the wave vector $k$. The corresponding group velocity is
\begin{equation}
v_{g}(k) = \frac{\partial \tilde{\omega}(k)}{\partial k} = k\, v_{\mathrm{ss}} \left[k^{2} - (2\tau_{\mathrm{ss}}v_{\mathrm{ss}})^{-2}\right]^{-1/2},
\end{equation}
and it approaches the undamped propagation velocity $v_{\mathrm{ss}}$ in the limit $\tau_{\mathrm{ss}} \to \infty$. Finally, we recall the distinction introduced by Enz \cite{enz1968one} and Hardy \cite{hardy1970phonon} between “drifting” and “driftless” second sound. The phenomenon described here belongs to the drifting category, since it arises from coupled balance equations for energy and crystal momentum derived from the LBTE.\\
The work by \textit{Huberman et al.} provides the first unambiguous experimental demonstration of second sound in graphite at temperatures exceeding $100\,\mathrm{K}$ \cite{huberman2019observation}, establishing phonon hydrodynamics as a relevant transport regime well beyond the cryogenic limit.
\begin{figure}[h!]
\centering
\includegraphics[width=0.8\textwidth]{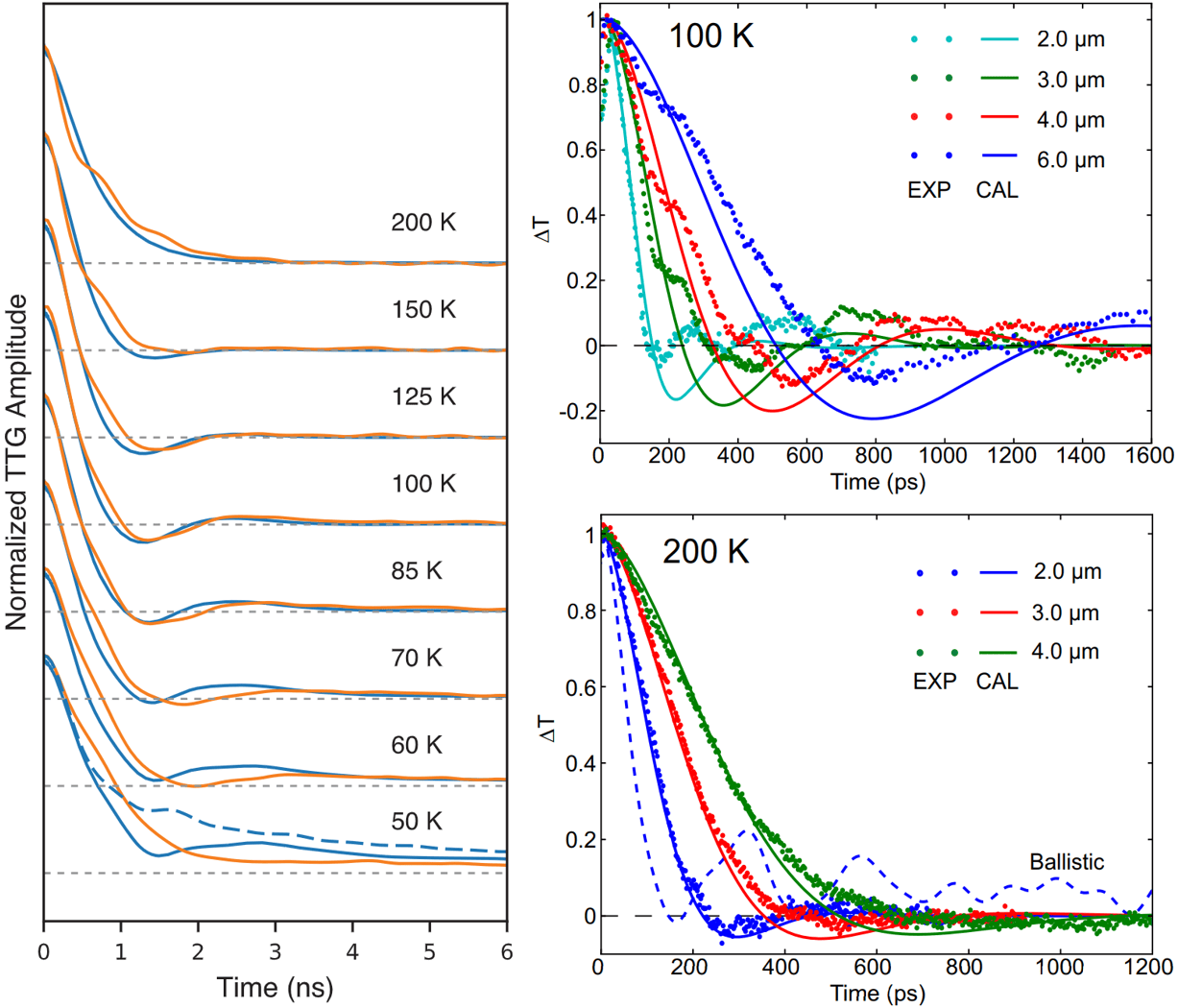}
\caption{\textbf{Temperature and transport length-scale evolution of transient thermal grating signals.} (Left panel) Time evolution of the transient thermal grating (TTG) temperature signal at several temperatures.  Experimental measurements (orange) and corresponding theoretical predictions (solid blue) are shown for several temperatures. The curve labeled 85 K refers to the measured response, while the simulation is performed at 80 K. Horizontal dashed lines mark the zero-signal reference for each temperature. The dashed blue curve at 50 K represents the calculated ballistic limit, obtained by suppressing phonon scattering in the transport model. \textit{Figure reproduced with permission from Ref. \cite{huberman2019observation}}. (Right panels) Normalized heterodyned TTG signals in graphite comparing experimental data (symbols) with first-principles simulations (solid lines) at 100 K (top) and 200 K (bottom). The appearance of a pronounced negative excursion in the signal indicates temperature waves propagation, with strength increasing at lower temperatures and longer transport length scales. \textit{Figure reproduced with permission from Ref. \cite{ding2022observation}}.}
\label{fig:ding_huberman_figure}
\end{figure}
Using transient thermal grating measurements, the authors observe a crossover from diffusive decay at high temperatures to damped oscillatory temperature dynamics at lower temperatures, accompanied by a phase inversion of the thermal grating—a hallmark of temperature waves dynamics. These oscillations exhibit a linear dispersion relation, with a propagation velocity of order $3\times10^{3}\,\mathrm{m/s}$, inconsistent with elastic acoustic modes and indicative of collective phonon flow.
\begin{figure}[h!]
\centering
\includegraphics[width=\textwidth]{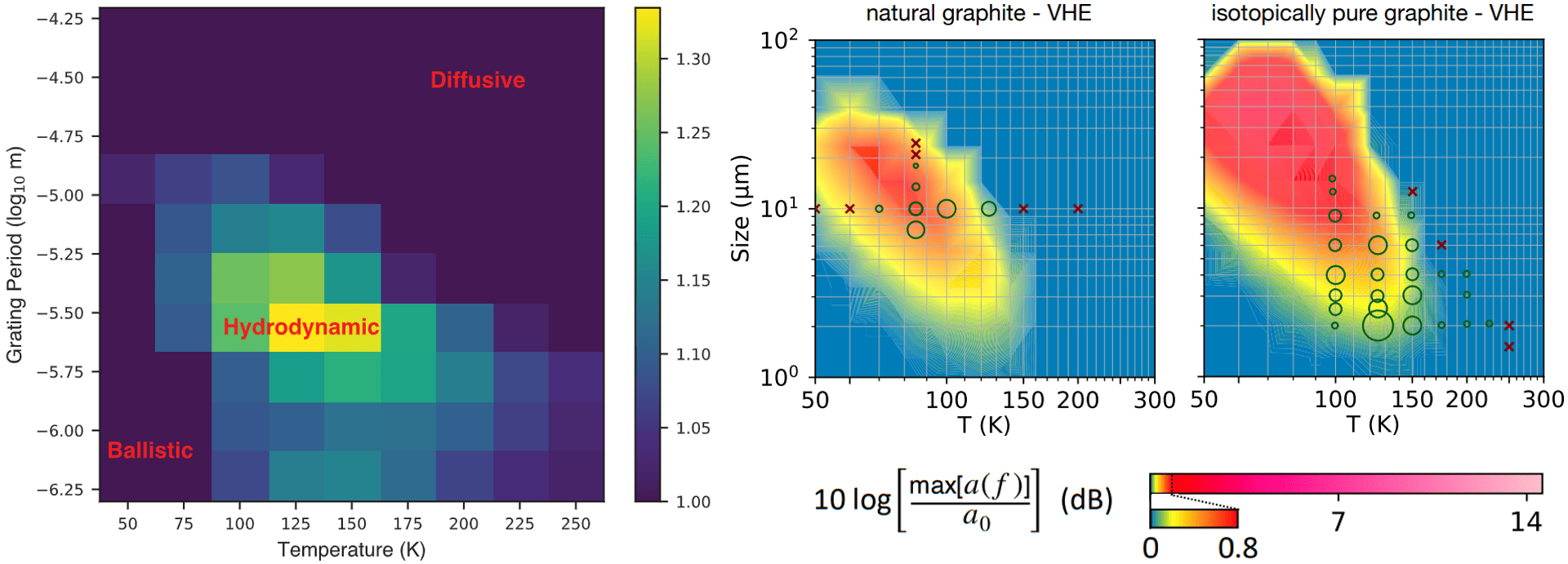}
\caption{\textbf{Temperature and characteristic length-scale window for phonon hydrodynamics in graphite.} (Left panel) Hydrodynamic regime of heat transport in graphite with natural isotope composition. The color map quantifies the relative strength of the temperature waves, defined as the ratio between the dominant resonant peak and the low-frequency background in the frequency-domain thermal response. A distinct region of temperature–length-scale space exhibits pronounced second-sound behavior, separating ballistic transport at low temperatures and short thermal-grating periods from a quasi-diffusive regime at higher temperatures and larger length scales, where the thermal response becomes purely relaxational. \textit{Figure reproduced with permission from Ref. \cite{huberman2019observation}}. (Central and right panels) Peak resonant enhancement of temperature waves as a function of device size, temperature, and isotopic composition for natural graphite (central panel) and isotopically purified graphite (right panel). The color maps display the maximum resonant response (see Ref. \cite{dragavsevic2023viscous} for definition) versus the characteristic length of a rectangular geometry and the average temperature about which the system is driven. Green open circles, with areas scaled to the experimentally inferred hydrodynamic strength, summarize measurements on natural and isotopically enriched samples reported by \textit{Huberman et al.} \cite{huberman2019observation} and \textit{Ding et al.} \cite{ding2022observation} while red crosses indicate regimes where transport remains diffusive. The viscous description accurately reproduces the temperature and length scales associated with the onset of hydrodynamic transport. The enhancement of hydrodynamic behavior predicted in isotopically purified graphite is also consistent with experimental observations. \textit{Figure reproduced with permission from Ref. \cite{dragavsevic2023viscous}}.}
\label{fig:huberman_figure}
\end{figure}
First-principles solutions of the time-dependent BTE quantitatively reproduce the observed dynamics, confirming that the effect originates from dominant momentum-conserving phonon-phonon scattering involving low-velocity transverse modes. \\
Moreover, the experimental observations reported by \textit{Huberman et al.} \cite{huberman2019observation} and by \textit{Ding et al.} \cite{ding2018phonon} provide a mutually consistent and progressively extended picture of phonon hydrodynamics in graphite. The left panel of Fig. \ref{fig:ding_huberman_figure} demonstrates that the damped oscillatory thermal response associated with second sound is confined to an intermediate temperature window, disappearing both at high temperatures, where diffusive transport dominates, and at low temperatures, where ballistic phonon propagation becomes prevalent. This nonmonotonic behavior establishes a clear hydrodynamic regime bounded by ballistic and diffusive limits. Building on this foundation, the ultrafast transient thermal grating measurements of \textit{Ding et al.} \cite{ding2022observation}, shown the right panels of Fig. \ref{fig:ding_huberman_figure}, reveal that the same temperature waves dynamics persists up to $\sim 200$--$225\,\mathrm{K}$ when the grating period is reduced to a few micrometers. The emergence of a pronounced phase inversion in the heterodyned signal at these elevated temperatures confirms that the hydrodynamic window shifts to higher temperatures as the characteristic transport length scale decreases. The combined experimental and theoretical results of Ref. \cite{huberman2019observation} are synthesized in the left panel of Fig. \ref{fig:huberman_figure}, which maps the hydrodynamic window in temperature--length-scale space: second sound emerges between ballistic and diffusive limits, persists up to $\sim150\,\mathrm{K}$ for micron-scale transport distances, and is predicted to extend to even higher temperatures at shorter length scales. Importantly, the phase diagram in the left panel of Fig. \ref{fig:huberman_figure} and results in Fig. \ref{fig:ding_huberman_figure} are consistent with simulations from Ref. \cite{dragavsevic2023viscous}, where simulated temperature waves are resonantly amplified under time-periodic heating. Solutions of the VHE exhibit a frequency-dependent response analogous to that of an underdamped oscillator, whereas Fourier’s law yields an overdamped, nonresonant behavior. This is summarized in the central and right panels of Fig. \ref{fig:huberman_figure}, which map the maximum resonant amplification as a function of temperature and device size. In natural graphite, the viscous hydrodynamic model predicts a well-defined resonant window that closely matches the temperature and length scales at which temperature waves dynamics was observed experimentally by \textit{Huberman et al.} \cite{huberman2019observation}.

\subsection{Guyer-Krumhansl equations} \label{GKE_section}

In this section we clarify how the Guyer-Krumhansl equations (GKE) \cite{guyer1966solution,sendra2021derivation} arise as a limiting case of the VHE. The GKE rely on two key assumptions: an isotropic and linear phonon dispersion, and a regime in which the drift-velocity contribution to the heat flux dominates over the part driven directly by temperature gradients.  
Under these conditions, the total heat flux takes the simplified form  
\begin{equation} 
\boldsymbol{Q}\equiv\boldsymbol{Q}^{D}=\alpha\boldsymbol{u},
\end{equation}
and the steady-state VHE reduce to
\begin{equation} \label{GKE1}
\nabla \cdot \boldsymbol{u} = 0, 
\end{equation}
\begin{equation} \label{GKE2}
\alpha\,\boldsymbol{u}(\boldsymbol{R}) = -\frac{\beta\alpha}{\gamma}\,\nabla T(\boldsymbol{R}) 
+ \frac{\eta}{\gamma}\alpha\nabla^{2}\boldsymbol{u}(\boldsymbol{R}),
\end{equation}
which coincide with the standard GKE form presented in e.g. Ref. \cite{sendra2021derivation}. Here, $\eta/\gamma$ defines an effective squared length scale governing nonlocal conduction, and $\beta\alpha/\gamma = \kappa^{M}$ corresponds to the momentum-driven part of the thermal conductivity (see Eq. \eqref{kappa_momentum}).
Taking the divergence of Eq. \eqref{GKE2} and using Eq. \eqref{GKE1} immediately yields the Laplace equation for the temperature field, $\nabla^{2} T(\boldsymbol{R}) = 0$, showing that, within the isotropic GKE, the steady-state temperature must always be harmonic.  A well-known consequence is that $T(\boldsymbol{R})$ cannot exhibit interior extrema; local maxima and minima can only occur at the boundaries. This behavior contrasts sharply with the predictions of the full VHE, as we will discuss in detail in the sections dedicated to thermal backflow (see sections \ref{tunnel_chamber} and \ref{strip_device_section}). Because the VHE do not assume a linear dispersion and retain both the drift and diffusive components of the heat flux, the resulting temperature field is not constrained to be harmonic. In fact, the VHE naturally allow $\nabla^{2} T \neq 0$ (see also Ref. \cite{coulter2025coupled}), similar to transport in rarefied-fluid systems \cite{sambasivam2014numerical}.

\subsection{Finite-size effects} \label{finite_size_effects}

Finite-size effects can be incorporated by applying Matthiessen-type corrections to the intrinsic diffusion-damped thermal conductivity, $\kappa^{D,\mathrm{bulk}}_{ij}$ \cite{simoncelli2020generalization,ziman2001electrons}, in close analogy with Bosanquet-type formulas used for rarefied-gas transport \cite{michalis2010rarefaction}:
\begin{equation}
\frac{1}{\kappa^{D}_{ij}}
=
\frac{1}{\kappa^{D,\mathrm{bulk}}_{ij}}
+
\frac{1}{\kappa^{D,\mathrm{ballistic}}_{ij}}.
\end{equation}
The intrinsic diffusion-damped conductivity $\kappa^{D,\mathrm{bulk}}_{ij}$ is defined in Eq. \eqref{diffusion_damped_kappa}, while its ballistic counterpart is given by
\begin{equation}
\kappa^{D,\mathrm{ballistic}}_{ij} =
\left[
\frac{1}{\mathcal{V}}\sum_{\nu}
\frac{(\hbar\omega_{\nu})^{2}\,\bar{n}_{\nu}\,(\bar{n}_{\nu}+1)}
{k_{\mathrm{B}}T^{2}}\frac{v^{i}_{\nu} v^{j}_{\nu}}{|\boldsymbol{v}_{\nu}|}\right]L_{S}\frac{\kappa^{D,\mathrm{bulk}}_{ij}}{\kappa_{ij}},
\end{equation}
where the term inside brackets corresponds to the prefactor $K^{S}_{ij}$ introduced in Ref. \cite{simoncelli2020generalization}, and $L_{S}$ denotes the characteristic length scale of the device. The final rescaling factor $\kappa^{D,\mathrm{bulk}}_{ij}/\kappa_{ij}$ ensures that only the non-hydrodynamic, diffusion-damped modes contribute to the ballistic correction, i.e.
\begin{equation}
\frac{\kappa^{D,\mathrm{bulk}}_{ij}}{\kappa_{ij}}=\frac{\kappa^{D,\mathrm{ballistic}}_{ij}}{K^{S}_{ij}\,L_{S}},
\end{equation}
which reflects the requirement that the hydrodynamic (momentum-conserving) portion of the conductivity remains excluded from the ballistic channel \cite{goblot2024imaging}. A completely analogous Matthiessen-type construction is used to treat boundary-induced corrections to the viscosity tensor. The ballistic viscosity is defined in Ref. \cite{simoncelli2020generalization} as (see also Eq. \eqref{slip_lenght}) 
\begin{equation}
M_{ijkl}=\frac{1}{\mathcal{V}}\sum_{\nu}\hbar^{2}q_{i} v^{j}_{\nu}q^{k} v^{l}_{\nu}\frac{\bar{n}_{\nu}(\bar{n}_{\nu}+1)}{k_{\mathrm{B}}T}\frac{1}{|\boldsymbol{v}v_{\nu}|}.
\end{equation}
In the basal plane of a system like graphite—where $M_{ijkl}$ is isotropic \cite{simoncelli2020generalization}—we can decompose the tensor into the same principal component ($\mathrm{cmp}$) used for the bulk viscosity: shear ($\mathrm{shr}$), rotational ($\mathrm{rot}$), and volumetric ($\mathrm{vol}$) \cite{dragavsevic2023viscous}. Finite-size corrections are then introduced for each principal component as
\begin{equation}
\frac{1}{\eta_{\mathrm{cmp}}}=\frac{1}{\eta^{\mathrm{bulk}}_{\mathrm{cmp}}}+\frac{1}{L_{S} M_{\mathrm{cmp}}},\qquad\mathrm{cmp}\in\{\mathrm{vol},\,\mathrm{shr},\,\mathrm{rot}\},
\end{equation}
where $M_{\mathrm{cmp}}$ denotes the corresponding principal component of the ballistic viscosity. Note that finite-size corrections arising from phonon–boundary scattering may be represented either as an effective further momentum-relaxation contribution to $\gamma_{ij}$, or, at the mesoscopic level, by explicitly incorporating drift-velocity boundary conditions together with size-dependent corrections to the transport coefficients.
\\\\
In the next two sections, we examine in detail how numerical solutions of the VHE enable the simulation of intriguing and counterintuitive hydrodynamic phenomena like steady-state thermal or heat backflow and time-dependent lattice cooling.

\subsection{Tunnel-chamber device for steady-state heat backflow} \label{tunnel_chamber}

\begin{figure}[h!]
\centering
\includegraphics[width=0.6\textwidth]{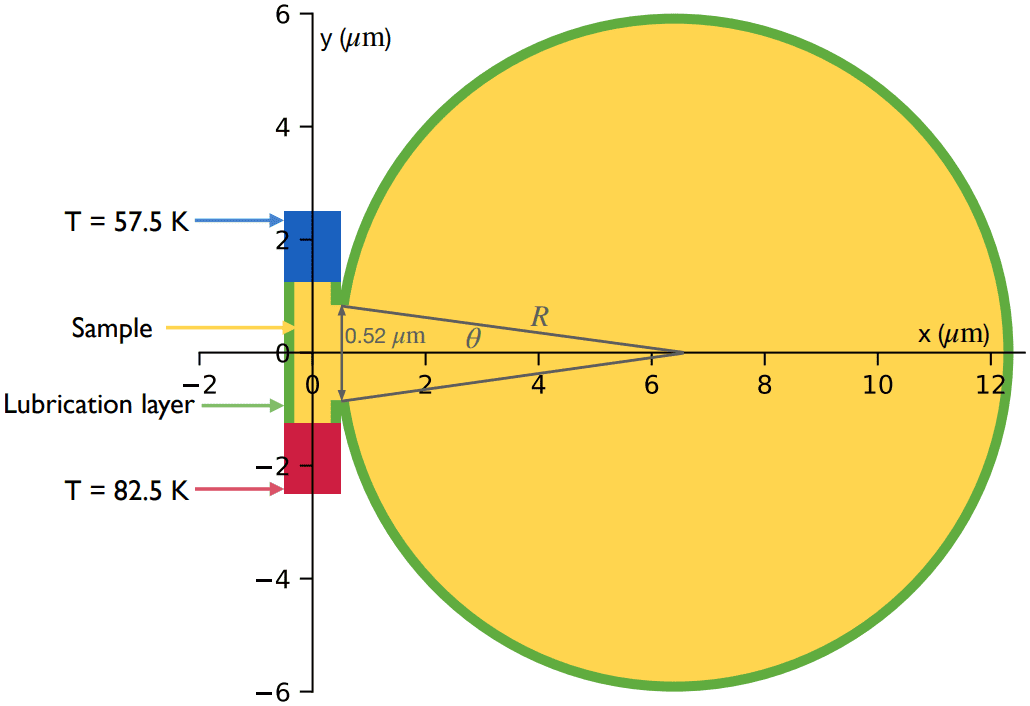}
\caption{\textbf{Tunnel-chamber geometry enabling viscous heat backflow.} The region in yellow represents the physical simulation domain in which all VHE coefficients take their actual material values, whereas the green region denotes the lubrication layer, where the viscosity is artificially reduced to model partially reflective boundary conditions (see section \ref{lubrication}). For visual clarity, both the lubrication-layer thickness (0.02\,$\mu$m) and the width of the tunnel–chamber opening (0.52\,$\mu$m) are depicted with exaggerated scale. The opening angle is $\theta=4.9672^{\circ}$. A temperature of 82.5\,K (57.5\,K) is imposed at the bottom (top) end of the tunnel, while all remaining boundaries are treated as adiabatic. \textit{Figure reproduced with permission from Ref. \cite{dragavsevic2023viscous}.}}
\label{fig:tunnel_chamber_setup}
\end{figure}

In this section we analyze a graphitic device featuring a tunnel-chamber structure \cite{aharon2022direct}, a layout known to facilitate hydrodynamic phonon flow and the emergence of vortical structures. In this configuration, a large temperature gradient is applied along the external tunnel, and the resulting heat flow is then studied within the adjoining chamber. Because the system operates in the hydrodynamic regime, the temperature bias imposed in the tunnel establishes both a well-defined local equilibrium temperature and a finite longitudinal phonon drift velocity at the tunnel–chamber interface, which in turn act as effective boundary conditions for the chamber.
\begin{figure}[h!]
\centering
\includegraphics[width=\textwidth]{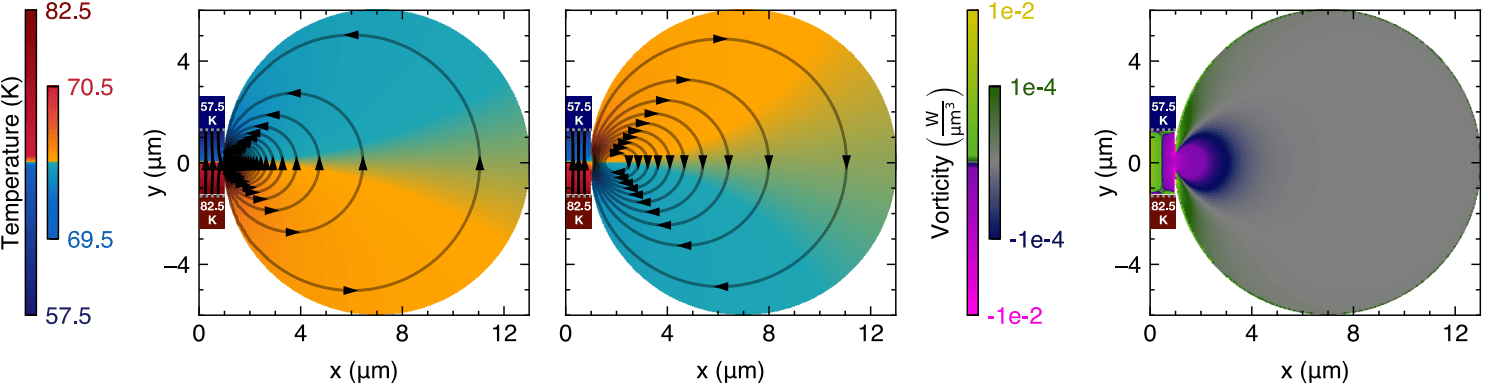}
\caption{\textbf{Viscous heat backflow and temperature inversion in a graphite tunnel-chamber geometry.} In-plane ($x$–$y$) heat-flow streamlines and temperature field in a graphite tunnel-chamber device. The left and central panels display, respectively, the solutions of Fourier’s law and VHE under a tunnel temperature bias of $T = 70 \pm 12.5\,\mathrm{K}$ at $y=\mp 1.25\,\mu\mathrm{m}$. All other boundaries are adiabatic ($\nabla T\!\cdot\!\hat{\boldsymbol{n}}=0$), and for the VHE we additionally impose $\boldsymbol{u}\!\cdot\!\hat{\boldsymbol{n}}=0$ and finite-slip conditions with slip length $0.4\,\mu\mathrm{m}$ \cite{bocquet2007flow}, representative of diffusive phonon-boundary scattering (see end notes). Fourier’s law solution produces a chamber temperature gradient aligned with that in the tunnel and cannot generate rotational heat motion. In contrast, the VHE solution exhibits a vortical heat-flow pattern that locally reverses the chamber temperature gradient, signaling viscous heat backflow. The right panel displays the vorticity of the VHE heat flux, $\nabla\times\boldsymbol{Q}_{\mathrm{tot}}$ (see also section \ref{disentanglement}); Fourier’s vorticity is identically zero and therefore not displayed. Streamline density is proportional to heat-flux magnitude: in the Fourier case streamlines in the chamber are $\sim$100 times denser than those in the tunnel, whereas for the VHE the ratio is $\sim$10. \textit{Figure reproduced with permission from Ref. \cite{dragavsevic2023viscous}.}}
\label{fig:tunnel_chamber_1}
\end{figure}
Fig. \ref{fig:tunnel_chamber_1} illustrates how thermal viscosity reshapes steady-state heat transport by comparing the numerical solution of Fourier’s inviscid equation (see left panel of Fig. \ref{fig:tunnel_chamber_1}) with that obtained from the viscous VHE (see central panel of Fig. \ref{fig:tunnel_chamber_1}). A key qualitative difference emerges: within the chamber, the VHE predicts a temperature profile that is inverted relative to the tunnel region—a behavior entirely opposite to the prediction of Fourier’s law. As shown in the right panel of Fig. \ref{fig:tunnel_chamber_1}, this temperature inversion—which could, in principle, be observed experimentally through thermal-imaging techniques \cite{menges2016nanoscale,cheng2022battery,cahill2014nanoscale,braun2022spatially,ziabari2018full,reihani2021quantitative,goblot2024imaging}—coincides with the formation of a vortex, where heat circulates locally and partially flows against the overall temperature gradient. To understand why the heat-backflow vortex in Fig. \ref{fig:tunnel_chamber_1} requires a nonzero thermal viscosity, we remember that graphite's in-plane transport is effectively isotropic, so that tensors proportional to the identity in the basal plane can be written without explicit Cartesian indices. In the inviscid limit, the second steady-state VHE \eqref{VHE2} reduces to
\begin{equation}
\beta \nabla T(\boldsymbol{R}) = -\gamma\,\boldsymbol{u}(\boldsymbol{R}),
\end{equation}
and inserting this into the first steady-state VHE \eqref{VHE1} leads to a Fourier-like, irrotational heat-transport equation, where the total heat flux is fully determined by the temperature gradient. Consequently, neither vorticity nor backflow can occur \cite{dragavsevic2023viscous,di2024vorticity,di2025vortices}. When thermal viscosity is finite, however, the second VHE \eqref{VHE2} no longer enforces proportionality between $\boldsymbol{u}$ and $\nabla T$ and the total heat flux, $\boldsymbol{Q}_{\mathrm{tot}}=-\kappa^{D}\nabla T + \alpha\,\boldsymbol{u}$ can no longer be rewritten as an irrotational field. Thus, viscosity is necessary for generating finite vorticity and enabling steady-state viscous heat backflow (as will be discussed below, transient temperature oscillations may also arise from nonlocal ballistic (BTE) effects). Nevertheless, viscosity by itself is not sufficient: the device must also possess a geometry and boundary conditions (see Fig. \ref{fig:tunnel_chamber_setup} for a schematic representation) that produce spatial variations in the drift velocity—i.e., nonzero $\nabla^{2}\boldsymbol{u}$—which in turn yield a heat flux with nonzero curl. 
A simple rectangular device, for instance, shows no appreciable differences between Fourier’s law and the VHE \cite{dragavsevic2023viscous} solutions, even though the applied temperature gradient can induce a non-zero drift velocity in the tunnel. Far from the thermalized boundaries, this drift velocity becomes nearly uniform and effectively sets a boundary condition for the attached chamber. Heat backflow emerges when the drifting component of the heat flux dominates and opposes the diffusive contribution, resulting in a local reversal of the temperature gradient. At the chamber entrance, the temperature–gradient flux behaves like a thermal dipole, producing counterclockwise, irrotational streamlines, while viscous effects cause the drifting heat flux to bend into clockwise, closed trajectories, thereby generating vortices and enabling temperature inversion. Frictionless boundaries (modeled via the lubrication layer technique discussed in section \ref{lubrication}) preserve the magnitude of the incoming drift velocity and thus strengthen these hydrodynamic signatures. Ultimately, vortex formation requires viscous forces to outweigh momentum-dissipation effects, giving rise to heat flux with nonzero vorticity. The interplay between the dipole-like diffusive flux and the vortex-forming drift flux depends on both material properties (e.g., isotopic composition) and device geometry (chamber radius and opening size in Fig. \ref{fig:tunnel_chamber_setup}) \cite{dragavsevic2023viscous}. The opening used here is chosen to be small enough ($L = 0.52\,\mu\mathrm{m}$ is also used as the characteristic length of the system, which enters the finite-size corrections discussed in section \ref{finite_size_effects}) to promote strong viscous backflow, yet large enough to ensure consistency between VHE and LBTE predictions (see section \ref{heat_backflow_LBTE}).\\
For the computational details underlying this study, the reader is referred to Ref. \cite{dragavsevic2023viscous}.

\subsubsection{Heat backflow from the LBTE} \label{heat_backflow_LBTE}

As discussed in the previous sections, the VHE arise from a coarse-grained formulation of the phonon LBTE, obtained by assuming spatially and temporally homogeneous linear response when extracting the transport coefficients---that is, by neglecting gradients of the non-equilibrium phonon population \cite{simoncelli2020generalization}. As a result, the VHE become increasingly reliable when such gradients are small. To assess the range of validity of the VHE for the geometries and conditions studied here, one can systematically compare their predictions with those obtained by solving the (much more expensive) full LBTE (including the full collision operator \cite{raya2022bte}), examining both steady-state and time-dependent regimes.
\begin{figure}[h!]
\centering
\includegraphics[width=\textwidth]{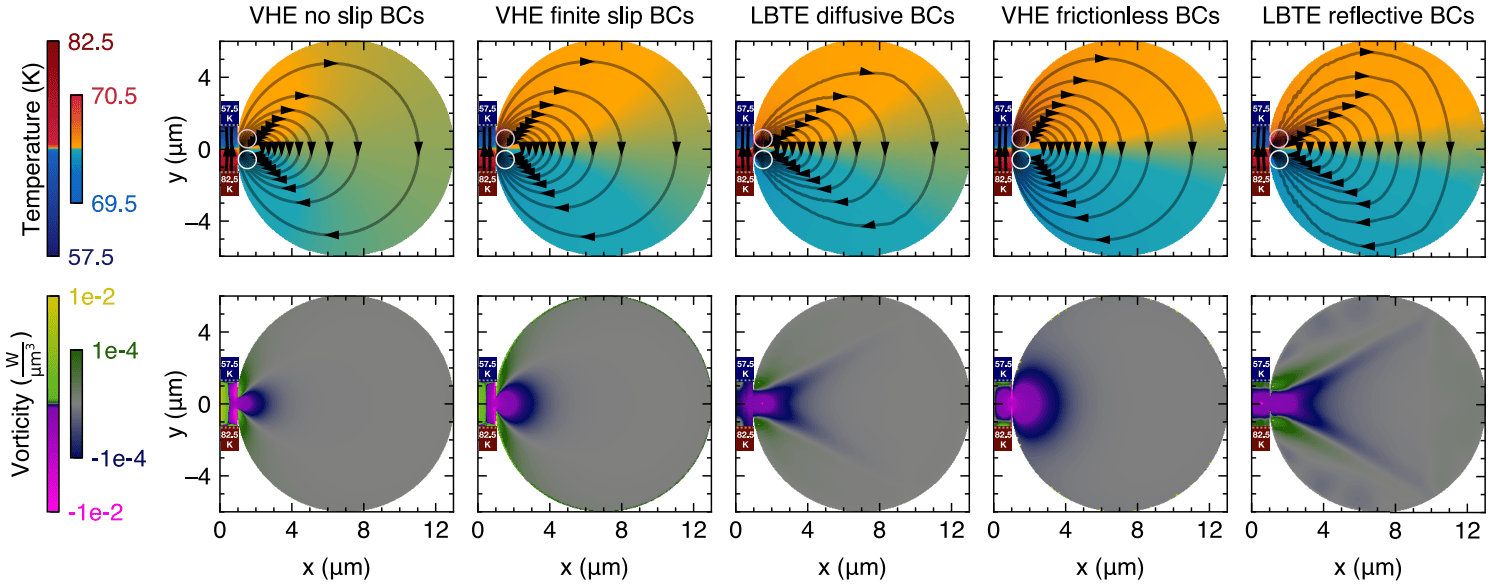}
\caption{\textbf{Viscous heat backflow and temperature inversion from VHE and LBTE, and sensitivity to boundary conditions.} Top panels display the in-plane ($x$–$y$) heat-flow streamlines together with the temperature field, while the bottom panels show the corresponding vorticity. Each column reports results for a tunnel--chamber structure made of natural graphite, where a temperature bias is applied across the tunnel (the boundaries at $y=\mp 1.25\,\mu\mathrm{m}$ are fixed at $70 \pm 12.5$\,K). All remaining boundaries are treated as adiabatic, $\boldsymbol{Q}_{\mathrm{tot}}\!\cdot\!\hat{\boldsymbol{n}}=0$, and the columns compare VHE and LBTE predictions under different assumptions for the tangential momentum transfer at the boundaries, that are, from left to right VHE with ideal no-slip conditions ($\boldsymbol{u}=0$);VHE results with a finite slip length of $0.4\,\mu\mathrm{m}$ (left), chosen to mimic the microscopic diffusive phonon-boundary scattering used in the corresponding LBTE calculations (right). VHE results with frictionless (zero-shear-stress) boundaries (center) and their microscopic counterpart obtained from LBTE with fully reflective phonon-boundary scattering (right). The thistle and beige circular markers (diameter $1\,\mu\mathrm{m}$), located at $(x,y)=(1.02\,\mu\mathrm{m},\pm 0.58\,\mu\mathrm{m})$, indicate the positions of the local temperature maximum (upper chamber) and minimum (lower chamber), respectively. These probes highlight the reversal of the temperature gradient inside the chamber—an unambiguous signature of viscous heat backflow—which becomes more pronounced as boundary friction decreases from left to right. Both VHE and LBTE predict heat-flow patterns with closed streamlines, indicating finite vorticity, as illustrated in the lower row. \textit{Figure reproduced with permission from Ref. \cite{dragavsevic2023viscous}.}}
\label{fig:tunnel_chamber_2}
\end{figure}
At steady state, Fig. \ref{fig:tunnel_chamber_2} demonstrates that the spatially resolved LBTE reproduces the viscous temperature inversion predicted by the VHE for the device of Fig. \ref{fig:tunnel_chamber_setup}. The simulations are performed using the deviational Monte Carlo algorithm implemented in the \texttt{BTE-Barna} package \cite{raya2022bte} with scattering rates computed across the full three-dimensional Brillouin zone of graphite \cite{dragavsevic2023viscous}. The device is modeled as translationally invariant along the out-of-plane direction, so that temperature and heat flux depend only on the in-plane coordinates $x$ and $y$, as in Fig. \ref{fig:tunnel_chamber_1}. Fig.  \ref{fig:tunnel_chamber_2} shows that the LBTE reproduces both the temperature inversion and the associated vorticity predicted by the VHE around $\bar{T}=70$\,K. In the LBTE simulations, the system is initialized at a uniform temperature $T(x,y)=\bar{T}$ and a $\Delta T=25$\,K bias is imposed across the tunnel (82.5\,K at $y=-1.25\,\mu$m and 57.5\,K at $y=1.25\,\mu$m). All remaining boundaries are taken as adiabatic, meaning no net heat crosses them. The results are obtained with steady-state quantities averaged over the interval $t\in[2,5.5]$\,ns. The adiabatic boundary condition alone does not fully specify the LBTE problem: a model for phonon-boundary scattering must also be chosen. Two limiting cases \cite{raya2022bte}: diffusive boundaries, which randomize the outgoing phonon direction following a Lambertian distribution \cite{raya2022bte,sabatti2017simulation} and (ii) specular/reflective boundaries, which preserve the tangential momentum component and flip the normal component \cite{raya2022bte}. As seen by comparing the third and fifth columns of Fig. \ref{fig:tunnel_chamber_2}, reflective boundaries enhance the magnitude of the temperature inversion and also produce larger heat-flux vorticities. In the VHE framework, adiabatic boundaries are enforced by setting to zero the normal components of (i) the temperature-gradient flux and (ii) the drift-velocity flux. The first condition, $\nabla T\cdot\hat{\boldsymbol{n}}=0$, is identical to the one used in Fourier theory. The second requires more care: our earlier work \cite{simoncelli2020generalization} employed a no-slip condition $\boldsymbol{u}=0$ at the boundary, corresponding to complete phonon-momentum loss. More generally, one may impose only $\boldsymbol{u}\cdot\hat{\boldsymbol{n}}=0$, allowing tangential drift (i.e., partial momentum retention), which more closely resembles realistic phonon-boundary interactions. To model arbitrary slip lengths $b$, the lubrication-layer strategy (see section \ref{lubrication}) is adopted. By varying $\eta_{\rm shr,ll}$ and $\eta_{\rm rot,ll}$ from their bulk values to zero, the boundary condition continuously sweeps from no-slip ($b=0$) to finite slip ($b=0.4\,\mu$m) to the frictionless limit ($b\to\infty$). This is physically meaningful: no-slip is appropriate when microscopic mean free paths are much smaller than device dimensions, whereas slip becomes essential in the opposite, rarefied regime \cite{laurent2011large}. Because the drift velocity is directly proportional to the average phonon momentum, the frictionless VHE condition is the mesoscopic analogue of specular reflection in the LBTE. This correspondence is confirmed by comparing the two top panels in the fourth and fifth columns of Fig. \ref{fig:tunnel_chamber_2}: the temperature fields agree quantitatively once the slip length is matched. The value $b=0.4\,\mu$m used in the second column of Fig. \ref{fig:tunnel_chamber_2} is obtained by reproducing the difference in temperature inversion observed between reflective and diffusive LBTE solutions \cite{dragavsevic2023viscous}. The comparison of vorticities shows a similar trend: finite-slip VHE more closely matches the diffusive LBTE case, while frictionless VHE and reflective LBTE display noticeable deviations near the chamber opening, where spatial variations in the heat flux are strongest. These discrepancies likely result from neglecting spatial gradients of the non-equilibrium phonon distribution when deriving the VHE \cite{simoncelli2020generalization}.\\
Although the VHE ignore these spatial gradients, they still encode a weak form of nonlocal constitutive behavior. For example, in section \ref{GKE_section} we demonstrate that the VHE reduce to the GKE under appropriate limits, and Ref. \cite{gurcan2013transport} shows that the GKE incorporate weakly nonlocal corrections governed by the length scale $\ell_\eta=\sqrt{\eta/\gamma}$. Such nonlocal effects are known to become important in nanoscale devices \cite{sendra2022hydrodynamic}. We remark that nonlocality could also arise independently of viscosity, e.g., in the ballistic regime through the term $\nabla n^{\delta}_\nu(\boldsymbol{R},t)$, which is not retained in the VHE derivation. For the conditions discussed in this review, viscous nonlocality alone suffices to reconcile VHE and LBTE predictions. In this sense, \textit{Hua and Lindsay} developed a direct eigendecomposition-based solution of the space–time–dependent BTE to investigate nonlocal thermal transport beyond the RTA \cite{hua2020space}. 
\begin{figure}[h!]
\centering
\includegraphics[width=0.8\textwidth]{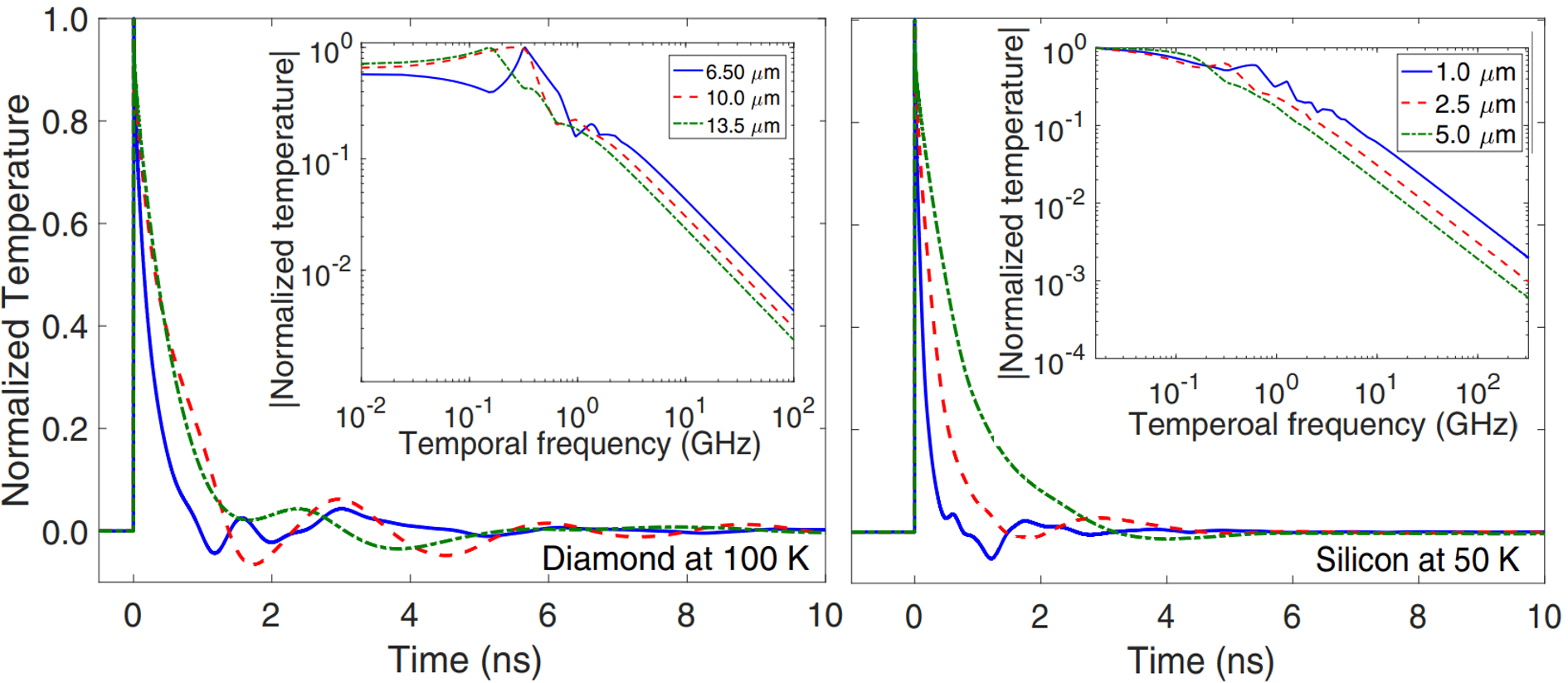}
\caption{\textbf{Time-domain evolution of transient thermal grating signals.} Time-domain transient thermal grating (TTG) response in diamond at 100 K (left panel) and in silicon at 50 K (right panel) for different grating periods. Insets show the corresponding temperature response amplitudes in the frequency domain as functions of temporal frequency. \textit{Figure reproduced with permission from Ref. \cite{hua2020space}.}}
\label{fig:lindsay_figure}
\end{figure}
One of their main results is that transient grating temperature oscillations can emerge in diamond and silicon when the experimental timescale becomes comparable to phonon relaxation times (see Fig. \ref{fig:lindsay_figure}), even in the absence of an explicit hydrodynamic viscosity term. As the grating period is reduced, the temperature response transitions from a monotonic decay to an oscillatory relaxation, reflecting the emergence of a space-time dependent effective thermal conductivity governed by long-mean-free-path phonons. The phase velocities associated with these oscillations are significantly smaller than those expected for hydrodynamic second sound, indicating that the effect is not driven by viscous momentum diffusion. This demonstrates that temperature waves dynamics in transient regimes can originate also from nonlocal thermal transport, and that a finite thermal viscosity is not a necessary condition for observing temperature oscillations.\\
As seen in section \ref{GKE_section} the VHE encompass, as a well-defined limiting case, the GKE, which constitutes a weakly nonlocal constitutive relation between heat flux and temperature gradient \cite{gurcan2013transport}. In this framework, the heat flux at position $\boldsymbol{R}$ depends on temperature gradients evaluated at neighboring points  $\boldsymbol{R}'$, reflecting the finite relaxation of momentum-carrying phonons. Starting from the VHE,
\begin{equation} \label{viscous_heat_T_appa}
C \frac{\partial T(\boldsymbol{R},t)}{\partial t} 
+ \sum_{i,j=1}^{3} W^i_{0j} \sqrt{T \bar{A}_jC} \frac{\partial u_j(\boldsymbol{R},t)}{\partial R_{i}} 
- \sum_{i,j=1}^{3} \kappa^{D}_{ij} \frac{\partial^2 T(\boldsymbol{R},t)}{\partial R_{i} \partial R_{j}} = 0,
\end{equation}
\begin{equation} \label{viscous_heat_U_appa}
A_i \frac{\partial u_i(\boldsymbol{R},t)}{\partial t} 
+ \sqrt{\frac{C A_i}{\bar{T}}} \sum_{j=1}^{3}  W^j_{i0} \frac{\partial T(\boldsymbol{R},t)}{\partial R_{j}} 
- \sum_{j,k,l=1}^{3} \eta_{ijkl} \frac{\partial^2 u_k(\boldsymbol{R},t)}{\partial R_{j} \partial R_{l}} 
= - \sum_{j=1}^{3} \sqrt{A_iA_j} D^U_{ij} u_j(\boldsymbol{R},t),
\end{equation}
and recalling that the total heat flux decomposes as $\boldsymbol{Q}^{\rm TOT}=\boldsymbol{Q}^\delta+\boldsymbol{Q}^D$, with $Q^{\delta}_i=-\kappa^{D}_{ij}\nabla_j T$ and $Q^{D}_i=\alpha_{ij}u_j$, one can focus on the GKE regime where $|\boldsymbol{Q}^D|\gg|\boldsymbol{Q}^\delta|$. In this limit, Eq. \eqref{viscous_heat_T_appa} reduces to $C\,\partial_t T+\alpha_{ij}\partial_i u_j=0$, while Fourier transforming Eq. \eqref{viscous_heat_U_appa} in space and time yields a closed relation between the drift velocity and the temperature field. Focusing on a one-dimensional geometry and performing the inverse transform, one obtains the weakly nonlocal constitutive law \cite{gurcan2013transport}
\begin{equation} \label{nonlocal}
\begin{split}
Q(R,t)
&=-\kappa^M\!\int\!dR'\!\int\!dt'\,
\frac{\sqrt{\tau_Q}}{2\lambda\sqrt{\pi(t-t')}}\,
\exp\!\left[-\frac{\tau_Q(R-R')^2}{4\lambda^2(t-t')}
-\frac{t-t'}{\tau_Q}\right]
\nabla_{R'}T(R',t'),
\end{split}
\end{equation}
where $\kappa^M=\alpha\beta/\gamma$ is the momentum-related contribution to the thermal conductivity, $\tau_Q=A/\gamma$ is the heat-flux relaxation time associated with resistive scattering, and $\lambda^2=\eta_{xxxx}/\gamma$ defines the characteristic nonlocal length scale \cite{sendra2021derivation}. Eq. \eqref{nonlocal} explicitly shows that, within the GKE limit of the VHE, steady-state nonlocality arises from a finite viscosity, which spatially spreads the response of the heat flux to temperature gradients. At the same time, nonlocal effects are not exclusively tied to viscosity: as demonstrated in Ref. \cite{allen2018analysis}, ballistic transport can also generate nonlocal behavior through spatial gradients of the out-of-equilibrium phonon distribution $\nabla n_\nu^\delta(\boldsymbol{R},t)$, a term neglected in the derivation of the VHE. Therefore, the VHE capture weakly nonlocal hydrodynamic effects encoded in viscosity \cite{gurcan2013transport,sendra2022hydrodynamic}, while omitting stronger nonlocalities characteristic of far-from-equilibrium ballistic regimes.\\
Finally, the derivation of the VHE—unlike the Navier--Stokes equations for ordinary fluids and the hydrodynamic equations used for electronic transport in, e.g., nanoscale conductors \cite{dagosta2006hydrodynamic}—is formulated strictly within linear response and therefore omits nonlinear advection terms. Nonlinearities of this kind have been explored only in simplified or approximate frameworks. For instance, they have been analyzed within the Callaway approximation for spatially uniform temperature waves in two-dimensional materials \cite{shang2022unified}. Earlier works \cite{banach2008chapman,banach1989irreducible} applied the Callaway approximation to construct nonlinear phonon hydrodynamics via a Chapman--Enskog expansion of the full nonlinear BTE \cite{guo2015phonon}, ordered in the small-Knudsen-number (long-wavelength) hydrodynamic limit. This expansion generates nonlinear constitutive relations in which macroscopic heat transport may acquire advective or amplitude-dependent corrections, and transport coefficients such as relaxation times or viscosities become temperature dependent. The structure closely parallels gas kinetic theory, where successive orders in the Knudsen number lead to Navier-Stokes, Burnett, and super-Burnett nonlinear hydrodynamics. However, the Callaway approximation itself suffers from several limitations (see Sec. \ref{callaway_section}).\\
A different nonlinear phonon hydrodynamics model was presented in Ref. \cite{nielsen1969heat}, although its nonlinearity arises solely from assuming a drifting local-equilibrium phonon gas—analogous to the nonlinearity of compressible Euler equations. It does not incorporate nonlinearities in the modern kinetic sense, nor does it include the full collision physics or viscous contributions. Additional remarks on nonlinear effects appear in models based on a weakly interacting phonon gas \cite{rogers1971transport} and in formulations employing Helmholtz potentials \cite{tzou2014longitudinal}.\\
Nonlinear extensions of the Maxwell-Cattaneo and GKE have also been proposed \cite{cimmelli2010nonequilibrium}, introducing an internal dynamical temperature, temperature-dependent coefficients, and quadratic terms in gradients or fluxes. These works examine direction-dependent heat-wave speeds in nonequilibrium steady states and define a “thermal Reynolds number” for phonon hydrodynamics. Nevertheless, such treatments remain essentially one-dimensional (see also Ref. \cite{melis2019indications}), weakly nonlinear, continuum-level approximations rather than microscopic descriptions based on the full LBTE. Another nonlinear framework, the thermomass theory \cite{dong2011generalized,wang2014theoretical,dong2016thermal}, interprets heat flow as a macroscopic fluid endowed with inertia and momentum, giving rise to Reynolds-type nonlinear terms. However, this approach is not derived from the phonon collision operator and lacks a quantum-mechanical foundation. Finally, recent work \cite{yadav2025derivation} employed Monte Carlo simulations within a relaxation-time approximation to investigate nonlinear phonon-transport regimes. To the best of our knowledge, incorporating these nonlinear effects while retaining the full LBTE collision matrix remains an open problem, and addressing it lies beyond the scope of the present review.

\subsection{Ring-shaped heater for lattice cooling in the transient domain} \label{lattice_cooling_section}

In this section, we summarize how the numerical solution of the VHE can be employed to model the hydrodynamic lattice cooling dynamics of a phonon device as a function of time.\\
As anticipated, recent time-resolved experiments in graphite have revealed instances of heat propagating against the applied temperature gradient, either in the form of second sound \cite{huberman2019observation,ding2022observation} or through transient lattice-cooling signatures \cite{jeong2021transient}. Early theoretical studies interpreted these observations using the microscopic LBTE \cite{huberman2019observation,ding2022observation,jeong2021transient,cepellotti2017transport,zhang2021transient}, but without explicitly resolving viscous contributions to heat transport, while other works relied on the inviscid DPLE  \cite{joseph1989heat,tzou1995unified,xu2002thermal,ordonez2010exact,kang2017method,gandolfi2019accessing,xu2021thermal,mazza2021thermal}. This naturally raises the question of how the hydrodynamic---and specifically viscous---mechanisms encoded in the VHE manifest in the time domain, and whether a direct connection exists between temperature-wave phenomena and transient viscous heat backflow.\\
This regime was addressed in Ref. \cite{dragavsevic2023viscous} by performing a time-resolved VHE simulation.
\begin{figure}[h!]
\centering
\includegraphics[width=\textwidth]{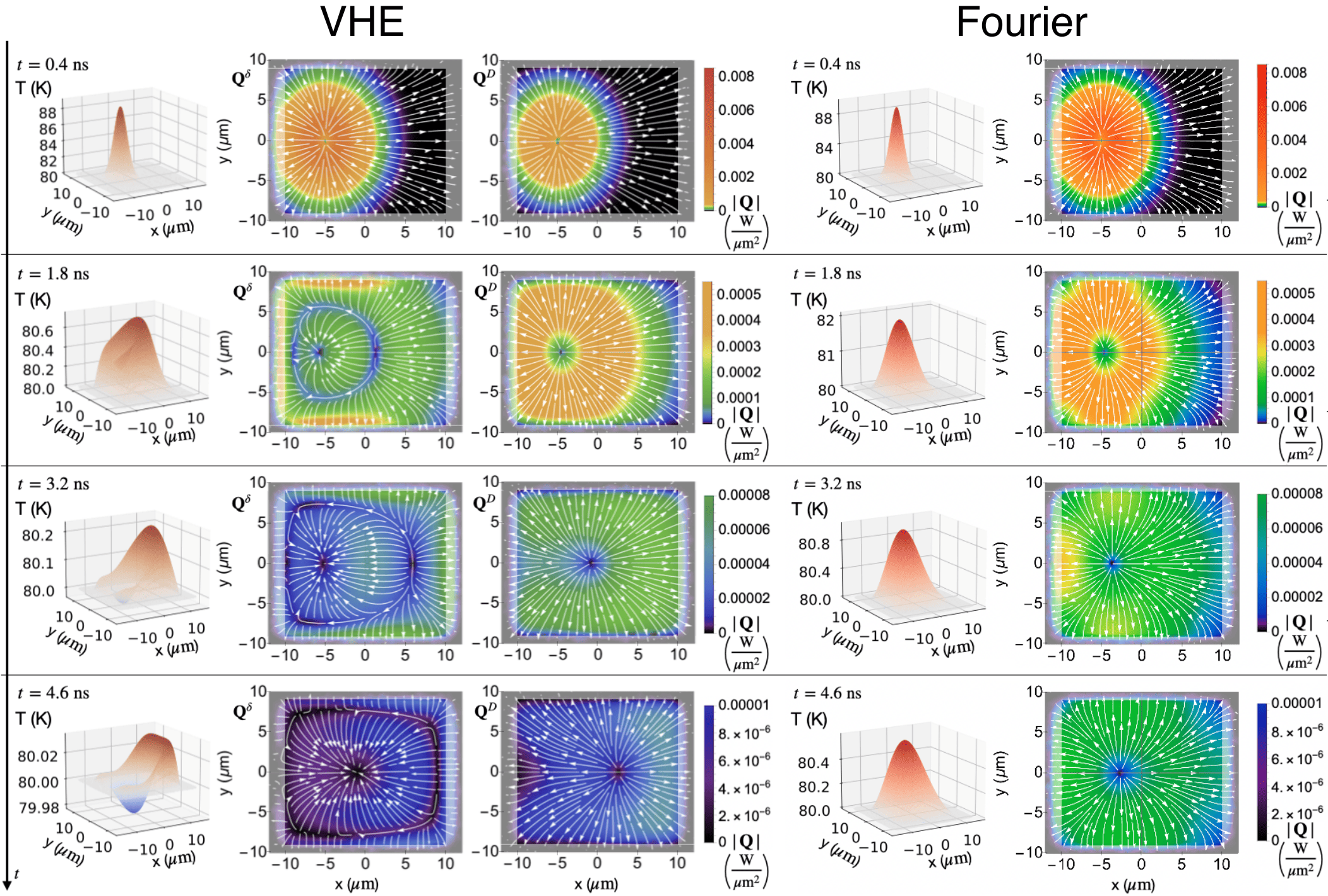}
\caption{\textbf{Transient viscous heat backflow.} Each row displays the system at successive times after the heating pulse is switched off at $t = 0.4\,\text{ns}$ (see main text). From left to right, we report the temperature field, the heat flux driven by the temperature gradient ($\boldsymbol{Q}^\delta$), the drift-induced heat flux ($\boldsymbol{Q}^D$) predicted by the VHE, and the temperature field and the heat flux predicted by Fourier's law, all obtained from time-dependent solutions in a graphite sample whose outer regions (shaded) are maintained at $80\,\text{K}$. In the two-dimensional maps, the color scale represents the magnitude of the corresponding heat-flux vector, while the white streamlines indicate its local direction. \textit{Figure reproduced with permission from Ref. \cite{dragavsevic2023viscous}.}}
\label{fig:lattice_cooling_1}
\end{figure}
Fig. \ref{fig:lattice_cooling_1} shows a rectangular sample initially at equilibrium at $T=80$ K with zero drift velocity ($u=0$). A localized heater, centered at $(x_c,0)$ with $x_c=5 \mu\mathrm{m}$, deposits heat for $0<t<t_{\mathrm{heat}}=0.4$ ns according to
\begin{equation}
\dot{q}(\boldsymbol{R},t)=\mathcal{H}\cdot\theta(t_{\mathrm{heat}}-t)\exp\!\left[-\frac{(x+x_c)^2}{2\sigma_x^2}-\frac{y^2}{2\sigma_y^2}\right],
\end{equation}
with parameters chosen such that the induced temperature excursion remains within 10\% of equilibrium. In particular, 
$\mathcal{H}=0.013\frac{\mathrm{W}}{\mathrm{\mu m^{3}}}$, $t_{\mathrm{heat}}=0.4\mathrm{ns}$, $\sigma_{x}=2\mu\mathrm{m}$ and $\sigma_{x}=2.8\mu\mathrm{m}$.
At $t=t_{\mathrm{heat}}$, the heater is switched off and the system relaxes back to equilibrium. Throughout the simulation, boundaries are maintained at $T=80$ K and $u=0$, mimicking experimentally realized thermalization conditions \cite{braun2022spatially}. \\
The time-dependent solution of the VHE (first column of Fig. \ref{fig:lattice_cooling_1}) displays an oscillatory relaxation of the temperature field, including the appearance of local temperatures below the initial equilibrium value. This contrasts sharply with predictions from Fourier's law: as shown in the fourth column of Fig. \ref{fig:lattice_cooling_1}, a positive temperature perturbation governed by the diffusion equation retains a strictly non-negative deviation from equilibrium, a consequence of the well-known smoothing property of diffusive models \cite{skinner2014mathematical}.\\
To identify the microscopic origin of such lattice cooling behavior, one can refer to the decomposition of the total heat flux $\boldsymbol{Q}$ into its temperature-gradient and drift components (see Eqs. \eqref{delta_flux} and \eqref{drifting_flux}, respectively). Viscous streamlines in Fig. \ref{fig:lattice_cooling_1} (second and third columns) reveal that during relaxation $\boldsymbol{Q}_{\delta}$ and $\boldsymbol{Q}_{\mathrm{D}}$ may point in opposite directions, generating transient heat backflow. This behavior arises from the finite time lag between the onset of a temperature gradient and the development of the corresponding drift response. Indeed, in the inviscid limit ($\eta=0$), the VHE reduce exactly to the lagged constitutive relation of the DPLE model \cite{tzou1995unified}:
\begin{equation}
\boldsymbol{Q}(\boldsymbol{R},t+\tau_Q)
= -\kappa \nabla T(\boldsymbol{R},t+\tau_T),
\end{equation}
where $\tau_Q = A/\gamma$ describes the delay between the emergence of a temperature gradient and the heat flux, and $\tau_T = \alpha\beta A/(\kappa\gamma^2)$ quantifies the inverse delay, that is, the time needed to generate the temperature gradient from an established heat flux. This demonstrates that, unlike in the steady-state case—where viscous stresses are strictly necessary for heat backflow—the transient regime can exhibit backflow even in the inviscid limit. Nevertheless, Ref. \cite{dragavsevic2023viscous} shows that incorporating viscous effects is essential to reconcile microscopic LBTE predictions and experimental trends quantitatively (see also section \ref{heat_backflow_LBTE}).\\
Here the characteristic size for finite-size scattering, $L$, is chosen to be the horizontal extent of the computational domain \cite{dragavsevic2023viscous}.

\subsubsection{Thermalization lengthscale}

In realistic experimental devices, thermalization does not occur abruptly at the boundary but instead takes place over a finite spatial region \cite{braun2022spatially}. This means that the transition from the device interior to the thermal reservoir is gradual rather than sharp. To capture this behavior, the boundaries are modeled using a compact sigmoid function \cite{dragavsevic2023viscous,tu2008manifolds}, which allows to interpolate smoothly between the thermal-bath region---where the temperature is fixed and the drift velocity vanishes (as required by the Bose-Einstein equilibrium distribution \cite{simoncelli2020generalization})---and the interior region, where temperature and drift velocity evolve according to the VHE. More precisely, the widely used smooth-step function is employed \cite{phillips2019smootherstep}, a fifth-order Hermite polynomial that interpolates monotonically between $0$ and $1$,
\begin{equation} \label{sigmoid}
f(R)=
\begin{cases}
0, & d(R)<0,\\[2pt]
1, & d(R)>1,\\[4pt]
6\,d(R)^5 - 15\,d(R)^4 + 10\,d(R)^3, & \text{otherwise},
\end{cases}
\end{equation}
where 
\begin{equation}
d(R)=\frac{r - R_{\mathrm{in}}}{R_{\mathrm{out}} - R_{\mathrm{in}}}, \qquad
r>0,\; R_{\mathrm{in}}>0,\; R_{\mathrm{out}}>0.
\end{equation}
This function is $C^2$-continuous and exactly equals $0$ and $1$ at $R_{\mathrm{in}}$ and $R_{\mathrm{out}}$, respectively. 
\begin{figure}[h!]
\centering
\includegraphics[width=0.6\textwidth]{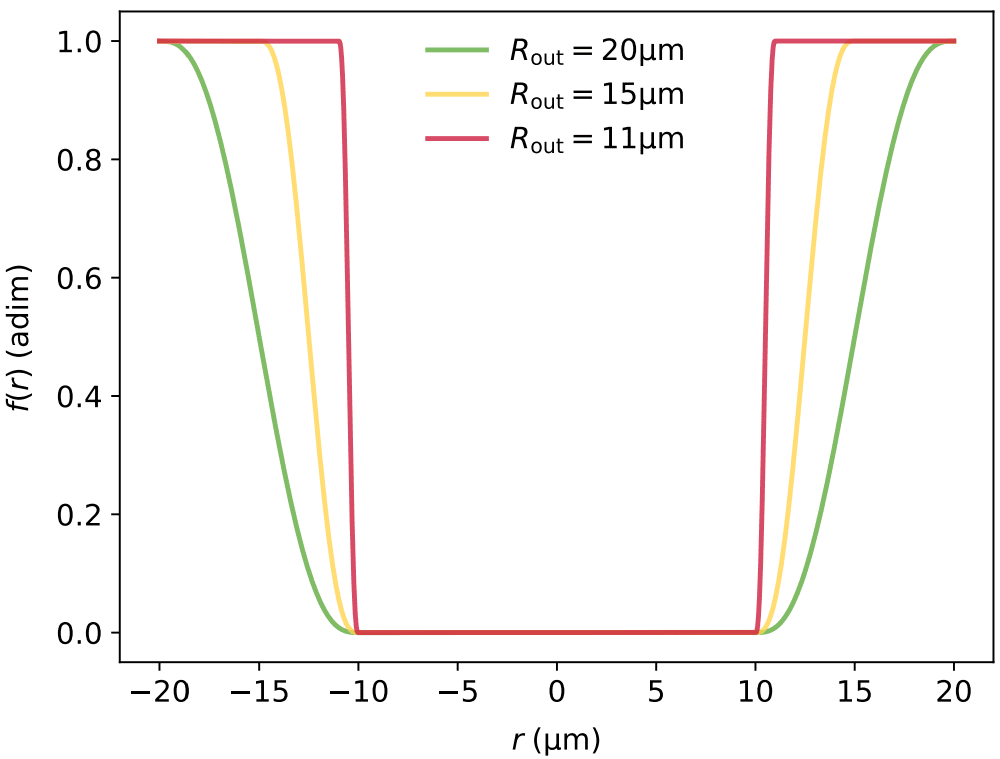}
\caption{\textbf{Sigmoid function for realistic thermalization.} The smooth-step function of Eq. \eqref{sigmoid} is shown for $R_{\rm in}=10\,\mu\text{m}$ and several choices of $R_{\rm out}$. A value $R_{\rm out}=11\,\mu\text{m}$ (red) corresponds to an almost instantaneous thermalization occurring over a very narrow region, $R_{\rm out}=15\,\mu\text{m}$ (orange) reflects a thermalization length consistent with experimental estimates \cite{braun2022spatially}, while $R_{\rm out}=20\,\mu\text{m}$ (green) models a substantially more gradual and less efficient coupling to the thermal reservoir. \textit{Figure reproduced with permission from Ref. \cite{dragavsevic2023viscous}.}}
\label{fig:hermite_poly}
\end{figure}
As shown in Fig. \ref{fig:hermite_poly}, choosing $R_{\mathrm{out}}\approx R_{\mathrm{in}}$ reproduces nearly ideal (spatially sharp) thermalization, whereas taking $R_{\mathrm{out}}\gg R_{\mathrm{in}}$ models inefficient thermalization that occurs over a large lengthscale. Using this sigmoid profile, the thermal-bath region ($f=1$) is enforced to satisfy $T(\boldsymbol{R},t)=T_{\mathrm{eq}}$ and $\mathbf{u}(\boldsymbol{R},t)=0$, while the device interior ($f=0$) evolves according to the full VHE. Combining these regimes yields the coupled equations
\begin{equation}
f(\boldsymbol{R})\,[T(\boldsymbol{R},t)-T_{\mathrm{eq}}] 
+ [1 - f(\boldsymbol{R})]
\left[
C\,\frac{\partial T(\boldsymbol{R},t)}{\partial t}
+ \alpha_{ij}\,\frac{\partial u_j(\boldsymbol{R},t)}{\partial r_i}
- \kappa_{ij}^{D}\,\frac{\partial^2 T(\boldsymbol{R},t)}{\partial r_i \partial r_j}
- \dot{q}(\boldsymbol{R},t)
\right]=0,
\end{equation}
and
\begin{equation}
f(\boldsymbol{R})\,u_i(\boldsymbol{R},t)
+ [1 - f(\boldsymbol{R})]
\left[
A_{ij}\,\frac{\partial u_j(\boldsymbol{R},t)}{\partial t}
+ \beta_{ij}\,\frac{\partial T(\boldsymbol{R},t)}{\partial r_j}
- \eta_{ijkl}\,\frac{\partial^2 u_k(\boldsymbol{R},t)}{\partial r_j \partial r_l}
+ \gamma_{ij} u_j(\boldsymbol{R},t)
\right]=0.
\end{equation}
These equations ensure a smooth and numerically stable transition between the free-evolving hydrodynamic region and the thermal reservoir, while preserving the correct physical behavior at the boundaries.

\subsubsection{Realistic thermalization}

\begin{figure}[h!]
\centering
\includegraphics[width=0.6\textwidth]{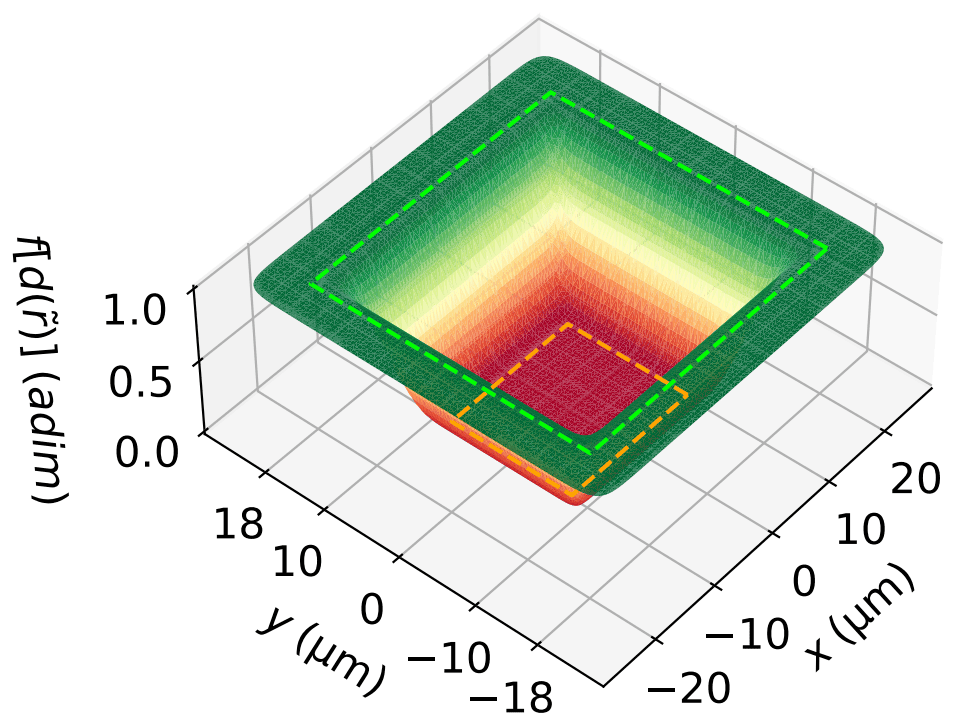}
\caption{\textbf{Realistic thermalization in a smoothed rectangular domain.} The green region (outside the dashed lime contour) corresponds to the fully thermalized portion of the device, where the sigmoid function $f[d(\tilde r)] = 1$ enforces a fixed temperature and vanishing drift velocity through coupling to the external thermal bath. The red region (inside the dashed orange contour) marks the domain where $f[d(\tilde r)] = 0$, meaning that temperature and drift velocity evolve solely according to the VHE. Between these two boundaries, the smooth transition produced by the sigmoid profile models a finite thermalization lengthscale, providing a realistic interpolation between the reservoir and the interior dynamics. \textit{Figure reproduced with permission from Ref. \cite{dragavsevic2023viscous}.}}
\label{fig:realistic_thermalization}
\end{figure}

To implement the realistic thermalization boundary conditions described in the previous section within the rectangular device of Fig. \ref{fig:lattice_cooling_1}, a smoothed rectangular geometry is adopted. This choice is motivated by numerical considerations: domains with rounded corners exhibited faster convergence with respect to mesh refinement and reduced numerical noise compared to sharp-edged rectangular domains \cite{dragavsevic2023viscous}. The smoothed rectangle used in Fig. \ref{fig:lattice_cooling_1} is defined by \cite{dragavsevic2023viscous}
\begin{equation}
\left(\frac{|x|}{a_{\rm out}}\right)^{2a_{\rm out}/\lambda}+\left(\frac{|y|}{b_{\rm out}}\right)^{2b_{\rm out}/\lambda}=1 ,
\end{equation}
where $a_{\rm out}$ and $b_{\rm out}$ represent the semi-axes of the outer region in which the thermal bath is imposed (outside the lime dashed contour in Fig. \ref{fig:realistic_thermalization}), and the parameter $\lambda$ controls the degree of corner smoothing.  
We set $\lambda = a_{\rm out}/3$, which provided a practical compromise between smoothness (beneficial for numerics) and fidelity to a rectangular shape. To incorporate the sigmoid thermalization profile of Eq. \ref{sigmoid} into a rectangular geometry, an effective distance function must replace the radial coordinate $r$. For this purpose we introduced the generalized distance
\begin{equation} \label{gen_distance}
\tilde{R}(x,y)=\left[
\left(\frac{|x|}{a_{\rm out}}\right)^{2a_{\rm out}/\lambda}+\left(\frac{|y|}{b_{\rm out}}\right)^{2b_{\rm out}/\lambda}
\right]^{1/p},
\end{equation}
where the exponent $p$ controls how sharply the two Cartesian contributions are combined. This $\tilde{R}(x,y)$ is then inserted into the mapping
\begin{equation}
d(\tilde{R}) = \frac{\tilde{R}(x,y)-R_{\rm in}}{R_{\rm out}-R_{\rm in}},
\end{equation}
which is subsequently used in the sigmoid of Eq. \eqref{sigmoid}. By construction, $d(\tilde{r})<0$ when $\tilde{r}(x,y)<R_{\rm in}$ and $d(\tilde{r})>1$ when $\tilde{r}(x,y)>R_{\rm out}$. Hence, choosing $R_{\rm in}=\tilde{r}(a_{\rm in},b_{\rm in})$ and $R_{\rm out}=\tilde{r}(a_{\rm out},b_{\rm out})$ guarantees that: (i) $f[d(\tilde{r})]=0$ inside the rectangle with semi-axes $(a_{\rm in}, b_{\rm in})$, where the VHE determine the fields $T$ and $\mathbf{u}$ (red region in Fig. \ref{fig:realistic_thermalization}) and (ii) $f[d(\tilde{r})]=1$ outside the rectangle with semi-axes $(a_{\rm out}, b_{\rm out})$, where the system is perfectly thermalized (green region). The exponent $p$ in Eq. \eqref{gen_distance} controls the shape of the transition region between these two boundaries.

\subsubsection{Boundary conditions for VHE lattice cooling}

In this section we examine how the magnitude of the viscous temperature oscillations—quantified through the lattice cooling strength (LCS)—is influenced by the thermalization lengthscale imposed at the device boundaries. The LCS descriptor \cite{dragavsevic2023viscous},
\begin{equation} \label{LCS}
\mathrm{LCS}=\frac{T_{\mathrm{eq}}-T_{\min}}{T_{\max}-T_{\min}},
\end{equation}
captures how far the temperature at the perturbation center drops below equilibrium during relaxation. Using the same heating protocol described previously, LCS is evaluated while varying only the spatial extent over which the boundary transitions from perfectly thermalized to free-evolving conditions. As the thermalization lengthscale increases, the boundaries act more gradually, reducing the amplitude of the resulting temperature oscillations \cite{dragavsevic2023viscous}. Nonetheless, lattice cooling remains robust: LCS stays appreciable even when thermalization occurs over distances comparable to several micrometers, and remains detectable for lengthscales up to roughly $9 \mu\mathrm{m}$, nearly half the size of the simulated device.

\subsubsection{Effects of isotopic disorder, temperature, and device size on VHE lattice cooling}

We now examine how lattice cooling, as quantified by the LCS, is influenced by three key factors in graphite: isotopic composition, operating temperature, and device dimensions \cite{dragavsevic2023viscous}.
\begin{figure}[h!]
\centering
\includegraphics[width=\textwidth]{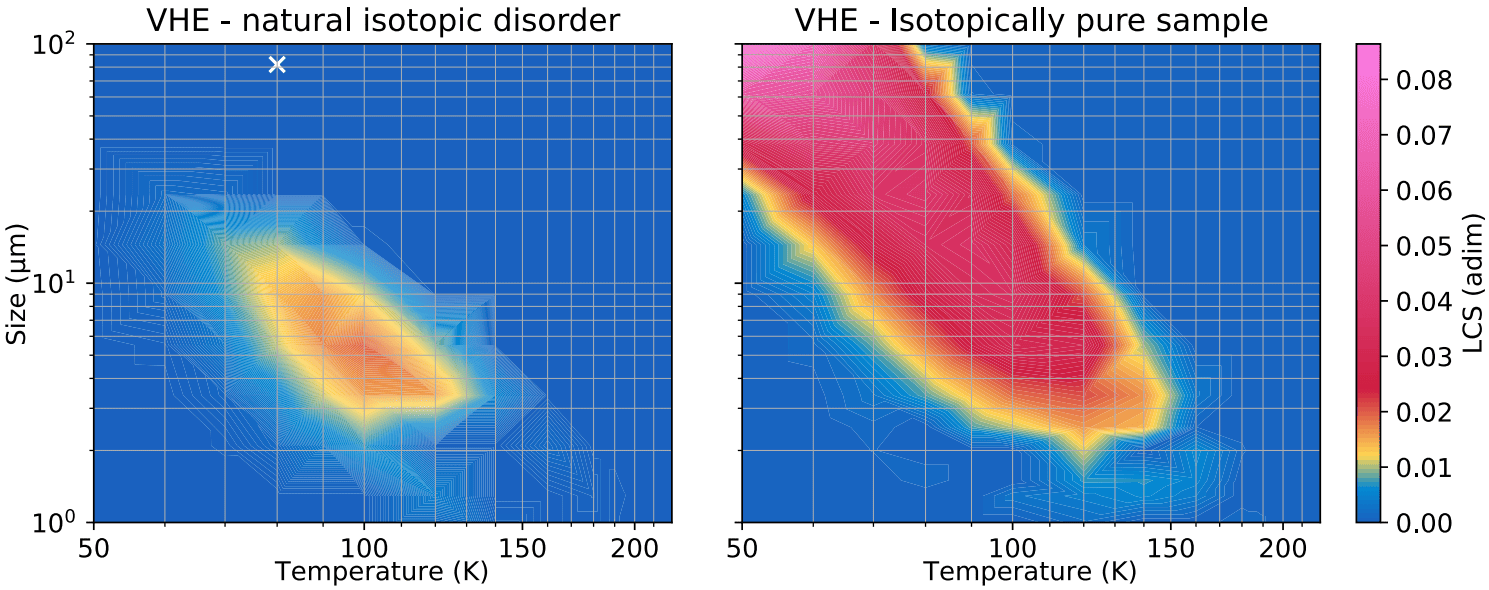}
\caption{\textbf{Lattice cooling strength as a function of system size, temperature, and isotopic composition.} For graphite with natural isotope concentrations (left panel), the lattice cooling strength (LCS, Eq. \eqref{LCS}) is noticeably reduced compared to isotopically enriched samples (right panel), reflecting the enhanced momentum-preserving scattering in the purified material. The white cross in the left panel marks the simulation parameters that most closely reproduce the single-spot pump--probe configuration described in the Supplementary Material of Ref. \cite{jeong2021transient}, where $\sigma_x = 6 \mu\mathrm{m}$ is used and the computational boundaries are placed far away ($80 \mu\mathrm{m}$). Under these conditions, the VHE predict no observable lattice cooling, in line with the experimental findings. \textit{Figure reproduced with permission from Ref. \cite{dragavsevic2023viscous}.}}
\label{fig:lattice_cooling_2}
\end{figure}
The impact of isotopic disorder (see Fig. \ref{fig:lattice_cooling_2}) enters the VHE through the temperature-dependent first-principles transport coefficients, for both naturally abundant graphite (98.9\% $^{12}$C, 1.1\% $^{13}$C) and isotopically enriched samples (99.9\% $^{12}$C, 0.1\% $^{13}$C).\\
The influence of equilibrium temperature was incorporated by using the tabulated temperature dependence of the VHE coefficients (Ref. \cite{simoncelli2020generalization}, Table III). This analysis shows that, for natural graphite, lattice cooling is strongest in the range $70$--$120 \mathrm{K}$ and remains substantial for device sizes between roughly $5$ and $20 \mu\mathrm{m}$. As the device dimensions exceed about $30 \mu\mathrm{m}$, the LCS rapidly diminishes, indicating that hydrodynamic signatures become too weak to generate observable temperature undershoots (see Fig. \ref{fig:lattice_cooling_2}). These conclusions are consistent with experimental trends. For example, the pump–probe measurements of Ref. \cite{jeong2021transient}, which used heaters of radius $6 \mu\mathrm{m}$, do not report temperature oscillations. By matching those conditions in the simulations—using $\sigma_x = 6 \mu\mathrm{m}$ and placing the boundaries far from the heating spot (an $80 \mu\mathrm{m}$ domain)—the LCS vanished, reproducing the experimental outcome. Furthermore, when simulating the ring-shaped geometry employed in Ref. \cite{jeong2021transient}, the VHE results align with the corresponding LBTE predictions and with the experimental observations.

\subsubsection{Comparison of time-dependent relaxation from VHE, DPLE and Fourier’s law}

In this section we compare the transient thermal response predicted by the viscous VHE, the inviscid DPLE, and diffusive Fourier’s law under identical simulation conditions as reported in Ref. \cite{dragavsevic2023viscous}.
\begin{figure}[h!]
\centering
\includegraphics[width=0.8\textwidth]{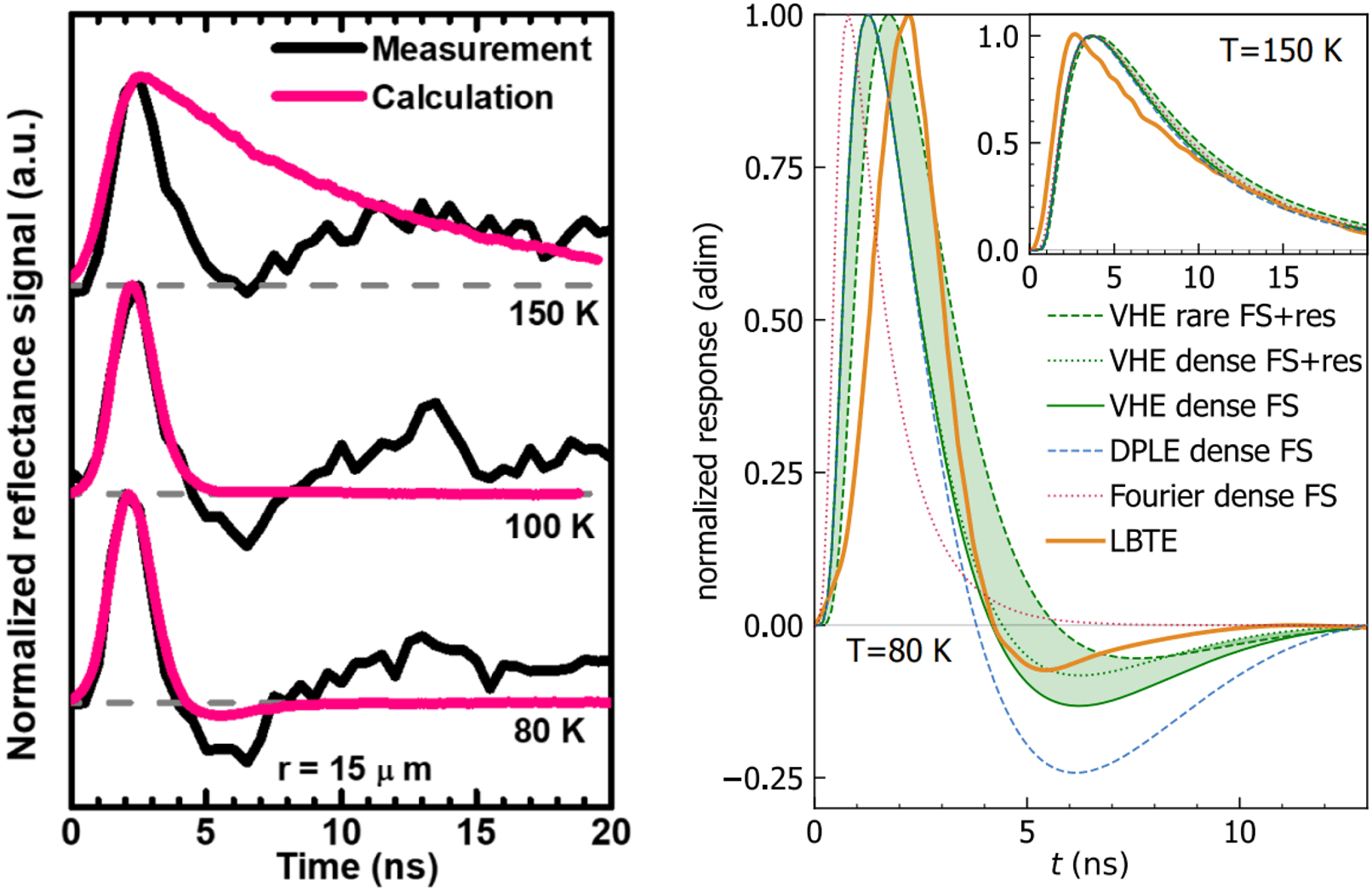}
\caption{\textbf{Time-resolved thermal responses and comparison of transport models.} (Left panel) Comparison between experimental measurements (normalized reflectance data) of Ref. \cite{jeong2021transient} for a pump-beam radius of $15\,\mu\mathrm{m}$ at three temperatures( $80\,\mathrm{K}$, $100\,\mathrm{K}$, and $150\,\mathrm{K}$) and calculated results also from Ref. \cite{jeong2021transient}. The calculations are obtained by simulating the heat-pulse experiment through a time-, real-space-, and reciprocal-space–resolved solution of the BTE employing the full three-phonon scattering matrix to account for strong normal phonon scattering. \textit{Figure reproduced with permission from Ref. \cite{jeong2021transient}.} (Right panel) Time-resolved temperature evolution at the center of a ring-shaped thermal excitation with radius $15\,\mu\mathrm{m}$ in isotopically pure graphite. At $T \simeq 80\,\mathrm{K}$, the lattice-cooling signal predicted by the LBTE (orange, from left panel) is well reproduced by the VHE (green). The inviscid DPLE (dashed blue) overestimates the cooling amplitude, while Fourier’s law yields a purely diffusive, strictly positive response. The effect of finite-size corrections within the VHE is illustrated by considering dense intrinsic scattering (solid green), enhanced Umklapp dissipation due to thermalized boundaries (dotted green), and rare intrinsic scattering with increased dissipation (dashed green). Inset: temperature response of the same device at $T = 150\,\mathrm{K}$, where no hydrodynamic lattice cooling is observed. \textit{Figure reproduced with permission from Ref. \cite{dragavsevic2023viscous}.}}
\label{fig:lattice_cooling_3}
\end{figure}
The simulated configuration closely mirrors the pump--probe experiment reported in Ref. \cite{jeong2021transient}, in which a ring-shaped heater of radius $15\,\mu\mathrm{m}$, modeled by a Gaussian power profile with a full width at half maximum (FWHM) of $3\,\mu\mathrm{m}$, is activated for $0.4\,\mathrm{ns}$ while the device boundaries are held at the equilibrium temperature $T_{\mathrm{eq}}$. The temperature at the center of the ring is monitored by convolving the local temperature field with a Gaussian probe of the same FWHM, consistent with the experimental protocol.\\
Both the viscous VHE and the inviscid DPLE predict a wavelike thermal relaxation at 80 K, as shown by the green and dashed blue curves in the right panel of Fig. \ref{fig:lattice_cooling_3} (see also Ref. \cite{zhang2025theoretical}). In these descriptions, the temperature transiently undershoots the equilibrium value and subsequently exhibits oscillations before relaxing back to $T_{\mathrm{eq}}$. The inviscid DPLE displays a noticeably faster relaxation with more pronounced oscillations, reflecting the absence of viscous damping. In contrast, Fourier’s law yields a purely diffusive response characterized by a monotonic decay without oscillatory features (red dotted curve), as expected for local heat transport. These trends are consistent with experimental observations. In particular, Ref. \cite{jeong2021transient} reports that, at temperatures around $150\,\mathrm{K}$, the time-resolved thermoreflectance signal measured at the center of a $15\,\mu\mathrm{m}$-diameter ring-shaped pump beam exhibits a conventional diffusive behavior: the signal rises rapidly to a maximum shortly after excitation and then decays monotonically to its baseline value (see left panel of Fig. \ref{fig:lattice_cooling_3}). Upon cooling the sample below approximately $120\,\mathrm{K}$, the temporal response changes qualitatively, with the emergence of a sharp positive peak at about $2\,\mathrm{ns}$, followed by a pronounced negative excursion around $6\,\mathrm{ns}$. The magnitude and clarity of this negative feature increase progressively as the temperature is further reduced, signaling the onset of hydrodynamic heat transport. The right panel of Fig. \ref{fig:lattice_cooling_3} further illustrates the role of finite-size effects by varying the relative importance of intrinsic phonon--phonon scattering and phonon--boundary scattering. The cases labeled “dense FS” and “rare FS” correspond to regimes in which intrinsic scattering events are, respectively, more frequent or less frequent than boundary interactions, using a characteristic domain length scale of $40\,\mu\mathrm{m}$. An additional scenario includes an explicit damping contribution to momentum transport, accounting for the finite thermalization length imposed by thermalized boundaries. The resulting variability in the predicted lattice-cooling signal is indicated by the shaded region. Overall, the influence of finite-size effects is found to be smaller than that associated with thermal viscosity. In particular, the depth of the lattice-cooling minimum is strongly controlled by viscous effects: incorporating viscosity within the VHE framework yields closer quantitative agreement with LBTE-based predictions, whereas inviscid descriptions such as the DPLE underestimate the cooling amplitude. In contrast, Fourier’s law predicts a strictly positive, diffusive response and fails to capture the oscillatory behavior observed both in LBTE calculations and in experiments \cite{jeong2021transient}. Recently, an analytical Green’s-function solution of the multidimensional BTE has been introduced to model temperature oscillations observed in ultrafast pump–probe experiments, while avoiding the computationally expensive inversion of the full phonon scattering matrix \cite{qian2025analytical}.\\
Finally, the VHE and LBTE exhibit good agreement over the time window relevant for lattice cooling, with discrepancies confined to very short times. These early-time differences arise from the coarse-graining procedure underlying the derivation of the VHE, which neglects rapid spatial and temporal variations of the nonequilibrium phonon distribution, but they do not affect the interpretation of the lattice-cooling phenomenon. At higher temperatures ($\bar{T} = 150\,\mathrm{K}$), both approaches predict an entirely positive thermal response (see inset in the right panel of Fig. \ref{fig:lattice_cooling_3}), confirming the absence of lattice cooling and indicating that thermal viscosity plays a negligible role in this regime. Overall, the VHE provide an efficient and accurate description of transient thermal transport, capturing the essential features of LBTE dynamics while clarifying the role of viscosity in low-temperature hydrodynamic heat flow.

\subsection{Limitations of the viscous heat equations}
\label{VHE_limitations}

Here, we summarize the main assumptions underlying the VHE and discuss their regime of applicability.\\
First, as discussed throughout this section, the VHE are obtained by coarse-graining the phonon LBTE. As such, they inherit from the LBTE an applicability restricted to the linear-response regime, where the deviation of the phonon population from equilibrium can be approximated as a linear functional of the temperature $T(\boldsymbol{R},t)$ and drift-velocity $\boldsymbol{u}(\boldsymbol{R},t)$ fields, and of their gradients \cite{simoncelli2020generalization}. Consequently, the VHE are expected to be accurate when temperature variations across the device are small compared with the average sample temperature, and sufficiently small that the transport coefficients entering the VHE can be regarded as temperature independent, i.e., evaluated at the average temperature of the sample. In the presence of very large temperature gradients or strong deviations from local equilibrium, nonlinear effects and higher-order corrections to the phonon distribution become important, and a nonlinear solution of the BTE may be required.\\
Second, the VHE neglect spatial gradients of the non-equilibrium part of the phonon population, $\nabla n_\nu^\delta(\boldsymbol{R},t)$, in the determination of the transport coefficients. This approximation is justified when the characteristic length scales over which the hydrodynamic fields vary are large compared to the phonon mean free paths. As device dimensions approach these microscopic scattering length scales, finite-size effects can be approximately incorporated through Bosanquet-type \cite{michalis2010rarefaction} rescalings of the transport coefficients, an approach commonly used to account for ballistic effects in rarefied fluids. A direct comparison between VHE and full space-dependent LBTE solutions, shown in Fig.~\ref{fig:tunnel_chamber_2}, indicates that, once finite-size effects are incorporated through suitable effective boundary conditions and Bosanquet-rescaled transport coefficients, the VHE reproduce hydrodynamic features such as temperature inversion and vortical heat-flow patterns within the parameter range considered here.\\
Third, the derivation of the VHE transport coefficients neglects the explicit time derivative of the non-equilibrium phonon population, $\partial_t n_\nu^\delta$, so that the transport coefficients are defined under static, frequency-independent conditions. This approximation is reasonably valid for the sub-GHz transient regimes considered in this work, as supported by comparison with time-dependent LBTE simulations in Fig.~\ref{fig:lattice_cooling_3}, which show that phenomena such as lattice cooling are reproduced within the relevant time window. Deviations are expected when the characteristic driving frequencies become comparable to intrinsic phonon scattering rates, so that temporal nonlocality and frequency-dependent transport coefficients become important \cite{chaput2013direct}.\\
Overall, the VHE aim to describe device-size effects in hydrodynamic and moderately nonlocal regimes, where phonon distributions remain close to local equilibrium, spatial and temporal gradients are sufficiently smooth, and transport coefficients are well defined. Near the boundary of this regime, additional phenomena may arise from the interplay between ballistic propagation, boundary scattering, and momentum-conserving collisions. One example is the phonon ``Knudsen minimum'' \cite{ding2018phonon}. The validity of the VHE in this crossover regime should therefore be assessed in future work, for example by systematic benchmarks against full space-dependent solutions of the LBTE. Outside this regime, namely for strongly nonlinear temperature fields, predominantly ballistic transport, or ultrafast dynamics, a full space- and time-dependent LBTE, a nonlinear BTE treatment, or higher-level atomistic approaches may be required.

\subsection*{Takeaways}

In this section we showed how the viscous heat equations emerge as a systematic mesoscopic coarse-graining of the phonon Boltzmann transport equation, obtained by projecting the dynamics onto the subspace spanned by the conserved energy and momentum collective modes (relaxons). This construction introduces the phonon drift velocity as an additional hydrodynamic field, enabling a unified continuum description of diffusive, hydrodynamic, and intermediate transport regimes. We demonstrated that thermal viscosity arises naturally from even relaxons and governs momentum diffusion, giving rise to hallmark hydrodynamic phenomena such as Poiseuille heat flow, finite heat-flux vorticity, and steady-state heat backflow in suitably confined geometries. The viscous heat equations were shown to reduce to well-known macroscopic models in appropriate limits: Fourier’s law at high temperatures and strong momentum dissipation, the Guyer–Krumhansl equations under isotropic, drift-dominated conditions, and the dual-phase-lag equation in the inviscid limit. Importantly, the viscous heat equations retain a direct microscopic foundation through first-principles transport coefficients while offering a computationally efficient framework capable of capturing boundary effects, nonlocality, and transient phenomena such as second sound and lattice cooling. Together, these results establish the viscous heat equations as the natural bridge between microscopic phonon kinetics and experimentally observable hydrodynamic heat transport.

\section{Analytical framework: modified Helmholtz and biharmonic equations}

As highlighted in earlier sections, the field of phonon hydrodynamics has seen accelerating progress, yet clear experimental signatures of viscous behavior at the macroscopic scale remain relatively scarce. Moreover, simple analytical techniques for diagnosing such effects are still limited, posing challenges for device-level modeling.\\
By coarse-graining the LBTE, one obtains the viscous heat equations (VHE) \cite{simoncelli2020generalization}, which play a role analogous to the Navier--Stokes equations for laminar fluids, with the crucial difference that they involve not only temperature but also a phonon drift-velocity field. This correspondence sharpens the analogy with pressure and velocity in classical fluids. The VHE faithfully describe the hydrodynamic, diffusive, and crossover regimes, dramatically reducing computational cost compared to the LBTE while retaining physical insight. Since solving the LBTE in nontrivial geometries remains extremely demanding, its direct usefulness at device scale is constrained; the VHE, instead, accommodate geometry and boundary conditions naturally, and have been validated against spatially resolved LBTE simulations in micron-scale systems \cite{dragavsevic2023viscous}.\\
In what follows, we demonstrate that the steady-state VHE can be recast into two decoupled modified biharmonic equations—one for a velocity potential and one for a stream function—thereby enabling analytic solutions in Fourier space. This procedure delivers a full closed-form expression for the temperature field and exposes its decomposition into compressible and vortical contributions. Compressibility, in particular, emerges as a defining property of viscous phonon flow, distinguishing it from most electronic hydrodynamic systems \cite{torre2015nonlocal,bandurin2016negative,levitov2016electron}, which are typically modeled as nearly incompressible \cite{levitov2016electron}. We identify the interplay of compressibility and vorticity as the microscopic source of thermal viscosity. Following the route of Ref. \cite{levitov2016electron}, we relate this mechanism to negative nonlocal thermal resistance in a two-dimensional strip—an unmistakable signature of hydrodynamic heat flow—arising from thermal vortices \cite{raya2022hydrodynamic,restuccia2023non,sykora2023multiscale,shang2020heat,zhang2021heat,tur2024microscopic} or thermal backflow \cite{dragavsevic2023viscous}.\\\\
We again consider the steady-state, isotropic version of the VHE (Eqs. \eqref{VHE1}–\eqref{VHE2}) and adopt the shorthand
\begin{equation} \label{easy_notation}
\alpha_{ij}=\alpha\delta_{ij}=W\delta_{ij}\sqrt{\bar{T}AC},\qquad
\beta_{ij}=\beta\delta_{ij}=W\delta_{ij}\sqrt{\frac{AC}{\bar{T}}},\qquad
\gamma_{ij}=\gamma\delta_{ij}=AD^{U}\delta_{ij}.
\end{equation}
Following the conventions of Ref. \cite{di2025vortices}, the steady-state VHE reduce to
\begin{equation} \label{VHE_preliminar}
\begin{cases}
\displaystyle
\sum_{i,j}\alpha\,\delta_{ij}\,\frac{\partial u_{j}}{\partial r_{i}}
-\sum_{i,j}\kappa^{D}\delta_{ij}\,\frac{\partial^{2}T}{\partial r_{i}\partial r_{j}}=0, \\[8pt]
\displaystyle
\sum_{j}\beta\,\delta_{ij}\frac{\partial T}{\partial r_{j}}
-\sum_{j,k,l}\eta^{ijkl}\frac{\partial^{2}u^{k}}{\partial r_{j}\partial r_{l}}
=-\sum_{j}\gamma\,\delta_{ij}u_{j}.
\end{cases}
\end{equation}
Assuming an isotropic diffusion-damped conductivity $\kappa^{D}$ and using Landau’s decomposition of viscous stresses \cite{landau2013fluid} into shear viscosity $\eta$ and volume viscosity $\zeta$, the above simplifies to
\begin{equation} \label{VHE}
\begin{cases}
\alpha\nabla\cdot\boldsymbol{u}=\kappa^{D}\nabla^{2}T,\\
\beta\nabla T-\eta\nabla^{2}\boldsymbol{u}-\left(\zeta+\frac{\eta}{3}\right)\nabla(\nabla\cdot\boldsymbol{u})
=-\gamma\boldsymbol{u}.
\end{cases}
\end{equation}
For incompressible fluids the term proportional to $\zeta$ vanishes, but phonon fluids are inherently compressible, in contrast to most electron hydrodynamic systems where compressibility is strongly suppressed \cite{levitov2016electron}.

\subsection{Decoupling of the steady-state VHE} \label{disentanglement}

We now isolate equations governing separately the divergence and curl of $\boldsymbol{u}$.  
Starting from Eq. \eqref{VHE}, the divergence of the first line gives
\begin{equation} \label{div}
\nabla\cdot(\alpha\boldsymbol{u}-\kappa^{D}\nabla T)=0.
\end{equation}
Defining for convenience
\begin{equation} \label{change}
\mu^{*}=\zeta+\frac{\eta}{3},
\end{equation}
the second line of Eq. \eqref{VHE} becomes
\begin{equation} \label{vhe_2_recast}
\nabla T=\frac{\eta\nabla^{2}\boldsymbol{u}+\mu^{*}\nabla(\nabla\cdot\boldsymbol{u})-\gamma\boldsymbol{u}}{\beta}.
\end{equation}
Substituting Eq. \eqref{vhe_2_recast} into Eq. \eqref{div} yields
\begin{equation}
\nabla\cdot\Big[\left(\tfrac{\kappa^{D}\gamma}{\beta}+\alpha\right)\boldsymbol{u}-\tfrac{\kappa^{D}\eta}{\beta}\nabla^{2}\boldsymbol{u}-\tfrac{\kappa^{D}\mu^{*}}{\beta}\nabla(\nabla\cdot\boldsymbol{u})\Big]=0.
\end{equation}
Introducing the thermal compressibility field \cite{di2025vortices,di2024vorticity}
\begin{equation} \label{compressibility}
\Phi=\nabla\cdot\boldsymbol{u},
\end{equation}
and noting that divergence and Laplacian commute, we arrive at
\begin{equation} \label{compressibility_eq}
\frac{\kappa^{D}(\eta+\mu^{*})}{\beta}\nabla^{2}\Phi-\left(\frac{\kappa^{D}\gamma}{\beta}+\alpha\right)\Phi=0.
\end{equation}
Next we define thermal vorticity as
\begin{equation} \label{vorticity}
\boldsymbol{\mathscr{W}}=\nabla\times\boldsymbol{u}.
\end{equation}
Taking the curl of the momentum-like VHE (second line of Eq. \eqref{VHE}) eliminates temperature and any gradient term, leaving
\begin{equation} \label{vorticity_eq}
\eta\nabla^{2}\boldsymbol{\mathscr{W}}-\gamma\boldsymbol{\mathscr{W}}=0.
\end{equation}
Thus both compressibility and vorticity satisfy modified Helmholtz equations:
\begin{equation} \label{VHE_compressibility_vorticity}
\begin{cases}
\displaystyle \frac{\kappa^{D}(\eta+\mu^{*})}{\beta}\nabla^{2}\Phi-\left(\frac{\kappa^{D}\gamma}{\beta}+\alpha\right)\Phi=0, \\[4pt]
\displaystyle \eta\nabla^{2}\boldsymbol{\mathscr{W}}-\gamma\boldsymbol{\mathscr{W}}=0.
\end{cases}
\end{equation}
Volume viscosity $\zeta$ influences the solution for $\Phi$ but drops out of the vorticity equation, which depends only on $\eta$.  
Given $\Phi$ and $\boldsymbol{\mathscr{W}}$, the velocity field follows from the Helmholtz decomposition
\begin{equation} \label{u_helmoltz_decomposition}
\boldsymbol{u}=-\nabla\phi+\nabla\times\boldsymbol{\Psi},
\end{equation}
with potentials given by
\begin{equation} \label{connection_vel_pot_strem_f_to_compress_and_vort}
\begin{split}
\phi(\boldsymbol{r})&=\frac{1}{4\pi}\int_{V}\frac{\Phi(\boldsymbol{r}')}{|\boldsymbol{r}-\boldsymbol{r}'|}\,dV'
-\frac{1}{4\pi}\int_{S}\hat{\boldsymbol{n}}'\frac{\boldsymbol{u}(\boldsymbol{r}')}{|\boldsymbol{r}-\boldsymbol{r}'|}\,dS',\\
\boldsymbol{\Psi}(\boldsymbol{r})&=\frac{1}{4\pi}\int_{V}\frac{\boldsymbol{\mathscr{W}}(\boldsymbol{r}')}{|\boldsymbol{r}-\boldsymbol{r}'|}\,dV'
-\frac{1}{4\pi}\int_{S}\hat{\boldsymbol{n}}'\frac{\boldsymbol{u}(\boldsymbol{r}')}{|\boldsymbol{r}-\boldsymbol{r}'|}\,dS'.
\end{split}
\end{equation}
Because $\Phi$ and $\boldsymbol{\mathscr{W}}$ lack direct experimental signatures, it is more convenient to work in Fourier space using $\phi$ and $\boldsymbol{\Psi}$, where boundary conditions on $\boldsymbol{u}$ are natural.\\
The above factorization does not apply for time-dependent fields. With temporal dynamics, the VHE read \cite{simoncelli2020generalization,di2025vortices}
\begin{equation} \label{VHE_time_dep}
\begin{cases}
C\frac{\partial T}{\partial t}+\alpha\nabla\cdot\boldsymbol{u}=\kappa^{D}\nabla^{2}T,\\
A\frac{\partial\boldsymbol{u}}{\partial t}+\beta\nabla T-\eta\nabla^{2}\boldsymbol{u}-\left(\zeta+\frac{\eta}{3}\right)\nabla(\nabla\cdot\boldsymbol{u})
=-\gamma\boldsymbol{u}.
\end{cases}
\end{equation}
Taking the curl yields for the vorticity:
\begin{equation} \label{vorticity_eq_time_dep}
A\frac{\partial\boldsymbol{\mathscr{W}}}{\partial t}-\eta\nabla^{2}\boldsymbol{\mathscr{W}}+\gamma\boldsymbol{\mathscr{W}}=0,
\end{equation}
while $\Phi$ couples directly to $\partial T/\partial t$ and therefore does not satisfy an analogous decoupled equation. The full transient behavior will be examined in Sec. \ref{transient_heat_backflow_section}.

\subsection{Biharmonic equations for the velocity potentials}

Using the decomposition \eqref{u_helmoltz_decomposition}, Eq. \eqref{VHE_compressibility_vorticity} becomes
\begin{equation} \label{potential_VHE}
\begin{cases}
\frac{\kappa^{D}(\eta+\mu^{*})}{\beta}\nabla^{2}\nabla\cdot[-\nabla\phi+\nabla\times\boldsymbol{\Psi}]
-\left(\frac{\kappa^{D}\gamma}{\beta}+\alpha\right)\nabla\cdot[-\nabla\phi+\nabla\times\boldsymbol{\Psi}]=0,\\[4pt]
\eta\nabla^{2}\nabla\times[-\nabla\phi+\nabla\times\boldsymbol{\Psi}]
-\gamma\nabla\times[-\nabla\phi+\nabla\times\boldsymbol{\Psi}]=0.
\end{cases}
\end{equation}
Invoking gauge freedom
\begin{equation} \label{gauge_freedom}
\boldsymbol{\Psi}\rightarrow\boldsymbol{\Psi}'=\boldsymbol{\Psi}+\nabla\mathcal{A},
\end{equation}
whose curl is unchanged, the first of Eqs. \eqref{potential_VHE} simplifies to
\begin{equation} \label{Phi_eq}
\frac{\kappa^{D}(\eta+\mu^{*})}{\beta}\nabla^{2}(\nabla^{2}\phi)
-\left(\frac{\kappa^{D}\gamma}{\beta}+\alpha\right)\nabla^{2}\phi=0.
\end{equation}
Similarly, fixing the gauge so that $\nabla\cdot\boldsymbol{\Psi}=0$ reduces the second equation to
\begin{equation} \label{Psi_eq}
\eta\nabla^{2}(\nabla^{2}\boldsymbol{\Psi})-\gamma\nabla^{2}\boldsymbol{\Psi}=0.
\end{equation}
In two dimensions, $\boldsymbol{\Psi}$ has only one nonzero component,  
\begin{equation}
\boldsymbol{\Psi}=(0,0,\psi),
\end{equation}
so that $\psi$ and $\phi$ obey the pair of modified biharmonic equations \cite{selvadurai2013partial,jiang2013second,biros2004embedded,greengard1998integral,di2025vortices}
\begin{equation} \label{potential_VHE_final}
\begin{cases}
\frac{\kappa^{D}(\eta+\mu^{*})}{\beta}\nabla^{2}(\nabla^{2}\phi)-\left(\frac{\kappa^{D}\gamma}{\beta}+\alpha\right)\nabla^{2}\phi=0,\\
\eta\nabla^{2}(\nabla^{2}\psi)-\gamma\nabla^{2}\psi=0.
\end{cases}
\end{equation}
The corresponding drift velocity follows directly from
\begin{equation} \label{potential_u}
\boldsymbol{u}
=-\nabla\phi+\nabla\times\boldsymbol{\Psi}
=\left(-\partial_{x}\phi+\partial_{y}\psi,\,-\partial_{y}\phi-\partial_{x}\psi\right).
\end{equation}
Equations \eqref{VHE_compressibility_vorticity} and \eqref{potential_VHE_final} represent the first complete analytical decoupling of heat flow in a compressible phonon fluid into independent compressible and vortical sectors. In the limit $\gamma\to 0$ the second of Eqs. \eqref{potential_VHE_final} reduces to the biharmonic equation of incompressible Stokes flow \cite{batchelor1967introduction}.\\
This analytical picture highlights the profound difference between Fourier conduction and hydrodynamic transport: while Fourier’s law yields harmonic equations unable to sustain interior extrema, the modified biharmonic structure of hydrodynamic heat flow naturally supports such features, enabling sign reversals and negative thermal response in confined geometries, as discussed in Sec. \ref{strip_device_section}.

\subsection{Analytical solution in a graphite strip} \label{strip_device_section}

\begin{figure}[!htb]
\centering
\includegraphics[width=0.8\textwidth]{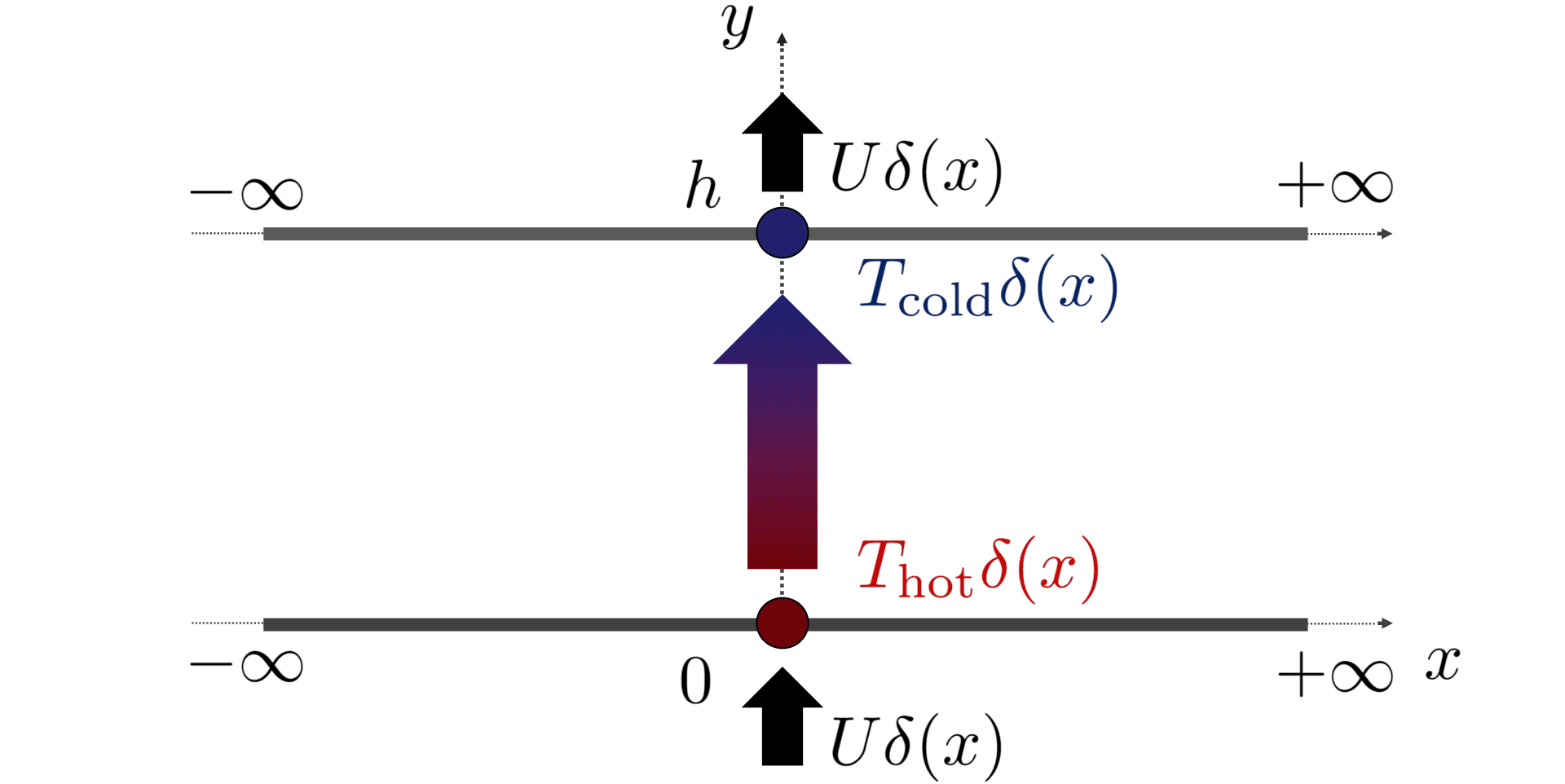}
\caption{\textbf{Geometry of the strip where the modified biharmonic equations \eqref{potential_VHE_final} are solved.} The strip extends infinitely along $x$ and has width $h$ along $y$. Point-like injection of drift velocity and temperature gradient is imposed at $x=0$ as specified in Eqs. \eqref{BCuy}–\eqref{temperature_BC_y=0,h}. In numerical simulations with finite contacts we set $h=5\,\mu{\rm m}$ and $L=20\,\mu{\rm m}$, ensuring convergence in the strip interior.}
\label{fig:strip_pic}
\end{figure}

\begin{figure}[!htb]
\centering
\includegraphics[width=0.7\textwidth]{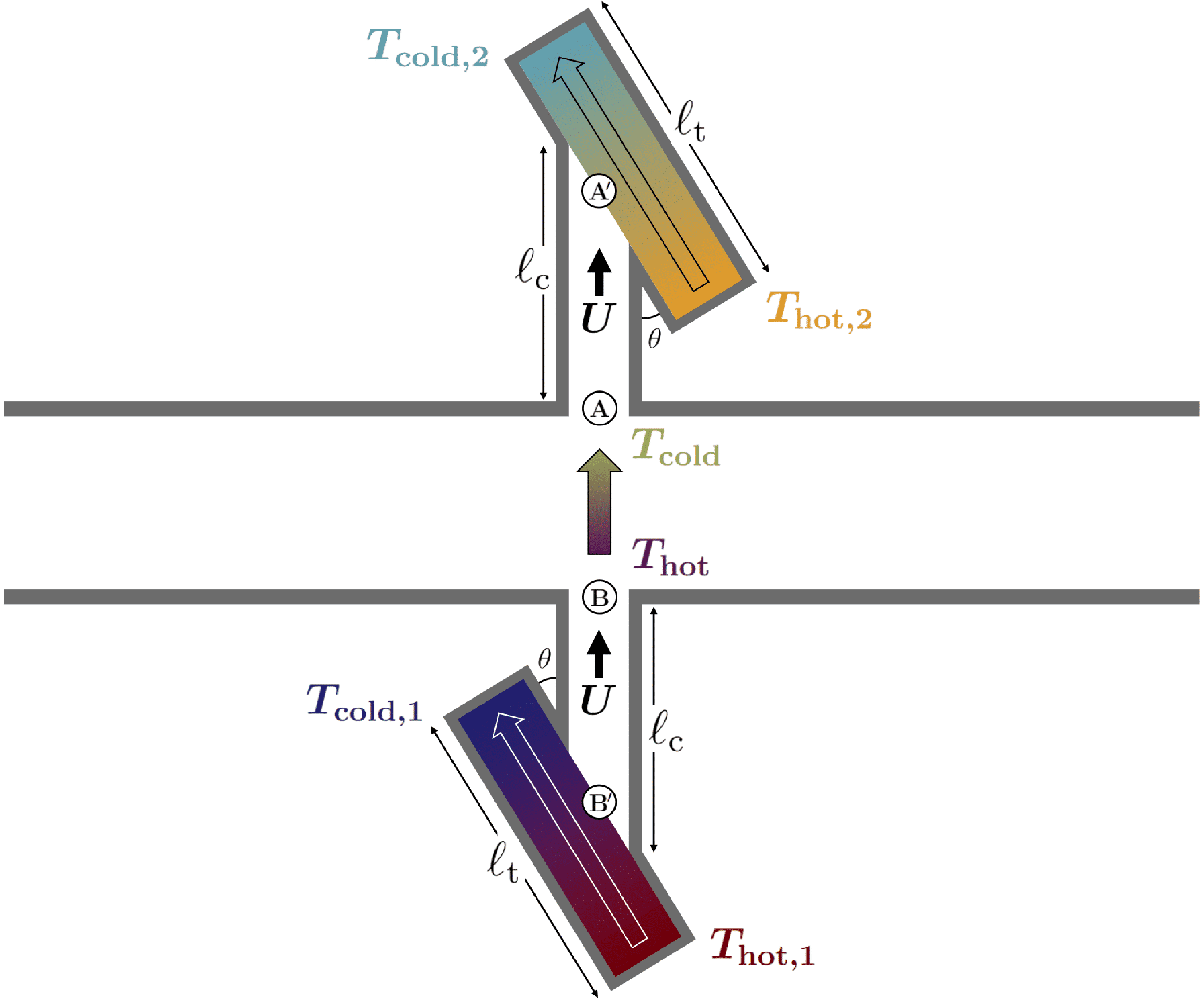}
\caption{\textbf{Experimental setup for phonon drift injection and thermal backflow in a 2D strip device.} Schematic of a two-channel thermal device in which vertically offset hot and cold reservoirs impose laterally displaced temperature gradients. This configuration generates a controlled transverse phonon drift velocity injected into a two-dimensional strip, while minimizing the longitudinal component through a small cantilever tilt. By tuning the relative channel lengths and temperature offsets, the setup enables directional control of the injected drift and the emergence of thermal backflow in the strip.
\textit{Figure reproduced with permission from Ref. \cite{di2025vortices}.}}
\label{fig:setup}
\end{figure}

We now apply this formalism to a two-dimensional graphite strip of width $h$ and infinite length \cite{levitov2016electron,di2025vortices}. A schematic is shown in Fig. \ref{fig:strip_pic}.
A mixed real–momentum representation proves convenient:
\[
f(x,y)=\frac{1}{2\pi}\!\int dk\,e^{ikx}f(k,y),\qquad
f(k,y)=\int dx\,e^{-ikx}f(x,y).
\]
Within this framework, Eqs. \eqref{potential_VHE_final} can be solved analytically subject to boundary conditions on the horizontal edges $y=0$ and $y=h$.  
The normal component of the drift velocity obeys
\begin{equation} \label{BCuy}
u_{y}(x,0)=u_{y}(x,h)=U\delta(x),
\end{equation}
while the tangential component satisfies no-slip conditions
\begin{equation} \label{BCux}
u_{x}(x,0)=u_{x}(x,h)=0.
\end{equation}
Temperature is pinned at the two boundaries via point-like hot and cold sources:
\begin{equation} \label{temperature_BC_y=0,h}
\begin{cases}
T(x,0)=T_{\rm hot}=\bar{T}+\Delta T\delta(x),\\
T(x,h)=T_{\rm cold}=\bar{T}-\Delta T\delta(x).
\end{cases}
\end{equation}
These conditions enforce a temperature gradient aligned with the injected phonon drift. The layout is illustrated in Fig. \ref{fig:strip_pic}. \\
In Fig. \ref{fig:setup} we outline a possible experimental configuration capable of inducing thermal backflow in in-plane graphite and other two-dimensional materials \cite{di2025vortices}. The arrangement employs two outer channels connected to the strip through adjustable cantilevers of length $\ell_{\rm c}$, chosen to guarantee rapid thermal equilibration—namely, that the temperatures at the ends of each cantilever are essentially the same ($T_{\rm cold}=T_{\rm A}=T_{\rm A'}$ and $T_{\rm hot}=T_{\rm B}=T_{\rm B'}$).
By shifting the temperature profiles of the upper and lower channels relative to one another and enforcing a gradient in the upper channel that is displaced with respect to the lower one,
\begin{equation}
\begin{cases}
T_{\rm hot,1}=T_{\rm hot,2}+\delta T,\\
T_{\rm cold,1}=T_{\rm cold,2}+\delta T,
\end{cases}
\end{equation}
the upper channel is kept at a higher temperature than the lower channel, thus establishing a vertical (top–bottom) temperature drop across the strip. The lengths of the channels ($\ell_{\rm t}$) may differ; varying $\ell_{\rm c}$ enables fine control of the required thermal conditions. Proper alignment of the imposed gradients generates a finite $y$-component of the injected drift velocity, while selecting a small cantilever tilt angle keeps the $x$-component minimal (small $\theta$). Owing to the symmetry of the design, the injected drift velocity closely reproduces the one emerging from the strip, and modifying the gradients in the channels provides directional control, thereby implementing the necessary boundary conditions for both the drift velocity and the temperature gradient.\\
To tackle the system analytically, we begin by restating the modified biharmonic equation in its generic scalar form:
\begin{equation} \label{f_equation}
\nabla^{2}(\nabla^{2}f)-c\nabla^{2}f=0.
\end{equation}
Performing a Fourier transform in $x$ (transitioning to $(k,y)$–space) converts this PDE into a fourth–order ODE of the form \cite{di2025vortices}
\begin{equation} \label{4th_order_ODE}
\left[k^{4}-2k^{2}\partial_{y}^{2}+\partial_{y}^{4}+ck^{2}-c\partial_{y}^{2}\right]f(k,y)=0
\;\;\Longrightarrow\;\;
(\partial_{y}^{2}-k^{2})(\partial_{y}^{2}-q^{2})f(k,y)=0,
\end{equation}
with the shorthand $q^{2}=k^{2}+c$. The general solution to Eq. \eqref{4th_order_ODE} consists of a linear combination of exponentials:
\begin{equation} \label{solution_f}
f(k,y)=a_{f-}e^{-ky}+a_{f+}e^{ky}+b_{f-}e^{-qy}+b_{f+}e^{qy},
\qquad q^{2}=k^{2}+c.
\end{equation}
Employing this template for the two potentials $\phi$ and $\psi$ appearing in Eqs. \eqref{potential_VHE_final}, one obtains \cite{di2025vortices}
\begin{equation} \label{solution_Phi}
\phi(k,y)=a_{\phi-}e^{-ky}+a_{\phi+}e^{ky}+b_{\phi-}e^{-q_{\phi}y}+b_{\phi+}e^{q_{\phi}y},
\qquad
q_{\phi}^{2}=k^{2}+\frac{\gamma}{\eta+\mu^{*}}+\frac{\alpha\beta}{\kappa^{D}(\eta+\mu^{*})},
\end{equation}
and
\begin{equation} \label{solution_psi}
\psi(k,y)=a_{\psi-}e^{-ky}+a_{\psi+}e^{ky}+b_{\psi-}e^{-q_{\psi}y}+b_{\psi+}e^{q_{\psi}y},
\qquad
q_{\psi}^{2}=k^{2}+\frac{\gamma}{\eta}.
\end{equation}
The next step is to enforce boundary conditions in order to determine the coefficients $a_{\phi/\psi,\pm}$ and $b_{\phi/\psi,\pm}$. Because the geometry and injection protocol (Fig. \ref{fig:strip_pic}) possess mirror symmetry about the midline $y=h/2$, the drift-velocity components obey the relations
\begin{equation} \label{u_symmetry_condition_k_space}
\begin{cases}
u_{x}(k,y)=-u_{x}(k,h-y),\\[2pt]
u_{y}(k,y)=u_{y}(k,h-y),
\end{cases}
\end{equation}
which directly constrain the coefficients appearing in Eqs. \eqref{solution_Phi}–\eqref{solution_psi}. Imposing Eq. \eqref{u_symmetry_condition_k_space} yields
\begin{equation} \label{reduced_coefficients}
\begin{cases}
a_{\phi-}=-a_{\phi+}e^{kh},\\
b_{\phi-}=-b_{\phi+}e^{q_{\phi}h},\\
a_{\psi-}=a_{\psi+}e^{kh},\\
b_{\psi-}=b_{\psi+}e^{q_{\psi}h}.
\end{cases}
\end{equation}
Using these relations, we express the potentials in a more compact form:
\begin{equation} \label{solution_psi_new}
\psi(k,y)=a_{\psi}\!\left(e^{ky}+e^{kh}e^{-ky}\right)
+b_{\psi}\!\left(e^{q_{\psi}y}+e^{q_{\psi}h}e^{-q_{\psi}y}\right),
\end{equation}
\begin{equation} \label{solution_Phi_new}
\phi(k,y)=a_{\phi}\!\left(e^{ky}-e^{kh}e^{-ky}\right)
+b_{\phi}\!\left(e^{q_{\phi}y}-e^{q_{\phi}h}e^{-q_{\phi}y}\right),
\end{equation}
where we have renamed $a_{\psi+}\equiv a_{\psi}$, $b_{\psi+}\equiv b_{\psi}$, $a_{\phi+}\equiv a_{\phi}$ and $b_{\phi+}\equiv b_{\phi}$.\\
Using Eqs. \eqref{solution_psi_new}–\eqref{solution_Phi_new} in the velocity reconstruction formula, we find
\begin{equation} \label{u_y_ky_general}
u_{y}(k,y)=
-ka_{\phi}^{\psi}\!\left(e^{ky}+e^{kh}e^{-ky}\right)
-q_{\phi}b_{\phi}\!\left(e^{q_{\phi}y}+e^{q_{\phi}h}e^{-q_{\phi}y}\right)
-ik\,b_{\psi}\!\left(e^{q_{\psi}y}+e^{q_{\psi}h}e^{-q_{\psi}y}\right),
\end{equation}
\begin{equation} \label{u_x_ky_general}
u_{x}(k,y)=
-ik\,a_{\phi}^{\psi}\!\left(e^{ky}-e^{kh}e^{-ky}\right)
-ik\,b_{\phi}\!\left(e^{q_{\phi}y}-e^{q_{\phi}h}e^{-q_{\phi}y}\right)
+b_{\psi}q_{\psi}\!\left(e^{q_{\psi}y}-e^{q_{\psi}h}e^{-q_{\psi}y}\right),
\end{equation}
with the definition $a_{\phi}^{\psi}=a_{\phi}+ia_{\psi}$. Also the temperature profile displays a specific symmetry relation. As was done for the drift–velocity field in Eq. \eqref{u_symmetry_condition_k_space}, we now examine the symmetry properties of the temperature in the strip geometry of Fig. \ref{fig:strip_pic}. Starting from the second VHE, Eq. \eqref{vhe_2_recast}, we may write
\begin{equation}
\nabla T=\frac{\eta\nabla^{2}\boldsymbol{u}
+\mu^{*}\,\nabla(\nabla\!\cdot\!\boldsymbol{u})
-\gamma\boldsymbol{u}}{\beta}.
\end{equation}
Inserting the Helmholtz decomposition of the drift velocity (Eq. \eqref{potential_u}) gives
\begin{equation} \label{relation_T_u}
\begin{split}
\nabla T=\left(-\frac{\gamma}{\beta}+\frac{\eta}{\beta}\nabla^{2}\right)
\!\left[-\frac{\partial\phi}{\partial x}
+\frac{\partial\psi}{\partial y}\,,\,
-\frac{\partial\phi}{\partial y}
-\frac{\partial\psi}{\partial x}\right]+\frac{\mu^{*}}{\beta}\,
\nabla\!\left(\nabla\!\cdot\!\left[-\frac{\partial\phi}{\partial x}
+\frac{\partial\psi}{\partial y}\,,\,
-\frac{\partial\phi}{\partial y}
-\frac{\partial\psi}{\partial x}\right]\right).
\end{split}
\end{equation}
Carrying out the derivatives and organizing terms leads to the explicit temperature expression in mixed $(k,y)$ space:
\begin{equation} \label{expression_for_temperature}
T(k,y)=\frac{\gamma}{\beta}
\left(e^{ky}-e^{kh}e^{-ky}\right)(a_{\phi}+ia_{\psi})
-\frac{\alpha}{\kappa^{D}}
\left(e^{q_{\phi}y}-e^{q_{\phi}h}e^{-q_{\phi}y}\right)b_{\phi},
\end{equation}
where the definitions of $q_{\phi}$ and $q_{\psi}$ from Eqs. \eqref{solution_Phi} and \eqref{solution_psi} have been used. We now exploit Eq. \eqref{expression_for_temperature} to deduce the symmetry of the temperature field. Evaluating it at the reflected coordinate $y\rightarrow h-y$ yields
\begin{equation} \label{symmetry_of_temperature}
\begin{split}
T(k,h-y)
&=\frac{\gamma}{\beta}\!\left(e^{kh}e^{-ky}-e^{ky}\right)(a_{\phi}+ia_{\psi})
-\frac{\alpha}{\kappa^{D}}\!\left(e^{q_{\phi}h}e^{-q_{\phi}y}-e^{q_{\phi}y}\right)b_{\phi}\\[4pt]
&=-\left[
\frac{\gamma}{\beta}\!\left(e^{ky}-e^{kh}e^{-ky}\right)(a_{\phi}+ia_{\psi})
-\frac{\alpha}{\kappa^{D}}\!\left(e^{q_{\phi}y}-e^{q_{\phi}h}e^{-q_{\phi}y}\right)b_{\phi}
\right]\\[4pt]
&=-T(k,y).
\end{split}
\end{equation}
Hence the temperature is strictly antisymmetric under reflection about the line $y=h/2$. This implies that only one temperature boundary condition — for instance at $y=0$ — is required to determine the full solution. Imposing the boundary value at $y=0$ gives
\begin{equation} \label{equation_T}
\bar{T}\delta(k)+\Delta T
=\frac{\gamma}{\beta}\left(1-e^{kh}\right)(a_{\phi}+ia_{\psi})
-\frac{\alpha}{\kappa^{D}}\left(1-e^{q_{\phi}h}\right)b_{\phi},
\end{equation}
where we have used
\begin{equation}
T(k,0)=\int_{-\infty}^{\infty}\!dx\,e^{-ikx}\,T(x,0)
=\int_{-\infty}^{\infty}\!dx\,e^{-ikx}\,[\bar{T}+\Delta T\,\delta(x)]
=\bar{T}\delta(k)+\Delta T.
\end{equation}
For simplicity, we absorb the resulting factor of $2\pi$ into the definition of the Dirac delta when expressing boundary conditions in momentum space. Finally, inserting Eqs. \eqref{u_y_ky_general}, \eqref{u_x_ky_general} and \eqref{expression_for_temperature} into the boundary conditions \eqref{BCuy}, \eqref{BCux} and \eqref{temperature_BC_y=0,h}, one obtains the closed system
\begin{equation} \label{final_system_coefficients_explicit_condensed_main}
\begin{cases}
\left(1+e^{kh}\right)a_{\phi}^{\psi}
+\dfrac{q_{\phi}}{k}\left(1+e^{q_{\phi}h}\right)b_{\phi}
+i\left(1+e^{q_{\psi}h}\right)b_{\psi}
=-\dfrac{U}{k},\\[6pt]
\left(1-e^{kh}\right)a_{\phi}^{\psi}
+\left(1-e^{q_{\phi}h}\right)b_{\phi}
+i\dfrac{q_{\psi}}{k}\left(1-e^{q_{\psi}h}\right)b_{\psi}
=0,\\[6pt]
\left(1-e^{kh}\right)a_{\phi}^{\psi}
-\dfrac{\alpha\beta}{\kappa^{D}\gamma}\left(1-e^{q_{\phi}h}\right)b_{\phi}
=\dfrac{\beta}{\gamma}T_{\mathrm{bc}},
\end{cases}
\end{equation}
where $T_{\mathrm{bc}}=\bar{T}\delta(k)+\Delta T$. Explicit expressions for $a_{\phi/\psi}$ and $b_{\phi/\psi}$ are reported in appendix \ref{system_of_equations}.

\subsubsection{Analytical temperature profile as a sum of compressibility and vorticity contributions}

We can now examine the temperature structure for both hydrodynamic and diffusive heat flow. The full solution can be decomposed as
\begin{equation} \label{T_as_sum_of_T_phi_and_T_psi_main}
T(x,y)=T_{\phi}(x,y)+T_{\psi}(x,y),
\end{equation}
with
\begin{equation} \label{T_profile}
\begin{split}
T_{\phi}(x,y)&=\frac{1}{2\pi}\int dk\,e^{ikx}
\Bigg[\frac{\gamma}{\beta}a_{\phi}(k)\!\left(e^{ky}-e^{kh}e^{-ky}\right)
-\frac{\alpha}{\kappa^{D}}b_{\phi}(k)\!\left(e^{q_{\phi}y}-e^{q_{\phi}h}e^{-q_{\phi}y}\right)\Bigg],\\[4pt]
T_{\psi}(x,y)&=\frac{i}{2\pi}\frac{\gamma}{\beta}\int dk\,e^{ikx}
\,a_{\psi}(k)\!\left(e^{ky}-e^{kh}e^{-ky}\right).
\end{split}
\end{equation}
This represents an important improvement over Ref. \cite{simoncelli2020generalization}, made possible by the present analytical treatment. Indeed, without this approach \cite{di2025vortices}, it would not have been possible to show that the temperature profile of a phonon system in the hydrodynamic regime (a phonon fluid) can be expressed as the sum of two distinct contributions: the compressibility contribution $T_{\phi}$ and the vorticity contribution $T_{\psi}$. This distinction introduced by our approach is crucial, as it makes it clear which part of the temperature profile gives rise to thermal backflow (namely $T_{\psi}$). In particular, by examining the compressibility and vorticity contributions to the temperature profile in Eq. \eqref{T_profile} in their most general form (i.e., with the velocity potential $\phi$ and spectral function $\psi$ coefficients implicit, before applying boundary conditions), it becomes clear that $T_{\psi}$ depends solely on the expression $e^{ky} - e^{kh} e^{-ky}$, which is antisymmetric under the transformation $y \to h-y$ and is entirely responsible for the sign-changing behavior of the temperature difference in regions immediately adjacent to the central heat flow region. In contrast, $T_{\phi}$ involves the expression $e^{q_{\phi}y} - e^{q_{\phi}h} e^{-q_{\phi}y}$, which disfavors thermal backflow and the associated negative thermal resistance. In fact, considering the general expression of the temperature difference across the strip (see Eq. \eqref{final_temperature_diff}) one can see that the first term on the right-hand side of Eq. \eqref{final_temperature_diff} arises from the part proportional to $e^{ky} - e^{kh} e^{-ky}$ in $T_{\phi}$ (related to the temperature gradient) and in $T_{\psi}$ (related to the injected drift velocity). The second term on the right-hand side of Eq. \eqref{final_temperature_diff} is due to the contribution proportional to $e^{q_{\phi}y} - e^{q_{\phi}h} e^{-q_{\phi}y}$ in $T_{\phi}$, which, as soon as one moves away from the $\xi \to 0$ limit, becomes the main factor suppressing backflow. This can be also seen analytically in a simpler manner by first taking the limit $\epsilon \to 0$ for a generic $\xi$, which corresponds to evaluating the solution when the momentum dissipation rate caused by Umklapp processes is minimal. In this case, $T_{\phi}$ is given by Eq. \eqref{T_phi_epsilon_to_0} and coincides with an irrotational temperature profile, i.e., it does not generate heat vortices (see Eq. \eqref{expression_for_temperature_irr_final} of the SI). This separation also enables the engineering of materials to maximize the transport coefficients present in $T_{\psi}$, making this contribution dominant and favoring thermal backflow (see later in the text). This level of insight is only possible through the analytical method, which allows the decoupling of the two contributions \cite{di2025vortices}.\\
To characterize deviations from Fourier transport, we introduce the analytical form of the Fourier deviation number (FDN) discussed in Eq. \eqref{FDN} \cite{simoncelli2020generalization} (see also appendix \ref{FDN_section}):
\begin{equation} \label{FDN_main}
\mathrm{FDN}=\frac{1}{\epsilon+\xi},
\qquad
\epsilon=\frac{\gamma h^{2}}{\eta},
\qquad
\xi=\frac{\kappa^{D}}{\alpha}\,\frac{\Delta T/h}{U}.
\end{equation}
Small $\epsilon$ and $\xi$ correspond to prominent hydrodynamic behavior.
\begin{figure}[!htb]
\centering
\includegraphics[width=0.8\textwidth]{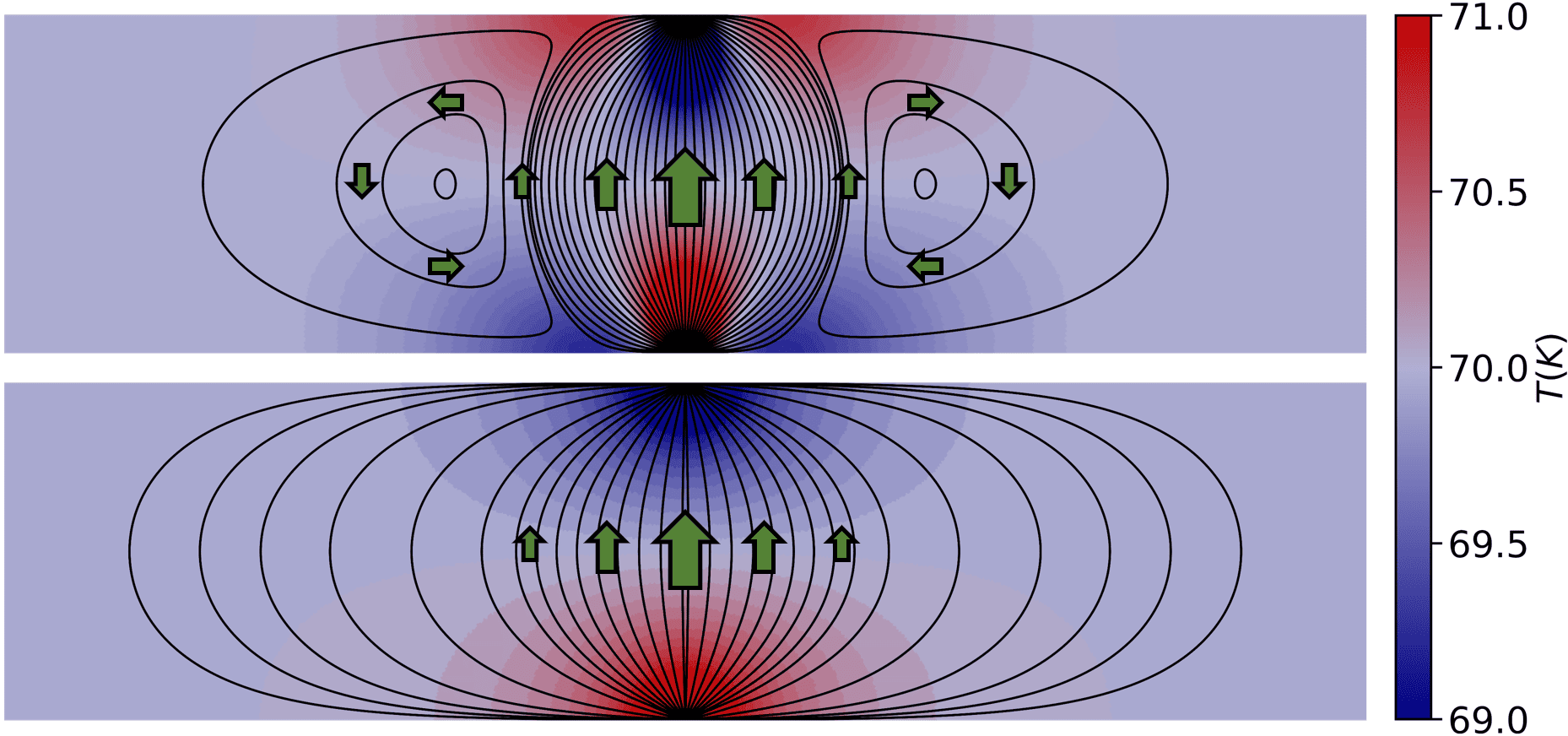}
\caption{\textbf{Temperature distribution in the 2D strip device obtained by analytically solving the modified biharmonic equation in both the viscous and diffusive regimes of thermal transport.} Temperature distribution in a two-dimensional graphite strip of height $h$, together with the corresponding streamlines for the viscous (upper panel) and diffusive (lower panel) regimes. In the viscous case, vortex formation generates a local reversal of heat flow, producing regions of negative thermal response and ultimately a negative thermal resistance across the device. The streamlines (black curves) divide the flow into a central conduction channel flanked by two recirculating side regions. They represent contour lines of the complex potential in Eq. \eqref{final_chi_xi_0}, while the associated temperature field is obtained from Eq. \eqref{T_xi_zero_and_epsilon_0}. These features provide clear experimental markers of compressible, rotational phonon hydrodynamics.\\ In contrast, heat transport in the diffusive limit follows the imposed temperature gradient, resulting in a monotonic resistance described by Eq. \eqref{final_temperature_Fourier}. Here, streamlines extend smoothly between source and drain without forming vortices or nodal structures, reflecting the absence of hydrodynamic backflow. Green arrows denote the direction of motion along each streamline family. Boundary conditions are $\bar{T}=70$ K, $\Delta T=1$ K, and $U=500$ m/s; transport coefficients are selected such that $\xi,\epsilon\to0$ in the viscous regime and $\xi,\epsilon\to\infty$ in the diffusive regime, highlighting the contrasting behaviors. \textit{Figure reproduced with permission from Ref. \cite{di2025vortices}.}}
\label{fig:backflow_main_1}
\end{figure}
In the hydrodynamic regime, the heat flux combines contributions from the temperature gradient and phonon drift velocity \cite{simoncelli2020generalization}, $\boldsymbol{Q}^{\delta}=-\kappa^{D}\nabla T$ and $\boldsymbol{Q}^{D}=\alpha\boldsymbol{u}$. Thus, $\xi\sim\langle \boldsymbol{Q}^{\delta}\rangle/\langle \boldsymbol{Q}^{D}\rangle$ quantifies the relative weight of diffusive versus drift flux, with averages set by the boundary conditions. When dissipation dominates, the temperature reverts to the Fourier expression (see appendix \ref{fourier_temperature_from_scratch_section}) \cite{di2025vortices}
\begin{equation} \label{final_temperature_Fourier}
T(x,y)=\frac{1}{2\pi}\int dk\,e^{ikx}\frac{T_{\mathrm{bc}}}{1-e^{kh}}
\left(e^{ky}-e^{kh}e^{-ky}\right),
\end{equation}
reproducing the bottom panel of Fig. \ref{fig:backflow_main_1}.

\subsubsection{Analytical profile of heat vortices}

\begin{figure}[h]
\centering
\includegraphics[width=0.8\textwidth]{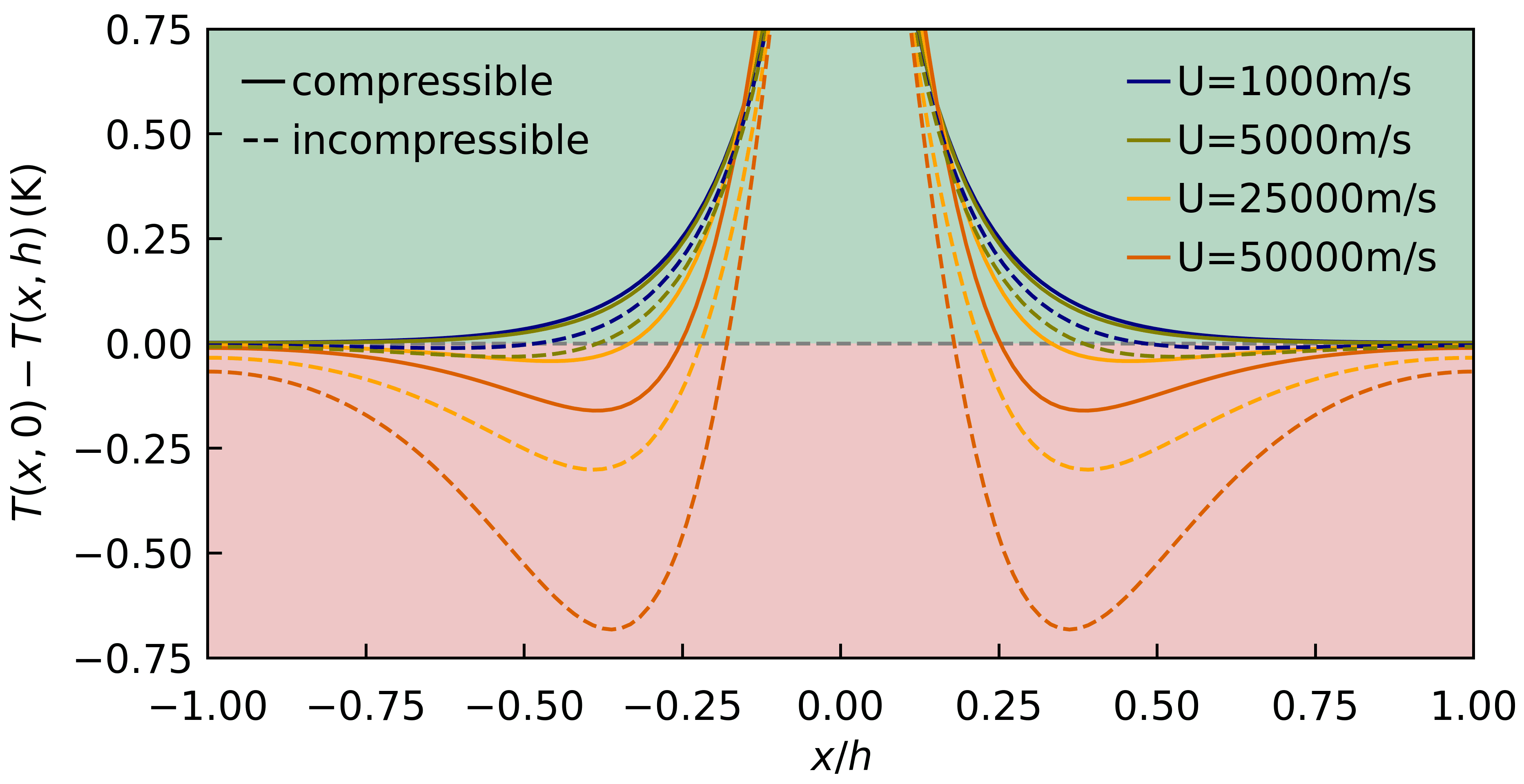}
\caption{\textbf{Comparison between compressible and incompressible thermal responses}. The temperature difference $T(x,0)-T(x,h)$ is shown as a function of $x$ for several injected drift velocities $U$. Solid curves correspond to the general compressible case, while dashed curves represent the incompressible limit (see Appendix \ref{incompressible_limit}). Blue and pink shading indicates positive and negative response, respectively. Here $\Delta T=1$ K, and the transport parameters are those of in-plane graphite obtained from first principles (Table I in Ref. \cite{simoncelli2020generalization}).}
\label{fig:backflow_main_3}
\end{figure}

We now turn to the analysis of the temperature distribution inside the strip in the ideal hydrodynamic regime, beginning with the limit where the drift-driven flux dominates over the flux induced by the temperature gradient, $\xi\to0$. In this case Eqs. \eqref{T_profile} reduce to (see also appendix \ref{xi_to_zero_T_profile_appendix}) \cite{di2025vortices}
\begin{equation} \label{T_xi_to_zero}
\resizebox{\textwidth}{!}{$
\begin{split}
T_{\phi}(x,y)&=\frac{1}{2\pi}\int dk\,e^{ikx}
\Bigg[\frac{T_{\mathrm{bc}}}{1-e^{kh}}\left(e^{ky}-e^{kh}e^{-ky}\right)
+\frac{1-e^{kh}}{1-e^{q_{\phi}h}}\,
\frac{U\gamma}{\beta}\frac{q_{\psi}}{k}G(q_{\psi},k)
\left(e^{q_{\phi}y}-e^{q_{\phi}h}e^{-q_{\phi}y}\right)\Bigg],\\
T_{\psi}(x,y)&=\frac{1}{2\pi}\frac{\gamma}{\beta}U\int dk\,e^{ikx}
\frac{q_{\psi}}{k}G(q_{\psi},k)
\left(e^{ky}-e^{kh}e^{-ky}\right),
\end{split}$}
\end{equation}
where the shorthand expression
\begin{equation} \label{condensed_notation_xi}
G(q_{\psi},k)=
\frac{1-e^{q_{\psi}h}}
{q_{\psi}(1-e^{q_{\psi}h})(1+e^{kh})
-k(1+e^{q_{\psi}h})(1-e^{kh})}
\end{equation}
has been introduced for compactness. Eq. \eqref{T_xi_to_zero} indicates a characteristic inverse-square spatial decay away from the injection points, as well as the emergence of two nodal lines along the directions $y=x$ and $y=-x$, which are associated with the primary vertical heat-transport channel. This becomes especially clear in the viscous limit, where $(q_{\psi}-k)\to0$ (corresponding to $\epsilon\to0$), yielding the expression (the full derivation can be found in Appendix \ref{xi_to_zero_T_profile_appendix}) \cite{di2025vortices}
\begin{equation} \label{T_xi_zero_and_epsilon_0}
\begin{split}
T(x,y)&=\frac{1}{2\pi}\int dk\,e^{ikx}\Bigg[
\sinh\!\left(ky-k\frac{h}{2}\right)
\left(
-\frac{8U\frac{\eta}{\beta}k\sinh\!\left(k\frac{h}{2}\right)}{kh+\sinh(kh)}
-\frac{T_{\rm bc}}{\sinh\!\left(k\frac{h}{2}\right)}
\right)\Bigg]\approx\\
&\approx \frac{4U}{\pi}\frac{\eta}{\beta}\frac{x^{2}-y^{2}}{(x^{2}+y^{2})^{2}}
-\frac{\Delta T}{\pi}\frac{y}{x^{2}+y^{2}}.
\end{split}
\end{equation}
As illustrated in the top panel of Fig. \ref{fig:backflow_main_1}, the spatial structure resulting from Eq. \eqref{T_xi_zero_and_epsilon_0} displays striking patterns. 
\begin{figure}[!htb]
\centering
\includegraphics[width=0.8\textwidth]{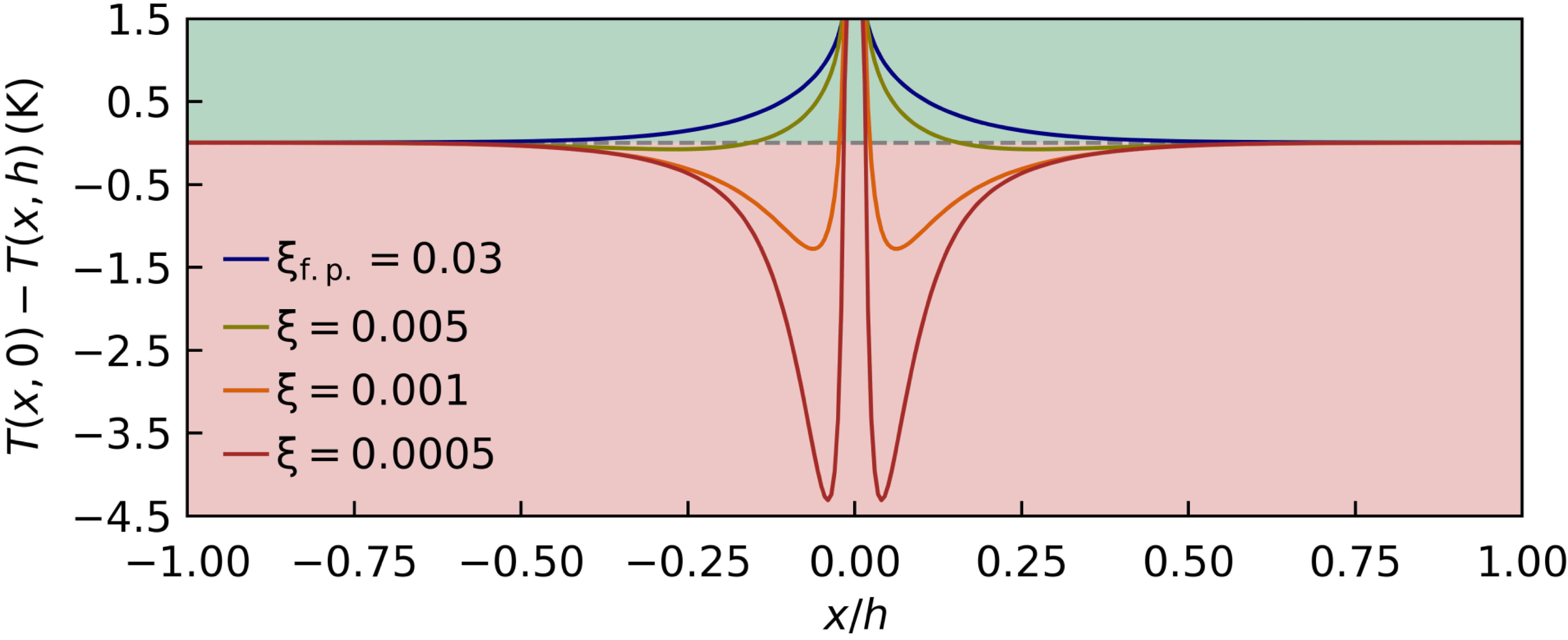}
\caption{\textbf{Non-local thermal response in viscous and diffusive regimes.} The temperature difference $T(x,0)-T(x,h)$ is shown as a function of the horizontal coordinate $x$ for several values of $\xi$, with $\epsilon=0.1$ fixed (see Eq. \eqref{temp_diff_xi_bis}). Far from the contacts (large $|x|$), the response is positive, characteristic of the diffusive regime, whereas near the leads—where viscous effects dominate—it becomes negative. The small positive bump near $x=0$ arises from the finite size of the contacts (approximately $0.05h$) used in the numerical implementation (see main text). A Fourier-like response persists up to relatively large $\xi$; $\xi_{\rm f.p.}$ corresponds to the first-principles value for in-plane graphite with natural isotope concentration. Green and pink regions highlight positive and negative response, respectively.}
\label{fig:backflow_main_2}
\end{figure}
Along the midline $y=h/2$, the temperature stabilizes due to symmetry, featuring multiple sign inversions that partition the plane into regions where $T>\bar{T}$ or $T<\bar{T}$. For small $x$ close to the dominant heat path ($x=0$), the temperature near the source is shifted upward relative to the midline equilibrium, whereas near the drain it becomes negative; this polarity reverses farther from the central axis. At the boundaries $y=0$ and $y=h$, excluding the contact zones, the sign of the temperature difference is opposite to that at the contacts. This behavior originates from the interaction between the main vertical conduction path and the adjacent regions in which thermal vortices emerge. These vortices are centered around $x\approx\pm h$, as confirmed numerically \cite{di2025vortices}. Crystal momentum spreads laterally away from the central conduction channel, and therefore any point outside the contact region on the upper boundary is connected, through a streamline, to a point on the lower boundary where the flow moves in the opposite direction (see Eq. \eqref{final_chi_xi_0}). This leads directly to a negative thermal resistance. Notably, the drift-velocity term in Eq. \eqref{T_xi_zero_and_epsilon_0} mirrors the electrical potential profile reported in Ref. \cite{levitov2016electron}. Approaching the contacts ($x\to0$), this term decays as $-y^{-2}$, overpowering the temperature-gradient contribution, which scales only as $-y^{-1}$ near $y=0,h$.\\
The temperature difference across the strip, obtained from Eqs. \eqref{T_xi_to_zero} (see Appendix \ref{xi_to_zero_T_profile_appendix}), is \cite{di2025vortices}
\begin{equation} \label{temp_diff_xi}
T(x,0)-T(x,h)=\frac{1}{\pi}\int dk\,e^{ikx}\left[
\Delta T+\frac{2U\gamma}{\beta}(1-e^{kh})\frac{q_{\psi}}{k}G(q_{\psi},k)
\right].
\end{equation}
In the limiting case $(q_{\psi}-k)\to0$, Eq. \eqref{temp_diff_xi} simplifies to
\begin{equation} \label{temp_diff_xi_bis}
\begin{split}
T(x,0)-T(x,h)&\approx
\frac{1}{\pi}\int dk\,e^{ikx}\left[
\Delta T+4U\frac{\eta}{\beta}
\frac{k\tanh\!\left(k\frac{h}{2}\right)\sinh(kh)}{kh+\sinh(kh)}
\right]\\
&\xrightarrow[|k|\gg1/h]{}2\Delta T\delta(x)-8U\frac{\eta}{\beta}\frac{1}{x^{2}},
\end{split}
\end{equation}
consistent with the electron-fluid result of Ref. \cite{levitov2016electron}. Fig. \ref{fig:backflow_main_2} exhibits this non-local response for several $\xi$ values: at large $\xi$, the behavior is diffusive and positive, while at small $\xi$ the thermal resistance becomes negative, as predicted by Eq. \eqref{temp_diff_xi_bis}. The origin of the negative response is twofold: reduced heat conductivity and weaker temperature gradients, along with the enhanced role of drift velocity. Conversely, modifying momentum dissipation through $\epsilon$ (Umklapp processes) has a comparatively minor impact (not shown), confirming that achieving sufficiently low $\xi$ is the key requirement for observing backflow, whereas decreasing momentum-relaxation rates alone is inadequate. This important point was not explicitly emphasized in Ref. \cite{simoncelli2020generalization} when discussing deviations from diffusive behavior. The analytical framework discussed here is essential for obtaining such an advance in understanding compared to the conventional picture of phonon hydrodynamics, which typically assumes only that Normal (momentum-conserving) processes must exceed Umklapp (momentum-nonconserving) processes. With this formulation, one can clearly demonstrate that backflow—a macroscopic hallmark of the hydrodynamic regime—depends explicitly on $\xi$, and that suppressing dissipative processes alone cannot sustain it \cite{di2025vortices}. These conclusions become even clearer when examining the incompressible and irrotational limits (see appendices \ref{incompressible_limit} and \ref{irrotational_limit}, respectively). \\
The strength of the thermal response is significantly influenced by the injected drift velocity $U$, which contributes directly to $\xi$. To disentangle its effect, all transport coefficients are fixed to their first-principles values for in-plane graphite with natural isotopic composition \cite{simoncelli2020generalization}, and only $U$ is varied (see Fig. \ref{fig:backflow_main_3}). For a 1 K applied temperature difference, a drift velocity above $\sim20{,}000$ m/s is required to generate even a $0.1$ K thermal backflow, posing a substantial experimental challenge. Using nearly isotopically pure graphite ($^{12}$C $99.95\%$, $^{13}$C $0.05\%$) can enhance the observable backflow up to $0.4$ K. The robustness of the negative response can be traced to the viscous term in the second VHE \eqref{VHE2}, which contains a Laplacian of the velocity and thus dominates at short spatial scales. As Fig. \ref{fig:backflow_main_2} indicates, this short-distance behavior is likely crucial for the detection of viscous heat transport in experiments.
Following Ref. \cite{levitov2016electron}, the delta-function boundary conditions \eqref{BCuy} and \eqref{temperature_BC_y=0,h} were simulated through Lorentzian functions, which generate sign reversals at the contact edges (equal to the HWHM) and negative response in the regions just outside the contacts.

\subsubsection{Complex potential and analytical streamlines equations}

To clarify the mechanism behind the heat vortices shown in the upper panel of Fig. \ref{fig:backflow_main_1}, we can compute the flow streamlines. Streamlines are a family of curves whose tangent vectors constitute the velocity vector field of the flow \cite{batchelor1967introduction}. These represent the trajectories of particles in a steady flow and therefore show the behavior of the phonon fluid. The general equation for streamlines is \cite{granger1995fluid}: $\frac{d\boldsymbol{r}_{s}}{ds}\times\boldsymbol{u}_{s}(\boldsymbol{r}_{s})=0$. If we make explicit the components of the velocity we deduce $\frac{dx_{s}}{u_{x}}=\frac{dy_{s}}{u_{y}}$, which shows that the set of curves $s$ are parallel to the velocity vector \cite{granger1995fluid}. In the incompressible case, since the velocity is simply $\boldsymbol{u}=\left(\frac{\partial\psi}{\partial y},-\frac{\partial\psi}{\partial x}\right)$ (see section \ref{incompressible_limit}), then streamlines are given by the equation $\psi=\text{const.}$. In the present case of a phonon fluid, the flow is described by a more complex drift velocity vector (see Eq. \eqref{potential_u}) in which the flow velocity components cannot be expressed as the derivatives of the scalar field $\psi$ but also include the derivatives of the field $\phi$ accounting for compressibility. For two-dimensional potential flow, streamlines are perpendicular to equipotential lines. So the idea is to use the two fields $\phi$ and $\psi$ to obtain the equation for the streamlines in the general compressible and rotational case. This may be done by defining a complex potential \cite{batchelor1967introduction,orlofftopic,meyer2012introduction,di2025vortices}. We define the complex potential $\chi$ as
\begin{equation} \label{complex_potential}
\chi(x,y)=i\psi(x,y)-\phi(x,y)
\end{equation}
and we show that this is related to the complex velocity vector $\boldsymbol{u}=u_{y}+iu_{x}$. From the Helmoltz decomposition \eqref{potential_u} we have 
\begin{equation} \label{u_{y}(x,y)+iu_{x}(x,y)_complex_pot}
\begin{split}
u_{y}(x,y)+iu_{x}(x,y)&=-\frac{\partial\phi(x,y)}{\partial y}-\frac{\partial\psi(x,y)}{\partial x}-i\frac{\partial\phi(x,y)}{\partial x}+i\frac{\partial\psi(x,y)}{\partial y}=\\
&=-\left(i\frac{\partial}{\partial x}+\frac{\partial}{\partial y}\right)\phi(x,y)+i\left(i\frac{\partial}{\partial x}+\frac{\partial}{\partial y}\right)\psi(x,y)=\\
&=\left(i\frac{\partial}{\partial x}+\frac{\partial}{\partial y}\right)\Big(i\psi(x,y)-\phi(x,y)\Big)
\left(i\frac{\partial}{\partial x}+\frac{\partial}{\partial y}\right)\chi(x,y),
\end{split}
\end{equation}
and $\chi(x,y)={\rm{const.}}$ gives streamlines expressions. \\
By evaluating the drift velocity components in the $\xi\to0$ limit (starting from Eqs. \eqref{u_y_ky_general} and \eqref{u_x_ky_general} and substituting coefficients from Eq. \eqref{final_coefficients_xi_to_zero_explicit}) we can evaluate the RHS of Eq. \eqref{u_{y}(x,y)+iu_{x}(x,y)_complex_pot}:
\begin{equation} \label{final_relation_complex_pot_xi}
\begin{split}
&\left(i\frac{\partial}{\partial x}+\frac{\partial}{\partial y}\right)\chi_{\xi\to0}(x,y)=\\
=&\,\frac{U}{2\pi}\int dk\,e^{ikx}G(k,q_{\psi})\left[-2q_{\psi}e^{kh}e^{-ky}-\frac{1 - e^{kh}}{1-e^{q_{\phi}h}}\Big((k-q_{\psi})e^{q_{\psi}y}+(k+q_{\psi})e^{q_{\psi}h}e^{-q_{\psi}y}\Big)\right]
\end{split}
\end{equation}
From this we can write
\begin{equation}
\left(i\frac{\partial}{\partial x}+\frac{\partial}{\partial y}\right)\chi(x,y)=\frac{1}{2\pi}\int dk\,e^{ikx}\left[-k+\frac{\partial}{\partial y}\right]\chi(k,y),
\end{equation}
that is
\begin{equation} \label{complex_potential_by_similarity}
\begin{split}
&-k\chi_{\xi\to0}(k,y)+\frac{\partial}{\partial y}\chi_{\xi\to0}(k,y)=\\
=&\,UG(q_{\psi},k)\left[-2q_{\psi}e^{kh}e^{-ky}-\frac{1 - e^{kh}}{1-e^{q_{\phi}h}}\Big((k-q_{\psi})e^{q_{\psi}y}+(k+q_{\psi})e^{q_{\psi}h}e^{-q_{\psi}y}\Big)\right]=\\
=&\,UG(q_{\psi},k)k\left[-\frac{q_{\psi}}{k}e^{kh}e^{-ky}-\frac{1-e^{kh}}{1-e^{q_{\phi}h}}\Big(e^{q_{\psi}y}+e^{q_{\psi}h}e^{-q_{\psi}y}\Big)\right]+\\
&-UG(q_{\psi},k)q_{\psi}\left[e^{kh}e^{-ky}-\frac{1-e^{kh}}{1-e^{q_{\phi}h}}\Big(e^{q_{\psi}y}-e^{q_{\psi}h}e^{-q_{\psi}y}\Big)\right].
\end{split}
\end{equation}
From this we get
\begin{equation} \label{obtained_chi_xi}
\begin{cases}
\chi_{\xi\to0}(k,y)=UG(q_{\psi},k)\left[\frac{q_{\psi}}{k}e^{kh}e^{-ky}+\frac{1-e^{kh}}{1-e^{q_{\phi}h}}\Big(e^{q_{\psi}y}+e^{q_{\psi}h}e^{-q_{\psi}y}\Big)\right],\\
\frac{\partial\chi_{\xi\to0}(k,y)}{\partial y}=-UG(q_{\psi},k)q_{\psi}\left[e^{kh}e^{-ky}-\frac{1-e^{kh}}{1-e^{q_{\phi}h}}\Big(e^{q_{\psi}y}-e^{q_{\psi}h}e^{-q_{\psi}y}\Big)\right],
\end{cases}
\end{equation}
and we clearly see that taking the derivative with respect to $y$ of the first of Eq. \eqref{obtained_chi_xi} we get the second of Eq. \eqref{obtained_chi_xi}. This allows to finally write 
\begin{equation} \label{complex_potential_result_xi}
\resizebox{\textwidth}{!}{$
\begin{split}
\chi_{\xi\to0}(x,y)&=\frac{U}{2\pi}\int dk\,e^{ikx}G(q_{\psi},k)\left[\frac{q_{\psi}}{k}\Big(e^{ky}+e^{kh}e^{-ky}\Big)+\frac{1-e^{kh}}{1-e^{q_{\phi}h}}\Big(e^{q_{\psi}y}+e^{q_{\psi}h}e^{-q_{\psi}y}\Big)\right]=\\
&=\frac{U}{2\pi}\int dk\,e^{ikx}\frac{\frac{q_{\psi}}{k}(1-e^{q_{\psi}h })\Big(e^{ky}+e^{kh}e^{-ky}\Big)+\frac{1-e^{kh}}{1-e^{q_{\phi}h}}(1-e^{q_{\psi}h })\Big(e^{q_{\psi}y}+e^{q_{\psi}h}e^{-q_{\psi}y}\Big)}{k(1 - e^{kh}) (1 + e^{q_{\psi}h })-q_{\psi}(1 + e^{kh}) (1 - e^{q_{\psi}h })}=\\
&=\frac{U}{2\pi}\int dk\,e^{ikx}\frac{\frac{q_{\psi}}{k}(1-e^{q_{\psi}h })\Big(e^{ky}+e^{kh}e^{-ky}\Big)+\frac{1-e^{kh}}{1-e^{q_{\phi}h}}(1-e^{q_{\psi}h })\Big(e^{q_{\psi}y}+e^{q_{\psi}h}e^{-q_{\psi}y}\Big)}{(q_{\psi}+k)(e^{q_{\psi}h}-e^{kh})+(q_{\psi}-k)(e^{(k+q_{\psi})h}-1)}.
\end{split}$}
\end{equation}
In the limits of $(q_{\psi}-k)\to0$ (that is $\epsilon\to0$) and $(q_{\phi}-k)\to0$, using similar algebra to that followed to manipulate Eq. \eqref{temp_diff_xi_to_0} (see appendix \ref{T_profile_appendix}), we can write \cite{di2025vortices}:
\begin{equation} \label{final_chi_xi_0}
\resizebox{\textwidth}{!}{$
\begin{split}
\chi_{\xi\to0}(x,y)&=\frac{U}{2\pi}\int dk\,e^{ikx}\frac{1}{k}\frac{q_{\psi}(1-e^{q_{\psi}h })\Big(e^{ky}+e^{kh}e^{-ky}\Big)+k(1-e^{kh})\Big(e^{q_{\psi}y}+e^{q_{\psi}h}e^{-q_{\psi}y}\Big)}{(q_{\psi}+k)(e^{q_{\psi}h}-e^{kh})+(q_{\psi}-k)(e^{(k+q_{\psi})h}-1)}\sim\\
&\sim\frac{U}{2\pi}\int dk\,e^{ikx}\frac{1}{k}\left[\frac{\cosh\left[k\left(y-\frac{h}{2}\right)\right]}{\cosh\left(k\frac{h}{2}\right)}+\frac{k\tanh\left(k\frac{h}{2}\right)}{kh+\sinh(kh)}\Big[y\sinh[k(h-y)]+(h-y)\sinh(ky)\Big]\right].
\end{split}$}
\end{equation}
Setting the condition $\chi(x,y)={\rm{const.}}$ to Eq. \eqref{final_chi_xi_0} allows to draw the streamlines for the compressible viscous case depicted in the upper panel of Fig. \ref{fig:backflow_main_1} \cite{di2025vortices}. Eq. \eqref{final_chi_xi_0} resembles the stream function of the electron fluid given in Eq. 8 of the Supplementary Information of Ref. \cite{levitov2016electron}.\\
Now the final expression of the complex potential $\chi$ (Eq. \eqref{final_relation_complex_pot_xi}) can be used to find the coordinates of the points $(x_{0},y_{0})$ within the strip around which the streamlines circulate. The $y_{0}$ coordinate is $y=\frac{h}{2}$, where the $u_{x}$ component of the drift velocity changes its direction by symmetry. Then, one can obtain the value of $x_{0}$ by imposing the condition that the $y$ component of the phonon drift velocity is zero, as it should be at the center of the vortices:
\begin{equation}
u_{y}\left(x_{0},\frac{h}{2}\right)=-\left.\frac{\partial\phi\left(x,\frac{h}{2}\right)}{\partial y}\right|_{x=x_{0}}-\left.\frac{\partial\psi\left(x,\frac{h}{2}\right)}{\partial x}\right|_{x=x_{0}}=0.
\end{equation}
This finally yields:
\begin{equation} 
-\frac{U}{\pi}\int dk\,\left.e^{ikx}G(q_{\psi},k)\left[q_{\psi}e^{k\frac{h}{2}}+k\frac{1 - e^{kh}}{1-e^{q_{\phi}h}}e^{q_{\psi}\frac{h}{2}}\right]\right|_{x=x_{0}}=0
\end{equation}
whose solution is obtained numerically and is given by \cite{di2025vortices}
\begin{equation}
x_{0}\approx\pm h.
\end{equation}
The points with coordinates $(x_{0},y_{0})=(\pm h,\frac{h}{2})$ are the centers of the vortices depicted in the upper panel of Fig. \ref{fig:backflow_main_1}. Most streamlines form open paths directly linking the source ($y=0$) and drain ($y=h$), but a subset of them close into loops, generating the vortices flanking the primary heat-transport channel.

\subsubsection{Irrotational and incompressible limits}

We can now separate the contributions to the temperature profile in Eq. \eqref{T_as_sum_of_T_phi_and_T_psi_main} into their irrotational and incompressible components to pinpoint which aspects predominantly drive vortex formation and thermal backflow.\\
In the irrotational limit ($\boldsymbol{\mathscr{W}}=\nabla\times\boldsymbol{u}=0$), the drift velocity originates entirely from the scalar potential $\phi$, i.e.\ $\boldsymbol{u}=-\nabla\phi=(-\partial\phi/\partial x,-\partial\phi/\partial y)$. The second of Eqs. \eqref{potential_VHE_final} becomes irrelevant (see Appendix \ref{irrotational_limit}). Solving the first equation using boundary conditions \eqref{BCuy} and \eqref{BCux} shows that drift injection alone does not generate thermal flow; however, in the presence of a temperature gradient, the irrotational temperature profile becomes
\begin{equation} \label{irr_temp}
T^{\boldsymbol{\mathscr{W}}=0}\equiv T_{\phi}^{\epsilon\to0}.
\end{equation}
Conversely, in the incompressible regime ($\Phi=\nabla\cdot\boldsymbol{u}=0$), the velocity is described solely by the stream function $\psi$, with $\boldsymbol{u}=\nabla\times\boldsymbol{\Psi}=(\partial\psi/\partial y,-\partial\psi/\partial x)$, rendering the first of Eqs. \eqref{potential_VHE_final} irrelevant (see Appendix \ref{incompressible_limit}). Solving the second equation under the same boundary conditions produces
\begin{equation} \label{inc_temp}
T^{\Phi=0}\equiv T_{\psi}^{\xi\to0}.
\end{equation}
This form matches the electrical potential obtained in Ref. \cite{levitov2016electron}, where electron flow is also incompressible. Thus, by injecting drift velocity, one can generate a thermal flow in the incompressible phonon fluid. The relations \eqref{irr_temp} and \eqref{inc_temp} emphasize that the $\epsilon\to0$ limit isolates the irrotational response, while the $\xi\to0$ limit isolates the incompressible contribution. In Fig. \ref{fig:backflow_main_3}, the incompressible temperature contrast is compared with the full compressible solution. The incompressible system displays enhanced thermal backflow and more pronounced vortex behavior, confirming that the $T_{\psi}$ component—associated with $\xi$—dominates the viscous response. Hence, the temperature field in Eq. \eqref{T_as_sum_of_T_phi_and_T_psi_main} can be decomposed naturally into the irrotational and incompressible contributions expected in the viscous limit (where $\text{FDN}\to\infty$).

\subsection{Numerical validation of thermal backflow}

The setup analyzed analytically in Fig. \ref{fig:backflow_main_1} is consistent with numerical solutions of the BTE.
\begin{figure}[!htb]
\centering
\includegraphics[width=\textwidth]{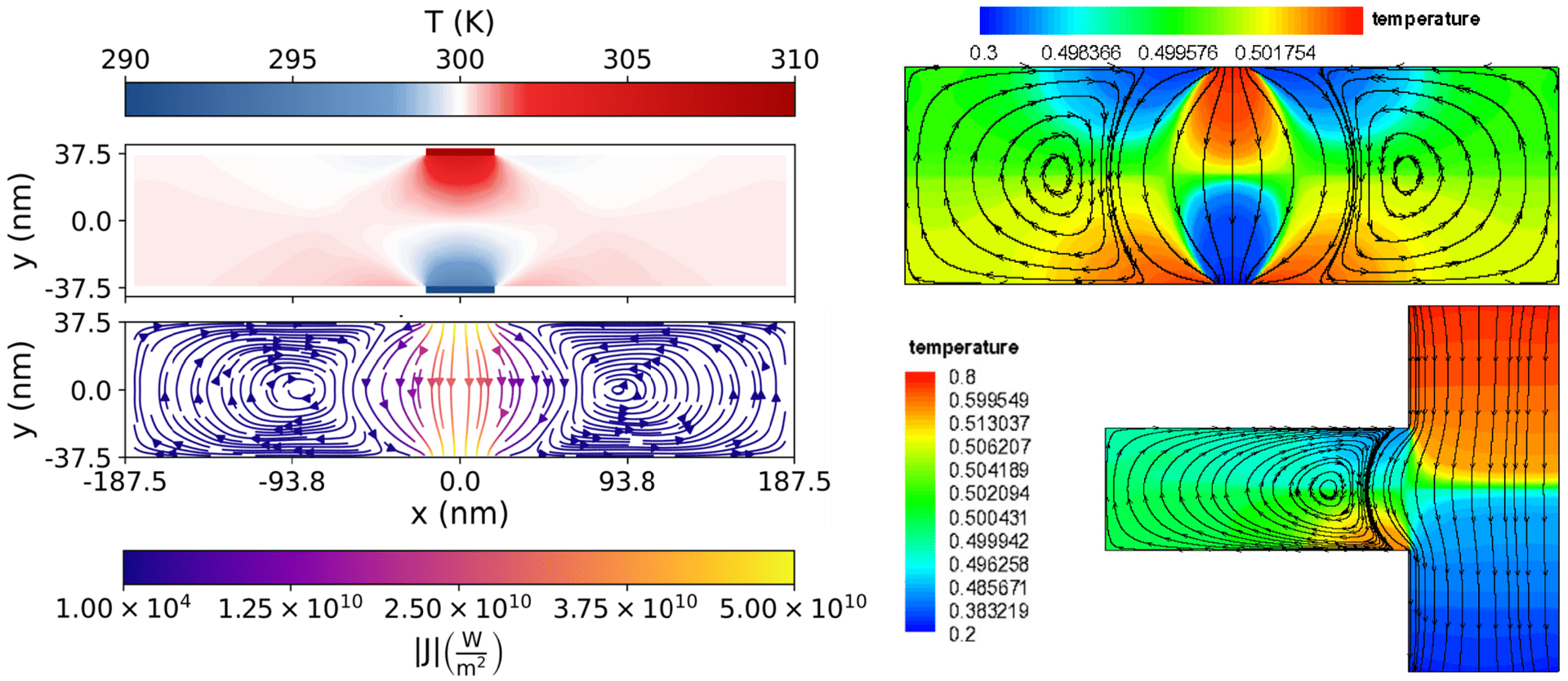}
\caption{\textbf{Numerical validation of thermal backflow in 2D strip devices.} (Left panel) Steady-state temperature distributions (top) and corresponding heat-flux patterns (bottom) obtained with beyond-RTA simulations as implemented in the \texttt{BTE-Barna} package \cite{raya2022bte}. The system dimensions are $W_{\mathrm{reservoir}} = 37.5\,\mathrm{nm}$, $W_{\mathrm{device}} = 375\,\mathrm{nm}$, and $H = 75\,\mathrm{nm}$, with boundary temperatures set to $T_{\mathrm{hot}} = 310\,\mathrm{K}$ and $T_{\mathrm{cold}} = 290\,\mathrm{K}$. In the temperature maps, the dark red and dark blue regions at the top and bottom denote the hot and cold isothermal reservoirs, respectively. \textit{Figure reproduced with permission from Ref. \cite{raya2022hydrodynamic}.} (Right panel) Representative examples of hydrodynamic phonon transport in graphene nanostructures. Phonon vortex patterns in rectangular a graphene ribbon of size $L = 3W = 15\,\mu\mathrm{m}$, at an average temperature of $100\,\mathrm{K}$ (top) and phonon cavity flow in a T-shaped graphene ribbon at $100\,\mathrm{K}$ for cavity aspect ratio $H/L_1 = 2$ (bottom). The color maps represent the dimensionless temperature $\theta = (T - T_c)/(T_h - T_c)$, while the streamlines indicate the direction and magnitude of the heat-flux field. \textit{Figure reproduced with permission from Ref. \cite{guo2021phonon}.}}
\label{fig:raya_guo_figure}
\end{figure}
For example, Ref. \cite{raya2022hydrodynamic} reproduces the same vortex-like temperature profiles using energy-based deviational Monte Carlo simulations (see the left panel of Fig. \ref{fig:raya_guo_figure}) to solve the position-dependent BTE as implemented in the \texttt{BTE-Barna} package \cite{raya2022bte}, while consistently finding the absence of vortices when the RTA is employed. That work also elucidates the mechanisms responsible for the emergence of hydrodynamic features, demonstrating that boundary scattering and the relation between the sample dimensions and the nonlocal length are the determining factors, largely independent of the relative importance of normal versus resistive scattering processes. From this perspective, the nonlocal length quantifies the ability of scattering to randomize the heat flux. These results highlight that approximations made in the scattering operator can have qualitative consequences for hydrodynamic signatures through their impact on the value of the nonlocal length.\\
Similarly, Ref. \cite{guo2021phonon} reports an analogous temperature profile using Callaway’s dual-relaxation-time model (see the top-right panel of Fig. \ref{fig:raya_guo_figure}). That study also considers a geometry closely related to the tunnel-chamber device discussed extensively in Sec. \ref{tunnel_chamber}, and finds that the resulting thermal backflow is qualitatively similar to that shown in Fig. \ref{fig:tunnel_chamber_1}. Once again, boundary scattering and the interplay between sample size and nonlocal length are identified as the key ingredients governing hydrodynamic transport, irrespective of the balance between normal and resistive scattering. This further supports the view that the nonlocal length provides a unifying framework for describing hydrodynamic behavior and underscores the sensitivity of hydrodynamic signatures to the microscopic treatment of scattering.

\subsection{Time-dependent domain}

In this section, we briefly show how the modified biharmonic equations—the analytical form of the VHE—can capture drifting second sound \cite{guyer1966thermal,hardy1970phonon,cepellotti2017transport,huberman2019observation}. Motivated by recent experiments in graphite, which observed heat temporarily flowing against the temperature gradient in the time-dependent regime, e.g., as second sound \cite{huberman2019observation,ding2022observation}, we present the analytical expressions of transient viscous heat backflow under different conditions. In particular, we investigate how viscous heat backflow, as predicted by the full analytical solutions of the modified-biharmonic equations, develops in the time domain.

\subsubsection{Second sound} \label{second_sound_analytical}

In order to observe second sound, temperature changes need to propagate following a damped wave equation \cite{simoncelli2020generalization}
\begin{equation} \label{ss_eq}
\frac{\partial^2 T(x,y,t)}{\partial t^2}+\frac{1}{\tau_{ss}}\frac{\partial T(x,y,t)}{\partial t}-v_{ss}^{2}\frac{\partial^{2}T(x,y,t)}{\partial y^{2}}=0,
\end{equation}
where $\tau_{ss}$ is the second-sound relaxation time, $v_{ss}^{2}=\frac{\kappa^{D}}{C\,\tau_{ss}(1-f)}$ is the squared second-sound velocity and $0<|f|<1$ is a constant, all to be determined. For simplicity, here we consider $\hat{y}$ as the direction of second-sound propagation. In an isotropic system, the drifting heat flux $\boldsymbol{Q}^{D}$ is collinear with the drift velocity, and the heat flux due to local temperature changes $\boldsymbol{Q}^{\delta}$ is collinear with the temperature gradient. Thus, it follows that the only nonzero component of the drift velocity must be along the second-sound propagation direction, $\boldsymbol{u}\equiv u_{y}=u$. With these conditions, and following the usual Fourier transform in the transverse direction $x$,
\begin{equation}
f(k,y,t) = \int_{-\infty}^{\infty} dx\, e^{-i k x} f(x,y,t),
\qquad
\partial_x \mapsto i k,\;\; \partial_x^2 \mapsto -k^2,
\end{equation}
Eq. \eqref{ss_eq} and the first VHE in mixed $(k,y,t)$-space respectively become
\begin{equation} \label{ss_eq_k}
\frac{\partial^2 T(k,y,t)}{\partial t^2}+\frac{1}{\tau_{ss}}\frac{\partial T(k,y,t)}{\partial t}-v_{ss}^{2}\frac{\partial^{2}T(k,y,t)}{\partial y^{2}}=0,
\end{equation}
and
\begin{equation}  \label{VHE_ss}
C\frac{\partial T(k,y,t)}{\partial t}+\alpha\frac{\partial u(k,y,t)}{\partial y} - \kappa^{D}\!\left(\frac{\partial^2}{\partial y^2} - k^2\right)T(k,y,t)=0,
\end{equation}
where the transverse wavenumber $k$ appears as an effective positive stiffness term. By substituting Eq. \eqref{VHE_ss} into Eq. \eqref{ss_eq_k}, one obtains the relation showing that the drift velocity and temperature are connected through \cite{simoncelli2020generalization}
\begin{equation} \label{u_ss_eq}
\alpha\frac{\partial u(k,y,t)}{\partial y}
= C\tau_{ss}(1-f)\frac{\partial^2T(k,y,t)}{\partial t^2}
- C f \frac{\partial T(k,y,t)}{\partial t}.
\end{equation}
By taking the derivative with respect to $y$ of \eqref{u_ss_eq} we have
\begin{equation} \label{condition_ss_u}
\alpha\frac{\partial^{2}u(k,y,t)}{\partial y^{2}}=\hat{O}\frac{\partial T(k,y,t)}{\partial y}.
\end{equation}
where the differential operator $\hat{O}$ is defined as
\begin{equation}
\hat{O}=\frac{\kappa^{D}}{v_{ss}^{2}}\frac{\partial^2}{\partial t^2}+\left(\frac{\kappa^{D}}{\tau_{ss}v_{ss}^{2}}-C\right)\frac{\partial}{\partial t}
\end{equation}
Now we apply the operator $\hat{O}$ to both sides of the second VHE. Using condition \eqref{condition_ss_u}, exploiting $v_{ss}^{2}$ and focusing on the close-to-equilibrium regime by neglecting higher-than-second order derivatives, we get (remember $\gamma=AD^{U}$)
\begin{equation} \label{u_ss_equation_done}
\frac{\partial^2 u(k,y,t)}{\partial t^2} + \underbrace{\frac{D^{U}f}{f-(1-f) \tau_{ss} D^{U}}}_{c_1} \frac{\partial u(k,y,t)}{\partial t} 
- \underbrace{\frac{W^2}{f - (1-f) \tau_{ss} D^{U}}}_{c_2} \frac{\partial^2 u(k,y,t)}{\partial y^2} = 0
\end{equation}
which shows that also the drift-velocity field follows a damped-wave equation provided that
\begin{equation}
f >\frac{\tau_{\mathrm{ss}}D^{U}}{1+\tau_{\mathrm{ss}}D^{U}},
\end{equation}
for which both constants $c_1$ and $c_2$ are positive. Remembering that $u\equiv u_{y}=-\partial\phi/\partial y-\partial\psi/\partial x$, and, for simplicity, focusing on the incompressible limit, we get the same damped-wave equation provided also for the stream function:
\begin{equation} \label{psi_ss_equation}
\frac{\partial^2 \psi(k,y,t)}{\partial t^2} + \underbrace{\frac{D^{U} f}{f-(1-f) \tau_{ss} D^{U}}}_{c_1} \frac{\partial \psi(k,y,t)}{\partial t} - \underbrace{\frac{W^2}{f - (1-f) \tau_{ss} D^{U}}}_{c_2} \frac{\partial^2 \psi(k,y,t)}{\partial y^2} = 0, \quad k \neq 0.
\end{equation}
By assuming plane waves along $y$ such that $f(k,y,t) = f(k,k',t) e^{ik'y}$, then $\partial_y \mapsto i k'$, $\partial_y^2 \mapsto -k'^2$, and the total in-plane wavenumber is $K^2 := k'^2 + k^2$. For harmonic time dependence $e^{-i \omega t}$, the dispersion relation is
\begin{equation}
-\omega^2 - \frac{i \omega}{\tau_{ss}} + v_{ss}^2 K^2 = 0, \quad v_{ss}^2 = \frac{\kappa^{D}}{C \tau_{ss} (1-f)}.
\end{equation}
and the damped frequency is
\begin{equation}
\bar{\omega}(K) = \sqrt{v_{ss}^2 K^2 - \frac{1}{4 \tau_{ss}^2}}.
\end{equation}
In this way, the solutions for each transverse mode $k$ are
\begin{equation}
\begin{split}
T(k,y,t) &= \frac{1}{2\pi} \int C_T(k,k')\, e^{-t/2\tau_{ss}} e^{i k' y - i \bar{\omega}(k,k') t} dk', \\
\psi(k,y,t) &= \frac{1}{i k} \frac{1}{2\pi} \int C_\psi(k,k')\, e^{-t/2\tau_{ss}} e^{i k' y - i \bar{\omega}(k,k') t} dk', \quad k \neq 0,
\end{split}
\end{equation}
with coefficients
\begin{equation}
\begin{split}
f &= D_U \tau_{ss},\\
\tau_{ss} &= \frac{C W^2}{\kappa^{D} D_U^2 + D_U C W^2},\\
v_{ss} &= \frac{\kappa^{D} D_U + C W^2}{C W}.
\end{split}
\end{equation}

\subsubsection{Transient viscous heat backflow} \label{transient_heat_backflow_section}

We concentrate on the incompressible limit, so only the stream function $\psi$ is involved. Rather than taking the steady-state limit, we consider the general time-dependent case. We demonstrate that it is still possible to derive a closed-form equation for the thermal vorticity (see Eq. \eqref{vorticity_eq_time_dep}). By applying the Helmholtz decomposition \eqref{u_helmoltz_decomposition} to this equation, we obtain the time-dependent extension of the modified-biharmonic equation for the stream function $\psi$, given by Eq. \eqref{potential_VHE_final}:
\begin{equation} \label{time_dep_psi_eq}
\frac{A}{\eta}\frac{\partial}{\partial t}(\nabla^{2}\psi)
-\nabla^{2}(\nabla^{2}\psi)
+\frac{\gamma}{\eta}\nabla^{2}\psi=0.
\end{equation}
Fourier transforming along $x$ we obtain
\begin{equation}
\left[(-k^{2}+\partial^{2}_{y})\frac{A}{\eta}\frac{\partial}{\partial t}
- \left(k^{2}-\partial^{2}_{y}\right)^2
+\frac{\gamma}{\eta}(k^{2}-\partial^{2}_{y})\right]\psi(k,y,t)=0.
\end{equation}
Rearranging:
\begin{equation}
\left(\partial^{2}_{y}-k^{2}\right)
\left(\partial^{2}_{y}-\frac{\gamma}{\eta}-\frac{A}{\eta}\frac{\partial}{\partial t}\right)\psi(k,y,t)=0.
\end{equation}
At this point we can exploit a Laplace transformation for the time domain. For transient problems it is often cleaner to use a Laplace transform ($e^{st}$, $\mathbb{R}\{s\}>0$) and then invert on a Bromwich contour; that picks the causal/retarded branch automatically and controls convergence of integrals. In this way we write
\begin{equation}
\psi(t)=\frac{1}{2\pi}\int ds\,e^{st}\psi(s),
\end{equation}
so that $\partial_t \to s$, giving
\begin{equation}
\left(\partial^{2}_{y}-k^{2}\right)
\left(\partial^{2}_{y}-q_\psi^{2}(s)\right)\psi(k,y,s)=0,
\end{equation}
with
\begin{equation}
q_{\psi}^{2}(s)=k^{2}+\frac{\gamma}{\eta}+\frac{A}{\eta}s.
\end{equation}
The general solution is
\begin{equation}
\psi(k,y,s)=a_{\psi-}e^{-ky}+a_{\psi+}e^{ky}
+b_{\psi-}e^{-q_{\psi}(s)y}+b_{\psi+}e^{q_{\psi}(s)y}.
\end{equation}
After imposing symmetry:
\begin{equation}
\psi(k,y,s)=a_{\psi}\left(e^{ky}+e^{kh}e^{-ky}\right)
+b_{\psi}\left(e^{q_{\psi}(s)y}+e^{q_{\psi}(s)h}e^{-q_{\psi}(s)y}\right).
\end{equation}
Proceeding as in the steady-state case, the temperature reads
\begin{equation}
T(x,y,t)=\frac{1}{(2\pi)^{2}}\frac{\gamma}{\beta}U
\int dk\,e^{ikx}
\int ds\,e^{st}
\frac{q_{\psi}(s)}{k}G(q_{\psi}(s),k)
\left(e^{ky}-e^{kh}e^{-ky}\right).
\end{equation}
We define
\begin{equation}
\tilde{G}(k,t)=\int ds\,e^{st}\frac{q_{\psi}(s)}{k}G(q_{\psi}(s),k).
\end{equation}
Changing variable:
\begin{equation}
s = \frac{\eta}{A}\left(q_\psi^2 - k^2 - \frac{\gamma}{\eta}\right),
\quad
ds = \frac{2\eta}{A}q_\psi dq_\psi,
\end{equation}
we obtain
\begin{equation} 
\tilde{G}(k,t)=
\frac{2\eta}{A}
e^{-\frac{\gamma}{A}t}
e^{-\frac{\eta}{A}k^2 t}
\int dq_\psi\, e^{-\frac{\eta}{A}q_\psi^2 t}
\frac{q_\psi^2}{k}G(q_\psi,k).
\end{equation}
For large $t$, the integral is Gaussian dominated, since the exponential suppresses large $q_\psi$. This allows us to focus on the small-$q_\psi$ limit. For $q_{\psi}\to0$ we have 
\begin{equation}
G\big(q_{\psi},k\big)\approx\frac{1-1-q_{\psi}h}{k\big(1-e^{kh}\big)-k\big(e^{kh}-1\big)}\approx-\frac{q_{\psi}h}{2k\big(1-e^{kh}\big)}.
\end{equation}
Thus, the integral becomes
\begin{equation}
g(k,t)=\int  dq_{\psi}  e^{-\frac{\eta}{A}q_{\psi}^2 t}\frac{q_{\psi}^2}{k} G(q_{\psi},k)
\approx-\frac{h}{2k^{2}\big(1-e^{kh}\big)}\int  dq_{\psi}  e^{-\frac{\eta}{A}q_{\psi}^2 t}q_{\psi}^{3}.
\end{equation}
This is a standard Gaussian integral ($q_{\psi}>0$):
\begin{equation}
\int_0^\infty dq\,q^n e^{- a q^2} = \frac{1}{2} a^{-\frac{n+1}{2}} \Gamma\left(\frac{n+1}{2}\right).
\end{equation}
For $n=3$ and $a=\frac{\eta}{A}t$, we obtain
\begin{equation}
\begin{split}
g(k,t)&\approx-\frac{h}{2k^{2}\big(1-e^{kh}\big)}\int_{0}^{\infty}  dq_{\psi}  e^{-\frac{\eta}{A}q_{\psi}^2 t}q_{\psi}^{3} \\
&= -\frac{h}{2k^{2}\big(1-e^{kh}\big)}\frac{1}{2} \left(\frac{\eta}{A}t\right)^{-2} \Gamma(2) \\
&= -\frac{h}{4k^{2}\big(1-e^{kh}\big)} \left(\frac{A}{\eta}\right)^2 t^{-2}.
\end{split}
\end{equation}
Substituting back, we obtain
\begin{equation}
\label{time_n_transform}
\begin{split}
\tilde{G}(k,t)
&= \frac{2\eta}{A}
e^{-\frac{\gamma}{A}t}
e^{-\frac{\eta}{A}k^2 t}
\, g(k,t) \\
&\approx 
\frac{2\eta}{A}
e^{-\frac{\gamma}{A}t}
e^{-\frac{\eta}{A}k^2 t}
\left[-\frac{h}{4k^{2}\big(1-e^{kh}\big)} \left(\frac{A}{\eta}\right)^2 t^{-2}\right] \\
&= -\frac{A}{\eta}\frac{h}{2k^{2}\big(1-e^{kh}\big)}
e^{-\left(\frac{\eta}{A}k^2+\frac{\gamma}{A}\right)t}
t^{-2}.
\end{split}
\end{equation}
Inserting Eq. \eqref{time_n_transform} into the temperature profile expression, we obtain
\begin{equation} \label{final_t_dep_sol}
\begin{split}
T(x,y,t)=-\frac{1}{(2\pi)^{2}}\frac{A\gamma}{2\eta\beta}hU\int dk\,e^{ikx}\frac{\left(e^{ky}-e^{kh}e^{-ky}\right)}{k^{2}\big(1-e^{kh}\big)}e^{-\left(\frac{\eta}{A}k^2+\frac{\gamma}{A}\right)t}t^{-2}.
\end{split}
\end{equation}
This solution is physically stable: all modes decay exponentially in time, with higher-$k$ components damped more rapidly due to viscosity. The transient dynamics is therefore governed by a redistribution of spectral weight rather than by any growth mechanism.
\begin{figure}[!htb]
\centering
\includegraphics[width=\textwidth]{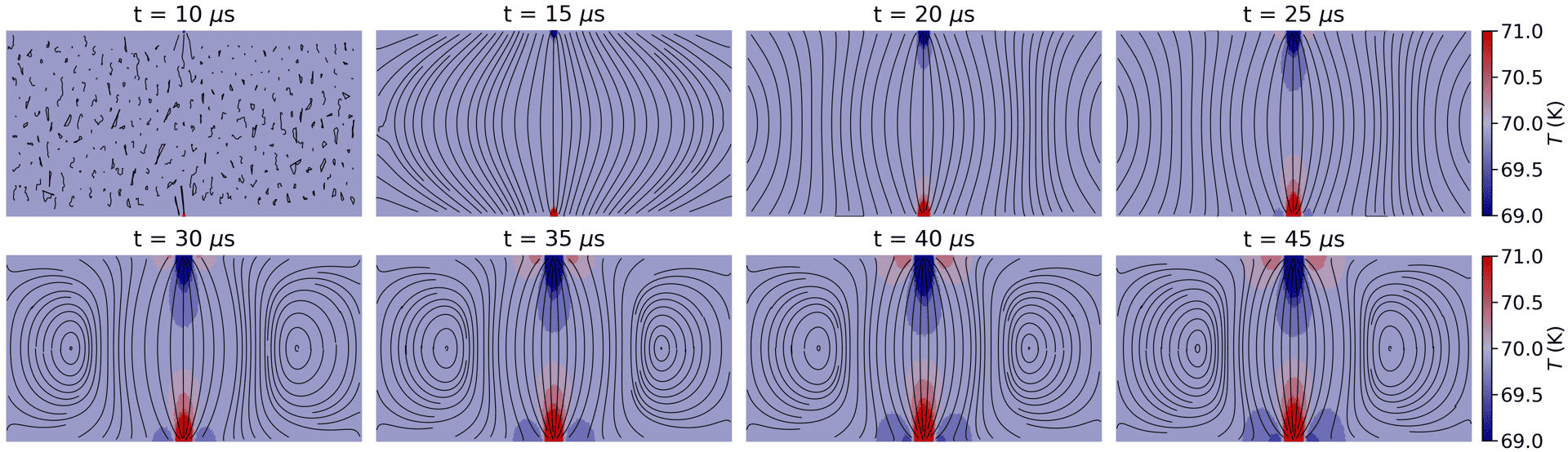}
\caption{\textbf{Transient viscous heat vortices.} Time evolution of the temperature-driven flow field. Transient vortex-like structures may appear at intermediate times due to the different decay rates of Fourier modes, but they do not correspond to any growing instability and progressively decay as the system relaxes.}
\label{fig:test3}
\end{figure}
At early times, the algebraic prefactor $t^{-2}$ suppresses the overall amplitude, resulting in a weak and diffuse temperature pattern. At intermediate times, the different decay rates of the Fourier components, set by $\exp[-(\eta k^2+\gamma)t/A]$, can produce transient spatial structures in the temperature field that resemble vortex-like features, as shown in Fig. \ref{fig:test3}. However, these structures do not grow indefinitely in time: they progressively decay and are eventually washed out at long times (not shown), as all modes remain exponentially damped and the system relaxes to a uniform state.

\subsection*{Takeaways}

In this section we developed an analytical framework for viscous phonon hydrodynamics by recasting the steady-state viscous heat equations into a pair of decoupled modified Helmholtz and biharmonic equations. This formulation enables a complete analytical separation of heat flow into compressible and vortical components, revealing the distinct physical roles played by phonon compressibility and vorticity. A key outcome of the analytical treatment is that it allows one to explicitly demonstrate that the temperature profile can be written as the sum of two additive contributions associated with compressibility and vorticity, a structural result that cannot be directly inferred from numerical solutions alone. We showed that the velocity field can be expressed in terms of a scalar velocity potential and a stream function, allowing the temperature profile to be decomposed unambiguously into irrotational and incompressible parts. This decomposition makes it possible to identify the microscopic origin of hydrodynamic signatures such as thermal vortices, heat backflow, and negative nonlocal thermal resistance, which are inaccessible within Fourier theory. By solving the resulting modified biharmonic equations in a confined strip geometry, we obtained closed-form expressions for the temperature and drift-velocity fields and established clear criteria for the emergence of hydrodynamic behavior. Finally, we demonstrated that this analytical structure extends naturally to the time-dependent domain, capturing transient phenomena such as drifting second sound and the dynamical formation of viscous heat vortices. Together, these results provide a transparent and computationally efficient bridge between mesoscopic hydrodynamic theory and experimentally observable signatures of phonon fluid behavior. More broadly, the generality of this analytical structure extends beyond phonon hydrodynamics. While compressibility is intrinsic to phonon fluids, the same framework becomes essential for electronic systems when drift velocities approach collective-mode velocities, where density fluctuations, plasmonic drag, or coherent plasmon dynamics emerge and incompressible descriptions break down. By retaining both compressibility and vorticity through a Helmholtz decomposition of the velocity field, the present approach provides a unified and transferable foundation for collective transport phenomena—including electrons, plasmons, electron–phonon bifluids, and other collective excitations—and offers a robust analytical reference for hydrodynamic transport across a wide range of quantum materials.

\section{Further theoretical and experimental advances in phonon hydrodynamics across new materials}

A systematic first-principles demonstration of phonon hydrodynamics in bulk crystalline solids was provided in 2020 by Ghosh \emph{et al.} through a comprehensive study of crystalline GeTe \cite{ghosh2020phonon,battaglia2022phonon}. This work is fully \emph{ab initio}, combining DFT with the direct solution of the LBTE, and relies on mode-resolved calculations of normal and Umklapp scattering rates over a wide temperature range. The emergence of a low-temperature hydrodynamic window was attributed to the rapid suppression of resistive scattering processes, while the kinetic--collective model and a Knudsen-number-based analysis were employed to quantitatively identify collective phonon transport. The interplay between characteristic size and collective phonon transport was clarified in 2021 through a follow-up first-principles investigation of crystalline GeTe by the same authors \cite{ghosh2021effect,battaglia2022phonon}. By combining density functional theory with the kinetic collective model, the authors demonstrated the existence of an intermediate-size regime between ballistic and diffusive transport, characterized by distinct scaling relations of thermal conductivity with size and controlled by the competition between momentum-conserving and resistive scattering processes.\\
Phonon hydrodynamics has subsequently been predicted and measured in a variety of additional and emerging materials, reflecting the continuing and growing interest in this field and, in particular, the search for new materials capable of exhibiting fluid-like phonon transport at temperatures compatible with technological applications. 
\begin{figure}[h!]
\centering
\includegraphics[width=0.8\textwidth]{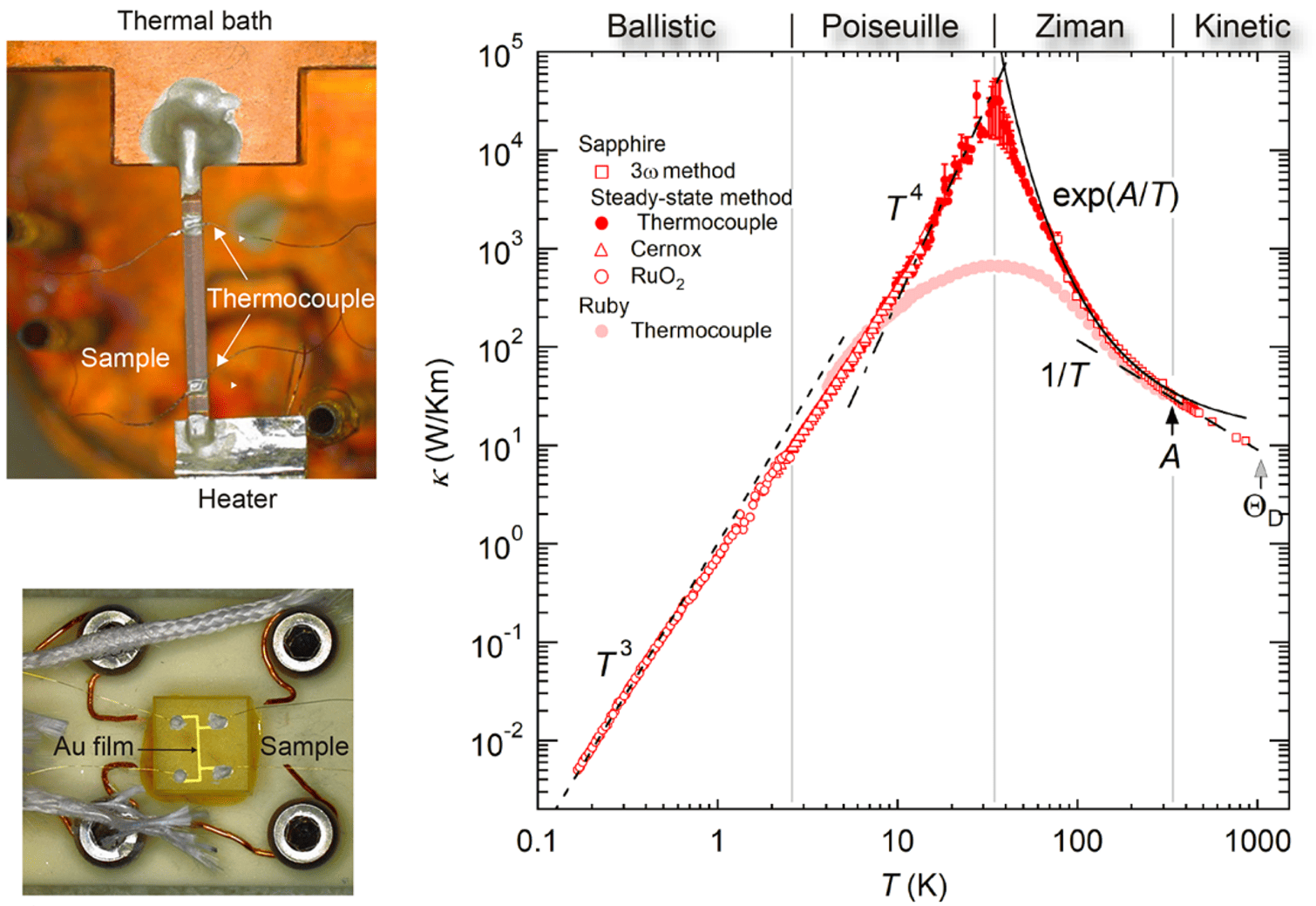}
\caption{\textbf{Phonon hydrodynamic regimes in sapphire.} (Left panel) Schematics of the experimental configurations performed in Ref. \cite{kawabata2025phonon} and used to measure thermal conductivity are illustrated for the steady-state method (top) and the 3$\omega$ method (bottom). (Right panel) Temperature dependence of the thermal conductivity for sapphire and ruby. In sapphire, $\kappa$ rises exponentially at intermediate temperatures, following $\kappa\propto\exp(A/T)$ with $A = 337$K, and exhibits a pronounced maximum near 40 K. Below this temperature, $\kappa$ decreases more rapidly than $T^{3}$, consistent with the onset of phonon Poiseuille transport. The functional form of $\kappa$ evolves from a $1/T$ behavior to an exponential dependence across the characteristic temperature $T = A$, allowing identification of four distinct transport regimes, indicated above in the panel. \textit{Figure reproduced with permission from Ref. \cite{kawabata2025phonon}.}}
\label{fig:kawabata_figure}
\end{figure}
In this context, very recently Kawabata \emph{et al.} reported a comprehensive experimental identification of hydrodynamic phonon transport regimes in sapphire \cite{kawabata2025phonon}. Their work is based on thermal-transport measurements spanning an exceptionally broad temperature interval (0.1--900~K), enabling the observation of the full sequence of four expected conductivity regimes upon cooling, including the Ziman and Poiseuille hydrodynamic regimes where momentum-conserving normal phonon collisions dominate (see Fig. \ref{fig:kawabata_figure}). Remarkably, sapphire was shown to display these regimes despite isotopic impurity, with an exponential rise of the thermal conductivity in the Ziman regime reaching values as large as $3.5\times 10^{4}$~W\,m$^{-1}$\,K$^{-1}$. By comparing against a proposed universal scaling of peak thermal conductivity with isotopic purity for ultrapure simple insulators, the authors found sapphire to exceed the expected trend by roughly an order of magnitude and argued that this enhancement may be linked to the proximity of optical and acoustic branches enabled by the large primitive-cell complexity. Importantly, we emphasize that the extension of phonon hydrodynamics to new two-dimensional and layered materials has been driven largely by first-principles studies in recent years. For example, Chen \emph{et al.} investigated bilayer black phosphorus under strain using density functional theory combined with the BTE \cite{chen2024modulating}. Their calculations revealed a nonmonotonic strain dependence of thermal conductivity arising from the interplay between optical phonon softening and the stabilization of the quadratic flexural acoustic mode through interlayer quasicovalent bonding, demonstrating that phonon hydrodynamics can be actively modulated by mechanical deformation. Moreover, Zhang \emph{et al.} studied phonon hydrodynamics in organic crystalline materials using first-principles phonon properties and BTE calculations \cite{zhang2020hydrodynamic}. Hydrodynamic phonon transport was predicted in bulk crystalline polymers such as polyacene and polyacetylene up to temperatures of approximately 50~K, while a weaker and unconventional hydrodynamic behavior was identified in polyethylene near 120~K and traced to torsional vibrational modes. This purely theoretical study proposed a modified quantitative criterion for the emergence of phonon hydrodynamics, validated against drifting phonon populations and thermal conductivity calculations.\\
In parallel, phonon hydrodynamics has witnessed a growing number of studies in systems with even more reduced dimensionality. Guo \emph{et al.} developed an efficient multiscale computational framework for anisotropic graphite ribbons, integrating full quantum-mechanical first-principles inputs within a kinetic theory description and complementing it with a macroscopic hydrodynamic approach \cite{guo2021size}. This study demonstrated pronounced size effects, including a phonon Knudsen minimum associated with the transition from ballistic to hydrodynamic transport, and established quantitative conditions on ribbon length and width for the emergence of collective phonon flow. Moreover, earlier indications of hydrodynamic-like behavior in confined silicon nanostructures were reported in 2019 by Melis \emph{et al.} using classical molecular dynamics simulations of telescopic silicon nanowires \cite{melis2019indications}. Although not based on first-principles force fields, the atomistic simulations revealed clear non-Fourier signatures such as vorticity, radial heat-flux profiles, and temperature discontinuities near contacts. Lattice-dynamics calculations were used to exclude confinement-induced mode localization, leading to the conclusion that a hydrodynamic description based on the GKE is required. A complementary multiscale approach was also introduced in 2019 by Beardo \emph{et al.}, who developed a finite-element implementation of hydrodynamic heat-transport equations and compared the model predictions directly with experimental thermal conductivity measurements in silicon thin films and holey membranes \cite{beardo2019hydrodynamic}. Quantitative agreement was obtained when the smallest characteristic size exceeded twice the intrinsic nonlocal length, providing indirect experimental validation of hydrodynamic phonon transport in nanostructured silicon. Most recently, Wang \emph{et al.} predicted anomalous temperature-dependent thermal conductivity in monolayer nitrogen-substituted diamane (NCCN). using first-principles BTE calculations that explicitly include four-phonon scattering processes \cite{wang2025unusual}. The dominance of momentum-conserving normal scattering over Umklapp processes across a broad temperature range leads to pronounced hydrodynamic behavior, persisting even above room temperature and over a wider temperature window than that observed in graphene. In parallel, Xu \emph{et al.} employed high-fidelity machine-learning interatomic potentials trained on density functional theory data to investigate ultrathin silicon nanowires \cite{xu2025critical}. While not strictly first-principles, these large-scale molecular dynamics simulations revealed a nonmonotonic dependence of thermal conductivity on nanowire diameter and showed that, at low frequencies, normal scattering dominates Umklapp scattering by several orders of magnitude, enabling fluid-like phonon transport.\\
The functional implications of phonon hydrodynamics were explored theoretically in 2024 by Zou \emph{et al.}, who demonstrated that asymmetric graphene microstructures can exhibit thermal rectification when phonon hydrodynamics is captured by numerically solving the phonon BTE within Callaway’s dual model \cite{zou2024thermal}. \begin{figure}[h!]
\centering
\includegraphics[width=0.8\textwidth]{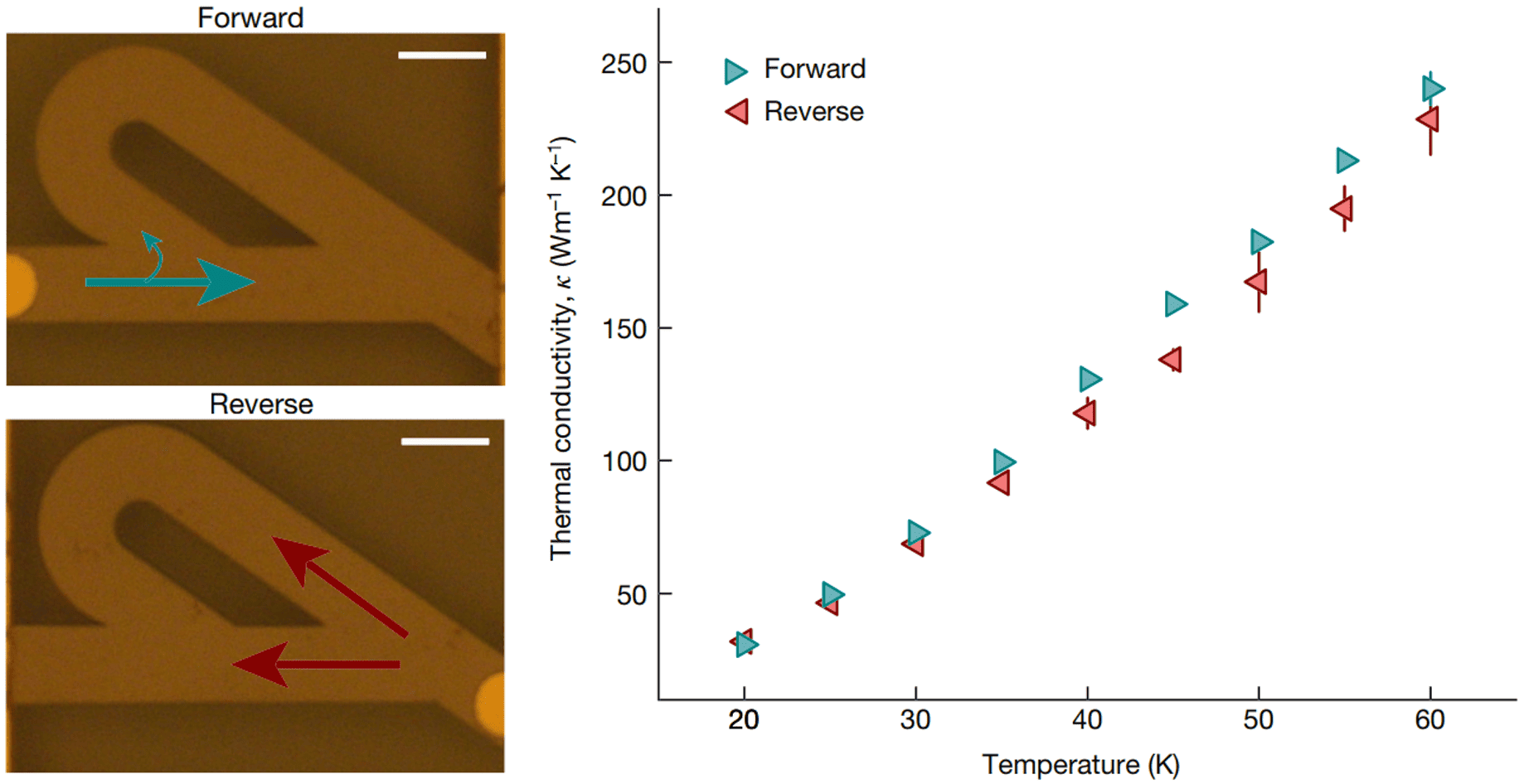}
\caption{\textbf{Graphite thermal Tesla valve.} Thermal transport in a graphite Tesla valve. (Left panel) Optical micrographs of the graphite Tesla valve illustrating the forward (top) and reverse (bottom) configurations, respectively. The arrows denote the imposed directions of heat flow. Scale bars correspond to 5 $\mu$m. (Right panel) Measured thermal conductivity of the graphite Tesla valve for heat flowing in the forward (cyan triangles) and reverse (red triangles) directions as a function of temperature. The error bars represent the standard deviation of the thermal conductivity values obtained from 3–10 independent measurements for each device orientation at each temperature. \textit{Figure reproduced with permission from Ref. \cite{huang2024graphite}.}}
\label{fig:tesla_valve_figure}
\end{figure}
Flexural acoustic phonons were identified as the dominant contributors to the rectification effect, with geometric asymmetry enabling direction-dependent heat flow. A decisive experimental breakthrough was achieved in 2024 by Huang \emph{et al.}, who realized a thermal Tesla valve in isotopically enriched graphite and directly measured a 15.2\% difference in thermal conductivity between opposite directions at 45~K \cite{huang2024graphite}. This experiment provides unambiguous evidence that collective phonon flow can be harnessed to achieve nonreciprocal heat transport in solid-state devices. The operating principle of the graphite-based Tesla valve realized by Huang \emph{et al.}, together with the resulting difference in effective thermal conductivity between the forward and reverse directions, is summarized in Fig. \ref{fig:tesla_valve_figure}. This directional asymmetry results in a lower thermal resistance in the forward direction, thereby enabling rectification and the overall operation of the device as a thermal diode.\\
Finally, a unifying computational perspective on nanoscale confinement effects was provided by Beardo \emph{et al.} in 2025, who critically compared ballistic and hydrodynamic formulations of the phonon BTE against experimental data and atomistic simulations \cite{dabo2025nanoscale}. This work highlighted the necessity of reconciling these approaches to establish a comprehensive theory of phonon transport beyond the diffusive regime.

\section{Conclusion and outlook}

We have reviewed the physics underlying thermal transport both at the quantum and continuum level, {i.e.}, at the levels where atomic vibrations can be resolved and only the evolution of statistical averages (e.g. temperature) are relevant, respectively. We have elucidated how the macroscopic Fourier's heat equation for the temperature and the microscopic BTE for the quantum atomic vibrational excitations (phonons) can be extended to enlarge their domain of applicability.
\\
Starting from the microscopic quantum treatment, we relied on the NEGF formalism for phonons to formally derive the semiclassical Peierls-Boltzmann transport equation, and extended it to account for collisional-broadening effects.  
We reviewed the hierarchy of ansatzes for Green's functions, tracking and discussing approximations involved in the derivation of the BTE.
Importantly, we showed that including collisional-broadening effects is essential for obtaining reliable results in two-dimensional materials with quadratic phonon dispersion relation.
As a perspective, it is worth mentioning that the formalism presented here can be further generalized to account for additional effects. First, it can naturally incorporate frequency lineshifts and pole renormalizations via the real part of the self-energy, which might be important in 2D systems with strong anharmonicity. 
Second, we recall that the quantum treatment presented focuses on the close-to-equilibrium regime, where only first-order terms in the gradient expansion of Green's function and phonon self energy are considered. Future work might extend this approximation by considering higher-order terms that would allow to describe effects beyond linear response.
\\
To review the theoretical description of heat transport at the mesoscopic level, we showed how to coarse-grain the linearized BTE into viscous heat equations, which describe not only heat diffusion but also the hydrodynamic regime where heat propagation resembles fluid dynamics. 
We reviewed known signatures of  hydrodynamic behavior for heat that cannot be described using Fourier's heat equation, specifically: second sound ({i.e.}, heat propagation as a temperature wave); Poiseuille heat flow ({i.e.}, heat flow having a parabolic-like profile akin to the profile of the velocity field of a fluid flowing in a pipe); heat vortices and temperature inversion.
We have relied on the recently introduced relaxon kinetic theory for thermal transport \cite{cepellotti2016thermal} to show that a viscosity for heat emerges from the solution of the microscopic Peierls-Boltzmann equation, and it has a complementary role to the thermal conductivity, since the former is determined exclusively from the even part of the microscopic solution, while the latter is determined exclusively from the odd part of the microscopic solution.
We have exploited microscopic (quasi) conservation laws to condense the physics described by the complex (integro-differential) microscopic BTE into a set of much simpler but equally-accurate mesoscopic differential equations, called “viscous heat equations”, which generalize Fourier's equation accounting for both heat diffusion and  hydrodynamics.
We have shown that for graphitic devices the hydrodynamic behavior for heat predicted from the viscous heat equations quantitatively rationalizes recent experimental measurements \cite{huberman2019observation,ding2022observation}, and to induce temperature inversion from steady-state heat vortices. We have also discussed how the appearance of signatures of  hydrodynamic heat transport in a device is influenced by its chemical composition, size, and temperature, demonstrating that these phenomena can be amplified by engineering the device’s geometry, boundary conditions, 
or exploiting resonance, paving the way for applications in next-generation technologies.
We have provided novel, fundamental insights on temperature waves, showing that the viscous temperature waves emerging from the VHE differ fundamentally from the inviscid heat waves emerging from the DPLE \cite{joseph_heat_1989,tzou_unified_1995,barletta_hyperbolic_1996,xu_thermal_2002,xu_thermal_2021}. 
From a technological perspective, we have discussed how the regime in which heat conduction violates the smoothing property of Fourier's equation can open new avenues to engineer thermal signals that mimic non-diffusive neuron transfer function for heat-based neuromorphic computing \cite{nataf_using_2024,torres_thermal_2023}.
\\
As a perspective, we discussed how these results share fundamental common underpinnings with other quasiparticle’s fluid-like transport phenomena in solids---involving, e.g., electrons \cite{e-vortexes,palm_observation_2024} coupled with phonons \cite{vool_imaging_2021,yang2021evidence,jaoui2022formation,huang2021electron,protik_elphbolt_2022,levchenko_transport_2020,coulter2025coupled}, magnons \cite{wei_giant_2022}, skyrmions \cite{PhysRevLett.130.106703}---thus may inspire analogous developments and applications.
We conclude by noting that both state-of-the-art Monte-Carlo solver for the exact LBTE \cite{raya2022bte} and the VHE focus on the close-to-equilibrium linear regime, and therefore neglect nonlinear advection terms. These have so far been discussed only (i) within the Callaway-approximated LBTE for homogeneous temperature waves in two-dimensional materials \cite{shang2022unified}, and (ii) in a recent study which benchmarked analytical predictions against a numerical Monte Carlo solution of the LBTE under the relaxation-time approximation \cite{yadav2025derivation}. Accounting for these nonlinear effects analytically and numerically while using the full LBTE collision matrix remains an open research problem which could be explored in future work.

\section*{Acknowledgments}
\addcontentsline{toc}{section}{Acknowledgments}

E.D. and N.M. acknowledge support from the Swiss National Science Foundation (SNSF), through Grant No. CRSII5\_189924 (“Hydronics” project). N.M. acknowledges support from NCCR MARVEL, a National Centre of Competence in Research, funded by the Swiss National Science Foundation (Grant No. 205602).

\section*{Disclosure statement}
\addcontentsline{toc}{section}{Disclosure statement}

\appendix

\section{Martin–Schwinger hierarchy and phonon self-energy} \label{self-energy_section}

In this section, we derive the functional form of the self-energy (based on the phonon Green's function) that captures the lowest order three-phonon scattering processes. Our approach builds upon the functional method first introduced by Martin and Schwinger \cite{martin1959theory,wehner1967phonon,wehner1966infra} in the context of cluster expansion, later refined and applied by Kadanoff \cite{kadanoff2018quantum}. This method complements Feynman’s diagrammatic approach, offering the additional advantage of deriving integral equations for the self-energy. Such equations enable precise perturbative analysis, allowing for direct handling of higher-order contributions and vertex corrections. We further clarify the connection between this Green’s-function framework and the more traditional kinetic-theory derivations based on the Liouville equation and the BBGKY hierarchy \cite{bogoliubov1946kineticI,bogoliubov1946kineticII,bogoliubov1947kinetic,born1946general,kirkwood1946statistical,kirkwood1947statistical,yvon1935theorie}. In particular, we show how the cluster expansion and the subsequent truncation of higher-order connected correlations provide the quantum analogue of the molecular-chaos (Stoßzahlansatz) assumption \cite{maxwell1867illustrationsI,maxwell1867illustrationsII,ehrenfest1990conceptual}, thereby establishing the link between the Martin--Schwinger hierarchy and the emergence of Boltzmann-type transport equations.

\subsection{Hierarchy truncation and emergence of Boltzmann transport}

This section briefly outlines how the emergence of Boltzmann transport can be understood as a consequence of truncating a hierarchy of correlations. We first recall the classical derivation based on the Liouville equation and the BBGKY hierarchy, where the BTE follows from the molecular-chaos assumption. We then discuss the quantum analogue in terms of the Liouville--von Neumann equation and correlation functions, highlighting how a similar closure—implemented through a Hartree--Fock–type factorization—leads to a kinetic equation for phonons and underlies the emergence of irreversibility and the $H$-theorem.

\subsubsection{Classical Liouville equation and BBGKY hierarchy}

In classical statistical mechanics, the starting point is the Liouville equation for the $N$-particle distribution function
\begin{equation}
\frac{\partial f_N}{\partial t} + \sum_{i=1}^{N}
\left(
\frac{\boldsymbol{p}_i}{m_i}\cdot \nabla_{\boldsymbol{r}_i}
+
\boldsymbol{F}_i \cdot \nabla_{\boldsymbol{p}_i}
\right)
f_N = 0,
\end{equation}
where $\boldsymbol{F}_i$ is the total force acting on particle $i$. 
By integrating over $(N-s)$ particles one defines the reduced $s$-particle distribution functions
\begin{equation}
f_s(\boldsymbol{r}_1,\dots,\boldsymbol{r}_s,
\boldsymbol{p}_1,\dots,\boldsymbol{p}_s,t)
=
\frac{N!}{(N-s)!}
\int d\Gamma_{s+1}\dots d\Gamma_{N}\, f_N,
\end{equation}
with $d\Gamma_i = d\boldsymbol{r}_i d\boldsymbol{p}_i$.  
The Liouville equation then generates the BBGKY hierarchy,
\begin{equation} \label{BBGKY_hierarchy}
\frac{\partial f_s}{\partial t} + \sum_{i=1}^{s}
\left(
\frac{\boldsymbol{p}_i}{m_i}\cdot \nabla_{\boldsymbol{r}_i}
+
\boldsymbol{F}_i^{(s)} \cdot \nabla_{\boldsymbol{p}_i}
\right)
f_s
=
\sum_{i=1}^{s}
\int d\Gamma_{s+1}\,
\boldsymbol{F}_{i,s+1}
\cdot \nabla_{\boldsymbol{p}_i}
f_{s+1},
\end{equation}
which couples the $s$-particle distribution function to the $(s+1)$-particle one.  
The BTE is then obtained by truncating this hierarchy at the level $s=1$ and invoking the molecular-chaos (Stoßzahlansatz) assumption, according to which the velocities of colliding particles are uncorrelated and independent of position. This means that the probability that a pair of particles with given velocities will collide can be calculated by considering each particle separately and neglecting any correlation between the probability of finding one particle with momentum $\boldsymbol{p}_{1}$ and the probability of finding another with momentum $\boldsymbol{p}_{2}$ in a small region $\boldsymbol{r}_{1}-\boldsymbol{r}_{2}$:
\begin{equation} \label{truncation}
f_2(\boldsymbol{r}_1,\boldsymbol{r}_2,
\boldsymbol{p}_1,\boldsymbol{p}_2,t)
\approx
f_1(\boldsymbol{r}_1,\boldsymbol{p}_1,t)
f_1(\boldsymbol{r}_2,\boldsymbol{p}_2,t).
\end{equation}
This step allows one to close the equation for $f_1$ in Eq. \eqref{BBGKY_hierarchy} and yields the  BTE for classical gases.

\subsubsection{Quantum Liouville equation and $H$-theorem}

In the quantum regime, the starting point is the Liouville--von Neumann equation for the density matrix, $\rho$,
\begin{equation}
i\hbar \frac{\partial \rho}{\partial t}=[H,\rho],
\end{equation}
which is the direct analogue of the classical Liouville equation and is strictly time-reversal invariant. The quantum analogue of the BBGKY hierarchy is obtained by deriving equations of motion for reduced density matrices or, equivalently, for correlation functions. In particular, one introduces the one-body (single-phonon) density matrix
\begin{equation}
\rho^{(1)}_{\nu\nu'}(t)=\langle \hat{a}^\dagger_{\nu'} \hat{a}_{\nu} \rangle
= \mathrm{Tr}\!\left[\rho(t)\,\hat{a}^\dagger_{\nu'} \hat{a}_{\nu}\right],\qquad \nu\equiv\boldsymbol{q}s,
\end{equation}
whose diagonal elements define the phonon occupation (distribution function)
\begin{equation}
\rho^{(1)}_{\nu\nu}(t)\equiv n_{\nu}(t)= \langle \hat{a}^\dagger_{\nu} \hat{a}_{\nu} \rangle.
\end{equation}
In complete analogy with the classical case, the resulting hierarchy does not close: the evolution of the one-body density matrix $\rho^{(1)}$ depends on higher-order correlators (e.g., three- and four-operator expectation values). Irreversibility emerges only after truncating this hierarchy through a suitable closure, which plays the role of the molecular-chaos (Stoßzahlansatz) assumption.\\
The key approximation that allows one to obtain a kinetic equation is therefore a factorization of higher-order correlation functions. In the context of interacting bosonic systems (such as phonons), this corresponds to a Hartree--Fock--type decoupling of operator averages. For instance, four-operator expectation values are approximated as products of two-point functions \cite{vasko2006quantum}. More explicitly, the equation for $\rho^{(1)}_{\nu\nu}(t)$ contains terms of the form
\begin{equation}
\langle \hat{a}_{\nu_1} \hat{a}_{\nu_2} \hat{a}_{\nu} \rangle,
\qquad
\langle \hat{a}^\dagger_{\nu_1} \hat{a}^\dagger_{\nu_2} \hat{a}_{\nu} \rangle,
\end{equation}
whose equations of motion involve four-operator expectation values such as
\begin{equation}
\langle \hat{a}^\dagger_{\nu_1} \hat{a}^\dagger_{\nu_2} \hat{a}_{\nu_3} \hat{a}_{\nu_4} \rangle,
\qquad
\langle \hat{a}_{\nu_1} \hat{a}_{\nu_2} \hat{a}_{\nu_3} \hat{a}_{\nu_4} \rangle,
\end{equation}
and analogous combinations. At this stage, the hierarchy clearly does not close. The Hartree--Fock–type closure is achieved by approximating the four-operator averages as products of two-point functions:
\begin{equation} \label{decoupling}
\begin{split}
\langle \hat{a}^\dagger_{\nu_1} \hat{a}^\dagger_{\nu_2}
\hat{a}_{\nu_3} \hat{a}_{\nu_4} \rangle
\;\approx\;&
\langle \hat{a}^\dagger_{\nu_1} \hat{a}_{\nu_4} \rangle
\langle \hat{a}^\dagger_{\nu_2} \hat{a}_{\nu_3} \rangle
+
\langle \hat{a}^\dagger_{\nu_1} \hat{a}_{\nu_3} \rangle
\langle \hat{a}^\dagger_{\nu_2} \hat{a}_{\nu_4} \rangle,
\end{split}
\end{equation}
while higher-order anomalous (non-number-conserving) correlators are neglected,
\begin{equation} \label{neglected_correlator}
\langle \hat{a}_{\nu_1} \hat{a}_{\nu_2}
\hat{a}_{\nu_3} \hat{a}_{\nu_4} \rangle \simeq 0,
\qquad
\langle \hat{a}^\dagger_{\nu_1} \hat{a}^\dagger_{\nu_2}
\hat{a}^\dagger_{\nu_3} \hat{a}^\dagger_{\nu_4} \rangle \simeq 0.
\end{equation}
In many-body terms, one distinguishes between number-conserving correlators, such as
\begin{equation}
\langle \hat{a}^\dagger \hat{a} \rangle,
\qquad
\langle \hat{a}^\dagger \hat{a}^\dagger \hat{a} \hat{a} \rangle,
\end{equation}
which contain equal numbers of creation and annihilation operators and are invariant under a global phase rotation $\hat{a} \to e^{i\phi} \hat{a}$ (i.e., they preserve the associated U(1) phase symmetry), and anomalous (phase-coherent) correlators, such as
\begin{equation}
\langle \hat{a} \hat{a} \rangle,
\qquad
\langle \hat{a} \hat{a} \hat{a} \hat{a} \rangle,
\label{eq:anomalous_correlators}
\end{equation}
which do not preserve this symmetry and encode phase coherence. It is therefore in this sense that the closure in Eq. \eqref{decoupling} reflects the absence of phase coherence \cite{stefanucci2013nonequilibrium,stefanucci2023and,stefanucci2024semiconductor} and the restriction to a normal (incoherent) state. It is important to stress that such notion of “coherences” should be clearly distinguished from the coherences introduced in the Wigner formulation of thermal transport \cite{simoncelli2019unified,simoncelli2022wigner}, which instead refer to off-diagonal elements of the one-body density matrix in the phonon eigenbasis,
\begin{equation}
n_{ss'}(\boldsymbol{q}) \equiv \langle \hat{a}^\dagger_{\boldsymbol{q}s} \hat{a}_{\boldsymbol{q}s'} \rangle, \qquad s \neq s'.
\end{equation}
These latter coherences are number-conserving and describe interband (i.e.\ $s \neq s'$) phase correlations responsible for wavelike transport and tunneling between quasi-degenerate vibrational modes. As already commented in the text, such interband coherences can be fully retained within the present work, since they originate from two-point functions.\\
Recently, more general theoretical frameworks have been developed that explicitly retain higher-order correlators such as those in Eq. \eqref{neglected_correlator}, in the context of electron, phonon, and electron-phonon transport \cite{stefanucci2024semiconductor}. In this context, in addition to the phonon occupations $n_{\nu}(t)=\langle \hat{a}^\dagger_{\nu}\hat{a}_{\nu}\rangle$, the coherent term 
\begin{equation} \label{coherent_term_stefanucci}
\Theta_{\nu}(t)=\langle \hat{a}_{\nu}\hat{a}_{-\nu}\rangle,
\end{equation}
is also retained, which account for correlations beyond a simple particle-number description. The resulting dynamics is governed by a coupled set of equations of motion,
\begin{equation}
\frac{d}{dt}n_{\nu}(t)=\mathcal{C}_{\nu}[n,\Theta],
\qquad
\frac{d}{dt}\Theta_{\nu}(t)+2i\omega_{\nu}\Theta_{\nu}(t)=\mathcal{S}_{\nu}[n,\Theta],
\end{equation}
where now there are two collision integrals (not only the $\text{CT}_{\nu}[n]$ described in the present work), $\mathcal{C}_{\nu}$ and $\mathcal{S}_{\nu}$, that explicitly couple occupations $\langle a^{\dagger}a\rangle$ and coherences $\langle aa\rangle$. Phase-coherent phonon states described by Eqs. \ref{eq:anomalous_correlators} and \eqref{coherent_term_stefanucci} are expected to be experimentally relevant only in specific driven conditions, e.g. under ultrafast excitation \cite{trovatello2020strongly}. For clarity, we refer to the phase coherence associated with anomalous correlators in Eq. \eqref{eq:anomalous_correlators} as \emph{intraband coherence}, to distinguish it from the \emph{interband coherence} appearing in the Wigner formulation of thermal transport (see below) which are present even when anomalous correlators are neglected.\\
Importantly, the Hartree--Fock–type approximation \eqref{decoupling} is conceptually distinct from the Markov approximation. The former concerns the truncation of correlations in the hierarchy, whereas the latter concerns the assumption of slow temporal variation of $n_{\nu}$, which allows one to evaluate time integrals and enforce energy conservation via $\delta$-functions. After performing the factorization in Eq. \eqref{decoupling} and, subsequently, the Markov approximation, one obtains a closed kinetic equation for the phonon occupation (not yet linearized here and written without imposing homogeneity that make spatial variations arise only from an applied temperature gradient),
\begin{equation}
\frac{\partial n_{\nu}(\boldsymbol{R},t)}{\partial t}+\boldsymbol{v}_{\nu}(\boldsymbol{R},t)\cdot\frac{\partial n_{\nu}(\boldsymbol{R},t)}{\partial\boldsymbol{R}}={\rm CT}_{\nu}(\boldsymbol{R},t)\equiv\left.\frac{\partial n_{\nu}(\boldsymbol{R},t)}{\partial t}\right|_{\text{coll.}},
\end{equation}
where the collision term ${\rm CT}_{\nu}$ is given by Eq. \eqref{final_BTE_scattering_qp_non_homo_main}. In summary, the emergence of irreversibility and the validity of Boltzmann’s $H$-theorem are thus directly tied to the Hartree--Fock–type closure of the hierarchy, which discards correlations built up during the evolution. Within this framework, the standard semiclassical BTE—derived via the GKBA ansatz—is obtained upon neglecting both interband (Wigner) and intraband (anomalous correlator) phase coherences.\\
The factorized structure of the occupation products reflects the same closure principle discussed for the molecular-chaos assumption: multi-phonon distributions entering the scattering probability are reduced to products of single-phonon occupations, neglecting connected higher-order correlations. The appropriate bosonic $\mathcal{H}$-functional is \cite{boltzmann1970weitere}
\begin{equation}
\mathcal{H}(t)=\sum_{\nu}\Big[n_{\nu}(t)\ln\big(n_{\nu}(t)\big)-\big(n_{\nu}(t)+1\big)\ln\big(n_{\nu}(t)+1\big)\Big],
\end{equation}
which is related to the entropy by $S=-k_{B}\mathcal{H}$ \cite{landau2013statistical,peierls1955quantum}. Taking the time derivative and using the BTE yields
\begin{equation}
\frac{d\mathcal{H}}{dt}=\sum_{\nu}\frac{\partial \mathcal{H}}{\partial n_{\nu}}\left.\frac{\partial n_{\nu}}{\partial t}\right|_{\text{coll.}}=\sum_{\nu}\frac{\partial \mathcal{H}}{\partial n_{\nu}}{\rm CT}_{\nu}=\sum_{\nu}\ln\!\left(\frac{n_{\nu}}{n_{\nu}+1}\right){\rm CT}_{\nu}.
\end{equation}
Substituting the explicit form of ${\rm CT}_{\nu}$ and exploiting microscopic detailed balance in momentum space,
\begin{equation}
|\mathcal{F}_{\nu-\nu'-\nu''}|^{2}=|\mathcal{F}_{-\nu\nu'\nu''}|^{2},
\end{equation}
together with the symmetry of the energy-conserving Dirac delta functions, the relabeling of dummy mode indices, and the permutation symmetry of the three-phonon matrix elements, one obtains after symmetrization
\begin{equation}
\frac{d\mathcal{H}}{dt}
=
-\frac{4\pi\hbar}{3}
\sum_{\nu\nu'\nu''}
|\mathcal{F}_{\nu-\nu'-\nu''}|^{2}
\delta(\omega_{\nu}-\omega_{\nu'}-\omega_{\nu''})
\left(X-Y\right)\ln\!\left(\frac{X}{Y}\right),
\end{equation}
where
\begin{equation}
X=(n_{\nu}+1)n_{\nu'}n_{\nu''},
\qquad
Y=n_{\nu}(n_{\nu'}+1)(n_{\nu''}+1).
\end{equation}
Since for positive $X,Y$ one has $(X-Y)\ln(X/Y)\ge 0$, it follows that
\begin{equation}
\frac{d\mathcal{H}}{dt}\le 0,
\qquad
\frac{dS}{dt}\ge 0.
\end{equation}
Entropy therefore increases monotonically until the equilibrium Bose-Einstein distribution is reached, where detailed balance ensures ${\rm CT}_{\nu}=0$. Here one can also clearly identify the apparent reversibility paradox (Umkehreinwand), emphasized by Loschmidt \cite{loschmidt1876sitzungsber}, whereby irreversible entropy growth is derived from time-reversal-invariant microscopic dynamics. This is solved by the fact that, although the underlying three-phonon Hamiltonian is time-reversal invariant, the derivation of monotonic entropy growth relies on a approximated factorization of occupations (see e.g. Eq. A25 in Ref. \cite{auerbach1984universal}) that discards the correlations dynamically produced by scattering: the resolution lies in the asymmetric treatment of correlations. As discussed above, the factorization implicit in the products $n_{\nu}n_{\nu'}$ assumes that incoming phonons are uncorrelated. However, three-phonon scattering can dynamically generates correlations among outgoing modes. If one were to evolve the system backward in time, these post-scattering correlations would need to be retained. By systematically neglecting them when reapplying the factorized form of the collision term, the derivation introduces an effective arrow of time. Irreversibility therefore does not originate from the three-phonon Hamiltonian itself, but from the neglect of dynamically generated higher-order correlations beyond the single-phonon distribution.

\subsection{Functional method}

We now reformulate the hierarchy discussed above within the Green’s-function framework, which provides a natural starting point for a diagrammatic and perturbative treatment of phonon interactions. In this approach, the hierarchy of equations of motion for correlation functions, known as the Martin--Schwinger hierarchy, plays a role directly analogous to the BBGKY hierarchy introduced in the previous subsection. The functional method allows one to systematically organize and truncate this hierarchy through a cluster expansion, thereby making explicit the connection between the Green’s-function formalism and the emergence of Boltzmann-type transport equations.\\
We start by considering the phonon Hamiltonian with its anharmonic (in general beyond third-order) part given by perturbation theory:
\begin{equation} \label{anharmonic_hamiltonian}
\begin{split}
H&=H_{\text{harm}}+\Delta H_{\text{anharm}}=\\
&=\sum_{\boldsymbol{q}_{1}s_{1}}\hbar\omega_{\boldsymbol{q}_{1}s_{1}}\left(a^{\dagger}
_{\boldsymbol{q}_{1}s_{1}}a_{\boldsymbol{q}_{1}s_{1}}+\frac{1}{2}\right)+\\
&\hspace{0.5cm}+\frac{1}{m!}\sum_{m\ge 3}\sum_{\boldsymbol{q}_{1}s_{1},...,\boldsymbol{q}_{m}s_{m}}\uppsi_{m}(\boldsymbol{q}_{1}s_{1},...,\boldsymbol{q}_{m}s_{m})\delta_{\boldsymbol{q}_{1}+...+\boldsymbol{q}_{m},\boldsymbol{K}}\, A_{\boldsymbol{q}_{1}s_{1}}\cdots A_{\boldsymbol{q}_{m}s_{m}}
\end{split}
\end{equation}
where $H_{\text{harm}}$ is the Hamiltonian in the harmonic approximation representing independent harmonic oscillators (free phonons) given in Eq. \eqref{harmonic_hamiltonian}, the squared frequencies $\omega_{\boldsymbol{q}_{1}s_{1}}^{2}$ are the eigenvalues of the dynamical matrix for wave number vector $\boldsymbol{q}_{1}$ and $\Delta H_{\text{anharm}}$ is the anharmonic part of the potential up to the $n$-th order of expansion around atomic equilibrium positions. In the interaction picture, the Hamiltonian \eqref{anharmonic_hamiltonian} leads to the following equation of motion for the phonon operator $ A$ \cite{wehner1967phonon,semwal1972thermal}
\begin{equation} \label{equation_of_motion_greens_function}
\hat{L}_{1} A(1)=\left[\frac{\partial}{\partial t_{1}^{2}}+\omega_{\boldsymbol{q}_{1}s_{1}}^{2}\right] A(1)=-2\omega_{\boldsymbol{q}_{1}s_{1}}\Big[\sum_{m\ge3}m\,\uppsi_{m}(-1,2,...,m) A(2)\cdots A(m)\Big],
\end{equation}
where we used the condensed notation
\begin{equation} \label{local_interaction}
\uppsi_{m}(1,2,...,m)=\uppsi_{m}(\boldsymbol{q}_{1}s_{1},\boldsymbol{q}_{2}s_{2},...,\boldsymbol{q}_{m}s_{m})\delta(t_{1}-t_{2})\cdots\delta(t_{m-1}-t_{m})
\end{equation}
with 
\begin{equation}
\begin{split}
k&=\boldsymbol{q}_{k}s_{k},t_{k},\\
-k&=-\boldsymbol{q}_{k}s_{k},t_{k}.
\end{split}
\end{equation}
Note that we have adopted the summation convention on the RHS of Eq. \eqref{equation_of_motion_greens_function}:
\begin{equation} \label{interaction_with_delta_on_time}
\uppsi_{m}(...,m)\cdots A(m)=\sum_{\boldsymbol{q}_{m}s_{m}}\int_{-\infty}^{\infty}dt_{m}\uppsi_{m}(...,\boldsymbol{q}_{m}s_{m})\cdots\delta(t_{m-1}-t_{m})\cdots A_{\boldsymbol{q}_{m}s_{m}}(t_{m}),
\end{equation}
where we wrote explicitly the time-dependence of the phonon operator $ A$. Dirac delta functions in time and wave vectors (see Eq. \eqref{delta_kroneker_self_energy}) inherently assume that interactions are both instantaneous and spatially localized, implying that phonons cannot change their positions during the scattering event. The impact of such quantum non-localities \cite{morawetz2017nonequilibrium} in scattering processes will be addressed in a separate article. All time-correlation functions can be derived from the time-ordered many-time Green's functions
\begin{equation}  \label{greens_function_Gn}
G_{n}(1,...,n)=-i\langle\hat{T}\left[A(1)\cdots A(n)\right]\rangle.
\end{equation}
The set of Green's functions \eqref{greens_function_Gn} in which we are interested is generated by the functional \cite{martin1959theory}
\begin{equation}
\mathcal{G}\left[\xi\right]=\langle \hat{T}e^{\xi(k)A(k)}\rangle=1+\sum_{n=1}^{\infty}\frac{1}{n!}\xi(1)\cdots\xi(n)G_{n}(1,...,n),
\end{equation}
with
\begin{equation} \label{Gn_as_functional_derivative}
G_{n}(1,...,n)=\left(\frac{\delta^{n}\mathcal{G}\left[\xi\right]}{\delta\xi(1)\cdots\delta\xi(n)}\right)_{\xi=0},
\end{equation}
where we have introduced the “source” function $\xi(k)$ which can be either real or complex \cite{wehner1966infra}. The following summation convention is used
\begin{equation}
\xi(k)A(k)=\sum_{\boldsymbol{q}_{k}s_{k}}\int_{-\infty}^{\infty}dt_{k}\xi(\boldsymbol{q}_{k}s_{k};t_{k})A(\boldsymbol{q}_{k}s_{k};t_{k}).
\end{equation}
The interaction picture is typically used to deal with the time evolution of phonon operators and in order to express correlation functions \cite{caldarelli2022many}. For the present purpose the nature of the $G$ functional is unimportant. However, as done by Kwok and Martin \cite{kwok1966unified} and Horie and Krumhansl \cite{horie1964boltzmann}, one can show that it coincides with the thermodynamic expectation value of the S-matrix \cite{klein1968linear}:
\begin{equation}
S(\infty)=\hat{T}{{\rm{exp}}}\left\lbrace-i\int_{-\infty}^{\infty}dt\,\Delta H^{I}_{\text{anharm}}(t)\right\rbrace,
\end{equation}
where the label $I$ indicates the interaction picture. We now apply the operator $\hat{L}_{1}$ on the Green's functions $G_{n}$ and, using Eq. \eqref{equation_of_motion_greens_function}, we get the equations of motion \cite{wehner1966infra,martin1959theory}
\begin{equation} \label{equation_of_motion_Gn}
\begin{split}
\hat{L}_{1}G_{n}(1,...,n)=&-2i\omega_{\boldsymbol{q}_{1}s_{1}}\sum_{k=2}^{n}\delta(1,k)G_{n-2}(2,...,k-1,k+1,...,n)+\\
&+2\omega^{2}_{\boldsymbol{q}_{1}s_{1}}\sum_{m\ge3}m\,\uppsi_{m}(-1,n+1,...,n+m-1)G_{n-m-2}(2,...,n+m-1)
\end{split}
\end{equation}
where 
\begin{equation} \label{delta_kroneker_self_energy}
\delta(1,k)=\delta_{\boldsymbol{q}_{1}+\boldsymbol{q}_{k},\boldsymbol{G}}\delta_{s_{1},s_{k}}\delta(t_{1}-t_{k}).
\end{equation}
At this stage, Eq. \eqref{equation_of_motion_Gn} defines a hierarchy of coupled equations of motion for the $n$-point Green’s functions, in which each $G_n$ is connected to higher-order correlators $G_{n+1}, G_{n+2}, \dots$. This structure is directly analogous to the BBGKY hierarchy discussed in the previous subsection, where the evolution of the $s$-particle distribution function depends on higher-order distributions. In this sense, the Martin--Schwinger hierarchy provides the Green’s-function counterpart of the BBGKY hierarchy within the quantum many-body formalism. \\
The functional method can be exploited to write the set of equations \eqref{equation_of_motion_Gn} as a single functional equation
\begin{equation} \label{functional_equation_of_motion}
\left[\hat{L}_{1}\frac{\delta}{\delta\xi(1)}+2i\omega_{\boldsymbol{q}_{1}s_{1}}\xi(-1)+2\omega_{\boldsymbol{q}_{1}s_{1}}\sum_{m\ge3}m\,\uppsi_{m}(-1,2,...,m)\frac{\delta^{m-1}}{\delta\xi(2)\cdots\delta\xi(m)}\right]\mathcal{G}[\xi]=0.
\end{equation}
The solution of Eq. \eqref{functional_equation_of_motion} can be determined by a cluster expansion of the Green's functions $G_{n}$ \cite{wehner1966infra,wehner1967phonon}. 

\subsection{Cluster expansion}

The hierarchy introduced above does not close, since the equation of motion for a given $G_n$ involves higher-order Green’s functions. In complete analogy with the BBGKY hierarchy, a closure scheme is therefore required. Within the Green’s-function formalism, this is achieved through the cluster expansion, which separates full correlation functions into connected and disconnected contributions. As we show below, this procedure provides a systematic way to express higher-order correlations in terms of lower-order ones.\\
The decoupling of the hierarchy of equations of motion \eqref{equation_of_motion_Gn} can be achieved by defining a new functional $\mathcal{F}[\xi]$ as
\begin{equation} \label{new_functional}
\mathcal{G}[\xi]=e^{\mathcal{F}[\xi]}
\end{equation}
where $\mathcal{F}[\xi]$ is defined as a power series in $\xi$ as
\begin{equation}
\mathcal{F}[\xi]=\sum_{n=1}^{\infty}\frac{1}{n!}\xi(1)\cdots\xi(n)f_{n}(1,...,n).
\end{equation}
At this stage we perform a cluster expansion on the exponential \eqref{new_functional} in terms of the new dynamical correlation functions $f_{n}(1,...,n)$. Then, the phonon Green's functions \eqref{greens_function_Gn} are given by the following relations
\begin{equation} \label{cluster_expansion_definitions}
\resizebox{\textwidth}{!}{$
\begin{split}
&G_{1}(1)=f_{1}(1)\\
&G_{2}(1,2)=f_{1}(1)f_{1}(2)+f_{2}(1,2)\\
&G_{3}(1,2,3)=f_{1}(1)f_{1}(2)f_{1}(3)+f_{1}(1)f_{2}(2,3)+f_{1}(2)f_{2}(3,1)+f_{1}(3)f_{2}(1,2)+f_{3}(1,2,3),\,\rm{etc.}
\end{split}$}
\end{equation}
These can be summarized in the general term
\begin{equation} \label{cluster_expansion}
\begin{split}
G_{n}(1,...,n)=\sum_{\substack{k,l,m\$n=k+2l+3m+...)}}\hat{\mathcal{S}}\Big[&f_{1}(1)\cdots f_{1}(k)\cdot\\
&f_{2}(k+1,k+2)\cdots f_{2}(k+2l-1,k+2l)\cdot\\
&f_{3}(k+2l+3m-2,k+2l+3m-1,k+2l+3m)\cdots\Big].
\end{split}
\end{equation}
The symmetrizing operator $\hat{\mathcal{S}}$ means that we have to sum over the $\frac{n!}{k!(2!)^{l}l!(3!)^{m}m!\cdots}$ different combinations of the variables $1,...,n$. Eq. \eqref{cluster_expansion} represents the usual cluster expansion of statistical mechanics \cite{landau2013statistical,mayer1941molecular,mayer1938statistical}. The inverse relations are found from the functional
derivatives of the logarithm of Eq. \eqref{new_functional},
\begin{equation} \label{fn_as_functional_derivative}
f_{n}(1,...,n)=\left(\frac{\delta^{n}\ln\left(\mathcal{G}\left[\xi\right]\right)}{\delta\xi(1)\cdots\delta\xi(n)}\right)_{\xi=0}.
\end{equation}
For the first $n$ indices of interest, we obtain
\begin{equation}
\resizebox{\textwidth}{!}{$
\begin{split}
&f_{1}(1)=G_{1}(1)\\
&f_{2}(1,2)=G_{2}(1,2)-G_{1}(1)G_{1}(2)\\
&f_{3}(1,2,3)=G_{3}(1,2,3)-G_{2}(1,2)G_{1}(3)-G_{2}(2,3)G_{1}(1)-G_{2}(3,1)G_{1}(2)+2G_{1}(1)G_{1}(2)G_{1}(3),\,\rm{etc.}
\end{split}$}
\end{equation}
which represent dynamical correlations beyond products of lower-order contributions. From a physical perspective, this decomposition corresponds to restricting the description to two-point correlations (the propagator), while neglecting genuine three- and higher-body connected correlations. Diagrammatically, this implies that the self-energy is constructed from two-particle irreducible diagrams built from the propagator $f_2$. These diagrams generate effective three-phonon scattering processes. This mirrors the closure of the BBGKY hierarchy at the two-particle level, leading to Boltzmann-type collision integrals. The functions $G_{n}$ and $f_{n}$ are generalized to functionals by the following definitions:
\begin{equation} \label{Gn_as_functional}
G_{n,[\xi]}(1,...,n)=-i\frac{\langle\hat{T}\big[e^{\xi(k)A(k)}A(1)\cdots A(n)\big]\rangle}{\langle\hat{T}\left[e^{\xi(k)A(k)}\right]\rangle}=\frac{1}{\mathcal{G}[\xi]}\frac{\delta^{n}\mathcal{G}\left[\xi\right]}{\delta\xi(1)\cdots\delta\xi(n)}
\end{equation}
and
\begin{equation} \label{fn_as_functional}
f_{n,[\xi]}(1,...,n)=\frac{\delta^{n}\ln\left(\mathcal{G}\left[\xi\right]\right)}{\delta\xi(1)\cdots\delta\xi(n)}=\frac{\delta^{n-1}f_{1,[\xi]}(1)}{\delta\xi(2)\cdots\delta\xi(n)},
\end{equation}
which respectively reduce to Eqs. \eqref{Gn_as_functional_derivative} and \eqref{fn_as_functional_derivative} if the source function $\xi$ approaches zero. \\
From the decomposition in Eq. \eqref{cluster_expansion_definitions} it is clear that the functions $f_n$ represent connected $n$-body correlations, while the products of lower-order terms describe statistically independent contributions. In particular, a truncation of the hierarchy at the level of $f_2$, obtained by neglecting $f_3$, $f_4$, etc., amounts to assuming that higher-order correlations are not independent degrees of freedom, but can be constructed from repeated two-body processes. This is precisely the quantum analogue of the molecular-chaos assumption in the BBGKY hierarchy, where the two-particle distribution function is factorized into a product of single-particle distributions.

\subsection{Dyson equation}

According to Eq. \eqref{Gn_as_functional}, the equations of motion for $G_{1,[\xi]}(1)$ is derived by simply operating with $\frac{1}{\mathcal{G}[\xi]}$ on Eq. \eqref{functional_equation_of_motion}. In this way we have
\begin{equation} \label{application_of_functional_G}
\left[\hat{L}_{1}\frac{1}{\mathcal{G}[\xi]}\frac{\delta}{\delta\xi(1)}+\frac{2i\omega_{\boldsymbol{q}_{1}s_{1}}}{\mathcal{G}[\xi]}\xi(-1)+\frac{2\omega_{\boldsymbol{q}_{1}s_{1}}}{\mathcal{G}[\xi]}\sum_{m\ge3}m\,\uppsi_{m}(-1,2,...,m)\frac{\delta^{m-1}}{\delta\xi(2)\cdots\delta\xi(m)}\right]\mathcal{G}[\xi]=0.
\end{equation}
From Eq. \eqref{Gn_as_functional} we respectively have
\begin{equation}
\begin{split}
&\frac{1}{\mathcal{G}[\xi]}\frac{\delta\mathcal{G}[\xi]}{\delta\xi(1)}=G_{1,[\xi]}(1),\\
&\frac{1}{\mathcal{G}[\xi]}\mathcal{G}[\xi]=1,\\
&\frac{1}{\mathcal{G}[\xi]}\frac{\delta^{m-1}\mathcal{G}[\xi]}{\delta\xi(2)\cdots\delta\xi(m)}=G_{m-1,[\xi]}(2,...,m).
\end{split}
\end{equation}
So Eq. \eqref{application_of_functional_G} becomes
\begin{equation}
\hat{L}_{1}G_{1,[\xi]}(1)+2i\omega_{\boldsymbol{q}_{1}s_{1}}\xi(-1)+2\omega_{\boldsymbol{q}_{1}s_{1}}\sum_{m\ge3}m\,\uppsi_{m}(-1,2,...,m)G_{m-1,[\xi]}(2,...,m)=0,
\end{equation}
which can be written as
\begin{equation}
\hat{L}_{1}G_{1,[\xi]}(1)=-2i\omega_{\boldsymbol{q}_{1}s_{1}}\xi(-1)-2\omega_{\boldsymbol{q}_{1}s_{1}}\sum_{m\ge3}m\,\uppsi_{m}(-1,2,...,m)G_{m-1,[\xi]}(2,...,m)=0,
\end{equation}
At this stage, we can use the cluster expansion definitions \eqref{cluster_expansion_definitions} to obtain the equation for $f_{1,[\xi]}(1)$:
\begin{equation} \label{equation_functional_f1_initial}
\resizebox{\textwidth}{!}{$
\begin{split}
&\hat{L}_{1}f_{1,[\xi]}(1)=\\
=&-2i\omega_{\boldsymbol{q}_{1}s_{1}}\xi(-1)-\\
&-2\omega_{\boldsymbol{q}_{1}s_{1}}\Big\lbrace3\Big[f_{1,[\xi]}(2)f_{1,[\xi]}(3)+f_{2,[\xi]}(2,3)\Big]\uppsi_{3}(-1,2,3)+\\
&\hspace{1.75cm}+4\Big[f_{1,[\xi]}(2)f_{1,[\xi]}(3)f_{1,[\xi]}(4)+f_{1,[\xi]}(2)f_{2,[\xi]}(3,4)+\\
&\hspace{2.5cm}+f_{1,[\xi]}(3)f_{2,[\xi]}(4,2)+f_{1,[\xi]}(4)f_{2,[\xi]}(2,3)+f_{3}(2,3,4)\Big]\uppsi_{4}(-1,2,3,4)\Big\rbrace=0,
\end{split}$}
\end{equation}
where we have retained only terms up to the 4th order of interactions. The $f_{1}f_{2}$ terms in the 4th order part of Eq. \eqref{equation_functional_f1_initial} give the same contribution, so we can write
\begin{equation} \label{equation_functional_f1}
\resizebox{\textwidth}{!}{$
\begin{split}
&\hat{L}_{1}f_{1,[\xi]}(1)=\\
=&-2\omega_{\boldsymbol{q}_{1}s_{1}}\Big\lbrace3\Big[f_{1,[\xi]}(2)f_{1,[\xi]}(3)+f_{2,[\xi]}(2,3)\Big]\uppsi_{3}(-1,2,3)+\\
&\hspace{1.75cm}+4\Big[f_{1,[\xi]}(2)f_{1,[\xi]}(3)f_{1,[\xi]}(4)+3f_{1,[\xi]}(2)f_{2,[\xi]}(3,4)+f_{3}(2,3,4)\Big]\uppsi_{4}(-1,2,3,4)\Big\rbrace=0.
\end{split}$}
\end{equation}
By differentiating Eq. \eqref{equation_functional_f1} the equations of motion of the higher functionals $f_{1,[\xi]}$, follow. Exploiting Eq. \eqref{fn_as_functional}, the first derivative gives
\begin{equation} \label{derivative_L1f1}
\begin{split}
&\frac{\delta}{\delta\xi(2)}\hat{L}_{1}f_{1,[\xi]}(1)=\\
=&\hat{L}_{1}f_{2,[\xi]}(1,2)=\\
=&-2i\omega_{\boldsymbol{q}_{1}s_{1}}\delta(1,2)-\\
&-2\omega_{\boldsymbol{q}_{1}s_{1}}\Bigg\lbrace3\uppsi_{3}(-1,2,3)\frac{\delta}{\delta\xi(2)}\Big(f_{1,[\xi]}(2')f_{1,[\xi]}(3)+f_{2,[\xi]}(2',3)\Big)+\\
&\hspace{1.75cm}+4\uppsi_{4}(-1,2,3,4)\frac{\delta}{\delta\xi(2)}\Big[f_{1,[\xi]}(2')f_{1,[\xi]}(3)f_{1,[\xi]}(4)+\\
&\hspace{6cm}+3f_{1,[\xi]}(2')f_{2,[\xi]}(3,4)+f_{3}(2',3,4)\Big]\Bigg\rbrace.
\end{split}
\end{equation}
So, from Eq. \eqref{fn_as_functional} the derivatives entering the 3rd order term are
\begin{equation} \label{fn_relations_1_a}
\begin{split}
\frac{\delta}{\delta\xi(2)}\Big[f_{1,[\xi]}(2')f_{1,[\xi]}(3)\Big]&=\frac{\delta f_{1,[\xi]}(2')}{\delta\xi(2)}f_{1,[\xi]}(3)+f_{1,[\xi]}(2')\frac{\delta f_{1,[\xi]}(3)}{\delta\xi(2)}=\\
&=f_{2,[\xi]}(2',2)f_{1,[\xi]}(3)+f_{1,[\xi]}(2')f_{2,[\xi]}(3,2)=2f_{2,[\xi]}(2',2)f_{1,[\xi]}(3)
\end{split}
\end{equation}
and
\begin{equation} \label{fn_relations_1_b}
\begin{split}
\frac{\delta f_{2,[\xi]}(2',3)}{\delta\xi(2)}=\frac{\delta^{2}f_{1,[\xi]}(2')}{\delta\xi(2)\delta\xi(3)}=f_{3,[\xi]}(2',2,3),
\end{split}
\end{equation}
while the derivatives in the 4th order term are
\begin{equation} \label{fn_relations_2_a}
\begin{split}
&\frac{\delta}{\delta\xi(2)}\Big[f_{1,[\xi]}(2')f_{1,[\xi]}(3)f_{1,[\xi]}(4)\Big]=\\
=&\frac{\delta}{\delta\xi(2)}\Big[f_{1,[\xi]}(2')f_{1,[\xi]}(3)\Big]f_{1,[\xi]}(4)+f_{1,[\xi]}(2')f_{1,[\xi]}(3)\frac{\delta f_{1,[\xi]}(4)}{\delta\xi(2)}=\\
=&\Big[f_{1,[\xi]}(3)f_{2,[\xi]}(2',2)+f_{1,[\xi]}(2')f_{2,[\xi]}(3,2)\Big]f_{1,[\xi]}(4)+f_{1,[\xi]}(2')f_{1,[\xi]}(3)f_{2,[\xi]}(4,2)=\\
=&f_{1,[\xi]}(3)f_{2,[\xi]}(2',2)f_{1,[\xi]}(4)+f_{1,[\xi]}(2')f_{2,[\xi]}(3,2)f_{1,[\xi]}(4)+f_{1,[\xi]}(2')f_{1,[\xi]}(3)f_{2,[\xi]}(4,2)=\\
=&3f_{1,[\xi]}(3)f_{2,[\xi]}(2',2)f_{1,[\xi]}(4),
\end{split}
\end{equation}
\begin{equation} \label{fn_relations_2_b}
\begin{split}
\frac{\delta}{\delta\xi(2)}\Big[f_{1,[\xi]}(2')f_{2,[\xi]}(3,4)\Big]&=\frac{\delta f_{1,[\xi]}(2')}{\delta\xi(2)}f_{2,[\xi]}(3,4)+f_{1,[\xi]}(2')\frac{\delta f_{2,[\xi]}(3,4)}{\delta\xi(2)}=\\
&=f_{2,[\xi]}(2',2)f_{2,[\xi]}(3,4)+f_{1,[\xi]}(2')f_{3,[\xi]}(3,2,4)
\end{split}
\end{equation}
and 
\begin{equation} \label{fn_relations_2_c}
\begin{split}
\frac{\delta f_{3}(2',3,4)}{\delta\xi(2)}&=\frac{\delta^{2}f_{2,[\xi]}(2',4)}{\delta\xi(2)\delta\xi(3)}=\frac{\delta}{\delta\xi(2)}\frac{\delta^{2}f_{1}(2')}{\delta\xi(3)\delta\xi(4)}=\\
&=\frac{\delta^{3}f_{1}(2')}{\delta\xi(2)\delta\xi(3)\delta\xi(4)}=f_{4,[\xi]}(2',2,3,4).
\end{split}
\end{equation}
In this way Eq. \eqref{derivative_L1f1} becomes
\begin{equation} \label{derivative_L1f1_bis}
\resizebox{\textwidth}{!}{$
\begin{split}
&\hat{L}_{1}f_{2,[\xi]}(1,2)=\\
=&-2i\omega_{\boldsymbol{q}_{1}s_{1}}\delta(1,2)-\\
&-2\omega_{\boldsymbol{q}_{1}s_{1}}\Bigg\lbrace3\uppsi_{3}(-1,2,3)\Big[2f_{2,[\xi]}(2',2)f_{1,[\xi]}(3)+f_{3,[\xi]}(2',2,3)\Big]+\\
&\hspace{1.75cm}+4\uppsi_{4}(-1,2,3,4)\Big[3f_{1,[\xi]}(3)f_{2,[\xi]}(2',2)f_{1,[\xi]}(4)+\\
&\hspace{5cm}+3\Big(f_{2,[\xi]}(2',2)f_{2,[\xi]}(3,4)+f_{1,[\xi]}(2')f_{3,[\xi]}(3,2,4)\Big)+f_{4,[\xi]}(2',2,3,4)\Big]\Bigg\rbrace.
\end{split}$}
\end{equation}
In practice it is convenient to eliminate correlation functions above the 2nd order. In this way, using Eqs. \eqref{fn_relations_1_b} and \eqref{fn_relations_2_c} for writing $f_{3,[\xi]}$ and $f_{4,[\xi]}$ in terms of $f_{2,[\xi]}$ respectively, we get
\begin{equation} \label{L1f2_eq}
\resizebox{\textwidth}{!}{$
\begin{split}
&\hat{L}_{1}f_{2,[\xi]}(1,2)=\\
=&-2i\omega_{\boldsymbol{q}_{1}s_{1}}\delta(1,2)-\\
&-2\omega_{\boldsymbol{q}_{1}s_{1}}\Bigg\lbrace3\uppsi_{3}(-1,2,3)\Bigg[2f_{2,[\xi]}(2',2)f_{1,[\xi]}(3)+\frac{\delta f_{2,[\xi]}(2',3)}{\delta\xi(2)}\Bigg]+\\
&\hspace{1.75cm}+4\uppsi_{4}(-1,2,3,4)\Bigg[3f_{1,[\xi]}(3)f_{2,[\xi]}(2',2)f_{1,[\xi]}(4)+\\
&\hspace{5.25cm}+3\Bigg(f_{2,[\xi]}(2',2)f_{2,[\xi]}(3,4)+f_{1,[\xi]}(2')\frac{\delta f_{2,[\xi]}(3,4)}{\delta\xi(2)}\Bigg)+\frac{\delta^{2}f_{2,[\xi]}(2',4)}{\delta\xi(2)\delta\xi(3)}\Bigg]\Bigg\rbrace.
\end{split}$}
\end{equation}
Furthermore, the functional derivatives of $f_{2,[\xi]}$ are expressed by
\begin{equation} \label{differentiating_f2_1}
\frac{\delta f_{2,[\xi]}(3,4)}{\delta\xi(2)}=\frac{\delta f_{2,[\xi]}(3,4)}{\delta f_{1,[\xi]}(2')}\frac{\delta f_{1,[\xi]}(2')}{\delta\xi(2)}=\frac{\delta f_{2,[\xi]}(3,4)}{\delta f_{1,[\xi]}(2')}f_{2,[\xi]}(2',2)
\end{equation}
and
\begin{equation} \label{differentiating_f2_2}
\frac{\delta^{2}f_{2,[\xi]}(2'',4)}{\delta\xi(2)\delta\xi(3)}=\left[\frac{\delta^{2}f_{2,[\xi]}(2'',4)}{\delta f_{1,[\xi]}(2')\delta f_{1,[\xi]}(2''')}f_{2,[\xi]}(2''',2)+\frac{\delta f_{2,[\xi]}(2'',4)}{\delta f_{1,[\xi]}(2''')}\frac{\delta f_{2,[\xi]}(2''',2)}{\delta f_{1,[\xi]}(2')}\right]f_{2,[\xi]}(2',2),
\end{equation}
where for convenience we wrote $2''$ in place of $2'$. Then, substituting in Eq. \eqref{L1f2_eq} we get
\begin{equation}
\resizebox{\textwidth}{!}{$
\begin{split}
&\hat{L}_{1}f_{2,[\xi]}(1,2)=\\
=&-2i\omega_{\boldsymbol{q}_{1}s_{1}}\delta(1,2)-\\
&-2\omega_{\boldsymbol{q}_{1}s_{1}}\Bigg\lbrace3\uppsi_{3}(-1,2,3)\Bigg[2f_{2,[\xi]}(2',2)f_{1,[\xi]}(3)+\frac{\delta f_{2,[\xi]}(3,4)}{\delta f_{1,[\xi]}(2')}f_{2,[\xi]}(2',2)\Bigg]+\\
&\hspace{1.75cm}+4\uppsi_{4}(-1,2,3,4)\Bigg[3f_{1,[\xi]}(3)f_{2,[\xi]}(2',2)f_{1,[\xi]}(4)+\\
&\hspace{5cm}+3\Bigg(f_{2,[\xi]}(2',2)f_{2,[\xi]}(3,4)+f_{1,[\xi]}(2')\frac{\delta f_{2,[\xi]}(3,4)}{\delta f_{1,[\xi]}(2')}f_{2,[\xi]}(2',2)\Bigg)+\\
&\hspace{5cm}+\left(\frac{\delta^{2}f_{2,[\xi]}(2'',4)}{\delta f_{1,[\xi]}(2')\delta f_{1,[\xi]}(2''')}f_{2,[\xi]}(2''',2)+\frac{\delta f_{2,[\xi]}(2'',4)}{\delta f_{1,[\xi]}(2''')}\frac{\delta f_{2,[\xi]}(2''',2)}{\delta f_{1,[\xi]}(2')}\right)f_{2,[\xi]}(2',2)\Bigg]\Bigg\rbrace.
\end{split}$}
\end{equation}
Factorizing $f_{2,[\xi]}(2',2)$ we get
\begin{equation}
\resizebox{\textwidth}{!}{$
\begin{split}
&\hat{L}_{1}f_{2,[\xi]}(1,2)=\\
=&-2i\omega_{\boldsymbol{q}_{1}s_{1}}\delta(1,2)-\\
&-2\omega_{\boldsymbol{q}_{1}s_{1}}\Bigg\lbrace3\uppsi_{3}(-1,2,3)\Bigg[2f_{1,[\xi]}(3)+\frac{\delta f_{2,[\xi]}(3,4)}{\delta f_{1,[\xi]}(2')}\Bigg]+\\
&\hspace{1.75cm}+4\uppsi_{4}(-1,2,3,4)\Bigg[3f_{1,[\xi]}(3)f_{1,[\xi]}(4)+3\Bigg(f_{2,[\xi]}(3,4)+f_{1,[\xi]}(2')\frac{\delta f_{2,[\xi]}(3,4)}{\delta f_{1,[\xi]}(2')}\Bigg)+\\
&\hspace{5cm}+\frac{\delta^{2}f_{2,[\xi]}(2'',4)}{\delta f_{1,[\xi]}(2')\delta f_{1,[\xi]}(2''')}f_{2,[\xi]}(2''',2)+\frac{\delta f_{2,[\xi]}(2'',4)}{\delta f_{1,[\xi]}(2''')}\frac{\delta f_{2,[\xi]}(2''',2)}{\delta f_{1,[\xi]}(2')}\Bigg]\Bigg\rbrace f_{2,[\xi]}(2',2).
\end{split}$}
\end{equation}
Then we can finally write
\begin{equation} \label{L1f2_eq_final}
\resizebox{\textwidth}{!}{$
\begin{split}
&\hat{L}_{1}f_{2,[\xi]}(1,2)=\\
=&-2i\omega_{\boldsymbol{q}_{1}s_{1}}\delta(1,2)-\\
&-2\omega_{\boldsymbol{q}_{1}s_{1}}\Bigg\lbrace6\uppsi_{3}(-1,2,3)f_{1,[\xi]}(3)+12\uppsi_{4}(-1,2,3,4)f_{1,[\xi]}(3)f_{1,[\xi]}(4)+12\uppsi_{4}(-1,2,3,4)f_{2,[\xi]}(3,4)+\\
&\hspace{1.75cm}+\Bigg[3\uppsi_{3}(-1,2,3)+12\uppsi_{4}(-1,2,3,4)f_{1,[\xi]}(2')+4\uppsi_{4}(-1,2,3,4)\frac{\delta f_{2,[\xi]}(2'',4)}{\delta f_{1,[\xi]}(2''')}\Bigg]\frac{\delta f_{2,[\xi]}(3,4)}{\delta f_{1,[\xi]}(2')}+\\
&\hspace{1.75cm}+4\uppsi_{4}(-1,2,3,4)\frac{\delta^{2}f_{2,[\xi]}(2'',4)}{\delta f_{1,[\xi]}(2')\delta f_{1,[\xi]}(2''')}f_{2,[\xi]}(2''',2)\Bigg\rbrace f_{2,[\xi]}(2',2).
\end{split}$}
\end{equation}
The solution of Eq. \eqref{L1f2_eq_final}, together with Eqs. \eqref{differentiating_f2_1} and \eqref{differentiating_f2_2}, is given by the Dyson equation:
\begin{equation} \label{dyson_eq_for_f2}
f_{2,[\xi]}(1,2)=f^{0}_{2,[\xi]}(1,2)-if^{0}_{2,[\xi]}(1,1')\Sigma(1',2')f_{2,[\xi]}(2',2)
\end{equation}
where $f^{0}_{2,[\xi]}$ is the general solution of Eq. \eqref{L1f2_eq} with vanishing interactions:
\begin{equation}
\hat{L}_{1}f^{0}_{2,[\xi]}(1,2)=-2i\omega_{\boldsymbol{q}_{1}s_{1}}\delta(1,2),
\end{equation}
and the term within curly brackets in Eq. \eqref{L1f2_eq_final} is the self-energy $\Sigma$ up to 4th order phonon-phonon interactions:
\begin{equation} \label{self_energy_before_vertices}
\resizebox{\textwidth}{!}{$
\begin{split}
&\Sigma(1,2)=\\
=&6\uppsi_{3}(-1,2,3)f_{1,[\xi]}(3)+12\uppsi_{4}(-1,2,3,4)f_{1,[\xi]}(3)f_{1,[\xi]}(4)+12\uppsi_{4}(-1,2,3,4)f_{2,[\xi]}(3,4)+\\
&+\Bigg[3\uppsi_{3}(-1,2,3)+12\uppsi_{4}(-1,2,3,4)f_{1,[\xi]}(2')+4\uppsi_{4}(-1,2,3,4)\frac{\delta f_{2,[\xi]}(2'',4)}{\delta f_{1,[\xi]}(2''')}\Bigg]\frac{\delta f_{2,[\xi]}(3,4)}{\delta f_{1,[\xi]}(2')}+\\
&+4\uppsi_{4}(-1,2,3,4)\frac{\delta^{2}f_{2,[\xi]}(2'',4)}{\delta f_{1,[\xi]}(2')\delta f_{1,[\xi]}(2''')}f_{2,[\xi]}(2''',2).
\end{split}$}
\end{equation}

\subsection{Vertex corrections} \label{vertex_corrections_section}

\begin{figure}[!htb]
\centering
\includegraphics[width=0.75\textwidth]{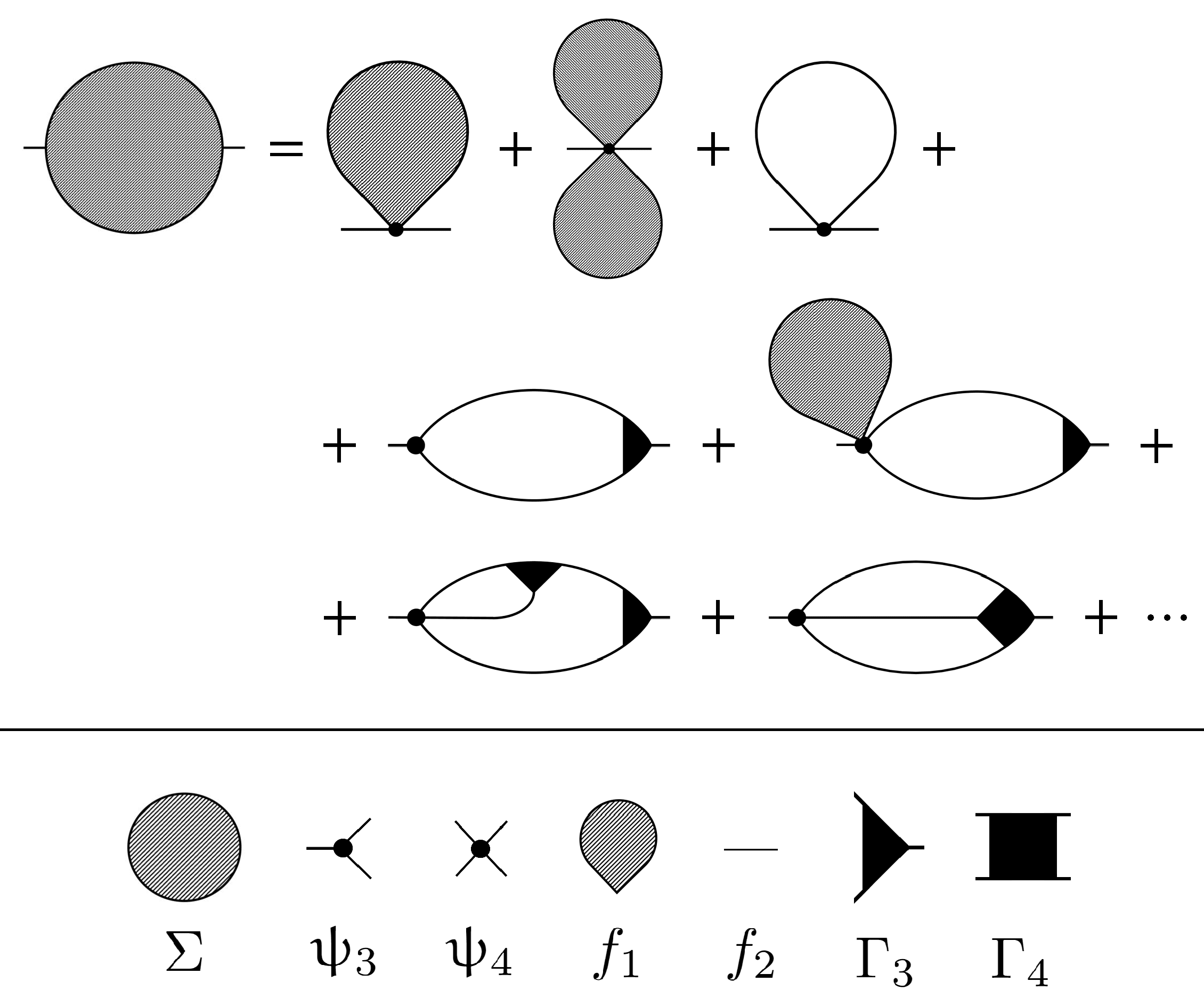}
\caption{\textbf{Diagrammatic interpretation of the phonon self-energy expansion in Eq. \eqref{self_energy_with_vertices_no_indices}.} We clearly see that the first two terms proportional to the two-point correlation function ($f_{2}$) are represented by the “loop” ($\uppsi_{4}f_{2}$) and “bubble” ($\uppsi_{3}f_{2}\Gamma_{3}f_{2}$) diagrams consistently with those of, e.g., Refs. \cite{maradudin1962scattering,paulatto2015first}.}
\label{fig:self_energy_diagrams_appendix2}
\end{figure}

\begin{figure}[!htb]
\centering
\includegraphics[width=0.6\textwidth]{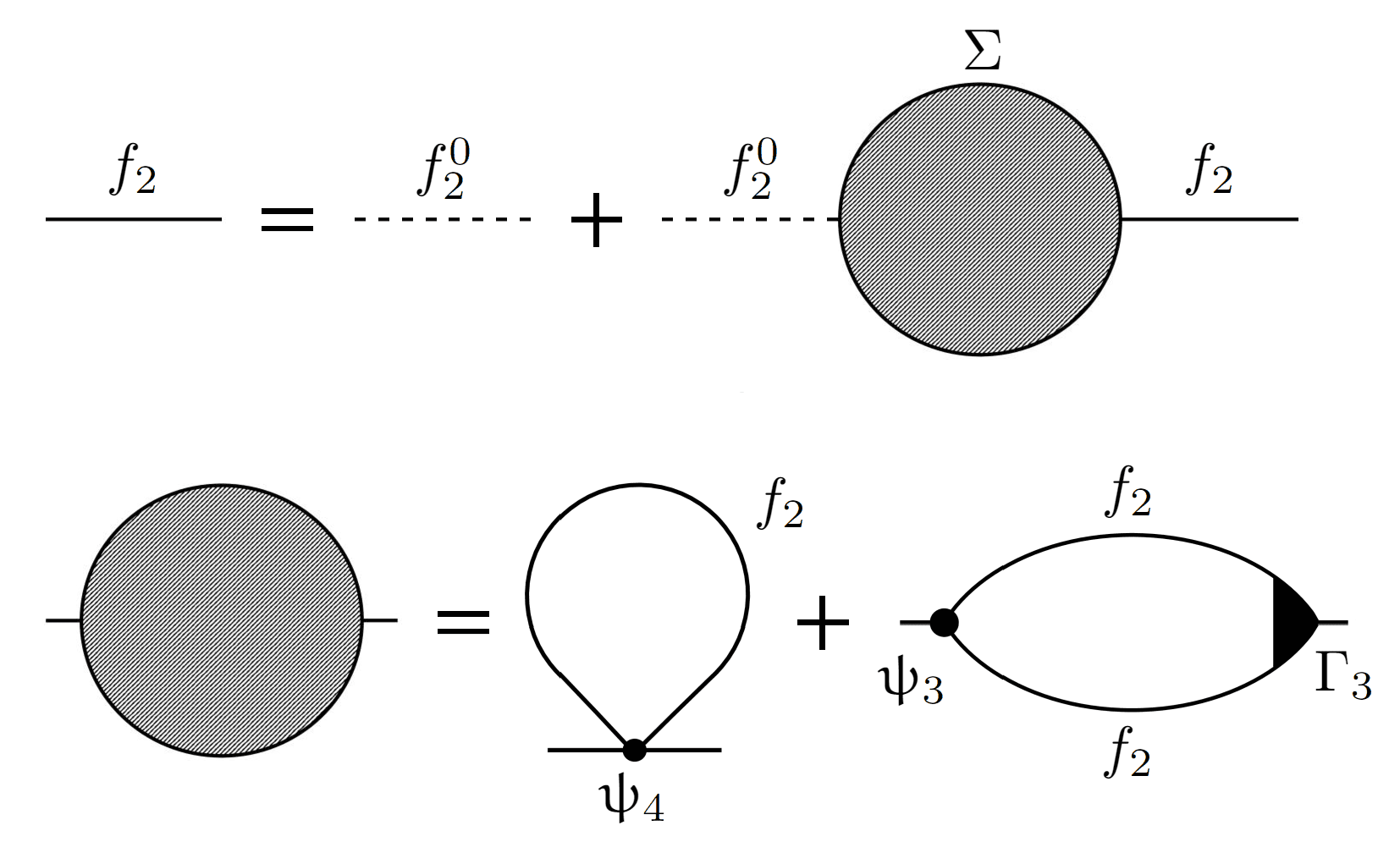}
\caption{\textbf{Diagrammatic representation of the phonon Dyson equation and self-energy.}
(Upper row) Feynman-diagram interpretation of the phonon Dyson equation given in Eq. \eqref{Dyson_eq_f2_final}, illustrating the renormalization of the phonon propagator.
(Lower row) Diagrammatic representation of the phonon self-energy with the leading bubble and loop diagrams, corresponding to Eq. \eqref{bubble_loop_self_energy_final}.}
\label{fig:self_energy_diagrams_appendix1}
\end{figure}

\begin{figure}[!htb]
\centering
\includegraphics[width=0.75\textwidth]{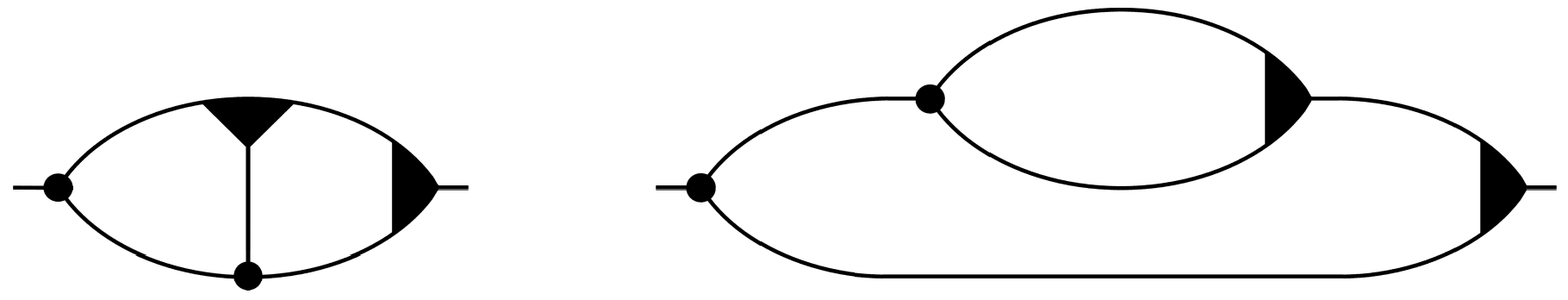}
\caption{\textbf{Comparison between vertex-correction and self-energy-insertion contributions to the three-phonon bubble self-energy involving cubic interactions.}
(Left) Self-energy diagram given by Eq. \eqref{sigma_vertex_correction_bubble_V3V3} obtained by accounting for vertex corrections to the three-phonon bubble self-energy generated by cubic interaction terms.
(Right) Self-energy diagram with a three-phonon bubble self-energy insertion generated through Dyson dressing of an internal propagator. Both diagrams contribute at $\mathcal{O}(\lambda^4)$ when all vertices are bare, see main text.}
\label{fig:vertex_correction_bubble_V3V3}
\end{figure}

In order to bring out the vertex contributions to the interaction, we need to manipulate the functional derivatives $\delta f_{2}/\delta f_{1}$ and $\delta^{2}f_{2}/\delta f_{1}^{2}$ given in the final epression of the self-energy \eqref{self_energy_before_vertices}. First, we can rewrite $\delta f_{2}/\delta f_{1}$ by defining a reciprocal correlation $f^{-1}_{2,[\xi]}$ as
\begin{equation} \label{f_2^-1_def}
f_{2,[\xi]}(1,-1)f^{-1}_{2,[\xi]}(-1,2)=\delta(1,-2).
\end{equation}
Differentiating the product in Eq. \eqref{f_2^-1_def} with respect to $f_{1,[\xi]}(3)$ and multiplying by $f_{2,[\xi]}(2,2')$, we arrive at
\begin{equation}
\begin{split}
&f_{2,[\xi]}(2,2')\frac{\delta}{\delta f_{1,[\xi]}(3)}\left[f_{2,[\xi]}(1,-1)f^{-1}_{2,[\xi]}(-1,2)\right]=\\
=&f_{2,[\xi]}(2,2')\left[\frac{\delta f_{2,[\xi]}(1,-1)}{\delta f_{1,[\xi]}(3)}f^{-1}_{2,[\xi]}(-1,2)+f_{2,[\xi]}(1,-1)\frac{\delta f^{-1}_{2,[\xi]}(-1,2)}{\delta f_{1,[\xi]}(3)}\right]=\\
=&f_{2,[\xi]}(2,2')\frac{\delta}{\delta f_{1,[\xi]}(3)}\left[\delta(1,-2)\right].
\end{split}
\end{equation}
So we are left with
\begin{equation}
\begin{split}
&f_{2,[\xi]}(2,2')\frac{\delta f_{2,[\xi]}(1,-1)}{\delta f_{1,[\xi]}(3)}f^{-1}_{2,[\xi]}(-1,2)+f_{2,[\xi]}(2,2')f_{2,[\xi]}(1,-1)\frac{\delta f^{-1}_{2,[\xi]}(-1,2)}{\delta f_{1,[\xi]}(3)}=\\
=&f_{2,[\xi]}(2,2')\frac{\delta}{\delta f_{1,[\xi]}(3)}\left[\delta(1,-2)\right].
\end{split}
\end{equation}
Finally having the Dirac delta on the RHS we can write the first needed quantity as
\begin{equation} \label{eq_Gamma3}
\frac{\delta f_{2,[\xi]}(1,2)}{\delta f_{1,[\xi]}(3)}=f_{2}(1,-1)\Gamma_{3}(-1,-2,3)f_{2}(-2,2),
\end{equation}
where
\begin{equation} \label{Gamma3}
\Gamma_{3}(-1,-2,3)=-\frac{\delta f^{-1}_{2,[\xi]}(-1,-2)}{\delta f_{1,[\xi]}(3)}
\end{equation}
is the third-order vertex function. \\
Now we can deal with $\delta^{2}f_{2}/\delta f_{1}^{2}$. Similarly to the procedure above, we can write the first derivative in the square brackets of Eq. \eqref{differentiating_f2_2} as follows:
\begin{equation}
\resizebox{\textwidth}{!}{$
\begin{split}
&\frac{\delta^{2}f_{2,[\xi]}(1,2)}{\delta f_{1,[\xi]}(3)\delta f_{1,[\xi]}(4)}=\\
=&\frac{\delta}{\delta f_{1,[\xi]}(4)}\left[\frac{\delta f_{2,[\xi]}(1,2)}{\delta f_{1,[\xi]}(3)}\right]=\frac{\delta}{\delta f_{1,[\xi]}(4)}\Big[f_{2,[\xi]}(1,-1)\Gamma_{3}(-1,-2,3)f_{2,[\xi]}(-2,2)\Big]=\\
=&\,\frac{\delta f_{2,[\xi]}(1,-1)}{\delta f_{1,[\xi]}(4)}\Big[\Gamma_{3}(-1,-2,3)f_{2,[\xi]}(-2,2)\Big]+f_{2,[\xi]}(1,-1)\frac{\delta}{\delta f_{1,[\xi]}(4)}\Big[\Gamma_{3}(-1,-2,3)f_{2,[\xi]}(-2,2)\Big]=\\
=&\,\frac{\delta f_{2,[\xi]}(1,-1)}{\delta f_{1,[\xi]}(4)}\Big[\Gamma_{3}(-1,-2,3)f_{2,[\xi]}(-2,2)\Big]+\\
&+f_{2,[\xi]}(1,-1)\left[\frac{\delta\Gamma_{3}(-1,-2,3)}{\delta f_{1,[\xi]}(4)}f_{2,[\xi]}(-2,2)+\Gamma_{3}(-1,-2,3)\frac{\delta f_{2,[\xi]}(-2,2)}{\delta f_{1,[\xi]}(4)}\right].
\end{split}$}
\end{equation}
Now we can substitute Eq. \eqref{eq_Gamma3} for the terms $\delta f_{2}/\delta f_{1}$ and Eq. \eqref{Gamma3} for $\Gamma_{3}$, getting
\begin{equation} \label{eq_Gamma3_Gamma4}
\begin{split}
\frac{\delta^{2}f_{2,[\xi]}(1,2)}{\delta f_{1,[\xi]}(3)\delta f_{1,[\xi]}(4)}=&\,f_{2,[\xi]}(1,-1)\Gamma_{3}(-1,1,4)f_{2,[\xi]}(1,-1)\Gamma_{3}(-1,-2,3)f_{2,[\xi]}(-2,2)+\\
&+f_{2,[\xi]}(1,-1)\Gamma_{4}(-1,-2,3,4)f_{2,[\xi]}(-2,2)+\\
&+f_{2,[\xi]}(1,-1)\Gamma_{3}(-1,-2,3)f_{2,[\xi]}(-2,2)\Gamma_{3}(2,-2,4)f_{2,[\xi]}(-2,2),
\end{split}
\end{equation}
where 
\begin{equation}
\Gamma_{4}(-1,-2,3,4)=\frac{\delta\Gamma_{3}(-1,-2,3)}{\delta f_{1,[\xi]}(4)}=-\frac{\delta^{2}f^{-1}_{2,[\xi]}(-1,-2)}{\delta f_{1,[\xi]}(3)\delta f_{1,[\xi]}(4)},
\end{equation}
is the fourth-order vertex function. So, we have finally obtained both the expressions of the two quantities $\delta f_{2}/\delta f_{1}$ and $\delta^{2}f_{2}/\delta f_{1}^{2}$ of Eq. \eqref{self_energy_before_vertices} in terms of the vertices $\Gamma_{3}$ and $\Gamma_{4}$. Now we can insert Eqs. \eqref{eq_Gamma3} and \eqref{eq_Gamma3_Gamma4} into Eq. \eqref{self_energy_before_vertices}, obtaining
\begin{equation}
\resizebox{\textwidth}{!}{$
\begin{split}
&\Sigma(1,2)=\\
=&\,\,6\uppsi_{3}(-1,2,3)f_{1,[\xi]}(3)+12\uppsi_{4}(-1,2,3,4)f_{1,[\xi]}(3)f_{1,[\xi]}(4)+12\uppsi_{4}(-1,2,3,4)f_{2,[\xi]}(3,4)+\\
&+3\uppsi_{3}(-1,2,3)f_{2,[\xi]}(3,-3)\Gamma_{3}(-3,-4,2')f_{2,[\xi]}(-4,4)+\\
&+12\uppsi_{4}(-1,2,3,4)f_{1,[\xi]}(2')f_{2,[\xi]}(3,-3)\Gamma_{3}(-3,-4,2')f_{2,[\xi]}(-4,4)+\\
&+4\uppsi_{4}(-1,2,3,4)f_{2,[\xi]}(2'',-2'')\Gamma_{3}(-2'',2''',-4)f_{2,[\xi]}(-4,4)f_{2,[\xi]}(3,-3)\Gamma_{3}(-3,-4,2')f_{2,[\xi]}(-4,4)+\\
&+4\uppsi_{4}(-1,2,3,4)f_{2,[\xi]}(2'',-2'')\Gamma_{3}(-2'',2'',2''')f_{2,[\xi]}(2'',-2'')\Gamma_{3}(-2'',-4,2')f_{2,[\xi]}(-4,4)f_{2,[\xi]}(2''',2)+\\
&+4\uppsi_{4}(-1,2,3,4)f_{2,[\xi]}(2'',-2'')\Gamma_{4}(-2'',-4,2',2''')f_{2,[\xi]}(-4,4)f_{2,[\xi]}(2''',2)+\\
&+4\uppsi_{4}(-1,2,3,4)f_{2,[\xi]}(2'',-2'')\Gamma_{3}(-2'',-4,2')f_{2,[\xi]}(-4,4)\Gamma_{3}(4,-4,2''')f_{2,[\xi]}(-4,4)f_{2,[\xi]}(2''',2).
\end{split}$}
\end{equation}
Adding together the terms proportional to $\Gamma_{3}\Gamma_{3}$ we finally get
\begin{equation} \label{self_energy_with_vertices}
\resizebox{\textwidth}{!}{$
\begin{split}
&\Sigma(1,2)=\\
=&\,\,6\uppsi_{3}(-1,2,3)f_{1,[\xi]}(3)+12\uppsi_{4}(-1,2,3,4)f_{1,[\xi]}(3)f_{1,[\xi]}(4)+12\uppsi_{4}(-1,2,3,4)f_{2,[\xi]}(3,4)+\\
&+3\uppsi_{3}(-1,2,3)f_{2,[\xi]}(3,-3)\Gamma_{3}(-3,-4,2')f_{2,[\xi]}(-4,4)+\\
&+12\uppsi_{4}(-1,2,3,4)f_{1,[\xi]}(2')f_{2,[\xi]}(3,-3)\Gamma_{3}(-3,-4,2')f_{2,[\xi]}(-4,4)+\\
&+12\uppsi_{4}(-1,2,3,4)f_{2,[\xi]}(2'',-2'')\Gamma_{3}(-2'',2''',-4)f_{2,[\xi]}(-4,4)f_{2,[\xi]}(3,-3)\Gamma_{3}(-3,-4,2')f_{2,[\xi]}(-4,4)+\\
&+4\uppsi_{4}(-1,2,3,4)f_{2,[\xi]}(2'',-2'')\Gamma_{4}(-2'',-4,2',2''')f_{2,[\xi]}(-4,4)f_{2,[\xi]}(2''',2).
\end{split}$}
\end{equation}
To facilitate the interpretation of the result in Eq. \eqref{self_energy_with_vertices}, we rewrite it omitting the indices,
\begin{equation} \label{self_energy_with_vertices_no_indices}
\begin{split}
\Sigma(1,2)=&\,\,6\uppsi_{3}f_{1}+12\uppsi_{4}f_{1}f_{1}+12\uppsi_{4}f_{2}+3\uppsi_{3}f_{2}\Gamma_{3}f_{2}+12\uppsi_{4}f_{1}f_{2}\Gamma_{3}f_{2}+\\
&+12\uppsi_{4}f_{2}\Gamma_{3}f_{2}f_{2}\Gamma_{3}f_{2}+4\uppsi_{4}f_{2}\Gamma_{4}f_{2}f_{2}.
\end{split}
\end{equation}
and depicting its diagrammatic representation in Fig. \ref{fig:self_energy_diagrams_appendix2}. Note that, for teasing out the connection, the order of the terms in Eq. \eqref{self_energy_with_vertices_no_indices} is consistent with that of the related diagrams in Fig. \ref{fig:self_energy_diagrams_appendix2}.
In order to assess the relative importance of the different contributions in Eq. \eqref{self_energy_with_vertices_no_indices}, it is useful to introduce a simple power-counting scheme following the arguments of Maradudin and Fein \cite{maradudin1962scattering} or Cowley \cite{cowley1963lattice}, which in turn build on the seminal work of Van Hove \cite{vanhove1959interactions,van1961problems}. Starting from the Taylor expansion of the interatomic potential around equilibrium, the $n$-th order term scales as
\begin{equation} \label{nth_order_potential}
V^{(n)} \sim \frac{V_{\mathrm{typ}}}{r_0^n} \, u^n,
\end{equation}
where $r_0$ is the nearest-neighbor distance, $u$ is a typical displacement, and $V_{\mathrm{typ}}$ is a characteristic energy scale of the potential. In a crystalline solid, the natural energy scale is set by the harmonic phonon energy. Factoring it out yields
\begin{equation}
V^{(n)} \sim V_{\mathrm{typ}}\left(\frac{u}{r_0}\right)^2 \left(\frac{u}{r_0}\right)^{n-2}.
\end{equation}
Since the quadratic fluctuations set the scale of the motion, the factor $V_{\mathrm{typ}}(u/r_0)^2$ is of order $\hbar\omega$ \cite{maradudin1962scattering,cowley1963lattice,van1961problems,vanhove1959interactions}, and the expectation value of the $n$-th order contribution to the potential energy scales as
\begin{equation}
\langle V^{(n)} \rangle \sim \hbar \omega \left(\frac{u}{r_0}\right)^{n-2}.
\end{equation}
Defining the small parameter $\lambda \equiv u/r_0 \ll 1$, the anharmonic Hamiltonian can be schematically written as
\begin{equation}
H_{\mathrm{anh}} \sim \lambda v_{3} + \lambda^{2} v_{4} + \cdots,
\end{equation}
where now the coefficients $v_{n}$ simply denote interaction operators of order $\hbar\omega$ and act as factors in the expansion in powers of $\lambda$. This immediately shows that the four-phonon interaction matrix element is of order $\lambda^{2}$, i.e.\ of second order with respect to the three-phonon interaction matrix element, and so on. Accordingly, the interaction vertices scale as
\begin{equation}
\uppsi_{3} \sim \mathcal{O}(\lambda), 
\qquad 
\uppsi_{4} \sim \mathcal{O}(\lambda^2),
\end{equation}
while the correlation functions $f_{2}$ are of order $\mathcal{O}(1)$, since they are determined by the harmonic dynamics (i.e. they correspond to expectation values evaluated with respect to the harmonic Hamiltonian). Within this scaling, the various terms entering the self-energy can be systematically organized according to their order in $\lambda$. In particular, the leading $\mathcal{O}(\lambda^2)$ contributions are given by the loop diagram, $\uppsi_{4} f_{2}$, and the bubble diagram, $\uppsi_{3} f_{2} \Gamma_{3} f_{2}$. Higher-order terms in Eq. \eqref{self_energy_with_vertices_no_indices}, such as $\uppsi_{4} f_{2} \Gamma_{3} f_{2} f_{2} \Gamma_{3} f_{2}$ and $\uppsi_{4} f_{2} \Gamma_{4} f_{2} f_{2}$, involve additional interaction vertices and scale as $\mathcal{O}(\lambda^4)$. These contributions correspond to more complex multi-phonon processes and are typically subleading in the weakly anharmonic regime. It is however important to emphasize that genuine four-phonon scattering processes—i.e. processes in which four phonons can be identified as colliding quasiparticles (decay or coalescence of one phonon into three, or scattering between two phonon pairs \cite{feng2016quantum,feng2017four,feng2018four})—originate from the diagram $\uppsi_{4} f_{2} \Gamma_{4} f_{2} f_{2}$, once it is combined with the additional $f_{2}$ entering the leading transport equation. This is not the only $\mathcal{O}(\lambda^4)$ diagram involving four-phonon interactions that contributes to the imaginary part of the self-energy (besides the $\mathcal{O}(\lambda^2)$ loop diagram $\uppsi_{4} f_{2}$—which contributes only to the real part and cannot be interpreted in terms of scattering of quasiparticles). One also has the higher-order contribution $\uppsi_{4} f_{2} \Gamma_{3} f_{2} f_{2} \Gamma_{3} f_{2}$. However, this term has a more complex structure: it involves one quartic vertex, two cubic vertices, and multiple internal propagators, and therefore does not correspond to a single elementary scattering event. Rather, this diagram should be interpreted as a composite process involving intermediate virtual states, i.e. a multi-step process of the form
\[
\nu \;\to\; (\text{intermediate state}) \;\to\; (\text{final state}).
\]
When extracting the imaginary part (e.g.\ via diagrammatic cutting rules), different cuts of the diagram correspond to different physical processes. As a result, one does not obtain a single well-defined scattering channel, but rather a sum of contributions that can include effective higher-order scattering processes, corrections to four-phonon scattering, and interference between different channels.\\
At this stage, it remains to explicitly determine the third- ($\Gamma_{3}$) and fourth- ($\Gamma_{4}$) order vertex functions. In order to do so, we multiply the Dyson equation \eqref{dyson_eq_for_f2} by $f_{2}^{-1}(2,2')$ and operate from the left with $\hat{L}_{1}$:
\begin{equation}
\hat{L}_{1}\delta(1,-2')=-i2\omega_{\boldsymbol{q}_{1}s_{1}}f_{2,[\xi]}^{-1}(-1,2')-2\omega_{\boldsymbol{q}_{1}s_{1}}\Sigma(-1,2').
\end{equation}
By differentiating with respect to $f_{1}$, we then obtain the integral equations
\begin{equation} \label{Gamma_3_derivative_of_sigma}
\Gamma_{3}(1,2,3)=-i\frac{\delta\Sigma(1,2)}{\delta f_{1,[\xi]}(3)}
\end{equation}
and
\begin{equation} \label{Gamma_4_derivative_of_sigma}
\Gamma_{4}(1,2,3,4)=-i\frac{\delta^{2}\Sigma(1,2)}{\delta f_{1,[\xi]}(3)\delta f_{1,[\xi]}(4)}.
\end{equation}
Using Eq. \eqref{self_energy_with_vertices_no_indices}, then Eqs. \eqref{Gamma_3_derivative_of_sigma} and \eqref{Gamma_4_derivative_of_sigma} become
\begin{equation} \label{final_Gamma_3}
\Gamma_{3}(1,2,3)=-i\Big[6\uppsi_{3}(1,2,3)+12\uppsi_{4}(1,2,3,4)f_{1,[\xi]}(4)\Big]+\cdots
\end{equation}
and
\begin{equation} \label{final_Gamma_4}
\Gamma_{4}(1,2,3,4)=-i12\uppsi_{4}(1,2,3,4)+\cdots,
\end{equation}
where we have only highlighted the first terms of the integral equations because they do not depend on $\Gamma_{3}$ and $\Gamma_{4}$. This is to say that, after the truncation in Eqs. \eqref{final_Gamma_3} and \eqref{final_Gamma_4}, more complex terms depending on both $\Gamma_{3}$ and $\Gamma_{4}$ would appear. Taking into account cubic and quartic anharmonicity we have the Dyson equation for $f_{2}$ 
\begin{equation} \label{Dyson_eq_f2_final}
f_{2,[\xi]}=f^{0}_{2,[\xi]}(1,2)-if^{0}_{2,[\xi]}(1,1')\Sigma(1',2')f_{2,[\xi]}(2',2)
\end{equation}
and the self-energy given by the “loop” and “bubble” diagrams \cite{lazzeri2003anharmonic,rousseau2010giant,calandra2007anharmonic}
\begin{equation} \label{bubble_loop_self_energy_final}
\Sigma(1,2)=12\uppsi_{4}(-1,2,3,4)f_{2,[\xi]}(3,4)+3\uppsi_{3}(-1,2,3)f_{2,[\xi]}(3,-3)\Gamma_{3}(-3,-4,2')f_{2,[\xi]}(-4,4),
\end{equation}
or equivalently,
\begin{equation} \label{bubble_loop_self_energy_final_in_G}
\Sigma(1,2)=12\uppsi_{4}(-1,2,3,4)G(3,4)+3\uppsi_{3}(-1,2,3)G(3,-3)\Gamma_{3}(-3,-4,2')G(-4,4).
\end{equation}
Finally, we can explicitly write Eq. \eqref{bubble_loop_self_energy_final} in its extended form: 
\begin{equation} \label{leading_self_energy_appendix}
\begin{split}
&\Sigma_{\nu_{1}\nu_{2}}(t_{1},t_{2})=\\
=&\sum_{\nu_{3}\nu_{4}}\Big[12\uppsi_{4,-\nu_{1}\nu_{2}\nu_{3}\nu_{4}}G_{\nu_{3}\nu_{4}}(t_{3},t_{4})+\\
&\hspace{0.75cm}+3\uppsi_{3,-\nu_{1}\nu_{2}\nu_{3}}(t_{1},t_{2},t_{3})G_{\nu_{3}-\nu_{3}}(t_{3},-t_{3})\Gamma_{3,\nu_{2}-\nu_{3}-\nu_{4}}(t_{2},t_{3},t_{4})G_{-\nu_{4}\nu_{4}}(-t_{4},t_{4})\Big],
\end{split}
\end{equation}
where, as usual, $\nu\equiv\boldsymbol{q}s$ and notations \eqref{interaction_with_delta_on_time} and \eqref{delta_kroneker_self_energy} hold. One should easily note that the time-ordered one-phonon Green function is obtained as decomposed into a correlated and an uncorrelated part (see Eq. \eqref{cluster_expansion_definitions}):
\begin{equation} \label{G_is_f_2_eq1}
G_{n}(1,2)=-i\langle\hat{T}\left[A(1)A(2)\right]\rangle=f_{1}(1)f_{1}(2)+f_{2}(1,2)
\end{equation}
where $f_{1}$ is time-independent and represents static strains \cite{klein1969derivation}, which we do not consider further
\begin{equation} \label{G_is_f_2_eq2}
f_{1}(1)=-i\langle A(1)\rangle=0.
\end{equation}
In this case, from Eq. \eqref{cluster_expansion_definitions}, it is evident how $G_{2}(1,2)=f_{2}(1,2)$ and the diagrammatic interpretation for the Green's function follows. Moreover, the diagrammatic representation of Eqs. \eqref{Dyson_eq_f2_final} and \eqref{bubble_loop_self_energy_final} are given in panel a and b of Fig. \ref{fig:self_energy_diagrams_appendix1}, respectively.
The vertex part $\Gamma_{3}$, to lowest order in the anharmonicity, is given by 
\begin{equation} \label{Gamma_3_0}
\Gamma_{3}^{0}=-i6\uppsi_{3}(1,2,3)
\end{equation}
When dealing with vertex corrections, we are mainly interested in the contributions to the self-energy which correspond to so called ladder diagrams \cite{sham1967equilibrium,sham1967temperature,kadanoff2018quantum,danielewicz1984quantum1,danielewicz1984quantum2,botermans1990quantum}. These stem from the second term on the RHS of Eq. \eqref{bubble_loop_self_energy_final} where $\Gamma_{3}$ is taken in a better approximation than that in Eq. \eqref{Gamma_3_0}. The equation for the vertex part $\Gamma_{3}$ is obtained by keeping further leading terms in Eq. \eqref{final_Gamma_3} \cite{klein1969derivation} (omitting terms proportional to $f_{1}$):
\begin{equation} \label{gamma_3_integral_eq_appendix}
\begin{split}
\Gamma_{3}(1,2,3)=&\,\,\Gamma_{3}^{0}(1,2,3)
-i12\uppsi_{4}(1,2,4,5)G(4,4')G(5,5')\Gamma_{3}(4',5',4)+\\
&+\Gamma_{3}^{0}(1,4,5)G(4,4')G(5,5')G(6,6')\Gamma_{3}(4',6,2)\Gamma_{3}(5',6',3)+\\
&+(-i12)^2\uppsi_{4}(1,2,4,5)G(4,4')G(5,5')\uppsi_{4}(4',5',6,7)G(6,6')G(7,7')\Gamma_{3}(6',7',3)-\\
&-i12\uppsi_{4}(1,4,5,6)G(4,4')G(5,5')G(6,6')G(7,7')G(8,8')\times\\
&\,\,\,\,\,\times\Gamma_{3}(4',7,2)\Gamma_{3}(5',8,3)\Gamma_{3}(6',7',8')+\\
&+\Gamma_{3}^{0}(1,4,5)
G(4,4')G(5,5')
G(6,6')G(7,7')
G(8,8')G(9,9')\times\\
&\,\,\,\,\,\times
\Gamma_{3}(4',6,7)
\Gamma_{3}(5',8,9)
\Gamma_{3}(6',8',2)
\Gamma_{3}(7',9',3)+\cdots
\end{split}
\end{equation}
If all vertex functions appearing on the right-hand side are replaced by their bare counterparts, Eq. \eqref{gamma_3_integral_eq_appendix} corresponds to an expansion up to order $\mathcal{O}(\lambda^{5})$. Restricting the expansion to the first two lines reproduces Eq. \eqref{gamma_3_integral_eq} of the main text, whose corresponding diagrams are shown in Fig. \ref{fig:vertex_diagrams}. It is important to note that every diagram contributing to the self-consistent equation for the three-phonon vertex $\Gamma_{3}$ satisfies the topological constraint
\begin{equation} \label{topological_relation_gamma3}
3N_{3}+4N_{4}=2I+E,
\end{equation}
where $N_{3}$ and $N_{4}$ denote the number of three-phonon (with three legs) and four-phonon (with four legs) interaction vertices, respectively, while $I$ and $E$ are the numbers of internal and external lines (or Green's functions). Since the vertex function $\Gamma_{3}$ possesses three external legs, in the present case one has $E=3$. Eq. \eqref{topological_relation_gamma3} can be verified graphically from the diagrams shown in Fig. \ref{fig:vertex_diagrams} of the main text. This relation also explains why the expansion of $\Gamma_{3}$ contains only odd powers of the anharmonicity parameter $\lambda$. Indeed, the quantity $2I$ is always even, and so the parity of the right-hand side of Eq. \eqref{topological_relation_gamma3} is determined entirely by the additional term $E=3$, which is odd. Consequently, the left-hand side must also be odd. Now, the contribution $4N_{4}$ is always even, therefore, the parity of the left-hand side is controlled exclusively by the term $3N_{3}$. Since the overall sum must be odd, $3N_{3}$ must itself be odd, which implies that $N_{3}$ must necessarily be odd as well. As a consequence, every diagram contributing to the three-point vertex function $\Gamma_{3}$ contains an odd number of cubic interaction vertices.\\
We can provide a diagrammatic interpretation of the first two contributions to the vertex corrections of the three-phonon bubble self-energy.
\begin{figure}[!htb]
\centering
\includegraphics[width=0.75\textwidth]{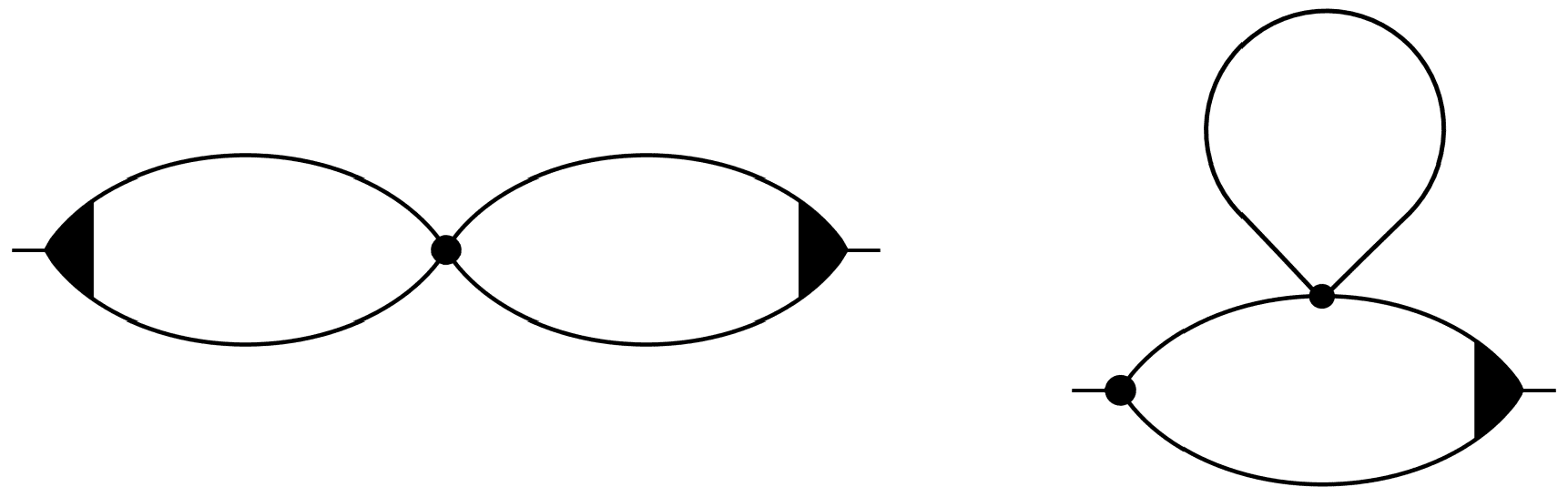}
\caption{\textbf{Comparison between vertex-correction and self-energy-insertion contributions to the three-phonon bubble self-energy involving quartic interactions.}
(Left) Self-energy diagram given by Eq. \eqref{sigma_vertex_correction_bubble_V4} obtained by accounting for vertex correction to the three-phonon bubble self-energy generated by the quartic interaction term. (Right) Self-energy diagram given by Eq. \eqref{loop_self_energy_insertion_correction} with a quartic loop self-energy insertion obtained through Dyson dressing of an internal propagator. Both diagrams contribute at $\mathcal{O}(\lambda^4)$ when all vertices are bare, see main text.}
\label{fig:vertex_correction_bubble_V4}
\end{figure}
These are the self-energy diagrams obtained by substituting the second and third terms on the RHS of Eq. \eqref{gamma_3_integral_eq_appendix} into the second term on the RHS of Eq. \eqref{bubble_loop_self_energy_final_in_G}. The first contribution that we obtain is
\begin{equation} \label{sigma_vertex_correction_bubble_V4}
-i36\uppsi_{3}(-1,2,3)G(3,-3)
\uppsi_{4}(-3,-4,4',5')G(4',4'')G(5',5'')
\Gamma_{3}(4'',5'',4')
G(-4,4).
\end{equation}
One can clearly see that this contribution is described by the diagram in the left panel of Fig. \ref{fig:vertex_correction_bubble_V4} by taking into account the arguments of the vertices and Green's functions. In a more compact notation, we define the three vertices
\begin{equation}
A=\psi_{3}(-1,2,3),\quad B=\psi_{4}(-3,-4,4',5'),\quad C=\Gamma_{3}(4'',5'',4').
\end{equation}
The connectivity of the diagram is determined by the Green's functions ($A$ and $C$ are vertices connected to external lines):
\begin{equation}
\begin{split}
&G(3,-3):\qquad A\leftrightarrow B,
\\
&G(4',4''):\qquad B\leftrightarrow C,
\\
&G(5',5''):\qquad B\leftrightarrow C,
\\
&G(-4,4):\qquad B\leftrightarrow A.
\end{split}
\end{equation}
Therefore, the corresponding diagram contains the edges
\begin{equation}
A-B,\qquad A-B,\qquad B-C,\qquad B-C.
\end{equation}
In this way, one can easily understand that the diagram consistent with these connections between the vertices is precisely the one shown in the left panel of Fig. \ref{fig:vertex_correction_bubble_V4}. In the right panel of Fig. \ref{fig:vertex_correction_bubble_V4} we show the bubble diagram with a quartic loop self-energy insertion, which has nothing to do with the one in the left panel obtained through vertex corrections. We decided to show it because globally it contains the same ingredients as $\propto\uppsi_{3}GG\uppsi_{4}GG\Gamma_{3}$, but one Green's function derives from a correlator evaluated at equal times and therefore represents a static correction corresponding to the standard thermal average of the loop diagram. Indeed, this diagram is represented by the edges
\begin{equation}
A-B,\qquad B-B,\qquad B-C,\qquad A-C,
\end{equation}
and is therefore inconsistent with Eq. \eqref{sigma_vertex_correction_bubble_V4}. This is to emphasize that one should carefully keep track of the time (or frequency) arguments of the vertex corrections to the three-phonon bubble, since they represent dynamical corrections. This diagram is instead obtained through self-energy insertion. In particular, this contribution is generated starting from the ordinary three-phonon bubble self-energy $\Sigma_{\mathrm{bubble}}=3\uppsi_{3}GG\Gamma_{3}$ and dressing one of its internal propagators through the Dyson equation $G=G_{0}+G_{0}\Sigma G$. By inserting the quartic loop self-energy $\Sigma_{\mathrm{loop}}=12\uppsi_{4}G$ into one propagator line of the bubble, one obtains
\begin{equation} \label{loop_self_energy_insertion_correction}
\Sigma^{(\mathrm{loop\ ins.})}=36\uppsi_{3}\,G\,\left(G\,\uppsi_{4}G(t,t)\,G\right)\Gamma_{3}.
\end{equation}
In this case, the Green's function of the quartic loop is evaluated at equal times and therefore represents a static Hartree-like correction corresponding to the thermal average of the loop diagram.
\\
Similarly, in the left panel of Fig. \ref{fig:vertex_correction_bubble_V3V3} we show the diagram corresponding to the second contribution:
\begin{equation} \label{sigma_vertex_correction_bubble_V3V3}
+3\uppsi_{3}(-1,2,3)G(3,-3)
\uppsi_{3}(-3,4',5') G(4',4'')G(5',5'')G(6',6'')\Gamma_{3}(4'',6',-4)
\Gamma_{3}(5'',6'',2')
G(-4,4).
\end{equation}
In this case we define the four cubic vertices
\begin{equation}
A=\psi_{3}(-1,2,3),\quad B=\psi_{3}(-3,4',5'),\quad C=\Gamma_{3}(4'',6',-4),\quad D=\Gamma_{3}(5'',6'',2').
\end{equation}
The connectivity of the diagram is then determined by the Green's functions:
\begin{equation}
\begin{split}
&G(3,-3):\qquad A\leftrightarrow B,
\\
&G(4',4''):\qquad B\leftrightarrow C,
\\
&G(5',5''):\qquad B\leftrightarrow D,
\\
&G(6',6''):\qquad C\leftrightarrow D,
\\
&G(-4,4):\qquad C\leftrightarrow A.
\end{split}
\end{equation}
Therefore, the corresponding diagram contains the edges
\begin{equation}
A-B,\qquad
B-C,\qquad
B-D,\qquad
C-D,\qquad
C-A,
\end{equation}
such that this topology is the one described in the left panel of Fig. \ref{fig:vertex_correction_bubble_V3V3}.
On the contrary, the diagram shown in the right panel of Fig. \ref{fig:vertex_correction_bubble_V3V3}, although composed of the same quantities (five Green's functions and four three-phonon vertices), has a topology inconsistent with Eq. \eqref{sigma_vertex_correction_bubble_V3V3} since its edges are
\begin{equation}
A-B,\qquad
B-C,\qquad
B-C,\qquad
C-D,\qquad
A-D.
\end{equation}
This diagram is instead obtained by inserting the ordinary bubble self-energy into one of the internal propagators of the three-phonon bubble. Starting from $\Sigma_{\mathrm{bubble}}=3\uppsi_{3}GG\Gamma_{3}$, one uses the Dyson expansion of an internal Green's function, $G=G_{0}+G_{0}\Sigma G$, and substitutes another bubble self-energy contribution, yielding
\begin{equation}
\begin{split}
\Sigma^{(\mathrm{bubble\ ins.})}=9\uppsi_{3}\,G\,\left(G\,\uppsi_{3}GG\Gamma_{3}\,G\right)\Gamma_{3}.
\end{split}
\end{equation}
In this case, differently from Eq. \eqref{loop_self_energy_insertion_correction}, we have not explicitly indicated the time dependence because the contribution is fully dynamical, i.e. all Green's functions are evaluated at different times and therefore represent propagating phonon correlations rather than static equal-time averages. In Ref. \cite{balkanski1983anharmonic} it is argued that the diagrams in Fig. \ref{fig:vertex_correction_bubble_V3V3} represent higher-order contributions to the proper self-energy of the Raman-active LO mode in silicon (both containing four three-phonon vertices, and therefore globally $\mathcal{O}(\lambda^4)$), together with the first diagram in the third row of Fig. \ref{fig:self_energy_diagrams_appendix2} (also of order $\mathcal{O}(\lambda^4)$ since it is composed of two three-phonon vertices and one four-phonon vertex). The latter derives from the diagrammatic expansion of the self-energy, the one in the left panel of Fig. \ref{fig:vertex_correction_bubble_V3V3} derives from the vertex corrections to the three-phonon bubble, whereas the one in the right panel derives from self-energy insertion. In particular, this is obtained by inserting the ordinary bubble self-energy itself, $\Sigma_{\mathrm{bubble}}=3\uppsi_{3}GG\Gamma_{3}$, into one propagator of the bubble.
\\
A similar approach can be followed to derive the integral equation for the vertex part $\Gamma_{4}$ (see Eq. 5.12 of Ref. \cite{wehner1967phonon}), where one can easily see that the lowest order of the $\Gamma_{4}$ vertex is simply:
\begin{equation} \label{Gamma_4_0}
\Gamma_{4}^{0}=-i12\uppsi_{4}(1,2,3,4)
\end{equation}
Importantly, as also noted in the main text, it is known from calculations of the self-energy \cite{semwal1972thermal,maradudin1962scattering,cowley1965anharmonic,cowley1966anharmonic} that the first term on the RHS of Eq. \eqref{bubble_loop_self_energy_final} (proportional to the four-phonon anharmonic potential) contributes to the same order of magnitude as the second one (proportional to the three-phonon anharmonic potential). So in principle four-phonon contributions should not be discarded; moreover if one goes beyond the 0th-order approximation given by $\Gamma_{3}^{0}$, we clearly see that taking four-phonon processes into account right from the beginning reveals to be important since those take part in the definition of the vertex part $\Gamma_{3}$ given in the second term on the RHS of Eq. \eqref{gamma_3_integral_eq_appendix} (or Fig. \ref{fig:vertex_diagrams}). At the present level of theory, we have already anticipated that frequency lineshifts are not taken into account given that we neglect the entire real part of the self-energy, $\text{Re}[\Sigma]=\Delta$, i.e. the contribution that affects also pole renormalizations in Eq. \eqref{z_pole_renormalization} and which is also responsible for lineshifts. This is consistent with what is shown in Ref. \cite{paulatto2015first}, where the three-phonon process described by the bubble diagram contributes to both linewidths and lineshifts, while the loop diagram has a two-phonon fourth-order vertex which contributes exclusively to lineshifts (the bubble contribution is the only self-energy term with an imaginary part). This has been explicitly discussed also in Ref. \cite{semwal1972thermal} (see Eqs. 32 and 33), where it is shown how the $\Delta$ contribution is given by a term proportional to $\uppsi_{3}$ and another term proportional to $\uppsi_{4}$ which contributes to the same order of magnitude. Here the entire $\Delta$ lineshifts contribution is neglected because a simple and better approximation is not easily achievable. \\
What can be demonstrated is that the frequency lineshifts could alleviate the issues arising from overdamped phonons in 2D materials \cite{bonini2012acoustic}. In fact, it can be shown that by including lineshifts effects, the final frequency of a phonon involved in a decay process where at least one out-of-plane (flexural) mode is present is no longer quadratic in the wave vector (which is the main reason why the quasiparticle picture breaks down and this phonon can no longer be defined as a well-behaved quasiparticle and becomes overdamped \cite{bonini2012acoustic}). Instead, it returns to being dominated by a linear contribution in the wave vector, similar to the other longitudinal and transverse acoustic modes \cite{aseginolaza2024bending}. Nevertheless, this result is closely related to the method employed in Ref. \cite{aseginolaza2024bending}, specifically the SSCHA approach \cite{SSCHA_0,SSCHA_1,SSCHA_2,SSCHA_3}, which allows for the renormalization of phonon frequencies, making them anharmonic by using fourth-order interactions and defining the concepts of physical and auxiliary frequencies. \\
Finally, the inclusion of 4th-order anharmonic interactions in the self-energy would lead to complications related to the definition of the heat flux beyond its harmonic formulation \cite{simoncelli2022wigner,caldarelli2022many,hardy1963energy}. In fact, to remain consistent with the perturbative order of the phonon Hamiltonian, the inclusion of 4th-order scattering processes would require adopting a third-order anharmonic heat-flux operator \cite{simoncelli2022wigner}, rather than the harmonic heat flux typically used when only third-order interactions are accounted for. A full derivation of this anharmonic heat flux is not yet available, and it remains unclear whether such a flux—combined with 4th-order anharmonic interactions—would yield additional contributions to phonon lifetimes that further reduce thermal conductivity, or whether it might instead lead to qualitatively different behavior.

\subsection{Tad-pole diagram}

The tad-pole diagram \cite{SSCHA_0,
calandra2007anharmonic,maradudin1962scattering,paulatto2015first} provides information about the atomic motion due to anharmonicity; at zero temperature, this is attributed to zero-point motion. Furthermore, this diagram is not relevant for a high-symmetry crystal because it only provides corrections at the Gamma point in the Brillouin zone \cite{lazzeri2003anharmonic}. The loop diagram depends on $\uppsi_{4}$ once, while the bubble depends on $\uppsi_{3}$ and $\Gamma_{3}$. If the vertex $\Gamma_{3}$ is not dressed, then the bubble depends on $\uppsi_{3}$ twice. Although the tad-pole also depends on $\uppsi_{3}$ twice, it is diagrammatically distinct from the bubble at zero-order in the vertex corrections. In the perturbative treatment above, we have assumed that the atoms oscillate around their harmonic positions, and under this assumption, the tad-pole is absent, as is the case in high-symmetry systems where the atomic positions are fixed by the symmetry group \cite{lazzeri2003anharmonic}.

\subsection{Phonon bubble self-energy in Wigner's mixed representation} \label{self_energy_wigner}

As reported in Ref. \cite{di2025broadening}, in order to write the bubble self-energy in Wigner's mixed representation we apply a Fourier transform to the self-energy in $\boldsymbol{q}\omega$ space, allowing it to be expressed in $\boldsymbol{q}t$ space (note that Eq. \eqref{bubble_self_energy_final_main} is equivalent to Eq. A23 in Ref. \cite{volz2020quantum} with $\omega_{2}=\omega-\omega_{1}$) \cite{di2025broadening}:
\begin{equation}
\begin{split}
\Sigma^{\lessgtr}_{s_{1}s_{2}}(\boldsymbol{q}_{1},t_{1};\boldsymbol{q}_{2},t_{2})=4i\pi\hbar\sum_{\substack{\nu_{1}'\nu_{2}'\\\nu_{1}''\nu_{2}''}}\mathcal{F}_{\nu_{1}-\nu_{1}'-\nu_{1}''}\mathcal{F}_{\nu_{2}-\nu_{2}'-\nu_{2}''}G^{\lessgtr}_{s_{1}'s_{2}'}(\boldsymbol{q}_{1}',t_{1};\boldsymbol{q}_{2}',t_{2})G^{\lessgtr}_{s_{1}''s_{2}''}(\boldsymbol{q}_{1}'',t_{1};\boldsymbol{q}_{2}'',t_{2}).
\end{split}
\end{equation}
Inverting the definition of Wigner's mixed representation \eqref{Wigner's_mixed_representation}, we can write
\begin{equation}
\begin{split}
G^{\lessgtr}_{s_{1}^{i}s_{2}^{i}}(\boldsymbol{q}_{1}^{i},t_{1};\boldsymbol{q}_{2}^{i},t_{2})&\equiv G_{s^{i}}^{\lessgtr}\left(\boldsymbol{q}^{i}+\frac{\boldsymbol{Q}^{i}}{2},t+\frac{\tau}{2};\boldsymbol{q}^{i}-\frac{\boldsymbol{Q}^{i}}{2},t-\frac{\tau}{2}\right)=\\
&=\iint\frac{d\omega_{1} d\boldsymbol{R}^{i}}{(2\pi)^{4}}e^{-i\omega_{1}\tau-i\boldsymbol{Q}^{i}\cdot\boldsymbol{R}^{i}}\tilde{G}_{\nu^{i}}^{\lessgtr}(\omega_{1};\boldsymbol{R}^{i},t),
\end{split}
\end{equation}
where $i=$ $',''$.
In this way can write:
\begin{equation}
\begin{split}
\Sigma^{\lessgtr}_{s_{1}s_{2}}(\boldsymbol{q}_{1},t_{1};\boldsymbol{q}_{2},t_{2})\equiv& \Sigma^{\lessgtr}_{s}\left(\boldsymbol{q}+\frac{\boldsymbol{Q}}{2},t+\frac{\tau}{2};\boldsymbol{q}-\frac{\boldsymbol{Q}}{2},t-\frac{\tau}{2}\right)=\\
=&\,4i\pi\hbar\sum_{\nu'\nu''}|\mathcal{F}_{\nu-\nu'-\nu''}|^{2}\iint\frac{d\omega_{1} d\boldsymbol{R}'}{(2\pi)^{4}}e^{-i\omega_{1}\tau-i\boldsymbol{Q}'\cdot\boldsymbol{R}'}\tilde{G}_{\nu'}^{\lessgtr}(\omega_{1};\boldsymbol{R}',t)\cdot\\
&\cdot\iint\frac{d\omega_{2} d\boldsymbol{R}''}{(2\pi)^{4}}e^{-i\omega_{2}\tau-i\boldsymbol{Q}''\cdot\boldsymbol{R}''}\tilde{G}_{\nu''}^{\lessgtr}(\omega_{2};\boldsymbol{R}'',t).
\end{split}
\end{equation}
So, finally, the self-energy in Wigner's mixed representation reads
\begin{equation}
\resizebox{\textwidth}{!}{$
\begin{split}
\tilde{\Sigma}_{\nu}^{\lessgtr}(\omega;\boldsymbol{R},t)=&\iint d\tau d\boldsymbol{Q}\,e^{i\omega\tau+i\boldsymbol{Q}\cdot\boldsymbol{R}}\Sigma_{s}^{\lessgtr}\left(\boldsymbol{q}+\frac{\boldsymbol{Q}}{2},t+\frac{\tau}{2};\boldsymbol{q}-\frac{\boldsymbol{Q}}{2},t-\frac{\tau}{2}\right)=\\
=&\,4i\pi\hbar\iint d\tau d\boldsymbol{Q}\,e^{i\omega\tau+i\boldsymbol{Q}\cdot\boldsymbol{R}}\sum_{\nu'\nu''}|\mathcal{F}_{\nu-\nu'-\nu''}|^{2}\iint\frac{d\omega_{1} d\boldsymbol{R}'}{(2\pi)^{4}}e^{-i\omega_{1}\tau-i\boldsymbol{Q}'\cdot\boldsymbol{R}'}\tilde{G}_{\nu'}^{\lessgtr}(\omega_{1};\boldsymbol{R}',t)\cdot\\
&\cdot\iint\frac{d\omega_{2} d\boldsymbol{R}''}{(2\pi)^{4}}e^{-i\omega_{2}\tau-i\boldsymbol{Q}''\cdot\boldsymbol{R}''}\tilde{G}_{\nu''}^{\lessgtr}(\omega_{2};\boldsymbol{R}'',t).
\end{split}$}
\end{equation}
We recall center-of-mass and relative variables,
\begin{equation} \label{center-of-mass_difference_variables_bis}
\begin{split}
&\boldsymbol{Q}^{i}=\boldsymbol{q}_{1}^{i}-\boldsymbol{q}_{2}^{i},\hspace{1cm}\tau=t_{1}-t_{2},\\
&\boldsymbol{q}^{i}=\frac{1}{2}(\boldsymbol{q}_{1}^{i}+\boldsymbol{q}_{2}^{i}),\hspace{1cm}t=\frac{1}{2}(t_{1}+t_{2})
\end{split}
\end{equation}
and
\begin{equation} \label{times_expressions_bis}
\begin{split}
&\boldsymbol{q}_{1}^{i}=\boldsymbol{q}^{i}+\frac{\boldsymbol{Q}^{i}}{2},\hspace{1cm}t_{1}=t+\frac{\tau}{2},\\
&\boldsymbol{q}_{2}^{i}=\boldsymbol{q}^{i}-\frac{\boldsymbol{Q}^{i}}{2},\hspace{1cm}t_{2}=t-\frac{\tau}{2},
\end{split}
\end{equation}
along with their related momentum and energy conservation laws: 
\begin{equation}
\begin{split}
&\boldsymbol{q}_{1}=\boldsymbol{q}_{1}'+\boldsymbol{q}_{1}''\\
&\boldsymbol{q}_{2}=\boldsymbol{q}_{2}'+\boldsymbol{q}_{2}''\\
&\boldsymbol{Q}=\boldsymbol{q}_{1}-\boldsymbol{q}_{2}=\boldsymbol{q}_{1}'+\boldsymbol{q}_{1}''-(\boldsymbol{q}_{2}'+\boldsymbol{q}_{2}'')=\boldsymbol{Q}'+\boldsymbol{Q}''\\
&\omega=\omega_{1}+\omega_{2}.
\end{split}
\end{equation}
In this way the self-energy in Wigner's mixed representation reads
\begin{equation} \label{almost_done_selfenergy_wigner}
\begin{split}
\tilde{\Sigma}_{\nu}^{\lessgtr}(\omega;\boldsymbol{R},t)=&\,4i\pi\hbar\int d\boldsymbol{Q}\sum_{\nu'\nu''}|\mathcal{F}_{\nu-\nu'-\nu''}|^{2}\iint\frac{d\omega_{1} d\boldsymbol{R}'}{(2\pi)^{4}}\iint\frac{d\omega_{2} d\boldsymbol{R}''}{(2\pi)^{4}}2\pi\delta(\omega-\omega_{1}-\omega_{2})\cdot\\
&\cdot e^{-i\boldsymbol{Q}\cdot(\boldsymbol{R}''-\boldsymbol{R})+i\boldsymbol{Q}'\cdot(\boldsymbol{R}''-\boldsymbol{R}')}\tilde{G}_{\nu'}^{\lessgtr}(\omega_{1};\boldsymbol{R}',t)\tilde{G}_{\nu''}^{\lessgtr}(\omega_{2};\boldsymbol{R}'',t)=\\
=&\,4i\pi\hbar\sum_{\nu'\nu''}|\mathcal{F}_{\nu-\nu'-\nu''}|^{2}\iint\frac{d\omega_{1} d\boldsymbol{R}'}{(2\pi)^{4}}\iint\frac{d\omega_{2} d\boldsymbol{R}''}{(2\pi)^{4}}2\pi\delta(\omega-\omega_{1}-\omega_{2})\cdot\\
&\cdot\int d\boldsymbol{Q}e^{i\boldsymbol{Q}\cdot\boldsymbol{R}}e^{-i\boldsymbol{Q}'\cdot\boldsymbol{R}'}e^{-i\boldsymbol{Q}''\cdot\boldsymbol{R}''}\tilde{G}_{\nu'}^{\lessgtr}(\omega_{1};\boldsymbol{R}',t)\tilde{G}_{\nu''}^{\lessgtr}(\omega_{2};\boldsymbol{R}'',t).
\end{split}
\end{equation}
Imposing $\boldsymbol{Q}'=\boldsymbol{Q}-\boldsymbol{Q}''$ we get
\begin{equation}
\begin{split}
\tilde{\Sigma}_{\nu}^{\lessgtr}(\omega;\boldsymbol{R},t)=&\,4i\pi\hbar\sum_{\nu'\nu''}|\mathcal{F}_{\nu-\nu'-\nu''}|^{2}\iint\frac{d\omega_{1} d\boldsymbol{R}'}{(2\pi)^{4}}\iint\frac{d\omega_{2} d\boldsymbol{R}''}{(2\pi)^{4}}2\pi\delta(\omega-\omega_{1}-\omega_{2})\cdot\\
&\cdot\int d\boldsymbol{Q}e^{i\boldsymbol{Q}\cdot(\boldsymbol{R}-\boldsymbol{R}')}e^{i\boldsymbol{Q}''\cdot(\boldsymbol{R}'-\boldsymbol{R}'')}\tilde{G}_{\nu'}^{\lessgtr}(\omega_{1};\boldsymbol{R}',t)\tilde{G}_{\nu''}^{\lessgtr}(\omega_{2};\boldsymbol{R}'',t).
\end{split}
\end{equation}
The term $(2\pi)^{-3}\int d\boldsymbol{Q}e^{i\boldsymbol{Q}\cdot(\boldsymbol{R}-\boldsymbol{R}')}$ is nothing but the exponential representation of the Dirac delta, so we get:
\begin{equation} \label{almost_done_selfenergy_wigner_1} 
\begin{split}
\tilde{\Sigma}_{\nu}^{\lessgtr}(\omega;\boldsymbol{R},t)=&\,4i\pi\hbar\sum_{\nu'\nu''}|\mathcal{F}_{\nu-\nu'-\nu''}|^{2}\iint\frac{d\omega_{1} d\boldsymbol{R}'}{(2\pi)^{4}}\iint\frac{d\omega_{2} d\boldsymbol{R}''}{(2\pi)^{4}}\frac{1}{(2\pi)^{2}}\delta(\omega-\omega_{1}-\omega_{2})\cdot\\
&\cdot\delta(\boldsymbol{R}-\boldsymbol{R}')e^{i\boldsymbol{Q}''\cdot(\boldsymbol{R}'-\boldsymbol{R}'')}\tilde{G}_{\nu'}^{\lessgtr}(\omega_{1};\boldsymbol{R}',t)\tilde{G}_{\nu''}^{\lessgtr}(\omega_{2};\boldsymbol{R}'',t).
\end{split}
\end{equation}
Similarly, we can use $\boldsymbol{Q}''=\boldsymbol{Q}-\boldsymbol{Q}'$ in Eq. \eqref{almost_done_selfenergy_wigner} obtaining
\begin{equation} \label{almost_done_selfenergy_wigner_2} 
\begin{split}
\tilde{\Sigma}_{\nu}^{\lessgtr}(\omega;\boldsymbol{R},t)=&\,4i\pi\hbar\sum_{\nu'\nu''}|\mathcal{F}_{\nu-\nu'-\nu''}|^{2}\iint\frac{d\omega_{1} d\boldsymbol{R}'}{(2\pi)^{4}}\iint\frac{d\omega_{2} d\boldsymbol{R}''}{(2\pi)^{4}}\frac{1}{(2\pi)^{2}}\delta(\omega-\omega_{1}-\omega_{2})\cdot\\
&\cdot\delta(\boldsymbol{R}-\boldsymbol{R}'')e^{-i\boldsymbol{Q}'\cdot(\boldsymbol{R}'-\boldsymbol{R}'')}\tilde{G}_{\nu'}^{\lessgtr}(\omega_{1};\boldsymbol{R}',t)\tilde{G}_{\nu''}^{\lessgtr}(\omega_{2};\boldsymbol{R}'',t),
\end{split}
\end{equation}
from which we find that Eq. \eqref{almost_done_selfenergy_wigner_1} equals Eq. \eqref{almost_done_selfenergy_wigner_2} only if we also have $\boldsymbol{R}'=\boldsymbol{R}''$. So, finally the self-energy in Wigner's mixed representation reads
\begin{equation} \label{final_selfenergy_wigner} 
\begin{split}
\tilde{\Sigma}_{\nu}^{\lessgtr}(\omega;\boldsymbol{R},t)&=4i\pi\hbar\sum_{\nu'\nu''}|\mathcal{F}_{\nu-\nu'-\nu''}|^{2}\int\frac{d\omega_{1}}{2\pi}\int\frac{d\omega_{2}}{2\pi}\delta(\omega-\omega_{1}-\omega_{2})\tilde{G}_{\nu'}^{\lessgtr}(\omega_{1};\boldsymbol{R},t)\tilde{G}_{\nu''}^{\lessgtr}(\omega_{2};\boldsymbol{R},t).
\end{split}
\end{equation}

\subsection{Resonant and off-resonant regimes}

We now clarify how the crystal response behaves both at and away from
resonance by examining the structure of the phonon Green's function. After
analytic continuation from the time-ordered correlation function to the retarded
response, the dressed retarded phonon Green's function is written as
\begin{equation}
G^{R}_{\nu}(\omega)
=
\frac{
2\omega_{\nu}
}{
\omega^{2}-\omega_{\nu}^{2}
-
2\omega_{\nu}\Sigma^{R}_{\nu}(\omega)
},
\qquad
\nu\equiv \boldsymbol{q}s ,
\label{eq:full_dressed_phonon_green_function}
\end{equation}
where, as already discusssed, we are neglecting the real part of the self-energy. Then, retaining the the full frequency dependence of the imaginary part of the retarded Green's function is
\begin{equation}
G^{R}_{\nu}(\omega)
=
\frac{
2\omega_{\nu}
}{
\omega^{2}-\omega_{\nu}^{2}
+
2i\omega_{\nu}\Gamma_{\nu}(\omega)
}.
\label{eq:full_dressed_phonon_green_function_no_shift}
\end{equation}
This form is the one required in the dipole-response diagrams \cite{fugallo2018infrared,cowley1963lattice}, since the external photon frequency probes the phonon propagator also away from the phonon pole. In principle, one could determine $\Gamma_{\nu}(\omega)$ from a fully
off-shell and self-consistent evaluation of the bubble self-energy, in which the
internal phonon lines are themselves dressed Green's functions. This would lead
to a frequency-dependent nonlinear problem of the schematic form
\begin{equation}
\Gamma_{\nu}(\omega)
=
\Gamma_{\nu}\!\left[\Gamma(\omega)\right],
\end{equation}
because the linewidth entering the internal propagators would affect the
self-energy that determines the linewidth itself. Such a treatment goes beyond the standard resonant approximation commonly used
for anharmonic phonon linewidths and phonon transport. In practice, one usually
adopts the lowest hierarchical approximation: the  Green's function in
Eq. \eqref{eq:full_dressed_phonon_green_function_no_shift} is kept
fully frequency dependent, while $\Gamma_{\nu}(\omega)$ is obtained from the
usual two-phonon bubble expression evaluated with two on-shell internal phonons, as
in standard implementations of anharmonic phonon scattering (see, e.g., Ref. \cite{togo2023first}). Equivalently, the energy-conservation factors are treated
through Dirac delta functions, their Gaussian regularizations, or a
self-consistent Lorentzian collisional broadening scheme
\cite{di2025broadening}. In this approximation the argument $\omega$ is retained as
the external frequency of the response, rather than being fixed from the outset
to the bare phonon frequency $\omega_{\nu}$ or to the frequency of the third
phonon entering the resonant bubble. The corresponding spectral function is \cite{togo2023first}
\begin{equation}
A_{\nu}(\omega)
=
\frac{1}{\pi}
\frac{
4\omega_{\nu}^{2}\Gamma_{\nu}(\omega)
}{
\left(\omega^{2}-\omega_{\nu}^{2}\right)^{2}
+
\left[2\omega_{\nu}\Gamma_{\nu}(\omega)\right]^{2}
}.
\label{eq:phonon_spectral_function_full_green}
\end{equation}
The commonly used resonant quasiparticle form is recovered only close to the
positive-frequency pole. Setting
\begin{equation}
\omega=\omega_{\nu}+\delta\omega,
\qquad
|\delta\omega|\ll \omega_{\nu},
\end{equation}
one obtains
\begin{equation}
\omega^{2}-\omega_{\nu}^{2}
=
2\omega_{\nu}\delta\omega
+
(\delta\omega)^{2}
\simeq
2\omega_{\nu}(\omega-\omega_{\nu}).
\label{eq:near_pole_expansion}
\end{equation}
Substituting this approximation into
Eq. \eqref{eq:full_dressed_phonon_green_function_no_shift} yields
\begin{equation}
G^{R}_{\nu}(\omega)
\simeq
\frac{1}{
\omega-\omega_{\nu}
+
i\Gamma_{\nu}(\omega)
}.
\label{eq:resonant_phonon_green_function}
\end{equation}
Eq. \eqref{eq:resonant_phonon_green_function} is the usual near-pole
propagator entering resonant one-phonon absorption and quasiparticle
phonon-scattering formulas. However, this near-pole form is not sufficient, for instance, to describe the
infrared response away from the main low-frequency resonances, where the
external photon frequency can lie far from any phonon pole and higher-order
phonon contributions become non-negligible in determining the dielectric
response. In this regime, one
must retain the full phonon propagator and the external-frequency dependence of
the linewidth, rather than evaluating the latter only at the resonant frequency
of the third phonon entering the bubble. Otherwise, the Lorentzian tails
generated by a purely resonant quasiparticle approximation may overestimate the
absorption intensity, or more generally fail to capture the correct
off-resonant frequency dependence of the spectrum
\cite{fugallo2018infrared}.

\section{Details on the analytical approach to phonon hydrodynamics}

\subsection{Final system of analytical equations for a 2D strip device} \label{system_of_equations}

Exploiting the boundary conditions \eqref{BCuy} and \eqref{BCux}and \eqref{temperature_BC_y=0,h} in the solutions \eqref{u_y_ky_general}, \eqref{u_x_ky_general} and \eqref{expression_for_temperature}, respectively, we get
\begin{equation} \label{final_system_coefficients_explicit_condensed}
\begin{cases}
\left(1+e^{kh}\right)a_{\phi}^{\psi}+\frac{q_{\phi}}{k}\left(1+e^{q_{\phi}h}\right)b_{\phi}+i\left(1+e^{q_{\psi}h}\right)b_{\psi}=-\frac{U}{k}
\\
\left(1-e^{kh}\right)a_{\phi}^{\psi}+\left(1-e^{q_{\phi}h}\right)b_{\phi}+i\frac{q_{\psi}}{k}\left(1-e^{q_{\psi}h}\right)b_{\psi}=0
\\
\left(1-e^{kh}\right)a_{\phi}^{\psi}-\frac{\alpha\beta}{\kappa\gamma}\left(1-e^{q_{\phi}h}\right)b_{\phi}=\frac{\beta}{\gamma}T_{{\rm{bc}}}
\end{cases}
\end{equation}
where $a_{\phi}^{\psi}=a_{\phi}+ia_{\psi}$ and $T_{{\rm{bc}}}=\bar{T}\delta(k)+\Delta T$. The system of equations \eqref{final_system_coefficients_explicit_condensed} is solved with \texttt{Mathematica} \cite{mathematica}. The results obtained for the coefficients are the following:
\begin{equation} \label{a_Phi^psi}
\resizebox{\textwidth}{!}{$
a_{\phi}^{\psi}=\frac{\frac{\beta}{\gamma} \big[(1 - e^{q_{\phi}h}) (1 + e^{q_{\psi}h }) - (1 + e^{q_{\phi}h}) (1 - e^{q_{\psi}h }) \frac{q_{\phi} q_{\psi}}{k^{2}}\big] T_{{\rm{bc}}} + \frac{\alpha\beta}{\kappa\gamma} (1 - e^{q_{\phi}h}) (1 - e^{q_{\psi}h })\frac{q_{\psi}}{k^{2}}U}{(1 - e^{kh}) \big[(1 - e^{q_{\phi}h}) (1 + e^{q_{\psi}h }) -  (1 + e^{q_{\phi}h}) (1 - e^{q_{\psi}h })  \frac{q_{\phi}q_{\psi}}{k^{2}}\big]+\frac{\alpha\beta}{\kappa\gamma} (1 - e^{q_{\phi}h}) \big[(1 - e^{kh}) (1 + e^{q_{\psi}h }) - (1 + e^{kh}) (1 - e^{q_{\psi}h }) \frac{q_{\psi}}{k}\big]},$}
\end{equation}
\begin{equation} \label{b_Phi}
\resizebox{\textwidth}{!}{$
b_{\phi}=\frac{\frac{\beta}{\gamma} k^{2} T_{{\rm{bc}}} (1+e^{q_{\psi}h})(1 - e^{kh}) -\frac{\beta}{\gamma} k q_{\psi} T_{{\rm{bc}}} (1 + e^{kh})(1 - e^{q_{\psi}h}) -q_{\psi} U (1 - e^{kh})(1- e^{q_{\psi}h})}{\left(\frac{\alpha\beta}{\kappa\gamma}+1\right)k^{2}(1 - e^{kh})(1-e^{q_{\phi}h})(1+e^{q_{\psi}h})-\frac{\alpha\beta}{\kappa\gamma}kq_{\psi}(1+e^{kh})(1-e^{q_{\phi}h})(1- e^{q_{\psi}h})- q_{\phi} q_{\psi}(1 - e^{kh})(1+e^{q_{\phi}h})(1-e^{q_{\psi}h})},$}
\end{equation}
\begin{equation} \label{b_psi}
\resizebox{\textwidth}{!}{$
b_{\psi}=ik\frac{\frac{\beta}{\gamma} T_{{\rm{bc}}} k(1 + e^{kh})(1- e^{q_{\phi}h}) - \frac{\beta}{\gamma} q_{\phi} T_{{\rm{bc}}} (1 - e^{kh})(1+e^{q_{\phi}h}) + \left(\frac{\alpha\beta}{\kappa\gamma}+1\right) U (1 - e^{kh})(1 - e^{q_{\phi}h})}{\left(\frac{\alpha\beta}{\kappa\gamma}+1\right)k^{2}(1 - e^{kh})(1 - e^{q_{\phi}h})(1+ e^{q_{\psi}h}) - \frac{\alpha\beta}{\kappa\gamma} k q_{\psi}(1 + e^{kh}) (1 - e^{q_{\phi}h}) (1 - e^{q_{\psi}h}) - q_{\phi} q_{\psi} (1 - e^{kh})(1 + e^{q_{\phi}h})(1 - e^{q_{\psi}h})}$}.
\end{equation}

\subsection{Deviations from diffusive behavior in the analytical approach} \label{FDN_section}

Recalling the definition of the parameters entering the VHE given in Eq. \eqref{easy_notation}, we can follow the same approach described in Ref. \cite{simoncelli2020generalization} to evaluate a descriptor which allows to capture the deviation from Fourier's diffusive behavior. We start by rewriting $q_{\phi}$ and $q_{\psi}$ in Eqs. \eqref{solution_Phi} and \eqref{solution_psi} by making explicit their physical content:
\begin{equation}
q_{\phi}^{2}=k^{2}+\frac{\gamma}{\eta+\mu^{*}}+\frac{\alpha\beta}{\kappa(\eta+\mu^{*})}=k^{2}+\frac{AD^{U}}{\eta+\mu^{*}}+\frac{W_{0}^{2}AC}{\kappa(\eta+\mu^{*})}
\end{equation}
and
\begin{equation}
q_{\psi}^{2}=k^{2}+\frac{\gamma}{\eta}=k^{2}+\frac{AD^{U}}{\eta}.
\end{equation}
Then, we define the dimensionless parameter:
\begin{equation} \label{parameter_epsilon}
\epsilon=\frac{\gamma h^{2}}{\eta}=\frac{AD^{U}h^{2}}{\eta}.
\end{equation}
This parameter characterizes the relative strength of the viscosity and thermal resistivity due to the momentum dissipation rate $D^{U}$. When $D^{U}\to0$ (\textit{i.e.} $\epsilon\to0$), we have
\begin{equation} \label{epsilon_to_0_limit}
q_{\psi}=\sqrt{k^{2}+\frac{\gamma}{\eta}}=\sqrt{k^{2}+\frac{\epsilon}{h^{2}}}
\xrightarrow[\epsilon\to0]{}k,
\end{equation}
and so also
\begin{equation} \label{q_Phi_limit_epsilon_to_0}
q_{\phi}=\sqrt{k^{2}+\frac{\gamma}{\eta+\mu^{*}}+\frac{\alpha\beta}{\kappa(\eta+\mu^{*})}}\xrightarrow[\epsilon\to0]{}\sqrt{k^{2}+\frac{\alpha\beta}{\kappa(\eta+\mu^{*})}}.
\end{equation}
On the other hand this limit makes also evident the definition of the parameter $\xi$ (see Eq. \eqref{final_temperature_viscous}):
\begin{equation} \label{parameter_xi}
\xi=\frac{\kappa}{\alpha}\frac{\frac{\Delta T}{h}}{U}=\frac{\kappa}{W_{0}\sqrt{\bar{T}AC}}\frac{\frac{\Delta T}{h}}{U},
\end{equation}
which takes into account the effect of boundary conditions with that of thermal conductivity and transport coefficient $\alpha$. Similarly we look at the viscous regime when $\xi\to0$.\\
Summing together $\epsilon$ in Eq. \eqref{parameter_epsilon} and $\xi$ in Eq. \eqref{parameter_xi} we find the Fourier deviation number (FDN) \cite{simoncelli2020generalization} as
\begin{equation} \label{FDN_apendices}
\text{FDN}=\left[\frac{\gamma h^{2}}{\eta}+\frac{\kappa}{\alpha}\frac{\frac{\Delta T}{h}}{U}\right]^{-1}=\left[\frac{AD^{U} h^{2}}{\eta}+\frac{\kappa}{W_{0}\sqrt{\bar{T}AC}}\frac{\frac{\Delta T}{h}}{U}\right]^{-1}=\frac{1}{\epsilon+\xi}.
\end{equation}
This descriptor is able to capture the conditions under which hydrodynamic heat transport is favored with respect to diffusive one. The higher the FDN (that is for both $\epsilon\to0$ and $\xi\to0$) the more dominant is the viscous character of the phonon fluid. Finally, we note that the FDN definition \eqref{FDN_apendices} obtained here is the analytical analogue of that of Ref. \cite{simoncelli2020generalization}, where the role of the characteristic velocity is played by the velocity $U$ injected into the strip. Finally, we rename the following polynomials to facilitate the discussion:
\begin{equation} \label{condensed_notation}
\begin{cases}
F(q_{\phi},k)=\frac{1-e^{q_{\phi}h}}{k(1-e^{q_{\phi}h })(1+e^{kh})-q_{\phi}(1+e^{q_{\phi}h })(1-e^{kh})},\\
G(q_{\psi},k)=\frac{1-e^{q_{\psi}h }}{k(1 - e^{kh}) (1 + e^{q_{\psi}h })-q_{\psi}(1 + e^{kh}) (1 - e^{q_{\psi}h })}.
\end{cases}
\end{equation}

\subsubsection{Limiting cases}

In the $\epsilon\to0$ limit, the coefficients of the solutions become:
\begin{equation} \label{final_coefficients_viscous}
\resizebox{\textwidth}{!}{$
\begin{cases}
\displaystyle{
a_{\phi}^{\psi}=\frac{\beta}{\gamma}\frac{T_{{\rm{bc}}}}{1-e^{kh}}+\frac{\alpha\beta}{\kappa\gamma}\frac{(1-e^{q_{\phi}h })U}{k(1-e^{q_{\phi}h })(1+e^{kh})-q_{\phi}(1+e^{q_{\phi}h })(1-e^{kh})}=\frac{\beta}{\gamma}\frac{T_{{\rm{bc}}}}{1-e^{kh}}+\frac{\alpha\beta}{\kappa\gamma}UF(q_{\phi},k)
}
\vspace{0.25cm}
\\
\displaystyle{
b_{\phi}=-\frac{(1 - e^{kh})U}{k(1-e^{q_{\phi}h })(1+e^{kh})-q_{\phi}(1+e^{q_{\phi}h })(1-e^{kh})}
=-U\frac{1-e^{kh}}{1-e^{q_{\phi}h}}F(q_{\phi},k)
}
\vspace{0.5cm}
\\
\displaystyle{
b_{\psi}=i\left[\frac{\beta}{\gamma}\frac{T_{{\rm{bc}}}}{1-e^{kh}}+\frac{(1 - e^{q_{\phi}h })\big(\frac{\alpha\beta}{\kappa\gamma}+1\big)U}{k(1-e^{q_{\phi}h })(1+e^{kh}) - q_{\phi}  (1+ e^{q_{\phi}h })(1 - e^{kh})}\right]=i\left[\frac{\beta}{\gamma}\frac{T_{{\rm{bc}}}}{1-e^{kh}}+\left(\frac{\alpha\beta}{\kappa\gamma}+1\right)UF(q_{\phi},k)\right]
},
\end{cases}$}
\end{equation}
and, knowing $a_{\phi}^{\psi}=a_{\phi}+ia_{\psi}$:
\begin{equation}
\begin{cases}
\displaystyle{
a_{\phi}=\frac{\beta}{\gamma}\frac{T_{{\rm{bc}}}}{1-e^{kh}}
}
\vspace{0.25cm}
\\
\displaystyle{a_{\psi}=-i\frac{\alpha\beta}{\kappa\gamma}UF(q_{\phi},k)
}.
\end{cases}
\end{equation}
In the $\xi\to0$ limit we have
\begin{equation} \label{final_coefficients_xi_to_zero_explicit}
\resizebox{\textwidth}{!}{$
\begin{cases}
\displaystyle{
a_{\phi}^{\psi}=\frac{\beta}{\gamma}\frac{T_{{\rm{bc}}}}{1-e^{kh}}+\frac{q_{\psi}}{k}\frac{(1 - e^{q_{\psi}h })U}{k(1 - e^{kh}) (1 + e^{q_{\psi}h }) - q_{\psi}(1 + e^{kh}) (1 - e^{q_{\psi}h })}=\frac{\beta}{\gamma}\frac{T_{{\rm{bc}}}}{1-e^{kh}}+\frac{q_{\psi}}{k}UG(q_{\psi},k)
}
\vspace{0.25cm}
\\
\displaystyle{
b_{\phi}=-\frac{1 - e^{kh}}{1-e^{q_{\phi}h}}\frac{\frac{\kappa\gamma}{\alpha\beta}\frac{q_{\psi}}{k} (1- e^{q_{\psi}h})U}{k(1 - e^{kh})(1+e^{q_{\psi}h})-q_{\psi}(1+e^{kh})(1- e^{q_{\psi}h})}=-\frac{\kappa\gamma}{\alpha\beta}U\frac{1 - e^{kh}}{1-e^{q_{\phi}h}}\frac{q_{\psi}}{k}G(q_{\psi},k)
}
\vspace{0.5cm}
\\
\displaystyle{
b_{\psi}=i\left(\frac{\kappa\gamma}{\alpha\beta}+1\right)\frac{  (1 - e^{kh})U}{k(1 - e^{kh})(1+ e^{q_{\psi}h}) - q_{\psi}(1 + e^{kh})  (1 - e^{q_{\psi}h})}=i\left(\frac{\kappa\gamma}{\alpha\beta}+1\right)U\frac{1 - e^{kh}}{1-e^{q_{\phi}h}}G(q_{\psi},k)
}
\end{cases}$}
\end{equation}
and, knowing $a_{\phi}^{\psi}=a_{\phi}+ia_{\psi}$:
\begin{equation}
\begin{cases}
\displaystyle{
a_{\phi}=\frac{\beta}{\gamma}\frac{T_{{\rm{bc}}}}{1-e^{kh}}
}
\vspace{0.25cm}
\\
\displaystyle{a_{\psi}=-i\frac{q_{\psi}}{k}UG(q_{\psi},k)
}.
\end{cases}
\end{equation}
Finally, in the $\epsilon\to\infty$ and/or $\xi\to\infty$ limits we get
\begin{equation} \label{final_coefficients_Fourier}
\begin{cases}
\displaystyle{
a_{\phi}=\frac{\beta}{\gamma}\frac{T_{{\rm{bc}}}}{1-e^{kh}}
}
\vspace{0.35cm}
\\
\displaystyle{
a_{\psi}=0
}
\vspace{0.35cm}
\\
\displaystyle{
b_{\phi}=0
}
\vspace{0.25cm}
\\
\displaystyle{
b_{\psi}=0
}
\end{cases}
\end{equation}
From these coefficients, it is evident that these limits lead to the diffusive temperature profile predicted by Fourier's law.

\subsection{Solution of the temperature profile} \label{T_profile_appendix}

By writing explicitly $a_{\phi}^{\psi}=a_{\phi}+ia_{\psi}$ in the solution of the temperature profile \eqref{expression_for_temperature}, we can show that it is given by the sum of two contributions, one related to the velocity potential and one related to the stream function:
\begin{equation} \label{T_as_sum_of_T_phi_and_T_psi}
\resizebox{\textwidth}{!}{$
\begin{split}
T(x,y)&=\frac{1}{2\pi}\int dk\,e^{ikx}\left[\frac{\gamma}{\beta}a_{\phi}\left(e^{ky}-e^{kh}e^{-ky}\right)-\frac{\alpha}{\kappa}b_{\phi}\left(e^{q_{\phi}y}-e^{q_{\phi}h}e^{-q_{\phi}y}\right)+i\frac{\gamma}{\beta}a_{\psi}\left(e^{ky}-e^{kh}e^{-ky}\right)\right]=\\
&=T_{\phi}(x,y)+T_{\psi}(x,y).
\end{split}$}
\end{equation}
where 
\begin{equation} \label{T_phi}
T_{\phi}(x,y)=\frac{1}{2\pi}\int dk\,e^{ikx}\left[\frac{\gamma}{\beta}a_{\phi}\left(e^{ky}-e^{kh}e^{-ky}\right)-\frac{\alpha}{\kappa}b_{\phi}\left(e^{q_{\phi}y}-e^{q_{\phi}h}e^{-q_{\phi}y}\right)\right]
\end{equation}
and 
\begin{equation} \label{T_psi}
T_{\psi}(x,y)=\frac{i}{2\pi}\int dk\,e^{ikx} \frac{\gamma}{\beta}a_{\psi}\left(e^{ky}-e^{kh}e^{-ky}\right).
\end{equation}
In summary, we have shown that the temperature profile can be seen as the sum of thermal compressibility and vorticity contributions (see Eqs. \eqref{connection_vel_pot_strem_f_to_compress_and_vort}) and we are now fully equipped to evaluate it in the viscous and diffusive limits.\\
Finally, we clearly see that the temperature difference across the strip is evaluated as
\begin{equation} \label{final_temperature_diff}
T(x,0)-T(x,h)=\frac{1}{2\pi}\int dk\,e^{ikx}\,2\left[\frac{\gamma}{\beta}\left(1-e^{kh}\right)a_{\phi}^{\psi}-\frac{\alpha}{\kappa}\left(1-e^{q_{\phi}h}\right)b_{\phi}\right].
\end{equation}

\subsubsection{Temperature profile for \texorpdfstring{$\xi\to0$}{xi to 0}} \label{xi_to_zero_T_profile_appendix}

The temperature profile for $\xi\to0$ becomes
\begin{equation} \label{final_temperature_xi_0}
\begin{split}
T_{\xi\to0}(x,y)=\frac{1}{2\pi}\int dk\,e^{ikx}\Bigg[&\left(e^{ky}-e^{kh}e^{-ky}\right)\left(\frac{T_{{\rm{bc}}}}{1-e^{kh}}+\frac{\gamma}{\beta}U\frac{q_{\psi}}{k}G(q_{\psi},k)\right)+\\
&+\frac{\gamma}{\beta}U\left(e^{q_{\phi}y}-e^{q_{\phi}h}e^{-q_{\phi}y}\right)\frac{1 - e^{kh}}{1-e^{q_{\phi}h}}\frac{q_{\psi}}{k}G(q_{\psi},k)\Bigg].
\end{split}
\end{equation}
In order to address the non-local thermal response we want to evaluate $T_{\xi\to0}(x,0)-T_{\xi\to0}(x,h)$. From Eq. \eqref{final_temperature_diff} we have
\begin{equation} \label{temp_diff_xi_to_0}
\begin{split}
&T_{\xi\to0}(x,0)-T_{\xi\to0}(x,h)=\\
=&\,\frac{1}{2\pi}\int dk\,e^{ikx}\,2\left[\Delta T+2\frac{\gamma}{\beta}U\left(1-e^{kh}\right)\frac{q_{\psi}}{k}G(q_{\psi},k)\right]=\\
=&\,\frac{1}{\pi}\int dk\,e^{ikx}\left[\Delta T+2\frac{\gamma}{\beta}U\frac{q_{\psi}}{k}\frac{(1-e^{kh})(1-e^{q_{\psi}h })}{(q_{\psi}+k)(e^{q_{\psi}h}-e^{kh})+(q_{\psi}-k)(e^{(k+q_{\psi})h}-1)}\right].
\end{split}
\end{equation}
The sign-changing behavior of the temperature difference across the strip given by Eq. \eqref{temp_diff_xi_to_0} is depicted in Fig. \ref{fig:backflow_main_2}. Finally, the second term in Eq. \eqref{temp_diff_xi_to_0} can be seen as the thermal analogue of the result obtained in Ref. \cite{levitov2016electron} for the electrical voltage (see Eq. 23 of the related Supplementary Information and see also section \ref{incompressible_limit}).\\
Moreover, we can further manipulate Eq. \eqref{temp_diff_xi_to_0} to analytically show the presence of a negative thermal response across the strip. We are interested in the viscous limit, which can can be analyzed by expanding the second term under the integral in Eq. \eqref{temp_diff_xi_to_0} in a small difference $q_{\psi}-k$, that is when we also take the $\epsilon\to0$ limit on top of the $\xi\to0$ solution. We start from the denominator:
\begin{equation}
\begin{split}
&\,k\Big[(q_{\psi}+k)(e^{q_{\psi}h}-e^{kh})+(q_{\psi}-k)(e^{(q_{\psi}+k)h}-1)\Big]=\\
=&\,k\Big[2k(e^{q_{\psi}h}-e^{kh})+(q_{\psi}-k)(e^{q_{\psi}h}-e^{kh})+(q_{\psi}-k)(e^{(q_{\psi}+k)h}-1)\Big]=\\
=&\,k\Big[2ke^{kh}(e^{(q_{\psi}-k)h}-1)+(q_{\psi}-k)e^{kh}(e^{(q_{\psi}-k)h}-1)+(q_{\psi}-k)e^{2kh}(e^{(q_{\psi}-k)h}-e^{-2kh})\Big]=\\
=&\,ke^{kh}\Big[2k(e^{(q_{\psi}-k)h}-1)+(q_{\psi}-k)(e^{(q_{\psi}-k)h}-1)+(q_{\psi}-k)e^{kh}(e^{(q_{\psi}-k)h}-e^{-2kh})\Big]\sim\\
\sim&\,ke^{kh}\Big[2kh(q_{\psi}-k)+(q_{\psi}-k)e^{kh}(1-e^{-2kh})\Big]=ke^{kh}(q_{\psi}-k)\Big[2kh+e^{kh}(1-e^{-2kh})\Big]=\\
=&\,2ke^{kh}(q_{\psi}-k)\left[kh+\frac{e^{kh}-e^{-kh}}{2}\right]=2ke^{kh}(q_{\psi}-k)\Big(kh+\sinh(kh)\Big),
\end{split}
\end{equation}
Then, knowing that
\begin{equation}
\begin{split}
\frac{\gamma}{\beta}&=\frac{\eta}{\beta}\frac{\gamma}{\eta}=\frac{\eta}{\beta}(q_{\psi}^{2}-k^{2})=\frac{\eta}{\beta}(q_{\psi}-k)(q_{\psi}+k)=\frac{\eta}{\beta}(q_{\psi}-k)(q_{\psi}-k+2k)=\\
&=\frac{\eta}{\beta}\Big[(q_{\psi}-k)^{2}+2k(q_{\psi}-k)\Big]\sim2\frac{\eta}{\beta}k(q_{\psi}-k)
\end{split}
\end{equation}
and
\begin{equation}
\begin{split}
&(1-e^{q_{\psi}h})(1-e^{kh})=-e^{kh}(1-e^{q_{\psi}h})(1-e^{-kh})=-e^{2kh}(e^{-kh}-e^{(q_{\psi}-k)h})(1-e^{-kh})\sim\\
\sim&-e^{2kh}(e^{-kh}-1)(1-e^{-kh})=e^{2kh}(1-e^{-kh})(1-e^{-kh})=e^{2kh}\frac{1-e^{-kh}}{1+e^{-kh}}(1-e^{-kh})(1+e^{-kh})=\\
=&\,e^{kh}\frac{1-e^{-kh}}{1+e^{-kh}}e^{kh}(1-e^{-2kh})=e^{kh}\frac{1-e^{-kh}}{1+e^{-kh}}(e^{kh}-e^{-kh})=2e^{kh}\frac{1-e^{-kh}}{1+e^{-kh}}\frac{e^{kh}-e^{-kh}}{2}=\\
=&\,2e^{kh}\tanh\left(k\frac{h}{2}\right)\sinh(kh),
\end{split}
\end{equation}
to the linear order in $q_{\psi}-k$, the numerator simply becomes
\begin{equation}
\begin{split}
&2\frac{\gamma}{\beta}q_{\psi}(1-e^{q_{\psi}h})(1-e^{kh})=4\frac{\eta}{\beta}k(q_{\psi}-k)q_{\psi}(1-e^{q_{\psi}h})(1-e^{kh})=\\
=&\,4\frac{\eta}{\beta}k(q_{\psi}-k)\Big[(q_{\psi}-k)(1-e^{q_{\psi}h})(1-e^{kh})+k(1-e^{q_{\psi}h})(1-e^{kh})\Big]=\\
=&\,8\frac{\eta}{\beta}k(q_{\psi}-k)^{2}e^{kh}\tanh\left(k\frac{h}{2}\right)\sinh(kh)+8\frac{\eta}{\beta}k^{2}(q_{\psi}-k)e^{kh}\tanh\left(k\frac{h}{2}\right)\sinh(kh)\sim\\
\sim&\,8\frac{\eta}{\beta}k^{2}(q_{\psi}-k)e^{kh}\tanh\left(k\frac{h}{2}\right)\sinh(kh).
\end{split}
\end{equation}
Putting everything together we get
\begin{equation}
\begin{split}
T(x,0)-T(x,h)\xrightarrow[(q_{\psi}-k)\to0]{}\frac{1}{\pi}\int dk\,e^{ikx}\left[\Delta T+4U\frac{\eta}{\beta}\frac{k\tanh\left(k\frac{h}{2}\right)\sinh(kh)}{kh+\sinh(kh)}\right].
\end{split}
\end{equation}
The temperature difference across the strip is a Fourier transform of an even function in $k$ and it only assumes real values. We can see that the second term of the integrated function vanishes at $k=0$. This means that $\int dx\big(T(x,0)-T(x,h)\big)$ vanishes, and so the temperature difference across the strip must go negative on a portion of the real $x$ axis. Furthermore, we have
\begin{equation} \label{identity}
\frac{k\tanh\left(k\frac{h}{2}\right)\sinh(kh)}{kh+\sinh(kh)}\xrightarrow[k\to\infty]{}|k|\,\,\,\,\leftrightarrow\,\,\,\,\int dk\,e^{ikx}|k|=-\frac{1}{(x-i0)^{2}}-\frac{1}{(x+i0)^{2}},
\end{equation}
which shows that $T(x,0)-T(x,h)$ goes negative for $x\ne0$ and small enough $x$ values. In fact, when $|x|\ll h$, $|k|\gg\frac{1}{h}$ and Eq. \eqref{identity} gives 
\begin{equation} \label{xi_0_temp_diff_limit}
T(x,0)-T(x,h)\xrightarrow[|k|\gg\frac{1}{h}]{}2\Delta T\delta(x)-8U\frac{\eta}{\beta}\frac{1}{x^{2}},
\end{equation}
which can lead to a negative thermal resistance. This result is consistent with that found in Ref. \cite{levitov2016electron} for the incompressible electron fluid. \\
Finally, in the following we show the inverse-square relationship with respect to the distance from the contacts of the temperature profile. From what we already discussed above, assuming small differences $q_{\phi}-k$, Eq. \eqref{final_temperature_xi_0} can be rewritten as
\begin{equation} \label{T_viscous_strong_limit}
T_{\xi\to0}(x,y)=\frac{1}{2\pi}\int dk\,e^{ikx}T_{\xi\to0}(k,y),
\end{equation}
where
\begin{equation}
T_{\xi\to0}(k,y)=\left(e^{ky}-e^{kh}e^{-ky}\right)\left(\frac{T_{{\rm{bc}}}}{1-e^{kh}}+2\frac{\gamma}{\beta}U\frac{q_{\psi}}{k}\frac{1-e^{q_{\psi}h }}{(q_{\psi}+k)(e^{q_{\psi}h}-e^{kh})+(q_{\psi}-k)(e^{(q_{\psi}+k)h}-1)}\right).
\end{equation}
The denominator reads
\begin{equation}
k\Big[(q_{\psi}+k)(e^{q_{\psi}h}-e^{kh})+(q_{\psi}-k)(e^{(q_{\psi}+k)h}-1)\Big]\sim2ke^{kh}(q_{\psi}-k)\Big(kh+\sinh(kh)\Big),
\end{equation}
Then, knowing $\frac{\gamma}{\beta}\sim2\frac{\eta}{\beta}k(q_{\psi}-k)$ and
\begin{equation}
\begin{split}
&(1-e^{q_{\psi}h})=e^{kh}(e^{-kh}-e^{(q_{\psi}-k)h})\sim e^{kh}(e^{-kh}-1)=1-e^{kh}=-2e^{k\frac{h}{2}}\frac{e^{kh}-1}{2e^{k\frac{h}{2}}}=\\
=&-2e^{k\frac{h}{2}}\sinh\left(k\frac{h}{2}\right)\equiv1-e^{kh},
\end{split}
\end{equation}
to the linear order in $q_{\psi}-k$, the numerator simply becomes
\begin{equation}
\begin{split}
&\,2\frac{\gamma}{\beta}q_{\psi}(1-e^{q_{\psi}h})=4\frac{\eta}{\beta}k(q_{\psi}-k)q_{\psi}(1-e^{q_{\psi}h})=\\
=&\,4\frac{\eta}{\beta}k(q_{\psi}-k)\Big[(q_{\psi}-k)(1-e^{q_{\psi}h})+k(1-e^{q_{\psi}h})\Big]=\\
=&\,4\frac{\eta}{\beta}k(q_{\psi}-k)\Big[-2(q_{\psi}-k)e^{k\frac{h}{2}}\sinh\left(k\frac{h}{2}\right)-2ke^{k\frac{h}{2}}\sinh\left(k\frac{h}{2}\right)\Big]\sim\\
\sim&-8\frac{\eta}{\beta}k^{2}(q_{\psi}-k)e^{k\frac{h}{2}}\sinh\left(k\frac{h}{2}\right).
\end{split}
\end{equation}
Merging altogether we get
\begin{equation}
T_{\xi\to0}(x,y)=\frac{1}{2\pi}\int dk\,e^{ikx}\left[\left(e^{ky}-e^{kh}e^{-ky}\right)e^{-k\frac{h}{2}}\left(-\frac{T_{{\rm{bc}}}}{2\sinh\left(k\frac{h}{2}\right)}-\frac{4U\frac{\eta}{\beta}k\sinh\left(k\frac{h}{2}\right)}{kh+\sinh(kh)}\right)\right].
\end{equation}
Then we also have 
\begin{equation}
\begin{split}
e^{ky}-e^{kh}e^{-ky}&=e^{k\frac{h}{2}}\left(e^{k\left(y-\frac{h}{2}\right)}-e^{-k\left(y-\frac{h}{2}\right)}\right)=e^{k\frac{h}{2}}2\frac{e^{k\left(y-\frac{h}{2}\right)}-e^{-k\left(y-\frac{h}{2}\right)}}{2}=\\
&=2e^{k\frac{h}{2}}\sinh\left[k\left(y-\frac{h}{2}\right)\right],
\end{split}
\end{equation}
so that the temperature profile can be written as
\begin{equation} \label{T_xi_zero_manipulation}
\begin{split}
T_{\xi\to0}(x,y)&=\frac{1}{2\pi}\int dk\,e^{ikx}\left[\sinh\left(ky-k\frac{h}{2}\right)\left(-\frac{T_{{\rm{bc}}}}{\sinh\left(k\frac{h}{2}\right)}-\frac{8U\frac{\eta}{\beta}k\sinh\left(k\frac{h}{2}\right)}{kh+\sinh(kh)}\right)\right]=\\
&=T_{\xi\to0}^{\Delta T}(x,y)+T_{\xi\to0}^{U}(x,y).
\end{split}
\end{equation}
Using the properties
\begin{equation} \label{properties_hyperbolic_functions}
\begin{cases}
\sinh\left(ky-k\frac{h}{2}\right)=\sinh\left(ky\right)\cosh\left(k\frac{h}{2}\right)-\cosh\left(ky\right)\sinh\left(k\frac{h}{2}\right),\\
\sinh\left(k\frac{h}{2}\right)\cosh\left(k\frac{h}{2}\right)=\frac{\sinh(kh)}{2},
\end{cases}
\end{equation}
we can write the second term of Eq. \eqref{T_xi_zero_manipulation} as
\begin{equation}
\begin{split}
T_{\xi\to0}^{U}(x,y)&=-\frac{1}{2\pi}\int dk\,e^{ikx}\left[\sinh\left(ky-k\frac{h}{2}\right)\frac{8U\frac{\eta}{\beta}k\sinh\left(k\frac{h}{2}\right)}{kh+\sinh(kh)}\right]=\\
&=\frac{1}{2\pi}U\frac{\eta}{\beta}\int dk\,e^{ikx}\left[\cosh\left(ky\right)\left(\frac{8k\sinh^{2}\left(k\frac{h}{2}\right)}{kh+\sinh(kh)}\right)-\sinh\left(ky\right)\left(\frac{4k\sinh\left(kh\right)}{kh+\sinh(kh)}\right)\right]=\\
&=\frac{1}{2\pi}U\frac{\eta}{\beta}\int dk\,e^{ikx}\Big[\cosh\left(ky\right)f(k)-\sinh\left(ky\right)g(k)\Big].
\end{split}
\end{equation}
We can see that for $k\to+\infty$ both $f(k)$ and $g(k)$ are $\sim4k$, while for $k\to-\infty$ we have $f(k)\sim-4k$ and  $g(k)\sim4k$.
\begin{figure}[!htb]
\centering
\includegraphics[width=0.99\textwidth]{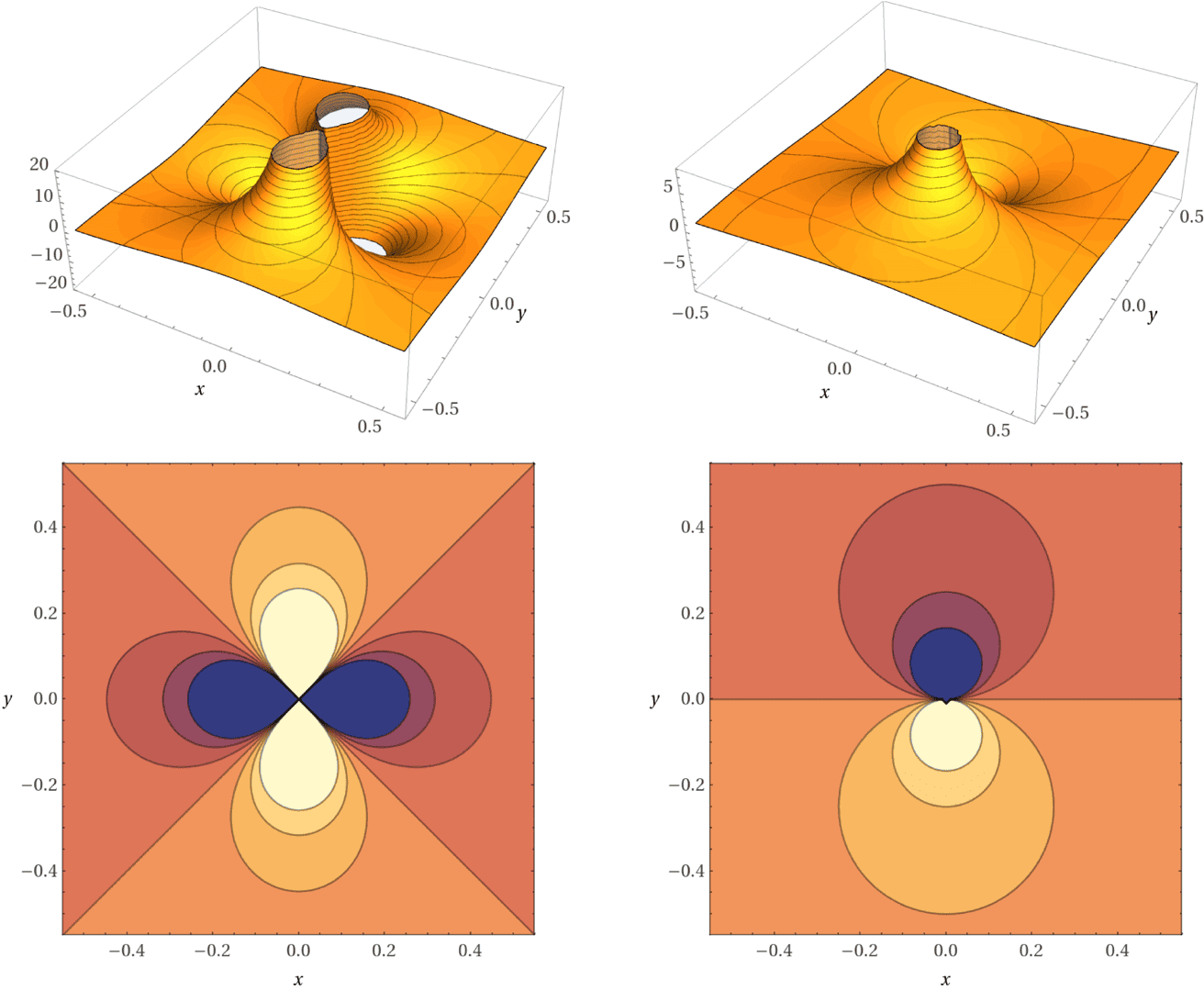}
\caption{\textbf{Analytical structure of the drift velocity and temperature gradient contribution to the temperature profile.} 3D plot of $\frac{x^{2}-y^{2}}{(x^{2}+y^{2})^{2}}$ appearing in the solution of $T_{\xi\to0}^{U}(x,y)$ (upper-left panel) and $\frac{\Delta T}{\pi}\frac{y}{x^{2}+y^{2}}$ appearing in the solution of $T_{\xi\to0}^{\Delta T}(x,y)$ (upper-right panel), together with their related 2D contours in (lower-left panel) and (lower-right panel), respectively (see Eqs. \eqref{T_U_xi_0} and \eqref{T_DeltaT_xi_0}). In the lower-left panel one can clearly spot the two nodal lines along $y=x$ and $y=-x$.}
\label{fig:backflow_SI_1}
\end{figure}
So we can split the integral as follows:
\begin{equation}
\begin{split}
T_{\xi\to0}^{U}(x,y)=\frac{1}{2\pi}U\frac{\eta}{\beta}\Bigg[&-4\int_{-\infty}^{0} dk\,e^{ikx}k\Big[\cosh\left(ky\right)+\sinh\left(ky\right)\Big]+\\
&+4\int_{0}^{\infty} dk\,e^{ikx}k\Big[\cosh\left(ky\right)-\sinh\left(ky\right)\Big]\Bigg].
\end{split}
\end{equation}
Using the substitution $ky=z$ (which also gives $ydk=dz$ and $k=\frac{z}{y}$) we get
\begin{equation}
\resizebox{\textwidth}{!}{$
\begin{split}
T_{\xi\to0}^{U}(x,y)&=\frac{U}{2\pi}\frac{\eta}{\beta}\frac{1}{y^{2}}\left[-4\int_{-\infty}^{0} dz\,e^{i\frac{x}{y}z}z\Big[\cosh\left(z\right)+\sinh\left(z\right)\Big]+4\int_{0}^{\infty} dz\,e^{i\frac{x}{y}z}z\Big[\cosh\left(z\right)-\sinh\left(z\right)\Big]\right]=\\
&=\frac{U}{2\pi}\frac{\eta}{\beta}\frac{1}{y^{2}}\left[4\int_{0}^{-\infty} dz\,e^{i\frac{x}{y}z}z\Big[\cosh\left(z\right)+\sinh\left(z\right)\Big]+4\int_{0}^{\infty} dz\,e^{i\frac{x}{y}z}z\Big[\cosh\left(z\right)-\sinh\left(z\right)\Big]\right]=\\
&=\frac{U}{2\pi}\frac{\eta}{\beta}\frac{1}{y^{2}}\left[4\int_{0}^{\infty}dz\,e^{-i\frac{x}{y}z}z\Big[\cosh\left(-z\right)+\sinh\left(-z\right)\Big]+4\int_{0}^{\infty} dz\,e^{i\frac{x}{y}z}z\Big[\cosh\left(z\right)-\sinh\left(z\right)\Big]\right].
\end{split}$}
\end{equation}
By further manipulating we get
\begin{equation} \label{T_U_xi_0}
\resizebox{\textwidth}{!}{$
\begin{split}
T_{\xi\to0}^{U}(x,y)&=\frac{4U}{2\pi}\frac{\eta}{\beta}\frac{1}{y^{2}}\int_{0}^{\infty}dz\Bigg[\left(e^{-i\frac{x}{y}z}+e^{i\frac{x}{y}z}\right)z\Big[\cosh\left(z\right)-\sinh\left(z\right)\Big]\Bigg]=\\
&=\frac{U}{2\pi}\frac{\eta}{\beta}\frac{1}{y^{2}}\int_{0}^{\infty}dz\left[8\cos\left(\frac{x}{y}z\right)z\Big[\cosh\left(z\right)-\sinh\left(z\right)\Big]\right]=\\
&=\frac{U}{2\pi}\frac{\eta}{\beta}\frac{1}{y^{2}}\left|\frac{8\Big[\cosh\left(z\right)-\sinh\left(z\right)\Big]\left[\frac{x}{y}\left(\frac{x^{2}}{y^{2}}z+z+2\right)\sin\left(\frac{x}{y}z\right)-\left(\frac{x^{2}}{y^{2}}(z-1)+z+1\right)\cos\left(\frac{x}{y}z\right)\right]}{\left(\frac{x^{2}}{y^{2}}+1\right)^{2}}\right|^{\infty}_{0}=\\
&=\frac{4U}{\pi}\frac{\eta}{\beta}\frac{1}{y^{2}}\frac{\frac{x^{2}}{y^{2}}-1}{\left(\frac{x^{2}}{y^{2}}+1\right)^{2}}=\frac{4U}{\pi}\frac{\eta}{\beta}\frac{1}{y^{2}}\frac{\frac{1}{y^{2}}(x^{2}-y^{2})}{\left[\frac{1}{y^{2}}(x^{2}+y^{2})\right]^{2}}=\frac{4U}{\pi}\frac{\eta}{\beta}\frac{x^{2}-y^{2}}{(x^{2}+y^{2})^{2}}.
\end{split}$}
\end{equation}
On the other hand, using Eq. \eqref{properties_hyperbolic_functions}, $T_{\xi\to0}^{\Delta T}(x,y)$ becomes
\begin{equation}
\begin{split} 
T_{\xi\to0}^{\Delta T}(x,y)&=-\frac{1}{2\pi}\int dk\,e^{ikx}T_{{\rm{bc}}}\frac{\sinh\left(ky-k\frac{h}{2}\right)}{\sinh\left(k\frac{h}{2}\right)}=\frac{1}{2\pi}\int dk\,e^{ikx}T_{{\rm{bc}}}\left[\cosh(ky)-\frac{\sinh\left(ky\right)}{\tanh\left(k\frac{h}{2}\right)}\right].
\end{split}
\end{equation}
Taking into account the explicit form of $T_{{\rm{bc}}}$ and approximating $\tanh\left(k\frac{h}{2}\right)\approx\text{sign}(kh)$ we can split the integral as
\begin{equation} \label{T_DeltaT_xi_0}
\resizebox{\textwidth}{!}{$
\begin{split}
T_{\xi\to0}^{\Delta T}(x,y)&=\frac{\Delta T}{2\pi}\left[\int_{-\infty}^{0} dk\,e^{ikx}\Big[\cosh(ky)+\sinh(ky)\Big]+\int_{0}^{\infty} dk\,e^{ikx}\Big[\cosh(ky)-\sinh(ky)\Big]\right]=\\
&=\frac{\Delta T}{2\pi}\left[-\int_{0}^{-\infty} dk\,e^{ikx}\Big[\cosh(ky)+\sinh(ky)\Big]+\int_{0}^{\infty} dk\,e^{ikx}\Big[\cosh(ky)-\sinh(ky)\Big]\right]=\\
&=\frac{\Delta T}{2\pi}\left[\int_{0}^{\infty}dk\,e^{-ikx}\Big[\cosh(-ky)+\sinh(-ky)\Big]+\int_{0}^{\infty} dk\,e^{ikx}\Big[\cosh(ky)-\sinh(ky)\Big]\right]=\\
&=\frac{\Delta T}{2\pi}\int_{0}^{\infty}dk\,\left(e^{-ikx}+e^{ikx}\right)\Big[\cosh(ky)-\sinh(ky)\Big]=\\
&=\frac{\Delta T}{\pi}\int_{0}^{\infty}dk\,\cos(kx)\Big[\cosh(ky)-\sinh(ky)\Big]=\\
&=\frac{\Delta T}{\pi}\left|\frac{\big[\cosh(ky)-\sinh(ky)\big]\big[x\sin(kx)-y\cos(kx)\big]}{x^{2}+y^{2}}\right|_{0}^{\infty}=-\frac{\Delta T}{\pi}\frac{y}{x^{2}+y^{2}}.
\end{split}$}
\end{equation}
So finally we have
\begin{equation} \label{final_trend_T_xi_to_0}
\begin{split}
T_{\xi\to0}(x,y)\approx\frac{4U}{\pi}\frac{\eta}{\beta}\frac{x^{2}-y^{2}}{(x^{2}+y^{2})^{2}}-\frac{\Delta T}{\pi}\frac{y}{x^{2}+y^{2}}.
\end{split}
\end{equation}
From Eq. \eqref{final_trend_T_xi_to_0} we see that, in the vicinity of the leads ($x\to0$), the drift velocity term $T_{\xi\to0}^{U}(x,y)$ behaves as $\sim-y^{-2}$. This dependence is much stronger than the temperature gradient one ($T_{\xi\to0}^{\Delta T}(x,y)\sim-y^{-1}$) when approaching e.g. the boundary $y=0$ (see Fig. \ref{fig:backflow_SI_1}). Moreover, we finally see that when the drift velocity term dominates over the temperature gradient one, that is when the effect of shear viscosity and boundary condition $U$ is maximized, then we find an inverse-square relationship with the distance from the contacts, also indicating the existence of two nodal lines along $y=x$ and $y=-x$ relative to the central heat flow path (see Fig \ref{fig:backflow_SI_1}b). This result is consistent with that found in Ref. \cite{levitov2016electron} for the incompressible electron fluid.\\ Furthermore, we can also consider the scenario where no drift velocity is injected transversely into the strip device, and relax the condition of vanishing $u_{x}$. In such cases one can show that the modified boundary conditions can alter the response near the leads while leaving it unchanged at greater distances \cite{levitov2016electron}.

\subsubsection{Compressibility and vorticity contributions to the temperature profile for \texorpdfstring{$\xi\to0$}{xi to 0}}

In the limit of $\xi\to0$ the compressibility and vorticity contributions to the temperature profile read:
\begin{equation} \label{T_phi_xi_to_0}
\resizebox{\textwidth}{!}{$
T_{\phi}(x,y)=\frac{1}{2\pi}\int dk\,e^{ikx}\left[\frac{T_{{\rm{bc}}}}{1-e^{kh}}\left(e^{ky}-e^{kh}e^{-ky}\right)+\frac{\gamma}{\beta}U\frac{1 - e^{kh}}{1-e^{q_{\phi}h}}\frac{q_{\psi}}{k}G(q_{\psi},k)\left(e^{q_{\phi}y}-e^{q_{\phi}h}e^{-q_{\phi}y}\right)\right]
$}
\end{equation}
and
\begin{equation} \label{T_psi_xi_to_0}
T_{\psi}(x,y)=\frac{1}{2\pi}\frac{\gamma}{\beta}U\int dk\,e^{ikx}\frac{q_{\psi}}{k}G(q_{\psi},k)\left(e^{ky}-e^{kh}e^{-ky}\right).
\end{equation}
Note again that $T_{\psi}$ resembles the expression of the electrical potential of an (incompressible) electron fluid \cite{levitov2016electron} (see Eq. 22 of the related Supplementary Information). Most importantly, as it will be evident later in the text, $T_{\psi}$ in Eq. \eqref{T_psi_xi_to_0} coincides with the temperature profile obtained in the incompressible limit (see Eq. \eqref{expression_for_temperature_final_inc}).

\subsubsection{Temperature profile for \texorpdfstring{$\epsilon\to0$}{epsilon to 0}} 

It is instructive to examine the temperature profile by first taking the limit $\epsilon\to0$, to clearly demonstrate the necessity of accounting for the parameter $\xi$ defined in Eq. \eqref{parameter_xi}. In this case, we have
\begin{equation} \label{final_temperature_viscous}
\begin{split}
T_{\epsilon\to0}(x,y)=\frac{1}{2\pi}\int dk\,e^{ikx}\Bigg[&\left(e^{ky}-e^{kh}e^{-ky}\right)\left(\frac{T_{{\rm{bc}}}}{1-e^{kh}}+\frac{\alpha}{\kappa}UF(q_{\phi},k)\right)+\\
&+\frac{\alpha}{\kappa}U\left(e^{q_{\phi}y}-e^{q_{\phi}h}e^{-q_{\phi}y}\right)\frac{1-e^{kh}}{1-e^{q_{\phi}h}}F(q_{\phi},k)\Bigg]
\end{split}
\end{equation}
It is worth noting that the terms proportional to $U$ in Eq. \eqref{final_temperature_viscous} can be rewritten as:
\begin{equation} \label{final_temperature_viscous_bis}
\frac{1}{\xi}F(q_{\phi},k)\left(e^{ky}-e^{kh}e^{-ky}\right)+\frac{1}{\xi}\frac{1-e^{kh}}{1-e^{q_{\phi}h}}F(q_{\phi},k)\left(e^{q_{\phi}y}-e^{q_{\phi}h}e^{-q_{\phi}y}\right),
\end{equation}
from which we clearly see that for $\xi\to0$ viscous effects are maximized (together with $\epsilon\to0$ this indeed translates into having $\rm{FDN}\to\infty$). On the other hand, for $\xi\to\infty$ we get
\begin{equation} \label{T_epsilon_to_0_xi_to_infinity}
T_{\substack{\epsilon\to0\\\xi\to\infty}}(x,y)=\frac{1}{2\pi}\int dk\,e^{ikx}\frac{T_{{\rm{bc}}}}{1-e^{kh}}\left(e^{ky}-e^{kh}e^{-ky}\right),
\end{equation}
that is precisely the temperature profile of Fourier's diffusive regime discussed in Appendix \ref{fourier_temperature_from_scratch_section}. This is to say that, in order to keep having a viscous contribution to thermal transport, both $\epsilon$ and $\xi$ have to be small, as prescribed by the FDN \eqref{FDN_apendices}. \\
The viscous behavior of the phonon fluid translates into a sign-changing thermal response in the vicinity of the strip leads. In this case, $T(x,0)-T(x,h)$ is given by the expression
\begin{equation} \label{temp_diff_epsilon_to_0}
\resizebox{\textwidth}{!}{$
\begin{split}
T_{\epsilon\to0}(x,0)-T_{\epsilon\to0}(x,h)&=\frac{1}{2\pi}\int dk\,e^{ikx}\,2\left[\Delta T+2\frac{\alpha}{\kappa}U(1-e^{kh})F(q_{\phi},k)\right]=\\
&=\frac{1}{2\pi}\int dk\,e^{ikx}\,2\left[\Delta T+2\frac{\alpha}{\kappa}U\frac{(1-e^{kh})(1-e^{q_{\phi}h})}{k(1-e^{q_{\phi}h })(1+e^{kh})-q_{\phi}(1+e^{q_{\phi}h })(1-e^{kh})}\right]=\\
&=\frac{1}{\pi}\int dk\,e^{ikx}\left[\Delta T-2\frac{\alpha}{\kappa}U\frac{(1-e^{kh})(1-e^{q_{\phi}h})}{(q_{\phi}+k)(e^{q_{\phi}h}-e^{kh})-(q_{\phi}-k)(e^{(q_{\phi}+k)h}-1)}\right].
\end{split}$}
\end{equation}
As done for the $\xi\to0$ limit discussed above, here we can write the denominator of the second term in Eq. \eqref{temp_diff_epsilon_to_0} as
\begin{equation}
\begin{split}
(q_{\phi}+k)(e^{q_{\phi}h}-e^{kh})-(q_{\phi}-k)(e^{(q_{\phi}+k)h}-1)=2e^{kh}(q_{\phi}-k)\Big(kh-\sinh(kh)\Big).
\end{split}
\end{equation}
Then we have
\begin{equation}
\begin{split}
-2\frac{\alpha}{\kappa}&=-2\frac{\alpha\beta}{\kappa(\eta+\mu^{*})}\frac{\eta+\mu^{*}}{\beta}=-2\frac{\eta+\mu^{*}}{\beta}(q_{\phi}^{2}-k^{2})=-2\frac{\eta+\mu^{*}}{\beta}(q_{\phi}-k)(q_{\phi}+k)=\\
&=-2\frac{\eta+\mu^{*}}{\beta}(q_{\phi}-k)(q_{\phi}-k+2k)=-2\frac{\eta+\mu^{*}}{\beta}\Big[(q_{\phi}-k)^{2}+2k(q_{\phi}-k)\Big]\sim\\
&\sim-4\frac{\eta+\mu^{*}}{\beta}k(q_{\phi}-k)
\end{split}
\end{equation}
which is true in the limit of $\epsilon\to0$. We also have
\begin{equation}
\begin{split}
(1-e^{q_{\phi}h})(1-e^{kh})\sim2e^{kh}\tanh\left(k\frac{h}{2}\right)\sinh(kh),
\end{split}
\end{equation}
so to linear order in $q_{\phi}-k$, the numerator of the second term in Eq. \eqref{temp_diff_epsilon_to_0} simply becomes
\begin{equation}
\begin{split}
-2\frac{\alpha}{\kappa}(1-e^{q_{\phi}h})(1-e^{kh})&=-4\frac{\eta+\mu^{*}}{\beta}k(q_{\phi}-k)(1-e^{q_{\phi}h})(1-e^{kh})=\\
&=-8\frac{\eta+\mu^{*}}{\beta}k(q_{\phi}-k)e^{kh}\tanh\left(k\frac{h}{2}\right)\sinh(kh)
\end{split}
\end{equation}
Merging everything together we get
\begin{equation} \label{diff_T_epsilon_0}
\begin{split}
T(x,0)-T(x,h)&\xrightarrow[(q_{\phi}-k)\to0]{}\frac{1}{\pi}\int dk\,e^{ikx}\left[\Delta T-4U\frac{\eta+\mu^{*}}{\beta}\frac{k\tanh\left(k\frac{h}{2}\right)\sinh(kh)}{kh-\sinh(kh)}\right].
\end{split}
\end{equation}
The temperature difference across the strip is a Fourier transform of an even function in $k$ and it only assumes real values. We can see that the integrated function is always positive $\forall k$.
\begin{figure}[!htb]
\centering
\includegraphics[width=0.75\textwidth]{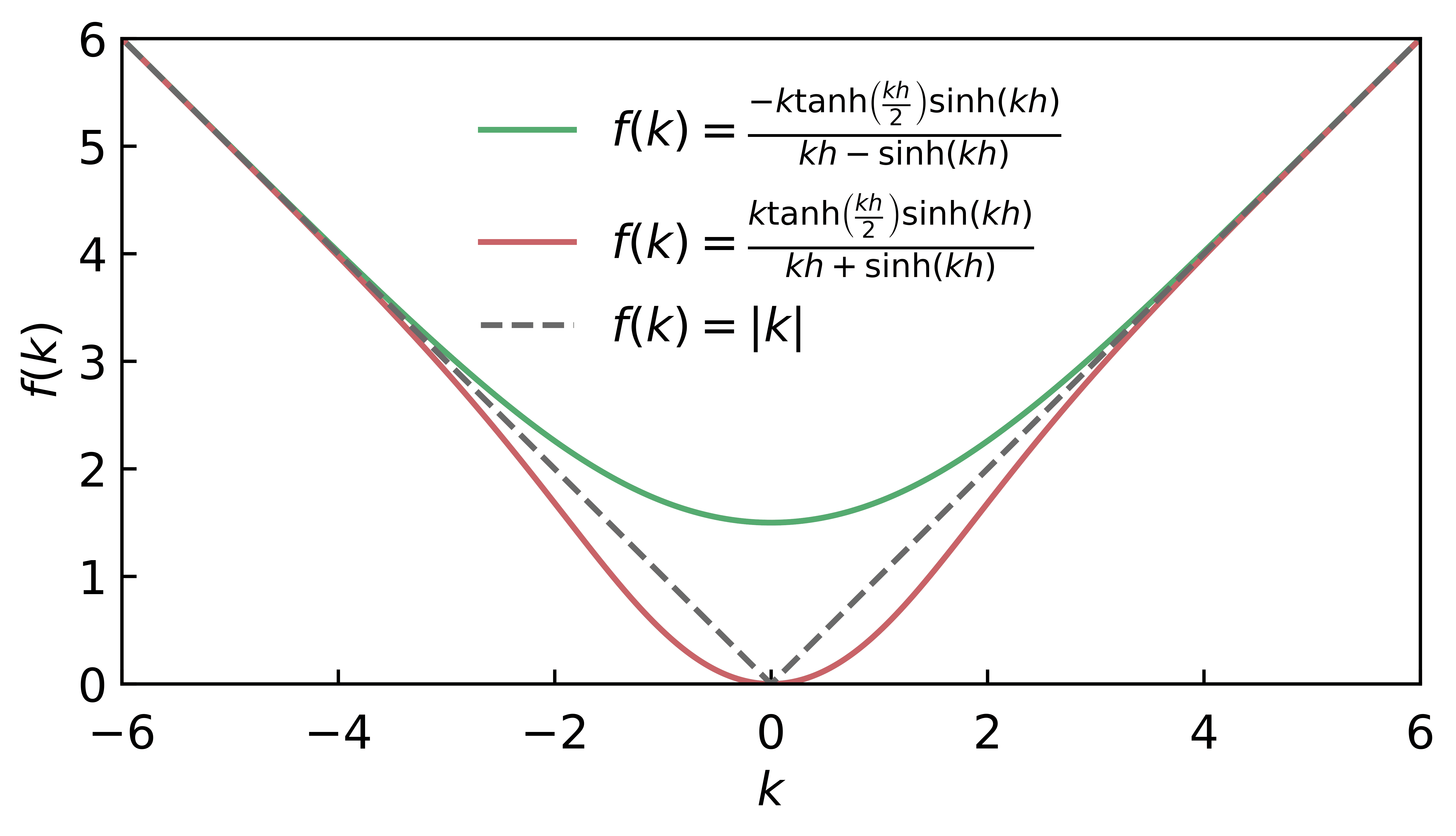}
\caption{\textbf{Integrands governing the temperature difference across the strip.} Plots of Eqs. \eqref{epsilon_to_zero_integrand} (green), \eqref{identity} (red), together with their $|k|$ asymptotic limit. In this example $h=2$.}
\label{fig:f_k_functions}
\end{figure}
Furthermore, the second term in square brackets of Eq. \eqref{diff_T_epsilon_0} is
\begin{equation}  \label{epsilon_to_zero_integrand}
\begin{split}
-\frac{k\tanh\left(k\frac{h}{2}\right)\sinh(kh)}{kh-\sinh(kh)}\xrightarrow[k\to\infty]{}|k|,
\end{split}
\end{equation}
In Fig. \ref{fig:f_k_functions} we compare the functions given in Eqs. \eqref{identity} and \eqref{epsilon_to_zero_integrand}. One can see that the same conclusions followed to derive Eq. \eqref{xi_0_temp_diff_limit} apply here, however, we can motivate the much weaker influence on the thermal response of the $\epsilon$ parameter as due to the larger deviations from the asymptote $|k|$ for sufficiently large $x$ values.

\subsubsection{Compressibility and vorticity contributions to the temperature profile for \texorpdfstring{$\epsilon\to0$}{epsilon to 0}}

In the limit of $\epsilon\to0$ the compressibility and vorticity contributions to the temperature profile read:
\begin{equation} \label{T_phi_epsilon_to_0}
\resizebox{\textwidth}{!}{$
T_{\phi}(x,y)=\frac{1}{2\pi}\int dk\,e^{ikx}\left[\frac{T_{{\rm{bc}}}}{1-e^{kh}}\left(e^{ky}-e^{kh}e^{-ky}\right)+\frac{\alpha}{\kappa}U\frac{1-e^{kh}}{1-e^{q_{\phi}h}}F(q_{\phi},k)\left(e^{q_{\phi}y}-e^{q_{\phi}h}e^{-q_{\phi}y}\right)\right]
$}
\end{equation}
and 
\begin{equation} \label{T_psi_epsilon_to_0}
T_{\psi}(x,y)=\frac{1}{2\pi}\frac{\alpha}{\kappa}U\int dk\,e^{ikx}F(q_{\phi},k)\left(e^{ky}-e^{kh}e^{-ky}\right).
\end{equation}
Importantly, as it will be evident later in the text, $T_{\phi}$ in Eq. \eqref{T_phi_epsilon_to_0} coincides with the temperature profile obtained in the irrotational limit (see Eq. \eqref{expression_for_temperature_irr_final}).

\subsection{Fourier's law solution} \label{fourier_temperature_from_scratch_section}

We can further validate our analysis by directly solving Fourier's law of heat conduction and showing that the result coincides with that of Eq. \eqref{final_temperature_Fourier}. The VHE reduce to Fourier's law when heat dissipation due to non-conservation of crystal momentum upon scattering dominates. In this case we are left with
\begin{equation}
\nabla^{2}T(x,y)=0.
\end{equation}
Fourier's law in $k$-space assumes the following form:
\begin{equation}
\left(-k^{2}+\frac{\partial^{2}}{\partial y^{2}}\right)T(k,y)=0.
\end{equation}
This is a simple second order differential equation with solution given by:
\begin{equation} \label{fourier_Tk(y)}
T(k,y)=A_{k}e^{ky}+B_{k}e^{-ky}.
\end{equation}
By recalling boundary conditions \eqref{temperature_BC_y=0,h} and the symmetry of the temperature profile \eqref{symmetry_of_temperature}
\begin{equation} \label{symmetry_of_temperature_bis}
T(x,h-y)=-T(x,y)
\end{equation}
we have
\begin{equation}
\begin{cases}
T(k,y)=A_{k}e^{ky}+B_{k}e^{-ky}\\
T(k,h-y)=A_{k}e^{kh}e^{-ky}+B_{k}e^{-kh}e^{ky}.
\end{cases}
\end{equation}
Now we see that Eq. \eqref{symmetry_of_temperature_bis} is verified only if 
\begin{equation} \label{relations_between_coefficients_fourier}
B_{k}=-A_{k}e^{kh}.
\end{equation}
So taking into account the symmetry of the solution through Eq. \eqref{relations_between_coefficients_fourier} we can write Eq. \eqref{fourier_Tk(y)} as follows:
\begin{equation} \label{fourier_Tk(y)_final}
T(k,y)=A_{k}\left(e^{ky}-e^{kh}e^{-ky}\right)
\end{equation}
Now we want to exploit the boundary condition at $y=0$:
\begin{equation} 
T(k,0)=\int_{-\infty}^{\infty} dx\,e^{-ikx}T(x,0)=\int_{-\infty}^{\infty} dx\,e^{-ikx}(\bar{T}+\Delta T\delta(x))=\bar{T}\delta(k)+\Delta T=T_{{\rm{bc}}}.
\end{equation}
By using this boundary condition on temperature we have
\begin{equation}
T(k,0)=A_{k}\left(1-e^{kh}\right)=T_{{\rm{bc}}},
\end{equation}
from which we get
\begin{equation}
A_{k}=\frac{T_{{\rm{bc}}}}{1-e^{kh}}.
\end{equation}
By plugging $A_{k}$ into the solution \eqref{fourier_Tk(y)_final} we have
\begin{equation} 
T(k,y)=\frac{T_{{\rm{bc}}}}{1-e^{kh}}\left(e^{ky}-e^{kh}e^{-ky}\right).
\end{equation}
So finally the temperature profile obtained by solving Fourier's law is given by
\begin{equation} \label{fourier_T_result}
T(x,y)=\frac{1}{2\pi}\int dk e^{ikx}\frac{T_{{\rm{bc}}}}{1-e^{kh}}\left(e^{ky}-e^{kh}e^{-ky}\right),
\end{equation}
which coincides with the result obtained in Eq. \eqref{final_temperature_Fourier}. Using Eq. \eqref{fourier_T_result} the temperature difference across the strip is:
\begin{equation}
T(x,0)-T(x,h)=\frac{1}{2\pi}\int dk e^{ikx}2\Delta T=2\Delta T\delta(x),
\end{equation}
from which it is evident that thermal backflow never arises in the diffusive Fourier's regime.

\subsection{Incompressible limit} \label{incompressible_limit}

In this section we show that in the incompressible limit the phonon fluid follows the same equation which describe charge transport in electronic viscous fluids. Due to the incompressible flow condition, $\Phi=\nabla\cdot\boldsymbol{u}=0$ (in electron systems charge transport is described by an incompressible flow when drift velocities are smaller than plasmonic velocities \cite{levitov2016electron}), we know that the drift velocity is defined by means of a single stream function $\psi$:
\begin{equation} \label{potential_u_incrompressible}
\boldsymbol{u}=\nabla\times\boldsymbol{\Psi}=\left(\frac{\partial\psi}{\partial y}\,,\,-\frac{\partial\psi}{\partial x}\right).
\end{equation}
In fact, this condition makes the first of Eq. \eqref{potential_VHE_final} irrelevant and the equation to solve is simply 
\begin{equation}
(\partial^{2}_{y}-k^{2})(\partial^{2}_{y}-q_{\psi}^{2})\psi(k,y)=0.
\end{equation}
The solution obtained by using boundary conditions on the drift velocity components is
\begin{equation}
\begin{split}
\psi(k,y)&=iU\frac{\frac{q_{\psi}}{k}(1-e^{q_{\psi}h})\left(e^{ky}+e^{kh}e^{-ky}\right)-(1-e^{kh})\left(e^{q_{\psi}y}+e^{q_{\psi}h}e^{-q_{\psi}y}\right)}{q_{\psi}(1-e^{q_{\psi}h})(1+e^{kh})-k(1+e^{q_{\psi}h})(1-e^{kh})}.
\end{split}
\end{equation}
We can use the expression of the stream function $\psi$ to find the coordinates of the points $(x_{0},y_{0})$ within the strip around which the streamlines circulate. The $y_{0}$ coordinate is $y=\frac{h}{2}$, where the $u_{x}$ component of the drift velocity changes its direction by symmetry. Then, we can obtain the value of $x_{0}$ by imposing the condition
\begin{equation}
\left.\frac{\partial\psi\left(x,\frac{h}{2}\right)}{\partial x}\right|_{x=x_{0}}=0
\end{equation}
on the stream function. We have
\begin{equation} 
\psi\left(x,\frac{h}{2}\right)=\frac{iU}{\pi}\int dk\,e^{ikx}G(q_{\psi},k)\left[\frac{q_{\psi}}{k}e^{k\frac{h}{2}}-\frac{1-e^{kh}}{1-e^{q_{\psi}h}}e^{q_{\psi}\frac{h}{2}}\right],
\end{equation}
and
\begin{equation} \label{equation_for_vortex_center}
\begin{split}
\left.\frac{\partial\psi\left(x,\frac{h}{2}\right)}{\partial x}\right|_{x=x_{0}}&=\left.\frac{U}{\pi}\int dk\,\cos(kx)G(q_{\psi},k)\left[k\frac{1-e^{kh}}{1-e^{q_{\psi}h}}e^{q_{\psi}\frac{h}{2}}-q_{\psi}e^{k\frac{h}{2}}\right]\right|_{x=x_{0}}=\\
&=\left.\frac{U}{\pi}\int dk\,\cos(kx)\frac{k(1-e^{kh})e^{q_{\psi}\frac{h}{2}}-q_{\psi}(1-e^{q_{\psi}h })e^{k\frac{h}{2}}}{k(1 - e^{kh}) (1 + e^{q_{\psi}h })-q_{\psi}(1 + e^{kh}) (1 - e^{q_{\psi}h })}\right|_{x=x_{0}}=\\
&=\left.\frac{U}{\pi}\int dk\,\cos(kx)\frac{k(1-e^{kh})e^{q_{\psi}\frac{h}{2}}-q_{\psi}(1-e^{q_{\psi}h })e^{k\frac{h}{2}}}{(q_{\psi}+k)(e^{q_{\psi}h}-e^{kh})+(q_{\psi}-k)(e^{(k+q_{\psi})h}-1)}\right|_{x=x_{0}}=0,
\end{split}
\end{equation}
whose solution is obtained numerically and is given by
\begin{equation}
x_{0}\approx\pm h.
\end{equation}

\subsubsection{Temperature in the incompressible limit}

According to the incompressible flow condition, the second VHE \eqref{VHE} becomes
\begin{equation}
\beta\nabla T-\eta\nabla^{2}\boldsymbol{u}=-\gamma\boldsymbol{u}. 
\end{equation}
So, using Eq. \eqref{potential_u_incrompressible} we have
\begin{equation} \label{relation_T_u_inc}
\begin{split}
\nabla T&=\left(-\frac{\gamma}{\beta}+\frac{\eta}{\beta}\nabla^{2}\right)\left[\frac{\partial\psi}{\partial y}\,,\,-\frac{\partial\psi}{\partial x}\right]=-\frac{\gamma}{\beta}\left[\frac{\partial\psi}{\partial y}\,,\,-\frac{\partial\psi}{\partial x}\right]+\frac{\eta}{\beta}\left[\frac{\partial^{3}\psi}{\partial x^{2}\partial y}+\frac{\partial^{3}\psi}{\partial y^{3}}\,,\,-\frac{\partial^{3}\psi}{\partial x^{3}}-\frac{\partial^{3}\psi}{\partial x\partial y^{2}}\right].
\end{split}
\end{equation}
Again, after performing the algebra we get
\begin{equation} \label{expression_for_temperature_inc}
\begin{split}
T(k,y)=-i\frac{\gamma}{\beta}\left(e^{kh}e^{-ky}-e^{ky}\right)a_{\psi+}.
\end{split}
\end{equation}
Note that from Eq. \eqref{expression_for_temperature_inc} we can uncover the symmetry of the temperature solution:
\begin{equation} \label{symmetry_of_temperature_inc}
\begin{split}
T(k,h-y)&=-i\frac{\gamma}{\beta}\left(e^{ky}-e^{kh}e^{-ky}\right)a_{\psi+}=-T(k,y),
\end{split}
\end{equation}
from which it is clearly seen that the temperature is antisymmetric with respect to the horizontal axis that divides the strip in half. By applying boundary conditions we have
\begin{equation} \label{expression_for_temperature_final_inc}
\begin{split}
T(x,y)&=\frac{1}{2\pi}\int dk\,e^{ikx}T(k,y)=\frac{i}{2\pi}\frac{\gamma}{\beta}\int dk\,e^{ikx}\left(e^{ky}-e^{kh}e^{-ky}\right)a_{\psi+}=\\
&=\frac{1}{2\pi}\frac{\gamma}{\beta}U\int dk\,e^{ikx}\frac{q_{\psi}}{k}G(q_{\psi},k)\left(e^{ky}-e^{kh}e^{-ky}\right).
\end{split}
\end{equation}
So we clearly see that by simply injecting a phonon drift velocity we manage to get a thermal flow inside the strip if the phonon flow is incompressible. From this result we easily see that Eq. \eqref{expression_for_temperature_final_inc} is identical to the $\xi\to0$ limit of the vorticity contribution $T_{\psi}$ given in Eq. \eqref{T_psi_xi_to_0} for the general compressible phonon fluid. Moreover, knowing that $a_{\psi-}=a_{\psi+}e^{kh}$ by symmetry, Eq. \eqref{expression_for_temperature_final_inc} can be also written as
\begin{equation}
\begin{split}
T(x,y)=\frac{i}{2\pi}\frac{\gamma}{\beta}\int dk\,e^{ikx}\left(e^{ky}a_{\psi+}-e^{-ky}a_{\psi-}\right).
\end{split}
\end{equation}
The form of this last result is identical to that obtained for the electrical potential of the electron fluid in Ref. \cite{levitov2016electron} (see Eq. 22 of the related Supplementary Information). Hence, this further validates our protocol and shows how the flow of an electron fluid can be seen as the analogue of the incompressible limit of the phonon fluid. As done in Eq. \eqref{temp_diff_xi_to_0} for the general compressible case in the $\epsilon\to0$ limit, here we have
\begin{equation} \label{temp_diff_inc}
\begin{split}
T(x,0)-T(x,h)&=\frac{1}{\pi}\frac{\gamma}{\beta}U\int dk\,e^{ikx}\frac{q_{\psi}}{k}G(q_{\psi},k)\left(1-e^{kh}\right)=\\
&=\frac{1}{\pi}\frac{\gamma}{\beta}U\int dk\,e^{ikx}\frac{q_{\psi}}{k}\frac{(1-e^{q_{\psi}h})(1-e^{kh})}{(q_{\psi}+k)(e^{q_{\psi}h}-e^{kh})+(q_{\psi}-k)(e^{(q_{\psi}+k)h}-1)},
\end{split}
\end{equation}
which is the analogue of Eq. 23 of the Supplementary Information of Ref. \cite{levitov2016electron}. \\
Finally, with the same reasoning followed to manipulate Eq. \eqref{temp_diff_xi_to_0} for the present incompressbile case we get
\begin{equation}
\begin{split}
T(x,0)-T(x,h)\xrightarrow[(q_{\psi}-k)\to0]{}\frac{1}{\pi}\int dk\,e^{ikx}2U\frac{\eta}{\beta}\frac{k\tanh\left(k\frac{h}{2}\right)\sinh(kh)}{kh+\sinh(kh)},
\end{split}
\end{equation}
leading to a negative thermal resistance 
\begin{equation}
T(x,0)-T(x,h)\xrightarrow[|k|\gg\frac{1}{h}]{}-4U\frac{\eta}{\beta}\frac{1}{x^{2}}.
\end{equation}

\subsection{Irrotational limit} \label{irrotational_limit}

By imposing the irrotational flow condition $\boldsymbol{\mathscr{W}}=\nabla\times\boldsymbol{u}=0$, we know that the drift velocity is defined through only one potential, this time it is the velocity potential $\phi$:
\begin{equation} \label{potential_u_irrotational}
\boldsymbol{u}=-\nabla\phi=\left(-\frac{\partial\phi}{\partial x}\,,\,-\frac{\partial\phi}{\partial y}\right).
\end{equation}
This condition makes the second of Eq. \eqref{potential_VHE_final} irrelevant and the equation to solve is simply 
\begin{equation}
(\partial^{2}_{y}-k^{2})(\partial^{2}_{y}-q_{\phi}^{2})\phi(k,y)=0.
\end{equation}
The solution obtained applying boundary conditions on drift velocity is
\begin{equation}
\begin{split}
\phi(k,y)=U\frac{(1-e^{kh})(e^{q_{\phi}y}-e^{q_{\phi}h}e^{-q_{\phi}y})-(1-e^{q_{\phi}h})(e^{ky}-e^{kh}e^{-ky})}{k(1+e^{kh})(1-e^{q_{\phi}h})-q_{\phi}(1+e^{q_{\phi}h})(1-e^{kh})},
\end{split}
\end{equation}

\subsubsection{Temperature in the irrotational limit} \label{section_temp_irr}

According to the irrotational flow condition, the second of Eq. \eqref{VHE} is irrelevant and we are simply left with the first
\begin{equation}
\nabla^{2}T=\frac{\alpha}{\kappa}\nabla\cdot\boldsymbol{u}
\end{equation}
So, using Eq. \eqref{potential_u_irrotational} we have
\begin{equation} \label{relation_T_u_irr}
\begin{split}
\nabla^{2}T=\frac{\partial^{2}T(x,y)}{\partial x^{2}}+\frac{\partial^{2}T(x,y)}{\partial y^{2}}=\frac{\alpha}{\kappa}\nabla\cdot\left[-\frac{\partial\phi}{\partial x}\,,\,-\frac{\partial\phi}{\partial y}\right]=\frac{\alpha}{\kappa}\left[-\frac{\partial^{2}\phi}{\partial x^{2}}-\frac{\partial^{2}\phi}{\partial y^{2}}\right]=-\frac{\alpha}{\kappa}\nabla^{2}\phi.
\end{split}
\end{equation}
Performing the algebra we have
\begin{equation} \label{expression_for_temperature_irr}
\begin{split}
T(k,y)=-\frac{\alpha}{\kappa}\phi(k,y)=-\frac{\alpha}{\kappa}\Big[\left(e^{ky}-e^{kh}e^{-ky}\right)a_{\phi+}+\left(e^{q_{\phi}y}-e^{q_{\phi}h}e^{-q_{\phi}y}\right)b_{\phi+}\Big],
\end{split}
\end{equation}
and applying boundary conditions on the drift velocity components we get
\begin{equation} 
\begin{split}
T(k,y)=\frac{U\alpha}{\kappa}F(q_{\phi},k)\left[\Big(e^{ky}-e^{kh}e^{-ky}\Big)-\frac{1-e^{kh}}{1-e^{q_{\phi h}}}\Big(e^{q_{\phi}y}-e^{q_{\phi}h}e^{-q_{\phi}y}\Big)\right].
\end{split}
\end{equation}
So we clearly see that, contrary to the incompressible case, by simply injecting a phonon drift velocity, when computing the thermal response across the strip we end up having the same temperature profile at $y=0$ and $y=h$, which translates into no signature of either thermal flow or backflow. This means that a different set of boundary conditions with at least one representing a temperature gradient is needed to have thermal flow.\\
From Eq. \eqref{expression_for_temperature_irr} we can show that the irrotational temperature reduces to the compressibility part of the temperature profile of the general compressible case given in Eq. \eqref{T_phi_epsilon_to_0}. In fact, in the limit of dominating viscous effects one can choose to further impose the boundary condition \eqref{temperature_BC_y=0,h}
\begin{equation}
\begin{split}
T_{{\rm{bc}}}=-\frac{\alpha}{\kappa}\left(1-e^{kh}\right)a_{\phi+}\,\,\,\,\rightarrow\,\,\,\,a_{\phi+}=-\frac{\kappa}{\alpha}\frac{T_{{\rm{bc}}}}{1-e^{kh}}.
\end{split}
\end{equation}
Then, substituting this in Eq. \eqref{expression_for_temperature_irr} leads to 
\begin{equation} \label{expression_for_temperature_irr_final}
T(x,y)=\frac{1}{2\pi}\int dk\,e^{ikx}\left[\frac{T_{{\rm{bc}}}}{1-e^{kh}}\left(e^{ky}-e^{kh}e^{-ky}\right)+\frac{\alpha}{\kappa}U\frac{1-e^{kh}}{1-e^{q_{\phi}h}}F(q_{\phi},k)\left(e^{q_{\phi}y}-e^{q_{\phi}h}e^{-q_{\phi}y}\right)\right].
\end{equation}
from which we clearly see that Eq. \eqref{expression_for_temperature_irr_final} coincides with the $\epsilon\to0$ limit of the compressibility contribution $T_{\phi}$ to the temperature given in Eq. \eqref{T_phi_epsilon_to_0} for the general compressible phonon fluid.\\
The temperature difference across the strip obtained using Eq. \eqref{expression_for_temperature_irr_final} is
\begin{equation}
T(x,0)-T(x,h)=\frac{1}{2\pi}\int dk\,e^{ikx}2\left[\Delta T+\frac{\alpha}{\kappa}U(1-e^{kh})F(q_{\phi},k)\right].
\end{equation}

\newpage
\section{List of symbols and notation}
\label{app:symbols}

This appendix provides a consolidated list of symbols and notation used throughout the manuscript. Unless otherwise stated, repeated Cartesian indices imply summation, and Fourier transforms follow the conventions specified in the main text.
\begin{longtable}{p{0.18\textwidth} p{0.50\textwidth} p{0.14\textwidth} p{0.18\textwidth}}
\label{tab:symbols}\\
\hline
\hline
\textbf{Symbol} & \textbf{Meaning / definition} & \textbf{Units} & \textbf{Where used} \\
\hline
\endfirsthead
\hline
\textbf{Symbol} & \textbf{Meaning / definition} & \textbf{Units} & \textbf{Where used} \\
\hline
\endhead
\hline
\endfoot
\hline
\endlastfoot
$\boldsymbol{R}$, $t$ &
Real-space position and time variables. &
m, s &
Throughout \\
$\boldsymbol{q}$, $\omega$ &
Wavevector and angular frequency (Fourier variables). &
m$^{-1}$, s$^{-1}$ &
Secs. 2--5 \\
$\nu\equiv\boldsymbol{q},s$ &
Composite phonon mode index (branch and wavevector). &
-- &
Secs. 2--5 \\
$\nabla$, $\partial_t$ &
Spatial gradient and time derivative. &
-- &
Secs. 4--6 \\

$\omega_{\nu}$ &
Phonon angular frequency. &
s$^{-1}$ &
Secs. 2--3 \\

$\boldsymbol{v}_{\nu}$ &
Phonon group velocity. &
m\,s$^{-1}$ &
Secs. 3--4 \\

$n_{\nu}$ &
Phonon distribution function. &
-- &
Secs. 3--4 \\

$G$ &
Phonon Green's function (matrix in mode indices). &
-- &
Sec. 2 \\

$G^{R}$, $G^{A}$ &
Retarded and advanced Green's functions. &
-- &
Sec. 2 \\

$G^{<}$, $G^{>}$ &
Lesser and greater Green's functions. &
-- &
Sec. 2 \\

$\Sigma$ &
Phonon self-energy. &
-- &
Sec. 2, App. A \\

$\Gamma$ &
Phonon linewidth, $\Gamma = -2\,\mathrm{Im}\{\Sigma^{R}\}$. &
s$^{-1}$ &
Sec. 2 \\

$\mathrm{A}$ &
Spectral function, $\mathrm{A}=i(G^{>}-G^{<})$. &
s &
Sec. 2 \\

$T(\boldsymbol{R},t)$ &
Local temperature field (linearized about a reference temperature where applicable). &
K &
Secs. 4--6 \\
$C$ &
Volumetric heat capacity. &
J\,m$^{-3}$\,K$^{-1}$ &
Secs. 4--5 \\
$\boldsymbol{Q}$ &
Heat (energy) flux &
W\,m$^{-2}$ &
Secs. 3--5 \\

$\kappa$ &
Thermal conductivity. &
W\,m$^{-1}$\,K$^{-1}$ &
Secs. 3--6 \\

$\kappa^{D}$ &
Diffusive (momentum-relaxing) contribution to thermal conductivity. &
W\,m$^{-1}$\,K$^{-1}$ &
Secs. 4--5 \\

$\tau$ &
Relaxation time (mode-dependent). &
s &
Secs. 3--4 \\

$\boldsymbol{u}(\boldsymbol{R},t)$ &
Phonon drift (hydrodynamic) velocity field. &
m\,s$^{-1}$ &
Secs. 4--6 \\

$\eta$ &
Shear viscosity of the phonon fluid. &
Pa\,s &
Secs. 4--6 \\

$\zeta$ &
Bulk viscosity of the phonon fluid. &
Pa\,s &
Secs. 4--6 \\

$\Phi$ &
Compressibility field, $\Phi=\nabla\!\cdot\!\boldsymbol{u}$. &
s$^{-1}$ &
Sec. 5 \\

$\boldsymbol{\mathscr{W}}$ &
Vorticity field, $\boldsymbol{\mathscr{W}}=\nabla\!\times\!\boldsymbol{u}$. &
s$^{-1}$ &
Sec. 5 \\

$\phi$ &
Velocity (scalar) potential associated with compressible flow. &
m$^{2}$\,s$^{-1}$ &
Sec. 5 \\

$\psi$ &
Stream function associated with vortical flow (2D). &
m$^{2}$\,s$^{-1}$ &
Sec. 5 \\

$T_{\phi}$, $T_{\psi}$ &
Additive compressibility and vorticity contributions to the temperature field. &
K &
Sec. 5 \\

$\mathrm{FDN}$ &
Fourier deviation number (measure of departure from Fourier transport). &
-- &
Secs. 4--5 \\

$\mathrm{Kn}$ &
Knudsen number. &
-- &
Secs. 1, 4--6 \\
\hline
\hline
\caption{List of symbols and notation used in the review.}
\end{longtable}


\end{document}